\documentclass[twocolumn]{aastex63}  % twocolumn, manuscript, preprint, preprint2, modern, linenumbers, tighten
\usepackage{amsmath}

\shortauthors{Law-Smith et al.}

\begin{document}

\title{Stellar Tidal Disruption Events with Abundances and Realistic Structures (STARS):\\Library of Fallback Rates} 

\author[0000-0001-8825-4790]{Jamie A.P. Law-Smith}
\affiliation{Department of Astronomy and Astrophysics, University of California, Santa Cruz, CA 95064, USA}
\affiliation{Niels Bohr Institute, University of Copenhagen, Blegdamsvej 17, 2100 Copenhagen, Denmark}

\author[0000-0003-4263-2228]{David A. Coulter}
\altaffiliation{NSF Fellow}
\affiliation{Department of Astronomy and Astrophysics, University of California, Santa Cruz, CA 95064, USA}

\author[0000-0002-9809-8215]{James Guillochon}
\noaffiliation

\author[0000-0001-6350-8168]{Brenna Mockler}
\affiliation{Department of Astronomy and Astrophysics, University of California, Santa Cruz, CA 95064, USA}
\affiliation{Niels Bohr Institute, University of Copenhagen, Blegdamsvej 17, 2100 Copenhagen, Denmark}

\author[0000-0003-2558-3102]{Enrico Ramirez-Ruiz}
\affiliation{Department of Astronomy and Astrophysics, University of California, Santa Cruz, CA 95064, USA}
\affiliation{Niels Bohr Institute, University of Copenhagen, Blegdamsvej 17, 2100 Copenhagen, Denmark}

\correspondingauthor{Jamie A.P. Law-Smith}
\email{lawsmith@ucsc.edu}

\begin{abstract}
We present the STARS library, a grid of tidal disruption event (TDE) simulations interpolated to provide the mass fallback rate ($dM/dt$) to the black hole for a main-sequence star of any stellar mass, stellar age, and impact parameter. We use a one-dimensional stellar evolution code to construct stars with accurate stellar structures and chemical abundances, then perform tidal disruption simulations in a three-dimensional adaptive-mesh hydrodynamics code with a Helmholtz equation of state, in unprecedented resolution: from 131 to 524 cells across the diameter of the star. The interpolated library of fallback rates is available on GitHub (\href{https://github.com/jamielaw-smith/STARS_library}{github.com/jamielaw-smith/STARS\_library}) and version 1.0.0 is archived on Zenodo; one can query the library for any stellar mass, stellar age, and impact parameter. We provide new fitting formulae for important disruption quantities ($\beta_{\rm{crit}},\Delta{M},\dot{M}_{\rm{peak}},t_{\rm{peak}},n_\infty$) as a function of stellar mass, stellar age, and impact parameter. Each of these quantities \replaced{vary}{varies} significantly with stellar mass and stellar age, but we are able to reduce all of our simulations to a single relationship that depends only on stellar structure, characterized by a single parameter $\rho_c/\bar\rho$, and impact parameter $\beta$. We also find that\added{,} in general, more centrally concentrated stars have steeper $dM/dt$ rise slopes and shallower decay slopes. For the same $\Delta{M}$, the $dM/dt$ shape varies significantly with stellar mass, promising the potential determination of stellar properties from the TDE light curve alone. The $dM/dt$ shape depends strongly on stellar structure and to a certain extent stellar mass, meaning that fitting TDEs using this library offers a better opportunity to determine the nature of the disrupted star and the black hole.
\end{abstract}


%%%%%%%%%%%%%%%%%%%%%%%%%%%%%%%%%%%%%%%%%%%%%%%%%%%%%%%%%%%%%%%%%%%%%%%%%%%%%%%%%%%%%%%%%%%%%%%%%%%%%%%%%%%%%%%%%%%%%%%%%%%%% 

\section{Introduction}\label{sec:introduction}

Near the region of influence of a galactic massive black hole (BH), a star scattered onto an orbit that brings it close to the BH can be tidally disrupted by the BH's gravitational field---this is a tidal disruption event (TDE).
The stellar material that remains bound to the BH produces a flare that is the signature of this event. 
TDEs were first explored theoretically with pioneering studies by \added{\citet{1979SvAL....5...16L},} \citet{1975Natur.254..295H}, \citet{1983A&A...121...97C}, \citet{1988Natur.333..523R}, and \citet{1989ApJ...346L..13E}.

Several dozen such flares have been observed at the centers of other galaxies \citep[for reviews of observations, see e.g.][]{2015JHEAp...7..148K,2017ApJ...838..149A, 2017ApJ...842...29H, 2020arXiv200101409V}, with observations now regularly capturing both the rise \citep[e.g.,][]{2019ApJ...883..111H} and decay \citep[e.g.,][]{2018MNRAS.480.5689H,2019ApJ...878...82V} of the transient in great detail, and even signatures of an accretion disk \citep[e.g.,][]{2019ApJ...880..120H,2020arXiv200309427H}. Fitting theoretical models to observed TDEs allows one\added{, in principle,} to extract the properties of the disruption: BH mass, BH spin, stellar mass, stellar age, impact parameter, and radiative efficiency. The first attempt to systematically extract BH masses from TDEs \citep{2019ApJ...872..151M} was remarkably successful, obtaining errors of order that of the $M$--$\sigma$ relationship. However, determination of other parameters is made difficult by degeneracies between stellar properties and BH properties. One needs better theoretical models of TDEs in order to extract more accurate information from observed events.

A combination of detailed theoretical modeling and high resolution observations can turn TDEs into unique tools to probe several astrophysical questions: 
(1) the BH mass function and in particular the possible existence of a cutoff in the BH mass function at low masses ($M_{\rm BH} \lesssim 10^5 M_\sun$),
(2) the BH spin distribution,
(3) the radiative efficiency of BH accretion and other questions of accretion physics,
(4) the stellar populations (stellar masses and ages) in galactic centers, as the stars at the centers of distant galaxies are exposed through their disruption and accretion,
(5) the dynamics operating in galactic centers; e.g., which mechanisms (two-body, resonant relaxation, secular effects, etc.) dominate how stars and BHs interact.

At the order-of-magnitude level, tidal disruption occurs when a star crosses the tidal disruption radius
\begin{equation}
r_{\rm t}=\left(\frac{M_{\rm BH}}{M_\star}\right)^{1/3} R_\star \propto \bar\rho_\star^{-1/3},
\label{eq:rt}
\end{equation}
at which point a star's self-gravity is smaller than the tidal acceleration across its radius.
The pericenter passage time of the star is approximately equal to the star's dynamical time, $t_{\rm p} \sim r_{\rm p} / v_{\rm p} \sim \sqrt{R_\star^3/GM_\star} = t_{\rm dyn, \star}$. The star is spun-up to a large fraction of its breakup angular velocity, and a quadrupole tidal distortion develops across its surface. These two effects are what tidally disrupt a star, and (again, at the order-of-magnitude level) they occur over the star's dynamical timescale, so the star does not have time to react hydrodynamically. 
The impact parameter of the encounter
\begin{equation}
\beta \equiv \frac{r_{\rm t}}{r_{\rm p}}
\end{equation}
is the ratio of the tidal radius to the pericenter distance; it is an order-of-magnitude measure of the ``strength'' of the tidal interaction.
The critical impact parameter $\beta_{\rm crit}$ is defined as the smallest impact parameter of full disruption (i.e., where the entire star's mass is disrupted), below which are partial disruptions and above which are ``post-critical'' encounters. 

The mass fallback rate $dM/dt \equiv \dot M_{\rm fb}$ of debris to pericenter is a central quantity of interest as it appears to track the \added{Optical/UV} luminosity evolution of observed TDEs closely \citep[e.g.,][]{2012Natur.485..217G, 2014ApJ...783...23G, 2019ApJ...872..151M}.\added{\footnote{In present model fitting, this is true provided a few additional free parameters (mostly related to the photosphere size).}}
For non-relativistic disruptions, one can scale mass fallback rate and time of return to pericenter with BH mass as follows:
\begin{align}
\dot M_{\rm peak} &\propto M_{\rm BH}^{-1/2}\ M_\star^{2}\ R_\star^{-3/2} \\
t_{\rm peak} &\propto M_{\rm BH}^{1/2}\ M_\star^{-1}\ R_\star^{3/2}
\label{eq:mdotpeaktpeakscalings}
\end{align}
where the $\dot M_\text{peak} \propto M_\star^{2}$ scaling results when we assume that a constant fraction of the star's mass is lost to the BH in the disruption.
Note that these formulae are for the peak quantities of the mass fallback rate, but can be applied to scale the entire $dM/dt$ curve.

Because the tidal radius depends inversely on the average density of the star (Eq.~\ref{eq:rt}), there is a maximum BH mass for disruption outside the innermost-bound spherical orbit for different types of stars. 
See Figure~1 of \citet{2017ApJ...841..132L} for the phase space in $M_{\rm object}$ and $M_{\rm BH}$ of stellar objects (from planets to evolved stars) that are expected to produce bright tidal disruption flares. 
We expect that most observed TDEs will be from MS stars, and that the mass function of TDEs is relatively flat for $M_\star \lesssim M_\sun$ \citep{2016MNRAS.461..371K} (see more detailed discussion in Section~\ref{sec:conclusion}). 
Thus, it is important to have a library of MS star simulations with which to fit observed events. This work is aimed to be the definitive library of the MS star parameter space. 
One can extend this library to include relativistic encounters, but these are rarer, and account for a small fraction of MS star disruptions (see discussion in Section~\ref{sec:conclusion}).

Our own Galactic Center's nuclear star cluster is host to young, massive stars, giant-branch stars, and an old population of main sequence stars \citep[e.g.,][]{2007A&A...469..125S}.
Stars in the Galactic Center also exhibit a range of metallicities, from metal-rich to metal-poor \citep{2017MNRAS.464..194F}.
In nuclear clusters outside our galaxy, there is evidence for a diversity of stellar ages and types \citep{2010ApJ...714..713S} and a wide range of star formation histories \citep{2014MNRAS.441.3570G}.
Additionally, TDEs appear to be observed preferentially in post-starburst galaxies \citep{2014ApJ...793...38A, 2016ApJ...818L..21F, 2017ApJ...850...22L, 2018ApJ...853...39G}.
Thus, we expect some TDEs to be sourced by more massive stars and also by stars of varying ages and compositions.
Observations of TDEs suggest that many disruptees are non-ZAMS stars  \citep[spectra show metal lines that evolve with time; e.g.,][]{2019ApJ...887..218L} as is expected \added{\citep[this was first argued for in][]{2016MNRAS.458..127K}}.
However, the theoretical modelling of TDEs thus far has largely ignored the stellar evolution aspect of the problem.

A brief (incomplete) summary of recent theoretical work, where we highlight features relevant to or differing in significant ways from this work, is below.
Before discussing the simulation work, we mention the so-called ``frozen-in'' approximation, studied in detail by \citet{2009MNRAS.392..332L}, in which the star arrives intact to pericenter. One can then integrate across the star in equal-orbital-energy slices and determine the mass fallback rate to the BH as a function of time analytically. In this framework, only the structure of the star matters. This approximation can only handle full disruptions (a single impact parameter).
\citet{2016MNRAS.458..127K} investigated abundance anomalies using this framework and 1D stellar evolution models from MESA.
\citet{2018ApJ...857..109G}, following the above two works, developed an analytic framework to study the composition as a function of time following disruption, using MESA stars.
In a recent significant extension to \citet{2009MNRAS.392..332L}, \citet{2019ApJ...883L..17C} developed an analytic framework to determine the asymptotic power-law slope of the fallback rate, finding $\propto t^{-5/3}$ for full disruptions and $\propto t^{-9/4}$ for partial disruptions, in general (but not detailed) agreement with simulations.

Despite the success and usefulness of the ``frozen-in'' approximation in calculating TDE properties, we have learned from hydrodynamical simulations that the star arrives at pericenter significantly distorted and spinning \citep[for a detailed discussion see][]{2019MNRAS.485L.146S}.
\citet{1982ApJ...263..377N} performed the first 3D tidal disruption simulations, of a $\gamma=5/3$ polytropic stellar structure.
\citet{1993ApJ...418..163K, 1993ApJ...418..181K} were the first to study stars with different stellar structures ($\gamma=(5/3, 1.5, 4/3)$) numerically.
\citet{2009ApJ...697L..77R} studied the tidal disruption and initial disk formation for $\gamma=5/3$ and $\gamma=1.4$ polytropes in 3D hydrodynamics.
\citet{2013ApJ...767...25G} performed the first systematic parameter-space study of the effect of both stellar structure and impact parameter, exploring $\gamma=4/3$ and $\gamma=5/3$ polytropes and a wide range of impact parameters. The effect of impact parameter made it clear that tidal disruptions are a \replaced{3-body}{3-dimensional} nonlinear hydrodynamical problem that needs to be studied by simulations, at least to a certain extent.
\citet{1993ApJ...410L..83L} were the first to explore relativistic tidal disruptions, on a Schwarzchild metric.
\citet{1997ApJ...479..164D} studied the disruption of a $\gamma=5/3$ polytrope by a rotating BH on a Kerr metric.
\citet{2012ApJ...749..117H} simulated the disruption of a white dwarf by a spinning BH.
\citet{2014PhRvD..90f4020C} and \citet{2017MNRAS.469.4483T} compared relativistic simulations to Newtonian simulations in detail, finding good agreement for non-relativistic encounters. 
\citet{2019MNRAS.487.4790G} performed a grid of general-relativistic simulations for a $\gamma=5/3$ polytrope, exploring higher impact parameters and spinning BH's, providing new fitting formulae for the relativistic regime, and again finding good agreement for non-relativistic encounters.
See also \citet{2019GReGr..51...30S} for a review of the status of TDEs in general relativity.
\citet{2008ApJ...679.1385R,2008CoPhC.179..184R,2009ApJ...695..404R} studied the tidal disruption of white dwarfs in detail for the first time, while \citet{2012ApJ...757..134M} were the first to study the tidal disruption of giant stars, whose highly segregated density profiles did not allow them to be fully disrupted.
\citet{2017ApJ...841..132L} performed a case-study simulating realistic stellar structures and compositions, with a hydrogen-envelope helium-core white dwarf.
\citet{2019ApJ...872..163G} studied the stellar spin dependence of fallback rates for a $\gamma=5/3$ polytrope.
\citet{2019MNRAS.487..981G} ran moving-mesh simulations of a ZAMS 1$M_\sun$ star with MESA stellar structure. Their results were consistent with the $\gamma=4/3$ result from \citet{2013ApJ...767...25G}.
\citet{2019ApJ...882L..26G} ran simulations with MESA stellar structures, for three stellar masses and ages, at one impact parameter $\beta=3$, and argued that the inferred BH mass from fitting TDEs to polytropic hydrodynamical simulations can be incorrect at the order-of-magnitude level.
\citet{2019ApJ...882L..25L} performed simulations with MESA stellar structures and a Helmholtz EOS, tracking chemical abundances for 49 elements in the 3D hydrodynamical simulations for the first time.
They found significant differences with the polytropic results of \citet{2013ApJ...767...25G} and also significant differences with the analytic predictions of the fallback-rate composition of \citet{2018ApJ...857..109G}.
\citet{2020arXiv200103501R,2020arXiv200103502R,2020arXiv200103503R,2020arXiv200103504R} recently posted results of a parameter-space study using a \added{fully} general-relativistic framework, for a range of stellar masses, impact parameters, and BH masses, and at a single stellar age \added{(see Sections~\ref{sec:methods} and \ref{sec:conclusion} for comparisons)}.

A number of theoretical studies have focused on the disk formation process
\citep[e.g.,][]{2000ApJ...545..772A, 2013MNRAS.434..909H, 2014ApJ...783...23G, 2015ApJ...804...85S, 2015ApJ...806..164P, 2015ApJ...809..166G, 2015ApJ...812L..39D, 2016MNRAS.461.3760H, 2016MNRAS.455.2253B, 2020MNRAS.492..686L, 2020MNRAS.495.1374B}. At present, the emission mechanism(s) responsible for the Optical/UV\deleted{, X-ray, and Radio components of} \added{emission in} TDE flares is an open question---the main candidates are (1) rapid disk circularization and accretion emission or (2) stream self-intersections and disk-formation-process emission---, but it is clear that the \added{Optical/UV} luminosity evolution of TDEs (the ``light curve'') tracks the mass fallback rate to the BH ($dM/dt$, the main output of the simulations in this work) very closely (see references above). This means that the fallback rate, which, as we show in this paper, is nearly solely determined by stellar structure, is a useful output.

In this paper, we study the disruption of main-sequence (MS) stars in a grid of stellar mass, stellar age, and impact parameter. Previously published systematic studies have used polytropic stellar structures, where $P=K \rho^\gamma = K \rho^{(n+1)/n}$. 
This work is a parameter space study using realistic internal stellar structures, chemical abundances, and equations of state (EOS).
We significantly expand upon the stellar structure study of \citet{2013ApJ...767...25G} by considering a wide range of density profiles derived from more accurate stellar models (as compared to polytropes): we study 14 distinct stellar structures, corresponding to different stellar masses and ages. The EOS is incorporated via the Helmholtz EOS, consistent with the MESA EOS of the stellar models.
We find (see Appendix) that the EOS contribution to the pressure support is small, and we argue that one can predict many of the properties of tidal disruption from stellar structure and impact parameter alone. 
In order to reduce our simulations into one relationship for various tidal disruption quantities, we parameterize the stellar structure by the single parameter $\rho_c/\bar\rho$, the ratio of the star's central density to its average density.
The simulations presented in this work have a full Helmholtz EOS tracking 49 elements; however, we study the chemical abundance of the fallback debris in a followup paper.

This paper is organized as follows: Section~\ref{sec:methods} describes our methods, Section~\ref{sec:results} describes our results, Section~\ref{sec:conclusion} concludes, and the Appendix describes the interpolated \texttt{STARS\_library} tool, as well as several other issues.

%%%%%%%%%%%%%%%%%%%%%%%%%%%%%%%%%%%%%%%%%%%%%%%%%%%%%%%%%%%%%%%%%%%%%%%%%%%%%%%%%%%%%%%%%%%%%%%%%%%%%%%%%%%%%%%%%%%%%%%%%%%%% 
%%%%%%%%%%%%%%%%%%%%%%%%%%%%%%%%%%%%%%%%%%%%%%%%%%%%%%%%%%%%%%%%%%%%%%%%%%%%%%%%%%%%%%%%%%%%%%%%%%%%%%%%%%%%%%%%%%%%%%%%%%%%% 

\section{Methods}\label{sec:methods}

We use the 1D stellar evolution code MESA \citep{2011ApJS..192....3P} to run a grid of models from 0.1$M_\sun$ to 10$M_\sun$, from pre-MS to zero-age main-sequence (ZAMS) to terminal-age main-sequence (TAMS). We define TAMS as a central hydrogen mass fraction of $10^{-3}$. We use the \texttt{mesa\_49.net} nuclear network, including 49 elements. See also \citet{2018ApJ...857..109G} and \citet{2019ApJ...882L..25L}  for details on the MESA setup. Table~\ref{tab:MESA_params} in the Appendix lists relevant or non-standard parameters for the MESA simulations. 
Stars with $M_\star \lesssim 0.8 M_\sun$ have a MS lifetime longer than the age of the universe; thus, for these stars, the oldest model we use in our tidal disruption calculations is at 10 Gyr, rather than TAMS. 

We then map the 1D stellar density profiles and chemical abundances into the 3D adaptive-mesh refinement (AMR) hydrodynamics code FLASH \citep{2000ApJS..131..273F}. 
We use the Helmholtz EOS and an extended Helmholtz table\footnote{Available at \url{http://cococubed.asu.edu/code_pages/eos.shtml}.}
that spans $10^{-12} \leq \rho\ {\rm [g/cm^3]} \leq 10^{15}$ and $10^3 \leq T\ {\rm [K]} \leq 10^{13}$.
This Helmholtz EOS is the backbone of the EOS module in MESA, so our mapping is self-consistent.
In FLASH, we use the sinks module to integrate trajectories and to track the position of the BH relative to the star, as in \citet{2017ApJ...834L..19G}. This is also an update from \citet{2013ApJ...767...25G}.
\added{We begin the simulations at $r = 10 r_{\rm t}$, where tidal effects are negligible.\footnote{For comparison, \citet{2019MNRAS.487..981G} begin at $r = 5 r_{\rm t}$.} }
In the multipole gravity solver, we use a maximum angular number of the multipole expansion of $l_m=20$. 
Our 3D box is 1000$R_\star$ on a side and we set the background density to $10^{-11}$ g/cm$^3$. This is such that the mass of the box is typically $\ll 0.1$\% of the stellar mass.
We use the following hydrodynamics parameters: the interpolation order is 3rd order, the slope limiter is the ``hybrid'' one, and we use the hybrid Riemann solver.  
We have no magnetic fields ($B=0$).
As in \citet{2013ApJ...767...25G}, in calculating the binding energy of each cell in the simulation with respect to the BH and stellar debris, we shift the center of mass of the star to be on a parabolic orbit (this is equivalent to centering the $dM/de$ distribution at $e=0$). 
\added{Note that in this paper $e$ denotes the specific orbital energy and not eccentricity.}
\citet{2013ApJ...767...25G} show that, because of the magnitude of this shift, it only affects the $dM/dt$ for $t>100$ yrs.
For the vast majority of events, the star is expected to approach on a parabolic or nearly-parabolic orbit \citep{2018ApJ...855..129H}.
See also \citet{2009ApJ...705..844G}, \citet{2011ApJ...732...74G}, \citet{2013ApJ...767...25G}, and \citet{2019ApJ...882L..25L} for details on the FLASH setup.
Table~\ref{tab:FLASH_params} in the Appendix lists several additional FLASH parameters, including the background grid values. 

The simulations in this paper are run with $1.5\times10^8$ maximum cells. This is a higher maximum cell count than in \citet{2019ApJ...882L..25L}.
We choose the maximum initial refinement based on the central concentration of the stellar density profile. We use 131 cells across the diameter of the star for less centrally concentrated stars ($\rho_c/\bar\rho \lesssim 150$) and 524 cells across the diameter of the star for more centrally concentrated stars ($\rho_c/\bar\rho \gtrsim 150$). 
In a few test simulations, results are nearly identical if we use 262 or 524 cells across the diameter of the star for the less centrally concentrated stars as well.
\added{See Section~\ref{sec:resolution-convergence} for our numerical convergence study.}
For comparison, the simulations in \citet{2013ApJ...767...25G} had $\approx$50 cells across the initial diameter, so our simulations have a factor of 2.6X to 10.5X higher initial resolution. 
The moving-mesh simulations of \citet{2019MNRAS.487..981G} have a maximum of $2.4\times10^5$ cells, thus initially $\approx$60 cells across the diameter of the star.
The simulations of \citet{2020arXiv200103501R,2020arXiv200103502R,2020arXiv200103503R,2020arXiv200103504R} initially have $\approx$50 cells across the diameter of the star and, in an important difference, the final box size is $17\times9\times14 R_\star$.

\begin{deluxetable*}{l l l l l l l}
\tablecaption{FLASH simulation grid in stellar mass, stellar age, and impact parameter. The simulations in this grid are interpolated in each of these dimensions in order to produce the library. 
Scaled results are indicated by $\ast$ for polytropes and $\dagger$ for our simulations (see text).
The ratio of the central to average density $\rho_c/\bar\rho$, the dynamical time of the star $t_{\rm dyn}$, and the corresponding $r_{\rm p}/r_{\rm g}$ for each $\beta$ is also listed. 
\label{tab:grid}}
\tablehead{
\colhead{$M_\star$} & \colhead{$R_\star$} & \colhead{stellar age} & \colhead{$\rho_c/\bar\rho$} & \colhead{$t_{\rm dyn}$} & \colhead{$\beta \equiv r_{\rm t}/r_{\rm p}$} & \colhead{$r_{\rm p}/r_{\rm g}$}
}
\startdata
0.1 $M_\sun$ & 0.1214 $R_\sun$ & 0 Gyr   &  5.5  & 213 s  & 0.5--1.2$^\ast$                     & 25--10\\
             & 0.1215 $R_\sun$ & 10 Gyr  &  5.5  & 213 s  & 0.5--1.2$^\ast$                     & 25--10\\
0.3 $M_\sun$ & 0.2814 $R_\sun$ & 0 Gyr   &  5.8 & 434 s  & 0.6, 0.7, 0.8, 0.9, 1.0, 2.0         & 33, 28, 25, 22, 20, 10\\
             & 0.2989 $R_\sun$ & 10 Gyr  &  5.8 & 475 s  & 0.6, 0.7, 0.8, 0.9, 1.0, 2.1         & 35, 30, 26, 23, 21, 10\\
0.5 $M_\sun$ & 0.4452 $R_\sun$ & 0 Gyr   &  11  & 669 s  & 0.6, 0.8, 1.0, 1.15, 1.4, 2.6        & 44, 33, 26, 23, 19, 10\\
             & 0.4564 $R_\sun$ & 10 Gyr  &  12  & 715 s  & 0.6, 0.8, 1.0, 1.2, 1.4, 2.8         & 46, 35, 28, 23, 20, 10\\
0.7 $M_\sun$ & 0.6485 $R_\sun$ & 0 Gyr   &  23  & 994 s  & 0.75, 1.0, 1.25, 1.5, 3.4            & 46, 34, 28, 23, 10\\
             & 0.6793 $R_\sun$ & 10 Gyr  &  36  & 1065 s & 0.8, 1.0, 1.15, 1.3, 1.5, 1.7, 3.6   & 45, 36, 31, 28, 24, 21, 10\\
1.0 $M_\sun$ & 0.9012 $R_\sun$ & 0 Gyr   &  42  & 1362 s & 1.0, 1.25, 1.5, 1.75, 2.0, 4.2       & 42, 34, 28, 24, 21, 10\\
             & 1.0455 $R_\sun$ & 4.8 Gyr &  138  & 1702 s & 1.0, 1.5, 2.0, 2.5, 3.0, 3.5, 4.9   & 49, 33, 24, 20, 16, 14, 10\\
             & 1.2872 $R_\sun$ & 8.4 Gyr &  756  & 2325 s & 1.0, 1.5, 2.0, 3.0, 4.0, 4.5, 5.0, 6.0   & 60, 40, 30, 20, 15, 13, 12, 10\\
1.5 $M_\sun$ & 1.6275 $R_\sun$ & 0 Gyr   &  128  & 2699 s & 1.0, 1.5, 2.0, 2.75, 6.7                 & 67, 51, 34, 24, 10 \\
             & 2.0805 $R_\sun$ & 2 Gyr   &  1697  & 3901 s & 2.0, 4.0, 6.0, 8.6                 & 43, 21, 14, 10 \\
3.0 $M_\sun$ & 1.8896 $R_\sun$ & 0 Gyr   &  73  & 2387 s & 0.6--4.0$^\ast$                   & 103--15\\
             & 3.3192 $R_\sun$ & 0.3 Gyr &  1198  & 5558 s & 1.5, 2.0, 3.0, 4.0, 4.5, 5.0, 7.0, 10.8        & 72, 54, 36, 27, 24, 22, 16, 10\\
10 $M_\sun$  & 3.6870 $R_\sun$ & 0 Gyr   &  38  & 3564 s & 1.0--4.2$^\dagger$      & 80--19 \\
             & 8.4232 $R_\sun$ & 0.02 Gyr&  1292  & 12307 s& 1.5--10.8$^\dagger$   & 123--17\\
\enddata
\end{deluxetable*}

Table~\ref{tab:grid} lists the parameter space in stellar mass, stellar age, and impact parameter studied in this work.
For each star, we run a range of impact parameters $\beta$ (the ratio of the tidal radius to the pericenter distance) from grazing partial disruptions to post-critical disruptions. 
The lowest $\beta$ for a given star corresponds to 1\% to 10\% mass lost (unbound) from the star in the encounter.
We run one very-post-critical (post-full-disruption) $\beta$ for each star such that $r_{\rm p}=10r_{\rm g}$ for this highest $\beta$ encounter, where $r_{\rm g}\equiv GM/c^2$ in this paper. This is for the purpose of interpolating our grid of $dM/dt$ curves in the maximum applicable range. We also plan to extend this library with relativistic simulations (applicable to the small fraction of very relativistic encounters) in future work.
For comparison, the most relativistic encounter in \citet{2013ApJ...767...25G} was the $\beta=4$ for the $\gamma=4/3$ star, which was $r_{\rm p}=11.8 r_{\rm g}$.

Table~\ref{tab:grid} also lists several other quantities, such as the ratio of the star's central density to average density $\rho_c/\bar\rho$ (a parameterization of the central concentration of the star that we use extensively later on), the dynamical time of the star, defined in this paper as $t_{\rm dyn} \equiv \sqrt{R_\star^3 / GM_\star}$, and the ratio of pericenter distance to gravitational radius, $r_{\rm p}/r_{\rm g}$, corresponding to each $\beta$.
In order to extend the range of the interpolated fallback rate library (see Appendix), we include a few results scaled from simulations where the stellar structures are nearly identical.
We scale $\gamma=5/3$ results for the 0.1$M_\sun$ stars, $\gamma=4/3$ results for the ZAMS 3$M_\sun$ star, our ZAMS 1$M_\sun$ results for the ZAMS 10$M_\sun$ star, and our TAMS 3$M_\sun$ results for the TAMS 10$M_\sun$ star.
Note that we do perform a 10$M_\sun$, $\beta=1.5$ simulation in Section~\ref{sec:stellar_structure_vs_EOS} in our ``stellar structure vs. EOS'' study.

\begin{figure*}[htp!]
\epsscale{0.55}
\plotone{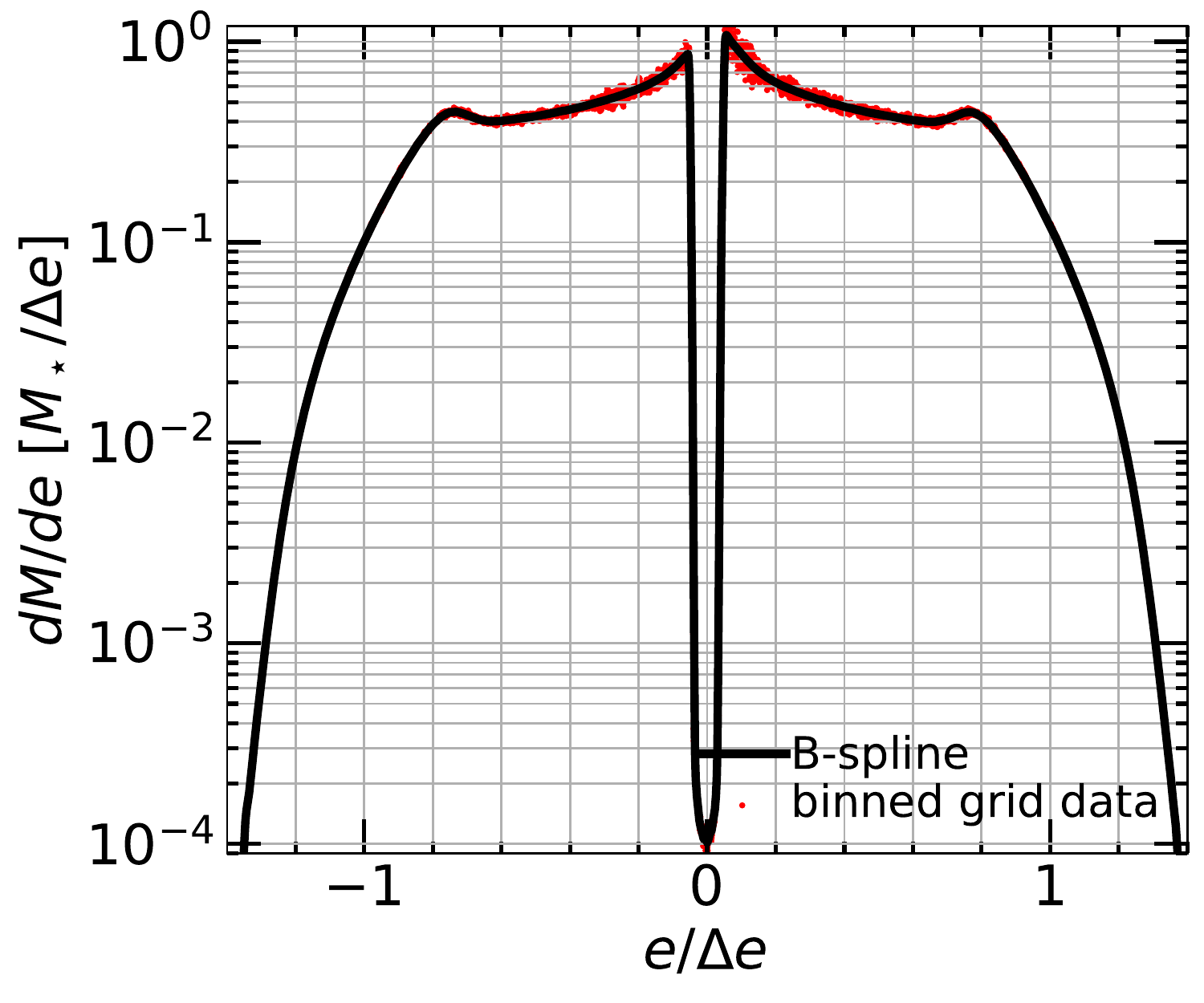} 
\plotone{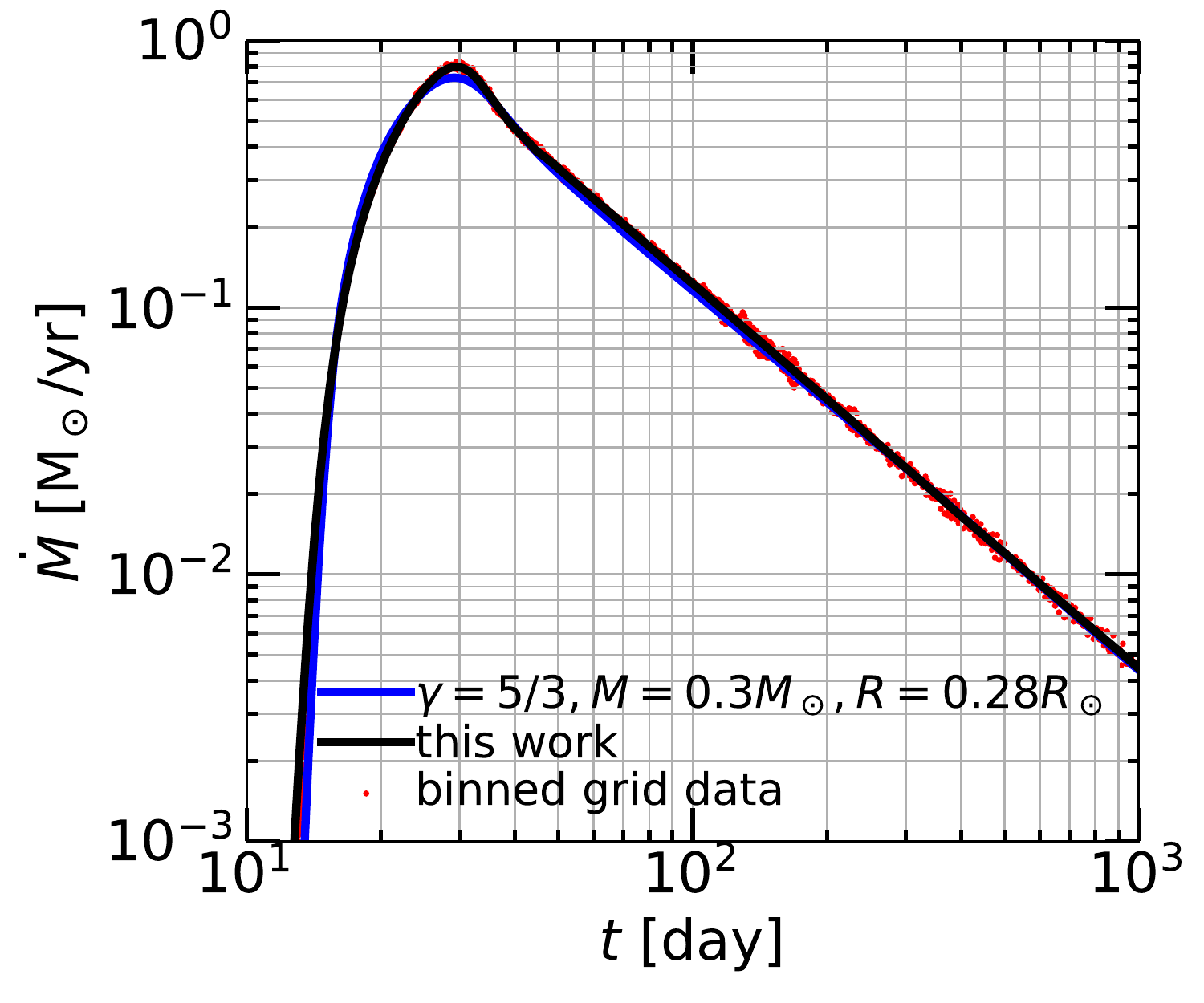} 
\caption{
Specific binding energy distribution, $dM/de$, and resulting mass fallback rate to the BH, $dM/dt$, for a $0.3 M_\sun$ ZAMS star constructed in MESA in a $\beta=0.9$ encounter with a $10^6 M_\sun$ BH; this is a full disruption. The finely binned hydrodynamical grid data is shown in red and the B-spline fit (this work) in black. The $x$- and $y$-axis of the $dM/de$ plot are normalized to the characteristic spread in binding energy, $\Delta e$ (see text). The $dM/dt$ plot also compares to a $\gamma=5/3$ polytropic simulation for this $\beta$ from \citet{2013ApJ...767...25G}, scaled to the same mass and radius as this star, in blue. It is expected to match quite closely as the stellar structure of a $0.3 M_\sun$ ZAMS star is well approximated by a $\gamma=5/3$ polytrope (see Figure~\ref{fig:MESA_profiles_vs_polytropes_norm}).
\label{fig:dmde}
}
\end{figure*}

We run at a single BH mass of $M_{\rm BH} = 10^6 M_\sun$. Our simulations are directly applicable to any non-relativistic encounter with a different BH mass\added{, to leading order in the ``tidal approximation'' ($R_\star/r \ll 1$),} because of the scaling of \replaced{disruption quantities}{$t$ and $dM/dt$} with $M_{\rm BH}$ in Eq.~(\ref{eq:mdotpeaktpeakscalings}).
See Figure~\ref{fig:mbh_beta_phase_space} in the Appendix for the range of applicability of our simulations. This grid of simulations applies to tidal disruptions from most stellar masses, ages, impact parameters, and BH masses. 
For $r_{\rm p}>10 r_{\rm g}$, the difference in $dM/dt$'s between Newtonian and relativistic simulations is $\lesssim10\%$ (see more detailed discussion in Section~\ref{sec:conclusion}).

At the last timestep before debris begins to leave the computational domain, we calculate the specific binding energy of every cell in the simulation relative to the BH and the star (if any self-bound mass remains), including only material bound to the BH and excluding material bound to the star or unbound from the BH. From this spread in specific binding energy distribution, or $dM/de$, we calculate the mass return rate to the BH (more accurately, to pericenter) as a function of time, using Kepler's third law:
\begin{equation}
\frac{dM}{dt}=\frac{dM}{de}\frac{de}{dt}=\frac{dM}{de}\frac{1}{3} (2 \pi G M_{\rm BH})^{2/3} t^{-5/3}.
\end{equation}

We run the simulations to a maximum time of 100$t_{\rm dyn}$ and the star reaches pericenter at approximately 20$t_{\rm dyn}$ into the simulation.
For moderate $\beta$'s, the debris remains inside the box for 100$t_{\rm dyn}$, but for high $\beta$'s the debris begins to leave the box earlier---for the most extreme $\beta$'s we study this is at approximately 50$t_{\rm dyn}$. So $dM/de$ distributions are calculated at 30--80$t_{\rm dyn}$ after pericenter. \citet{2013ApJ...767...25G} studied fallback rates at up to 550$t_{\rm dyn}$ after pericenter, showing that the resulting shape is identical to those calculated at earlier times as we do. By not having a large enough box to follow the entire debris for $t>100 t_{\rm dyn}$, we do not sacrifice any accuracy in the final result, but instead gain precision as a smaller box allows for higher resolution.
Note that however, unlike in \citet{2020arXiv200103501R,2020arXiv200103502R,2020arXiv200103503R,2020arXiv200103504R}, where debris leaves the computational domain after a few dynamical timescales, the debris in our simulations remains in the box for many dynamical timescales after pericenter.
\added{The effect this difference has on fallback rate determinations is unclear.} 

As an example of our analysis method, Figure~\ref{fig:dmde} shows the specific binding energy distribution and resulting mass return rate to the BH for a $0.3M_\sun$ ZAMS star in a $\beta=0.9$ encounter with a $10^6 M_\sun$ BH. $dM/de$ is plotted in units of $M_\star/\Delta e$, where $\Delta e= G M_\star^{2/3} M_{\rm BH}^{1/3} / R_\star$, an order-of-magnitude estimate of the range in fluid binding energies \citep{2013MNRAS.435.1809S}. 
The $x$-axis is normalized similarly.
The hydrodynamical grid data from the simulation is binned and then fit with a B-spline. The $dM/dt$ is extended by finding the average slope of the last 10--20\% of the $dM/dt$. The extended section is not visible on this plot as it is for $t\gtrsim10^4$ s. The result for a $\gamma=5/3$ polytrope, for this $\beta$, scaled to the same mass and radius as this star, is also shown. It is expected to match quite closely as the stellar structure of a $0.3 M_\sun$ ZAMS star is well approximated by a $\gamma=5/3$ polytrope (see Figure~\ref{fig:MESA_profiles_vs_polytropes_norm} in the Appendix). 

The small differences between the polytropic and MESA initial condition simulations are likely due to differences in resolution, numerical method, and smoothing algorithm between the two works. The B-splines applied to smooth the $dM/de$ distributions in this work have been examined and calibrated in detail for the $dM/de$ result from each simulation, and so the resulting $dM/dt$ curves in this work have higher fidelity to the raw hydrodynamical grid data.

%%%%%%%%%%%%%%%%%%%%%%%%%%%%%%%%%%%%%%%%%%%%%%%%%%%%%%%%%%%%%%%%%%%%%%%%%%%%%%%%%%%%%%%%%%%%%%%%%%%%%%%%%%%%%%%%%%%%%%%%%%%%% 

\section{Results}\label{sec:results}

\subsection{Stellar structure}

\begin{figure*}[htp!]
\epsscale{0.57}
\plotone{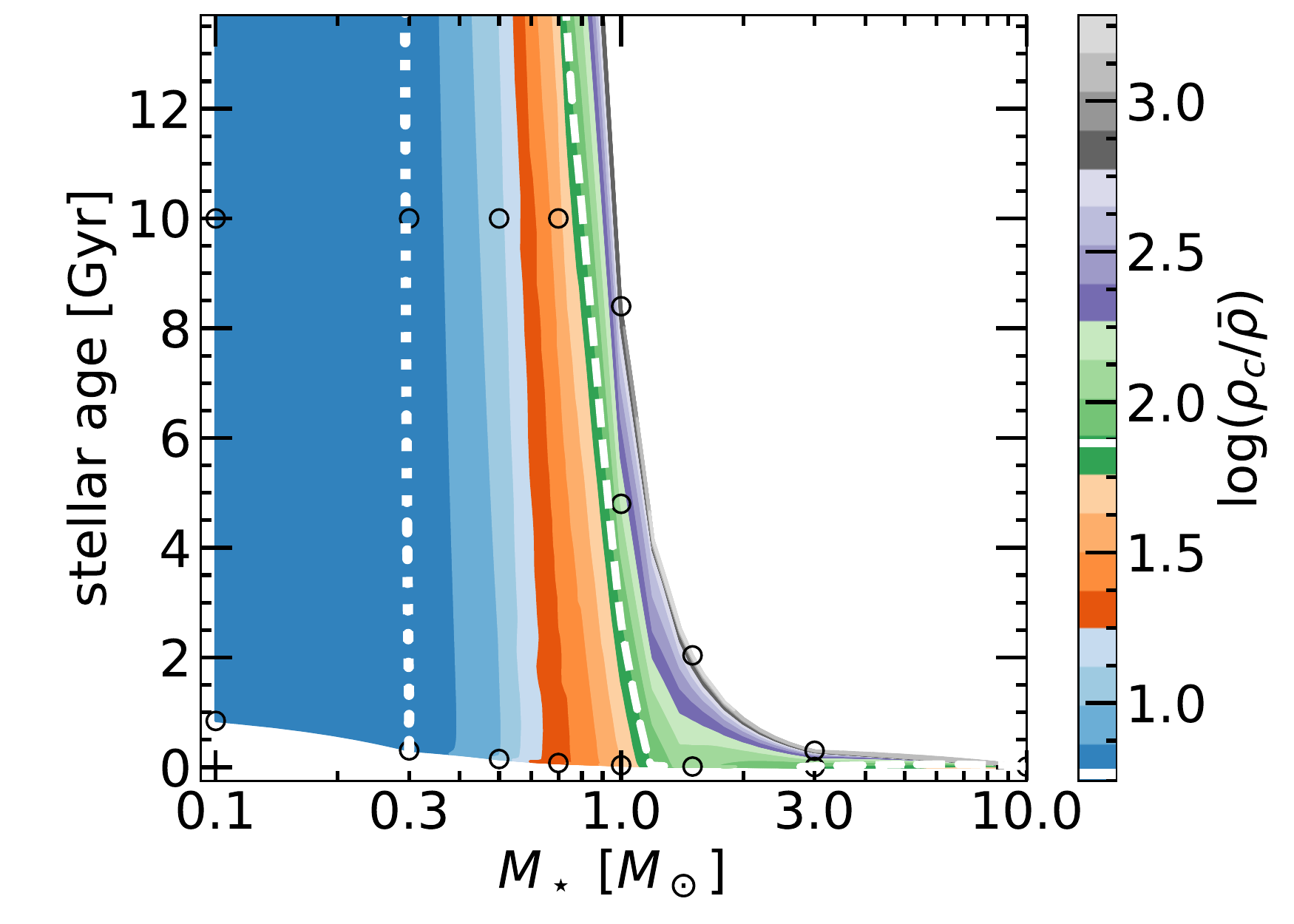}
\epsscale{0.58}
\plotone{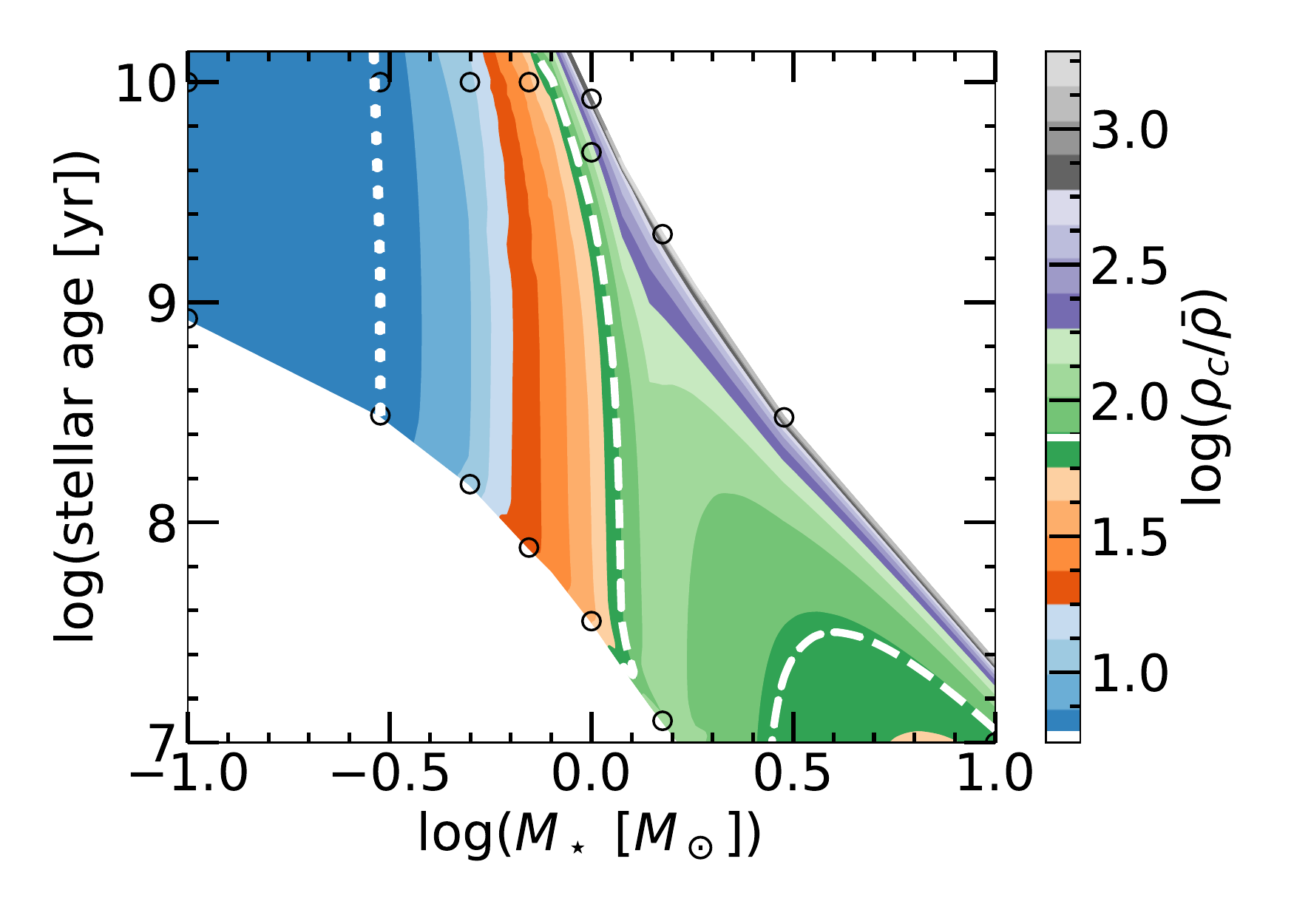} 
\epsscale{0.52}
\plotone{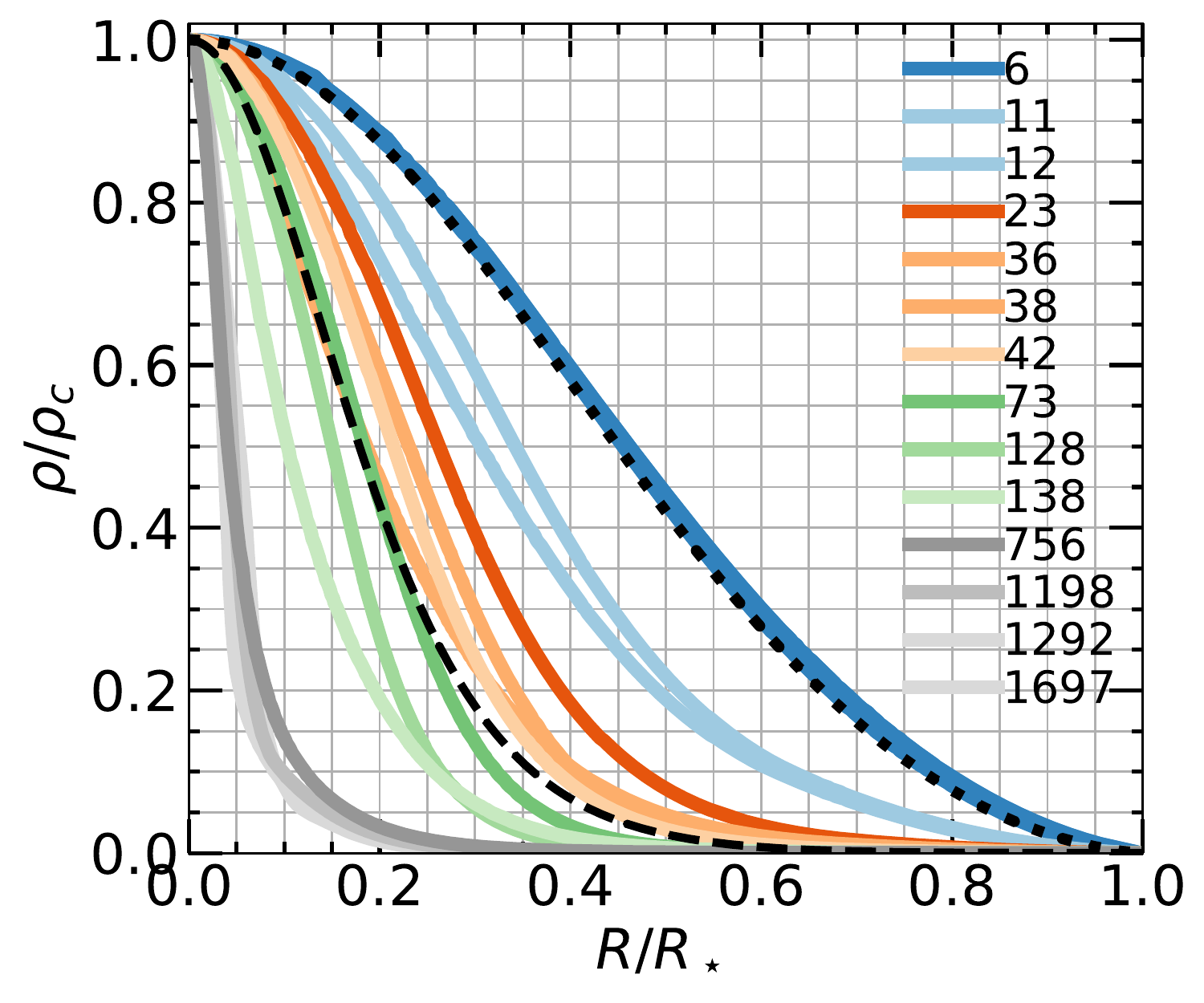}
\epsscale{0.59}
\plotone{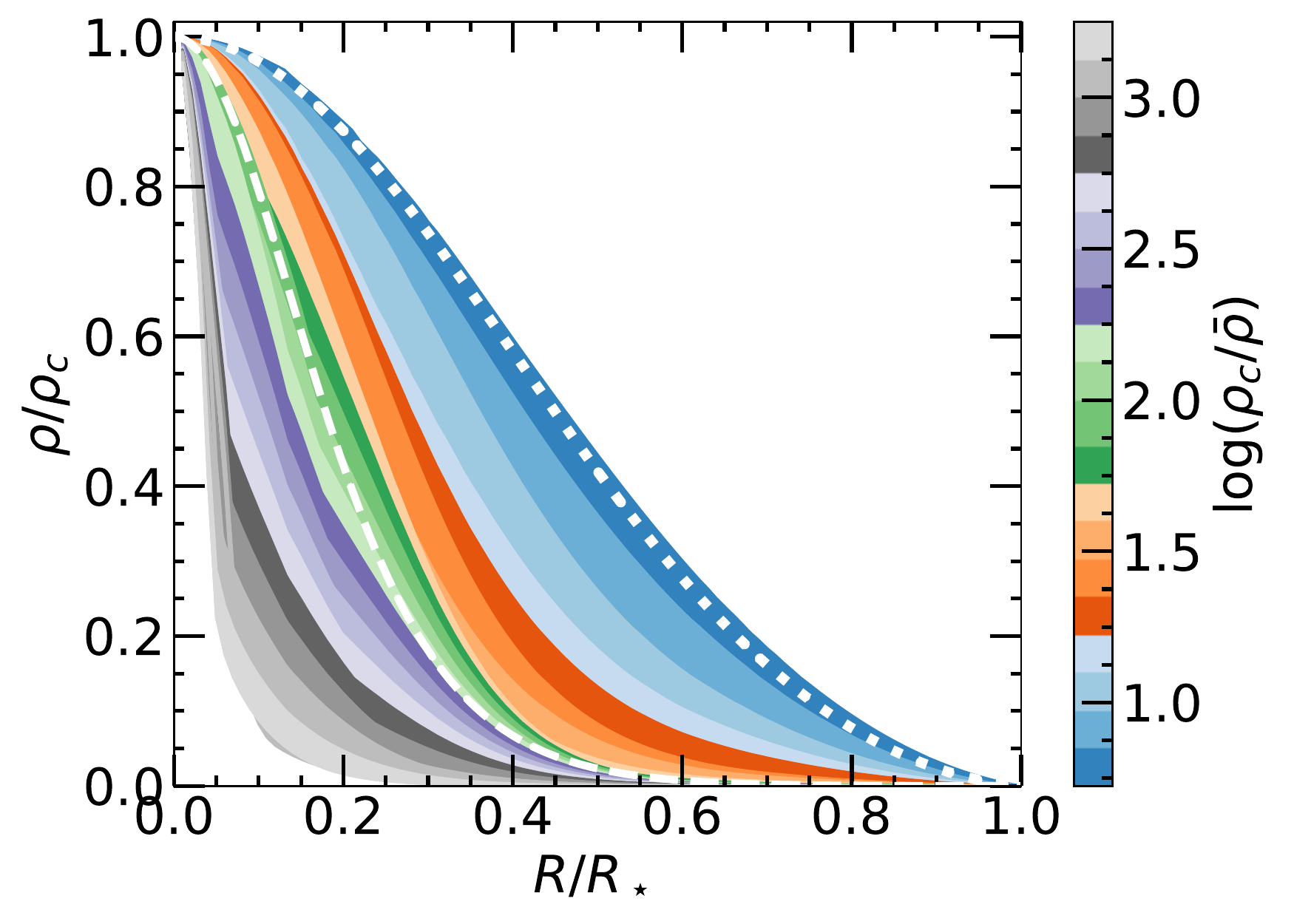}
\caption{
Stellar structure, parameterized by the ratio of central density to average density. 
Top: as a function of stellar mass and stellar age for MS stars, based on finely-spaced MESA grid. $x$-axis is log scale. $y$-axis is linear scale on left panel and log scale on right panel and its maximum is 13.8 Gyr. Dotted line corresponds to $\gamma\approx5/3$ polytrope (\added{$\rho_c/\bar\rho=5.8$, from a ZAMS 0.3 $M_\sun$ star} \deleted{using $M$, $R$, $\rho_c$ of a ZAMS 0.3 $M_\sun$ star}) and dashed line corresponds to $\gamma\approx4/3$ polytrope (\added{$\rho_c/\bar\rho=73$, from a ZAMS 3 $M_\sun$ star} \deleted{using $M$, $R$, $\rho_c$ of a ZAMS 3 $M_\sun$ star}). Black circles are stars for which we have done tidal disruption simulations in FLASH (note we run multiple $\beta$'s for each star).
Bottom: density profiles for MS stars, normalized to central density and stellar radius. Left panel is stars we have disrupted in FLASH (numbers in legend are $\rho_c/\bar\rho$) and right panel is based on finely-spaced MESA grid. Dotted and dashed lines are $\gamma=5/3$ and $\gamma=4/3$ polytropes respectively.
\label{fig:structure}
}
\end{figure*}

Figure~\ref{fig:structure} shows the ratio of central density to average density $\rho_c/\bar\rho$  for MS stars, interpolated based on a finely-spaced MESA grid (more finely spaced than for our TDE simulations; Table~\ref{tab:grid}). This shows the range of stellar structures on the main sequence. We perform TDE simulations in FLASH for structures at the extremes and interpolate the resulting $dM/dt$'s in stellar mass and stellar age in between these extremes. For some regions one can use polytropic stellar structures rather than MESA initial conditions (e.g., for $M_\star \lesssim 0.3 M_\sun$). 

The top panels show $\rho_c/\bar\rho$ in the space of stellar age vs. stellar mass. Here one sees that lower-mass stars ($\lesssim 0.8 M_\sun$), whose main-sequence lifetimes are $\gtrsim$ the age of the universe, evolve slowly and have roughly constant stellar structures over 10 Gyr. More massive stars ($\gtrsim 0.8 M_\sun$) have shorter lifetimes and evolve through different stellar structures more rapidly. 
We expect a wide variety of stellar masses and stellar ages to source observed TDEs; see Section~\ref{sec:conclusion} for discussion.

The bottom panels show normalized density profiles colored by $\rho_c/\bar\rho$, both individually for the stars we simulate in FLASH and interpolated on a more finely-spaced grid of MESA stars. One can see that $\rho_c/\bar\rho$ maps nearly 1-1 onto the stellar density profile, at least for the main sequence. This makes it a very good single-value parameter to describe the stellar structure, and we use it to reduce our simulations into a single relationship and to provide fitting formulae for each TDE quantity.

\begin{figure*}[htp!]
\epsscale{1.15}
\plotone{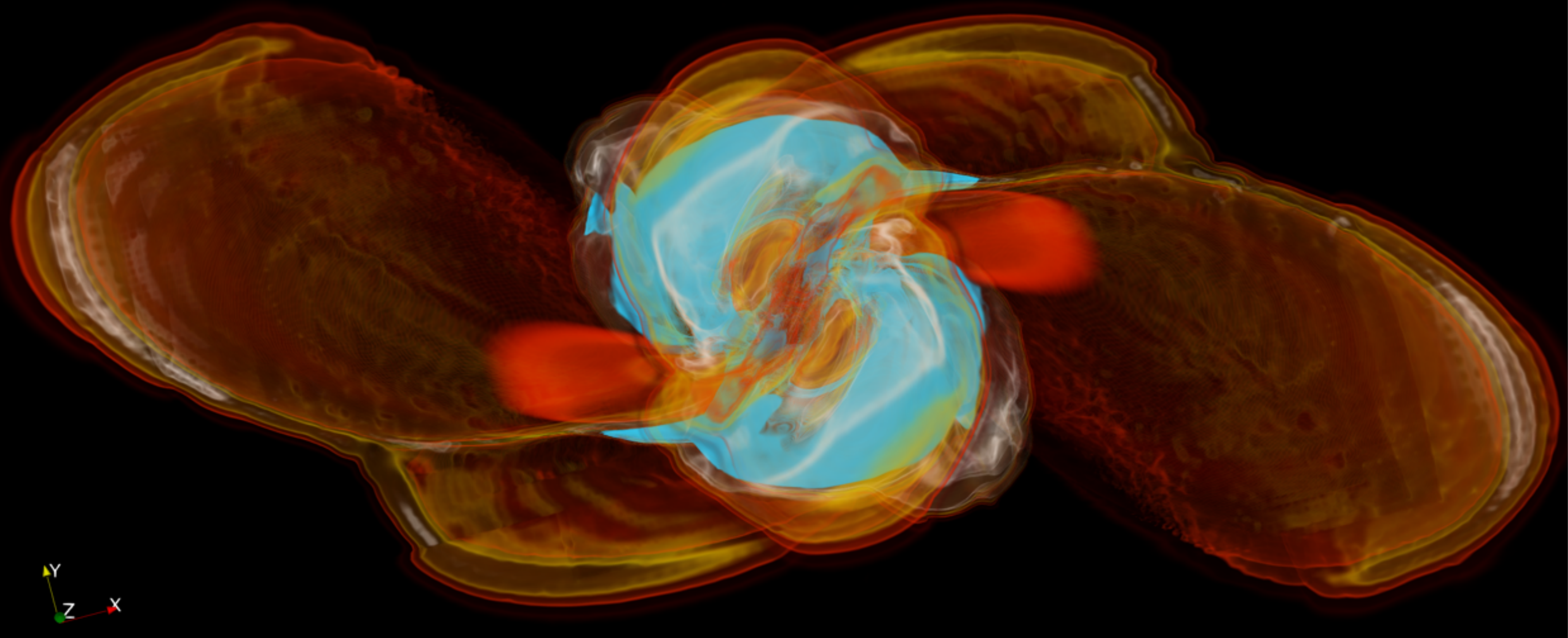}
\caption{
Volume rendering of a $1M_\sun$ ZAMS star at $t-t_{\rm p} \approx 3 t_{\rm dyn}$ in a $\beta=1$ encounter with a $10^6 M_\sun$ BH. Color corresponds to density, and the colorbar and transfer function are chosen to qualitatively highlight the stratified density structure of the debris. Videos of the simulations are available at \url{https://www.youtube.com/channel/UCShahcfGrj5dOZTTrOEqSOA}.
\label{fig:vol_rend}
}
\end{figure*}

For comparison, for giant stars, $\rho_c/\bar\rho \gtrsim 10^6$. In \citet{2012ApJ...757..134M}, due to numerical limitations, the authors use $\rho_{\rm core}/\bar\rho \approx 2\times10^3$ for the hydrodynamical simulations. They are unable to fully disrupt the giant stars, due to the highly segregated density profile (much more centrally concentrated than that of MS stars despite the artificial value of $\rho_{\rm core}/\bar\rho$) and extended envelope structure.

Figure~\ref{fig:MESA_profiles_vs_polytropes_norm} in the Appendix shows density profiles for each star we simulate as compared to $\gamma=5/3$ and $\gamma=4/3$ stellar structures. Also see Figure~1 in \citet{2019ApJ...882L..25L} for the stellar structure evolution of a $1M_\sun$ and $3M_\sun$ star over their MS lifetimes (also compared to polytropic stellar profiles). Only a few stars correspond closely to polytropic stellar structures; for all others, MESA initial conditions are significantly more accurate.

\subsection{Qualitative tidal disruption results}

As an example of one of our FLASH simulations, Figure~\ref{fig:vol_rend} shows a volume rendering of the disruption of a $1M_\sun$ ZAMS star at $t-t_{\rm p} \approx 3 t_{\rm dyn}$ in a $\beta=1$ encounter with a $10^6 M_\sun$ BH (where $t_{\rm p}$ is the pericenter time and $t_{\rm dyn}$ is the dynamical time of the star). Color corresponds to density. One can see the stratified structure following the disruption of a star with an accurate internal stellar structure. This encounter is a partial disruption in which a core survives.
Note that this snapshot is zoomed in on the star in order to highlight the density structure of the debris a few dynamical times after pericenter, but that our computational domain is 1000$R_\star$ on a side (roughly 100 times the size of this volume rendering), and the debris eventually expands to fill this.

\begin{figure*}[htp!]
\epsscale{1.15}
\plotone{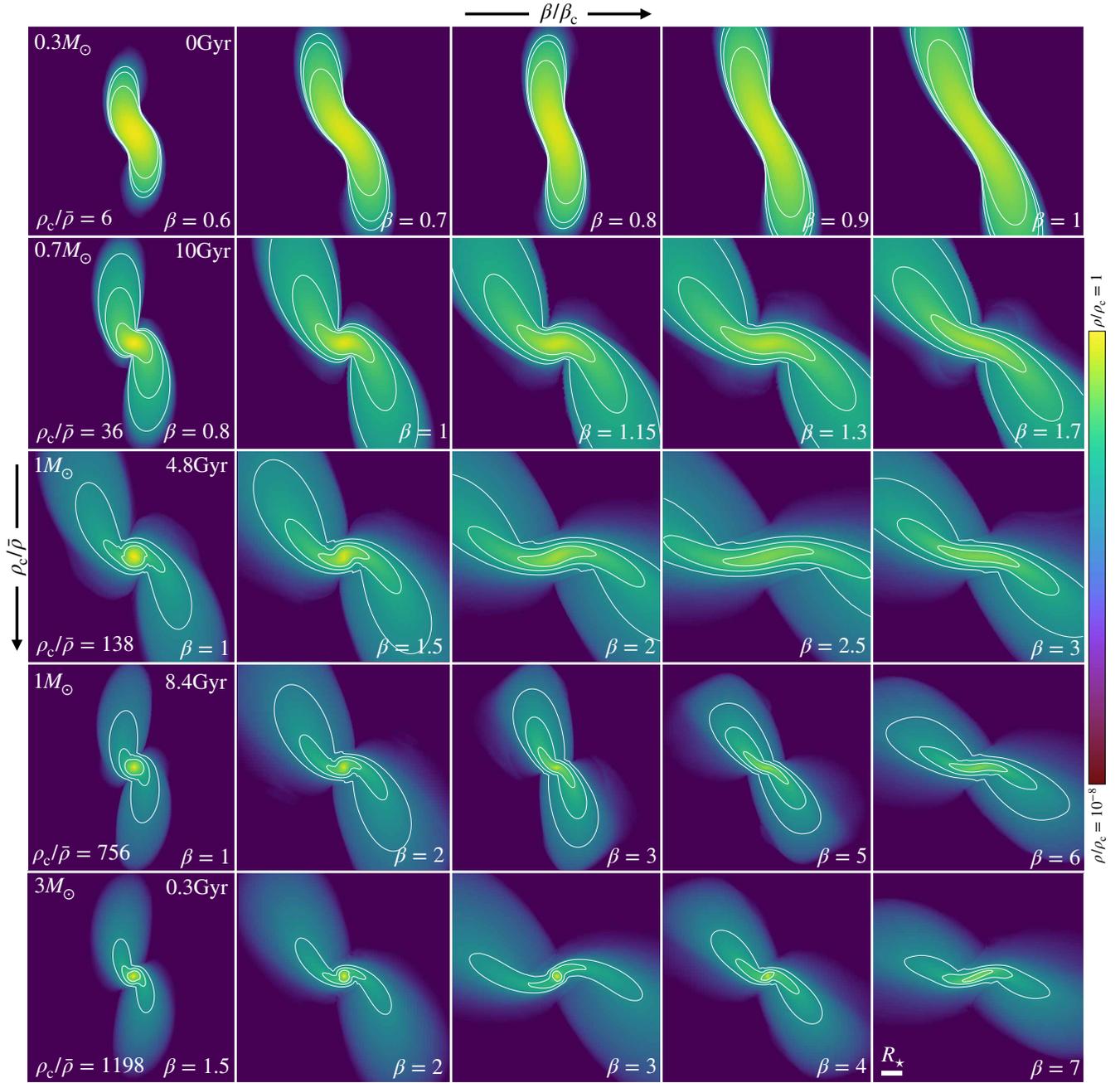}
\caption{
2D density slices of a few different simulations.
The axes of this plot grid are central concentration ($\rho_c/\bar\rho$) vs. impact parameter ($\beta/\beta_c$). 
Each row is a different star and each panel is a different impact parameter. 
$\beta$ increases from left to right and $\rho_c/\bar\rho$ increases from top to bottom. 
Color is the logarithm of density and is normalized to the initial central density of the star, extending to $10^{-8}$ of this value. 
Each panel has width $10R_\star$. All panels are at $t- t_{\rm p} \approx 2t_{\rm dyn}$. White contours correspond to densities of 1, $10^{-1}$, and $10^{-2}$ g/cm$^3$.
\label{fig:butterfly}
}
\end{figure*}

As an example of the range of vulnerability to tidal disruption of our grid, Figure~\ref{fig:butterfly} shows 2D density slices of the disruption of several different stars at different impact parameters, all at $\approx 2 t_{\rm dyn}$ after pericenter. The axes are arranged such that $\beta$ increases from left to right, and central concentration ($\rho_c/\bar\rho$) increases from top to bottom. 
The white contours correspond to absolute values of density (1, $10^{-1}$, and $10^{-2}$ g/cm$^3$), illustrating the different stellar structures and also the amount the different layers of the star are spun up.
Increasing $\beta$ both increasingly distorts the star and spins it up. The more centrally concentrated stars have ``layers'' that are more differentiated. 
This leads to the outer layers being torqued more than the inner layers, and the core remaining sequestered and undisturbed at higher $\beta$'s for more centrally concentrated stars.
Increasing central concentration allows the star to survive higher-$\beta$ encounters. For example, a ZAMS 0.3$M_\sun$ star has a larger fraction of its mass at larger radii, and is thus fully disrupted by a $\beta=1$ encounter, whereas a TAMS 1$M_\sun$ star has only a small fraction of its mass at larger radii from its sequestered core, and is thus relatively undisturbed by a $\beta=1$ encounter.

\subsection{Mass lost}

\begin{figure*}[tp!]
\epsscale{0.55}
\plotone{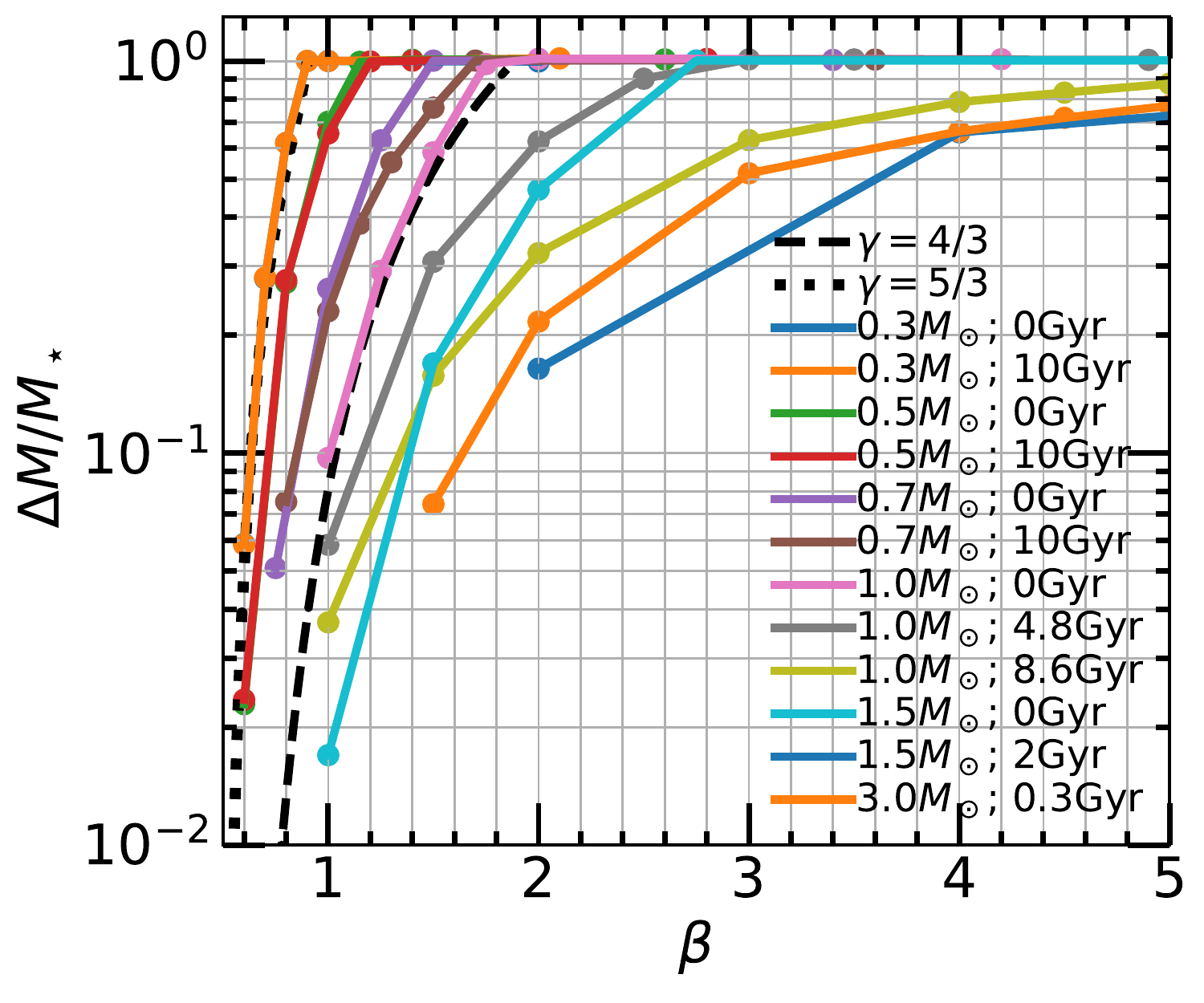}
\plotone{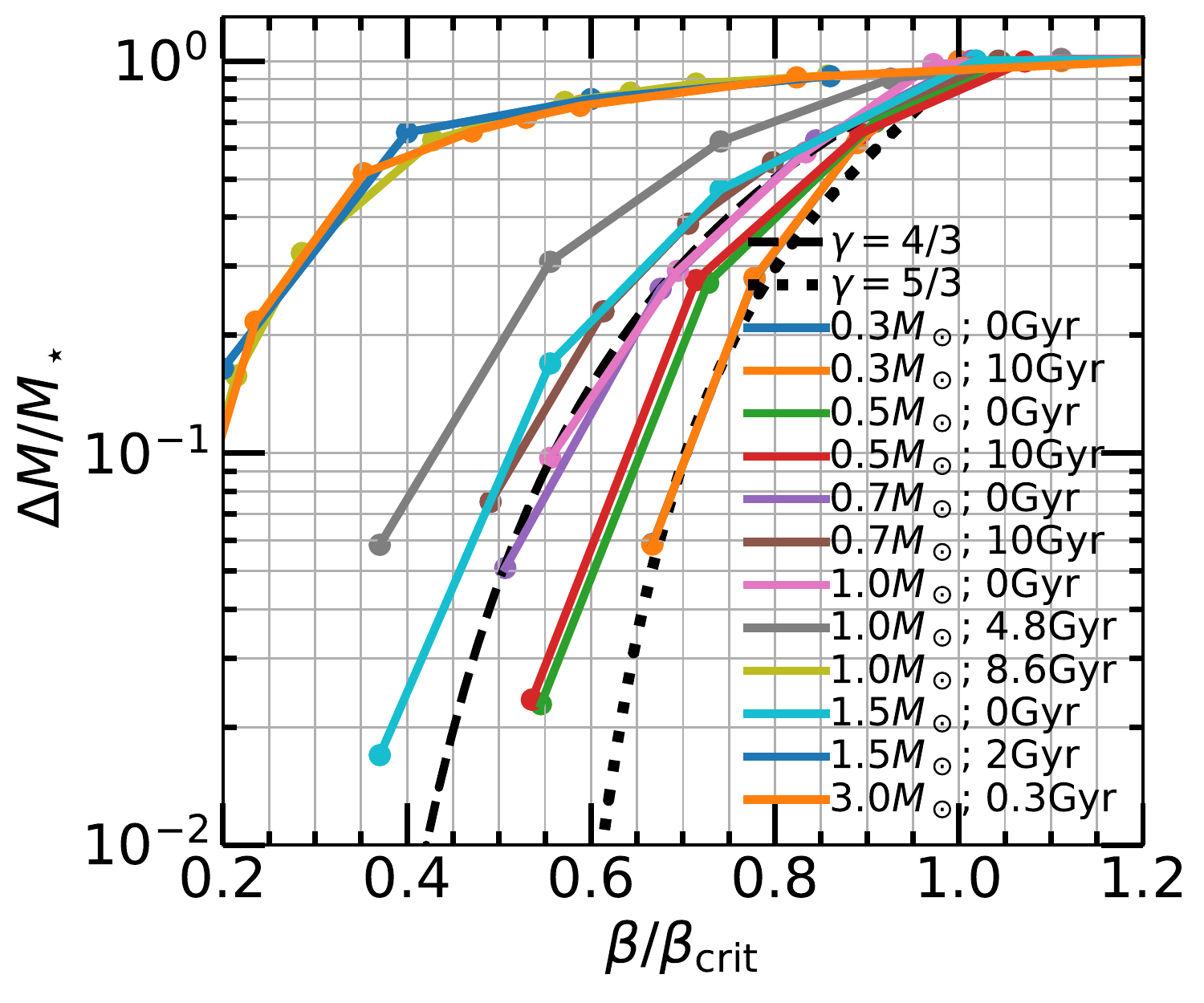}
\plotone{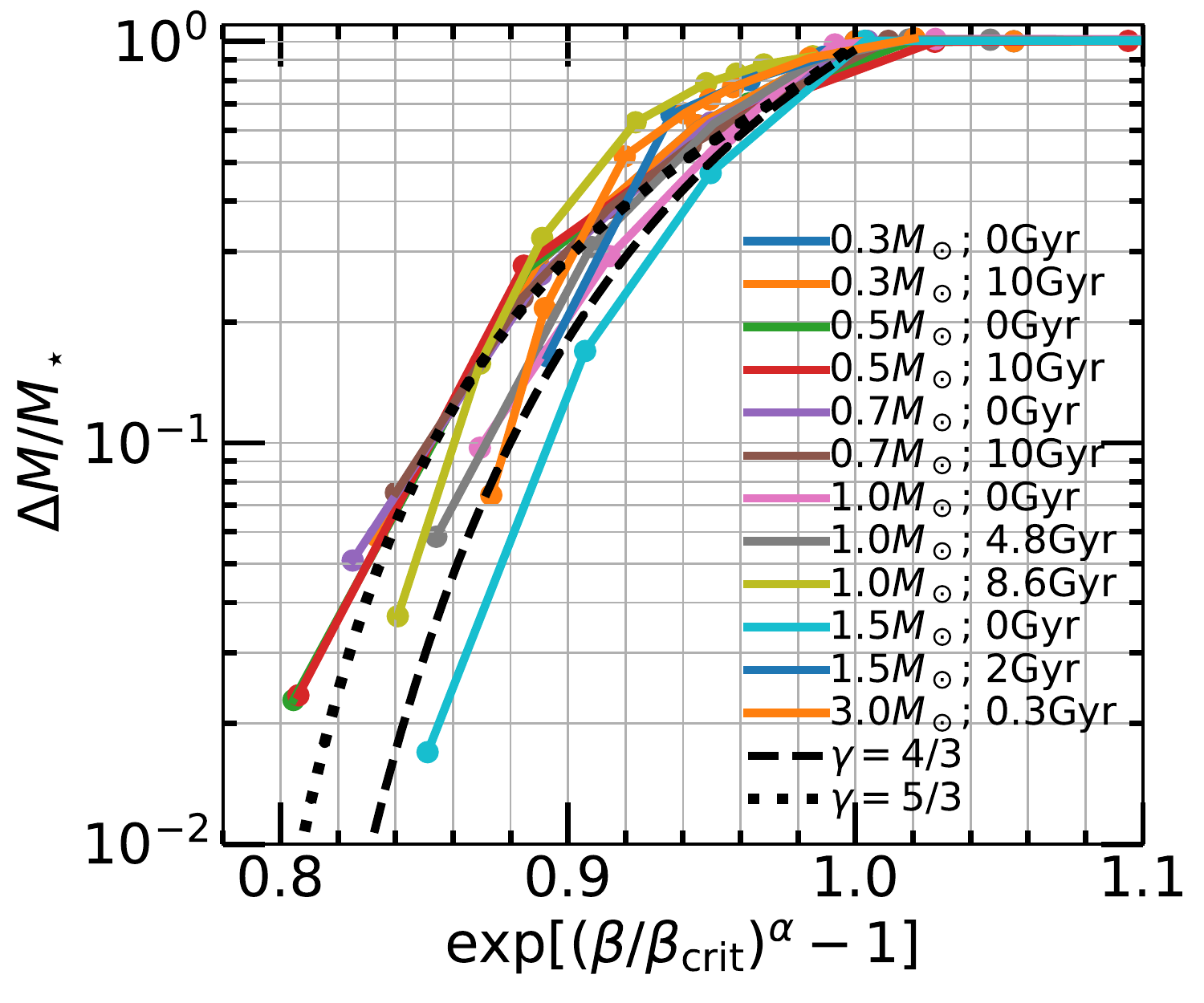}
\plotone{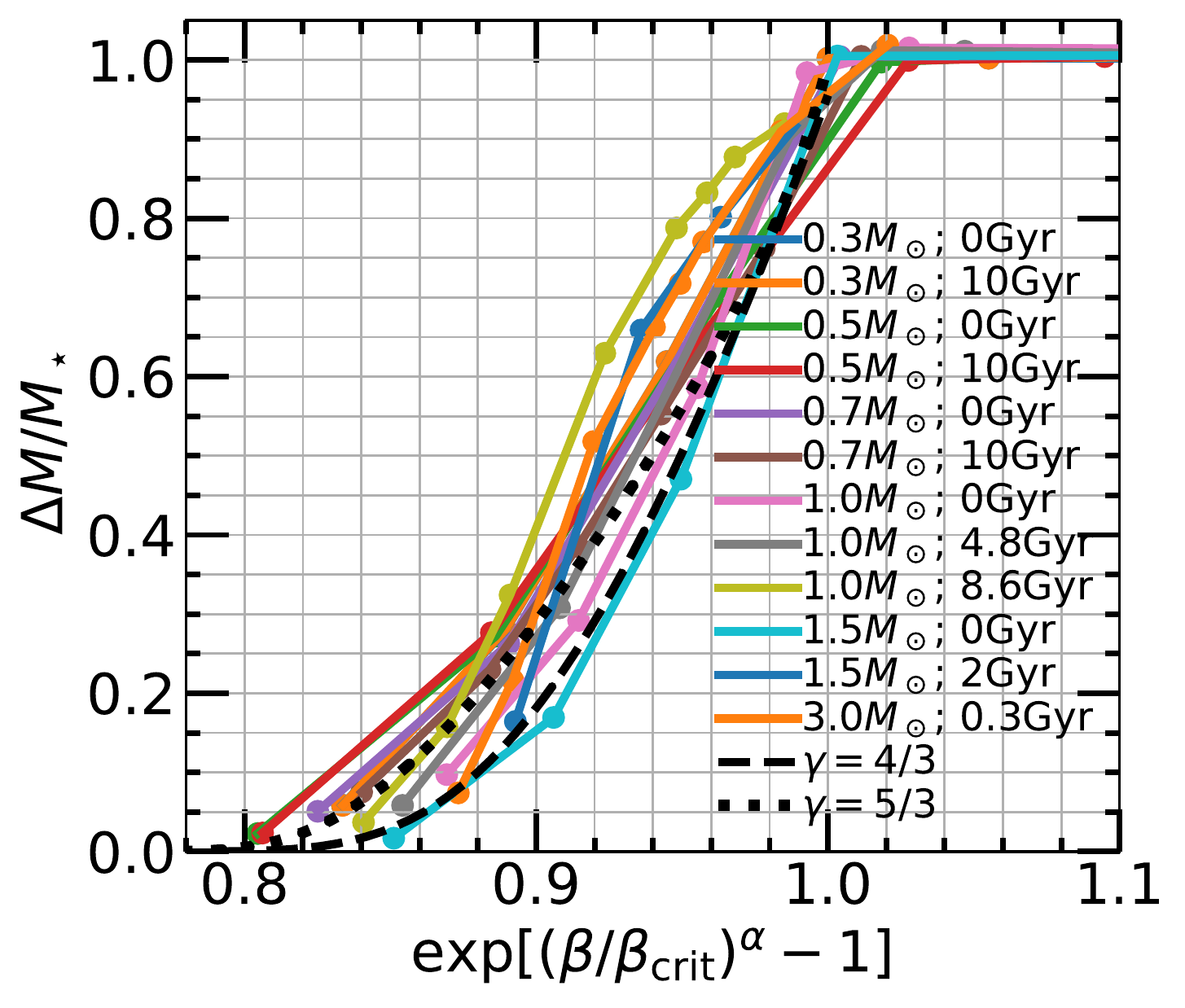}
\caption{
Top left: fractional mass lost $\Delta M/M_\star$ vs. impact parameter $\beta$. 
Top right: $x$-axis is normalized to the critical impact parameter for each star (see Table~\ref{tab:beta_crit}).
Bottom left: $x$-axis is scaled with a structural parameter, $\alpha = (\rho_c/\bar\rho)^{-1/3}$ (see text).
Bottom right: linear $y$-axis.
Results from \citet{2013ApJ...767...25G} for a $\gamma=5/3$ and $\gamma=4/3$ polytrope are shown by the dotted and dashed lines respectively.
\label{fig:deltaM_vs_beta}
}
\end{figure*}

Figure~\ref{fig:deltaM_vs_beta} shows fractional mass lost from the star $\Delta M/M_\star$ as a function of impact parameter $\beta$. The $x$-axis in the top left panel is the raw $\beta$ and the $x$-axis in the top right panel is normalized to the critical $\beta$ for full disruption for each star. Generally, more massive stars and stars further along in their MS evolution are more centrally concentrated, and thus must be disrupted deeper relative to their nominal ``tidal radii'' in order to lose the same amount of fractional mass. One can see that the mass-loss prescriptions for $\gamma=5/3$ and $\gamma=4/3$ polytropes are inadequate to describe the more centrally concentrated stars, which have critical impact parameters of $\beta_{\rm crit}>2$.

The bottom panels have a scaled $x$-axis, 
\begin{equation}
x = \exp \left[ \left( \beta/\beta_{\rm crit} \right)^\alpha -1 \right],\ \ \ \alpha = (\rho_c/\bar\rho)^{-1/3},
\end{equation}
constructed \added{by trial-and-error} in order to reduce all of the simulations into a single relationship. 
\added{The physical meaning of the functional form of $x$ is unclear at present, but the sharp $1/3$ factor suggests at least a partial physically-motivated origin as this number appears in basic TDE theory.}
This formula accounts for the dependence of $\Delta M$ vs. $\beta$ on stellar structure.
The fact that we are able to express all of the simulations in a single relationship implies that $\Delta M$ depends only on $\rho_c/\bar\rho$ and $\beta$.
Analytic and B-spline fits for this relation, allowing one to obtain the mass lost for any stellar mass, stellar age, and impact parameter, are provided in the Appendix.
The simulations in this work all reduce to a single relation for other disruption quantities as well (see below in this section).

\subsection{Critical impact parameter}

\begin{figure}[t!]
\epsscale{1.1}
\plotone{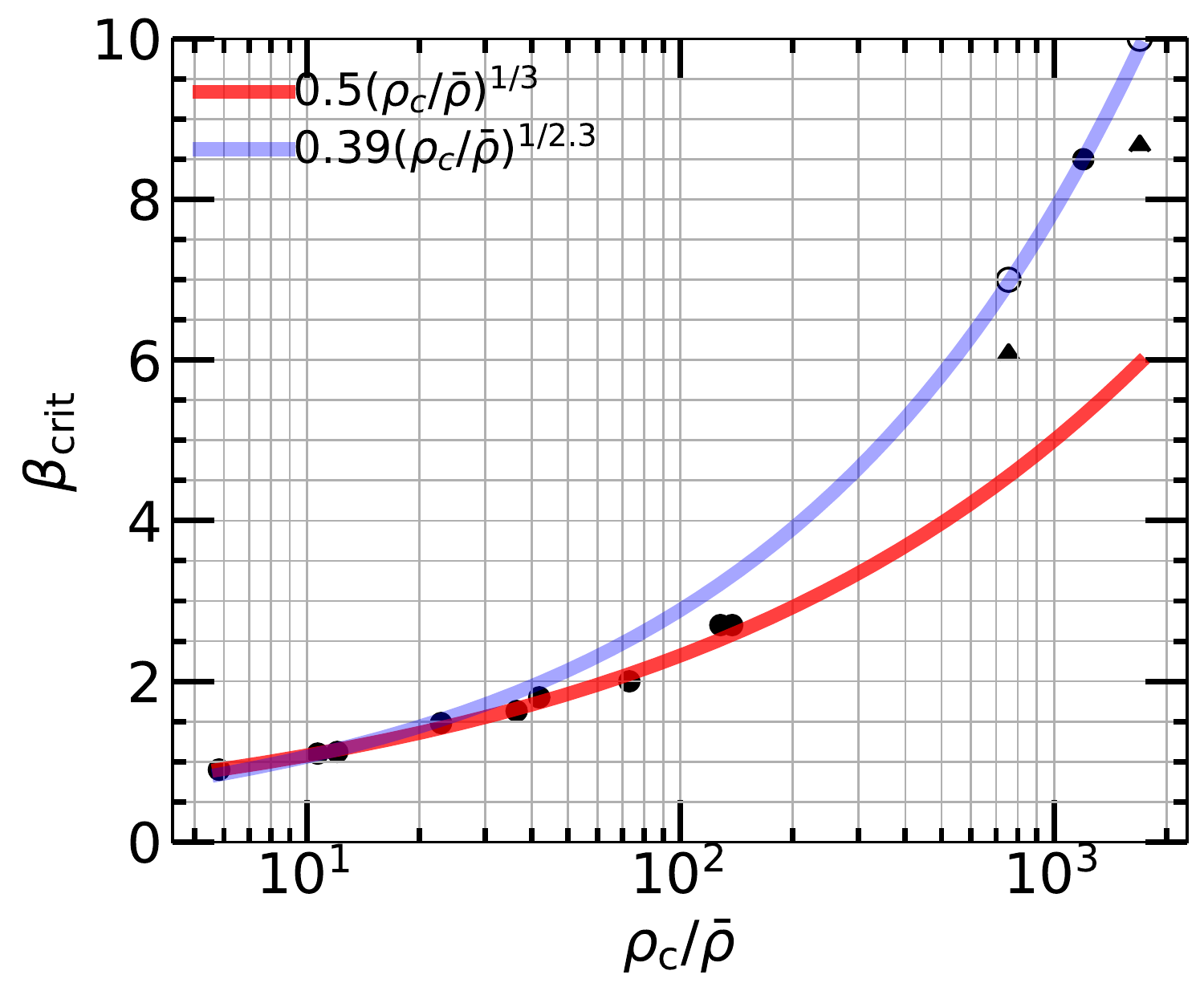}
\caption{
Critical impact parameter for full disruption $\beta_{\rm crit}$ vs. ratio of stellar central density to average density $\rho_{\rm c}/\bar\rho$. Triangle markers indicate lower limits and open circles indicate extrapolation. Simple fits in red and purple. See Table~\ref{tab:beta_crit} as well as Figure~\ref{fig:beta_crit_vs_Mstar2} in Appendix for dependence of $\beta_{\rm crit}$ and $(r_p/r_g)_{\rm crit}$ with stellar mass.
\label{fig:beta_crit_vs_Mstar}
}
\end{figure}

We did not sample finely enough in $\beta$ to find the exact critical impact parameter for full disruption for each star, but we estimate this within the bounds associated with our spacing in $\beta$ and the mass lost as a function of $\beta$ for the star.
Figure~\ref{fig:beta_crit_vs_Mstar} shows the critical $\beta$ as a function of $\rho_c/\bar\rho$. For nearly all stars, the critical $\beta$ is well-fit by a simple relation,\added{\footnote{We note that this formula (for stars with lower central concentrations, $\rho_c/\bar\rho \lesssim 500$), is very similar to that found by \citet{2020arXiv200103502R}.}}
\begin{equation}
\beta_{\rm crit} \approx 0.5 \left(\frac{\rho_c}{\bar\rho} \right)^{1/3},\ \ \ \rho_c/\bar\rho \lesssim 500.
\label{eq:beta_crit}
\end{equation}
At the highest central concentrations ($\rho_c/\bar\rho \gtrsim 500$), the critical $\beta$ is higher than predicted with this relation, and is instead better fit by a steeper relation,
\begin{equation}
\beta_{\rm crit} \approx 0.39 \left(\frac{\rho_c}{\bar\rho} \right)^{1/2.3},\ \ \ \rho_c/\bar\rho \gtrsim 500.
\label{eq:beta_crit2}
\end{equation}
\replaced{This}{Equation (\ref{eq:beta_crit}) and Equation (\ref{eq:beta_crit2})} allow one to predict the approximate critical impact parameter for full disruption for any main-sequence star.
\added{We note that already for $\rho_c/\bar\rho \gtrsim 100$, Equation (\ref{eq:beta_crit}) slightly underpredicts $\beta_{\rm crit}$, and that there is likely a smooth transition between the two power-law relationships, but we fit only two equations for simplicity.}
All one needs is $\rho_c/\bar\rho$ for that star, obtained from, e.g., a MESA model directly or pre-computed grids such as the MIST models \citep{2016ApJ...823..102C}.
So the ability to fully disrupt a star is a simple function of this ratio of densities.   

Note that for 2 points at the highest central concentrations, the quoted $\beta_{\rm crit}$ is obtained through extrapolation, as the maximum $\beta$ simulated for that star did not fully disrupt it. 
The lower limits obtained from our simulations are also shown.
We note also that the resolution (in term of maximum number of cells in the simulation) required to precisely determine $\beta_{\rm crit}$ becomes significantly higher for these most centrally concentrated stars. At the end of the simulation when the debris has expanded, and thus the linear resolution has decreased from the initial maximum level of refinement (in order to resolve the same number of cells in the simulation box), the highest density ``core'' of the star---for the stars with $\rho_c/\bar\rho \gtrsim 10^3$---becomes smaller than a grid cell and so is not resolved. However, the aim of this study is not to precisely determine the critical impact parameter for these most centrally concentrated stars, but rather to determine the mass fallback rate to the BH following their disruption.

Table~\ref{tab:beta_crit} lists the critical impact parameter $\beta_{\rm crit}$ for full disruption for all stars studied. 
Errors represent the grid spacing and so are overestimates.
We also list the corresponding physical pericenter distance in gravitational radii $r_{\rm p}/r_{\rm g}$. 
For stars that do not evolve significantly in structure over 10 Gyr, their $\beta_{\rm crit}$'s remain the same or very similar. For stars that evolve significantly over $<10$ Gyr, the $\beta_{\rm crit}$ increases dramatically over a star's lifetime. For example, for a 1$M_\sun$ star, the $\beta_{\rm crit}$ increases from $\approx$1.8 to $\approx$7.0\added{, a factor of 3.9X,} over its main-sequence lifetime, a result of its $\rho_c/\bar\rho$ increasing from 42 to 756. This corresponds to the critical $r_{\rm p}/r_{\rm g}$ for full disruption decreasing by a factor of 2.7X, from 23.4 to 8.6. Thus, a TAMS 1$M_\sun$ star must approach 2.7X closer the BH in order to be fully disrupted compared to its ZAMS self. Similar trends are seen for the 1.5$M_\sun$ and 3$M_\sun$ stars, where the critical $r_{\rm p}/r_{\rm g}$ decreases by a factor $>$2X from ZAMS to TAMS. Thus, for higher-mass stars ($M_\star \gtrsim 0.8 M_\sun$), at the same pericenter distance, the ZAMS star is roughly 2X as vulnerable to tidal disruption and associated mass loss as the TAMS star. For lower-mass stars ($M_\star \lesssim 0.8 M_\sun$), the critical $r_{\rm p}/r_{\rm g}$ is roughly constant over 10 Gyr. See the Appendix for the dependence of $\beta_{\rm crit}$ and the associated $r_{\rm p}/r_{\rm g}$ with stellar mass.

\begin{deluxetable}{l l l l l}
\tablecaption{Critical impact parameter for full disruption $\beta_{\rm crit}$ and the corresponding physical pericenter distance in gravitational radii $r_{\rm p}/r_{\rm g}$. $\dagger$ indicates extrapolation and $\ast$ indicates polytropic result. See also Figure~\ref{fig:beta_crit_vs_Mstar}, Figure~\ref{fig:beta_crit_vs_Mstar2}. \label{tab:beta_crit}}
\tablehead{
\colhead{$M_\star$} & \colhead{$R_\star$} & \colhead{stellar age} & \colhead{$\beta_{\rm crit}$} & \colhead{$r_{\rm p}/r_{\rm g}$}
}
\startdata
0.3 $M_\sun$ & 0.2814 $R_\sun$ & 0 Gyr   & $0.9 \pm 0.1$  & 22.0\\
             & 0.2989 $R_\sun$ & 10 Gyr  & $0.9 \pm 0.1$  & 23.4\\
0.5 $M_\sun$ & 0.4452 $R_\sun$ & 0 Gyr   & $1.1 \pm 0.1$ & 24.0\\
             & 0.4564 $R_\sun$ & 10 Gyr  & $1.1 \pm 0.1$ & 24.7\\
0.7 $M_\sun$ & 0.6485 $R_\sun$ & 0 Gyr   & $1.5 \pm 0.2$  & 23.2\\
             & 0.6793 $R_\sun$ & 10 Gyr  & $1.6 \pm 0.1$  & 22.1\\
1.0 $M_\sun$ & 0.9012 $R_\sun$ & 0 Gyr   & $1.8 \pm 0.1$  & 23.4\\
             & 1.0455 $R_\sun$ & 4.8 Gyr & $2.7 \pm 0.2$  & 18.1\\
             & 1.2872 $R_\sun$ & 8.4 Gyr & $7.0^\dagger$  & 8.6\\
1.5 $M_\sun$ & 1.6275 $R_\sun$ & 0 Gyr   & $2.7 \pm 0.5$ & 23.5\\
             & 2.0805 $R_\sun$ & 2 Gyr   & $10^\dagger$  & 8.6\\
3.0 $M_\sun$ & 1.8896 $R_\sun$ & 0 Gyr   & $2.0^\ast \pm 0.25$ & 30.8\\
             & 3.3192 $R_\sun$ & 0.3 Gyr & $8.5 \pm 1.5$  & 12.7\\
\enddata
\end{deluxetable}

As $\rho_c/\bar\rho$ increases, the star's expansion in response to mass loss decreases.
As a result, the star becomes less vulnerable to tidal \replaced{dissipation}{disruption} and thus the critical impact parameter for full disruption monotonically increases with increasing $\rho_c/\bar\rho$.
For example, note the transition in critical impact parameter from $\beta_{\rm crit} = 0.9 < 1$ for the 0.3$M_\sun$ star to $\beta_{\rm crit} = 1.1 > 1$ for the 0.5$M_\sun$ star. 
The 0.3$M_\sun$ star corresponds to a $\gamma=5/3$ polytrope (see Figure~\ref{fig:MESA_profiles_vs_polytropes_norm}) and the 0.5$M_\sun$ star has an intermediate structure in between a $\gamma=5/3$ and $\gamma=4/3$ polytrope.
In the most extreme cases, for example for giant stars \citep{2012ApJ...757..134M} with a sequestered core-envelope structure, the remaining envelope contracts and thus its disruption requires increasingly higher $\beta$'s for the same amount of mass loss (in fact, these stars are unable to be fully disrupted by massive BHs).

\subsection{Mass fallback rates}

\subsubsection{All encounters}

\begin{figure*}[htp!]
\epsscale{0.38}
\plotone{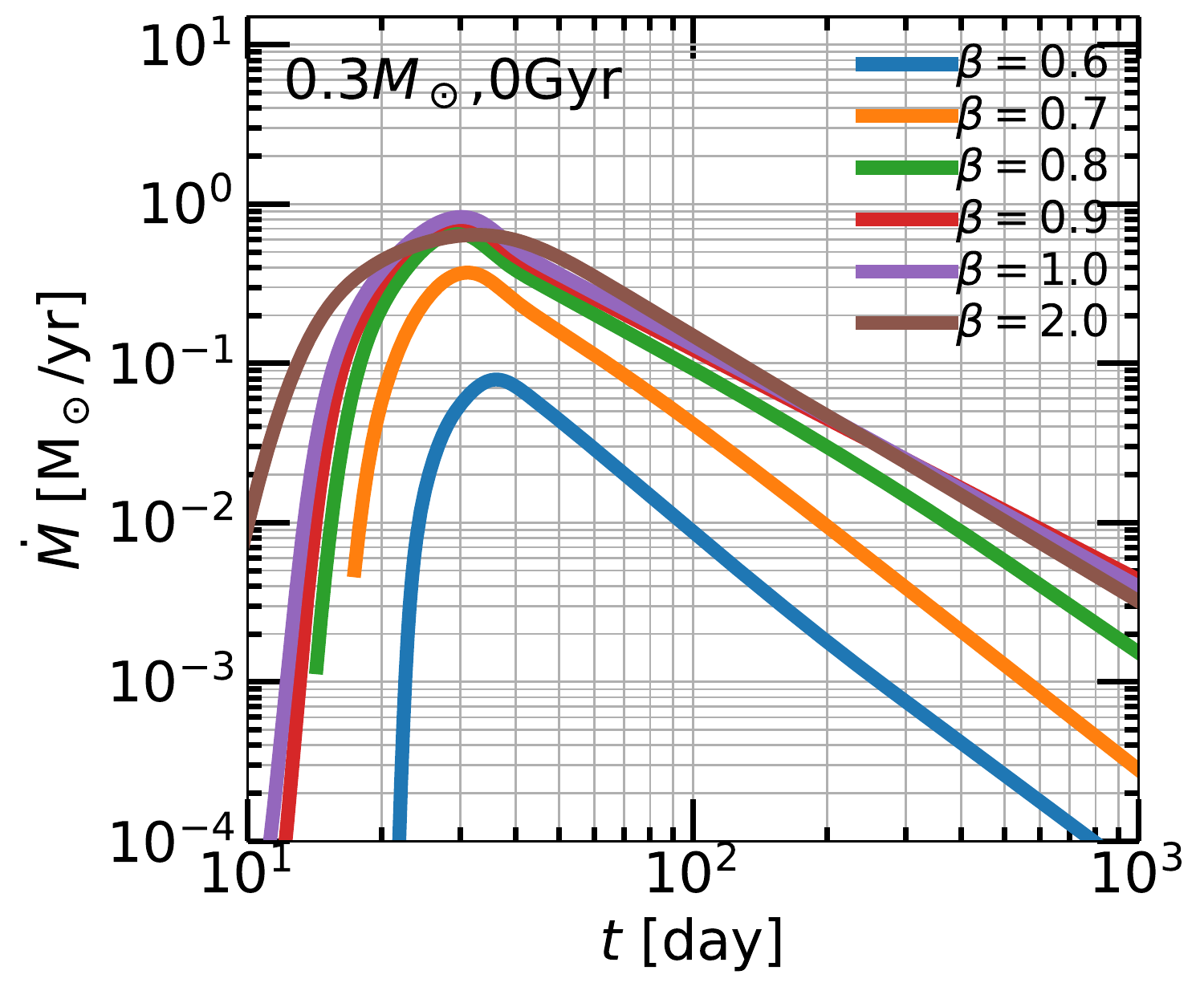}
\plotone{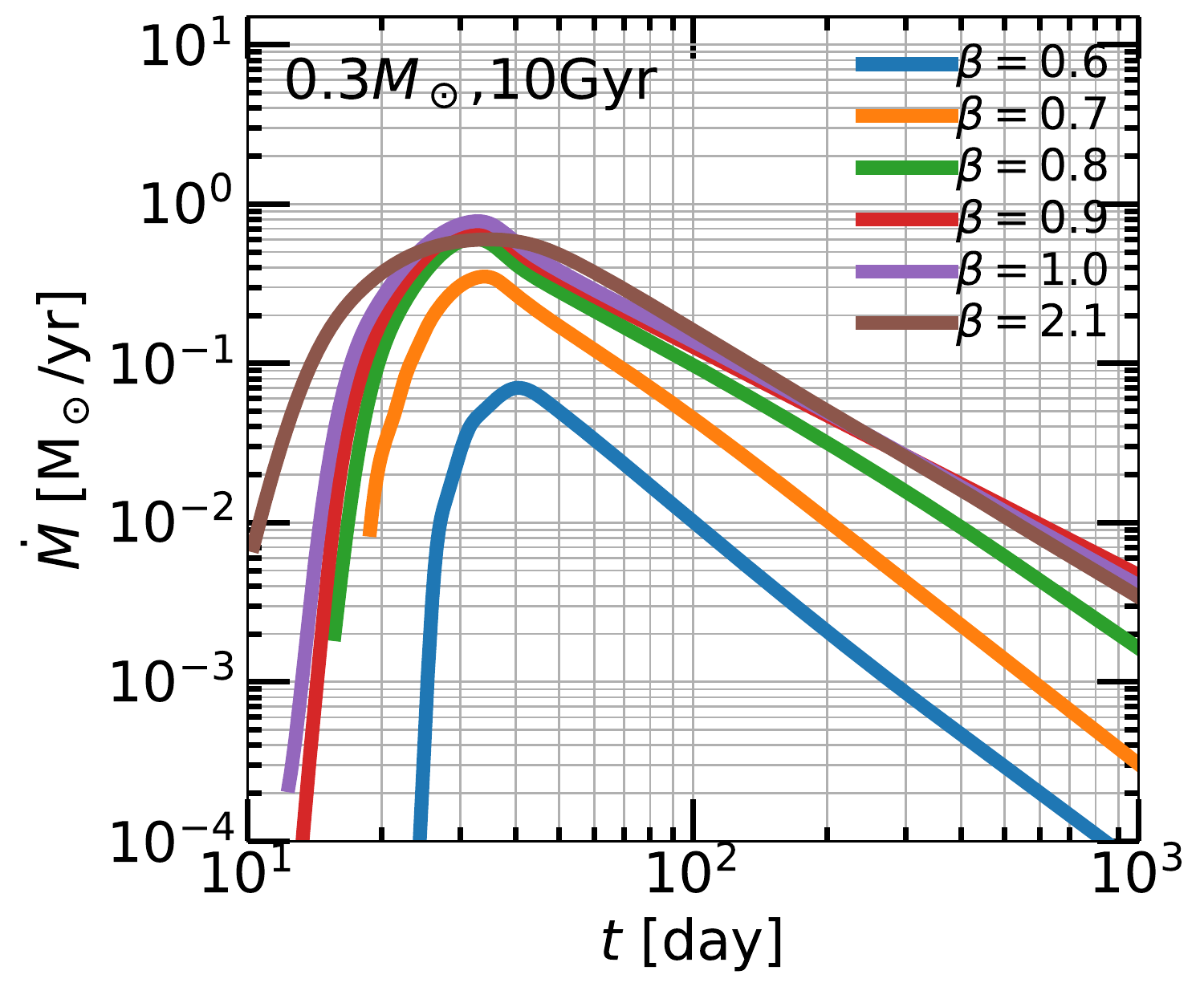}
\plotone{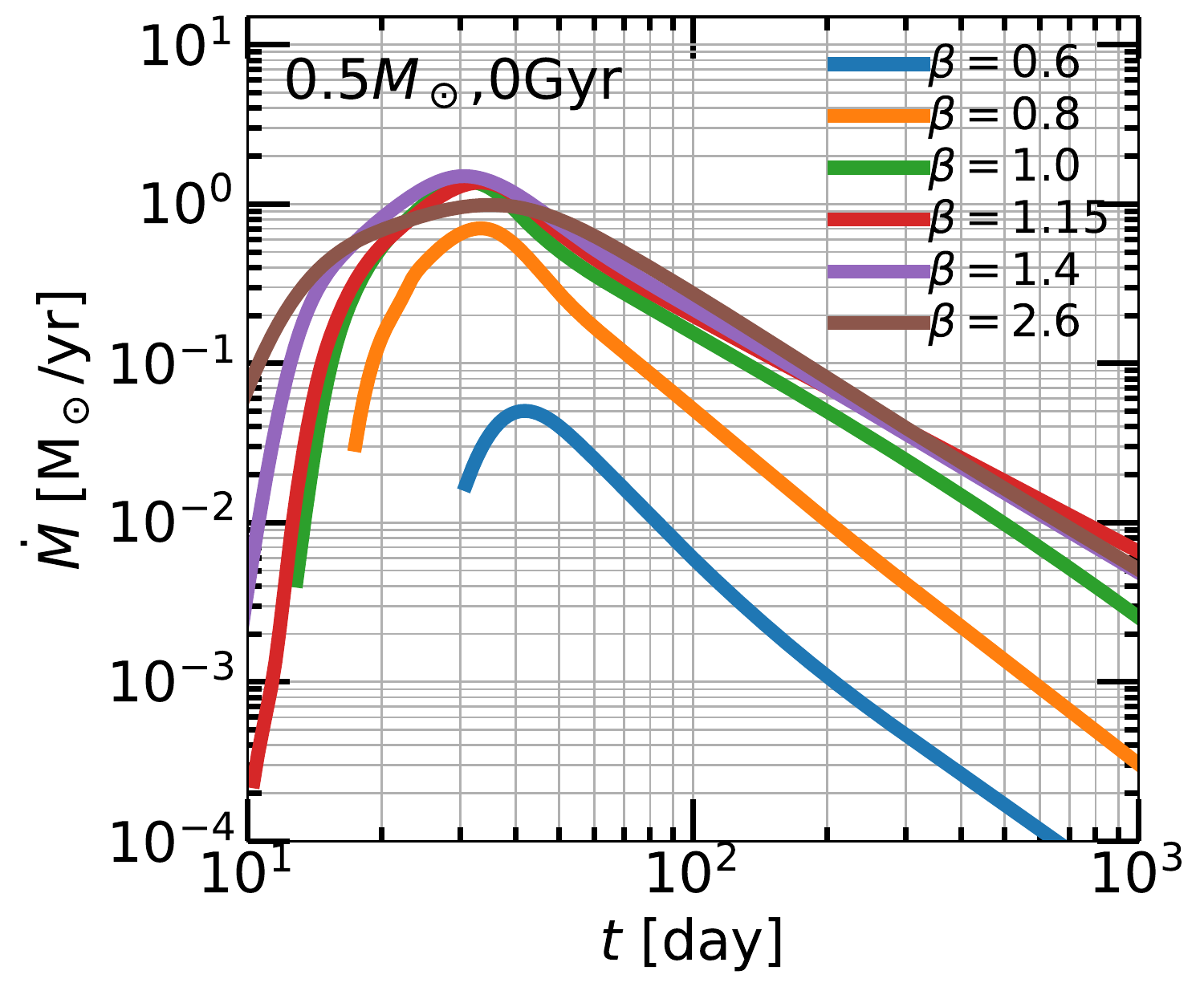}
\plotone{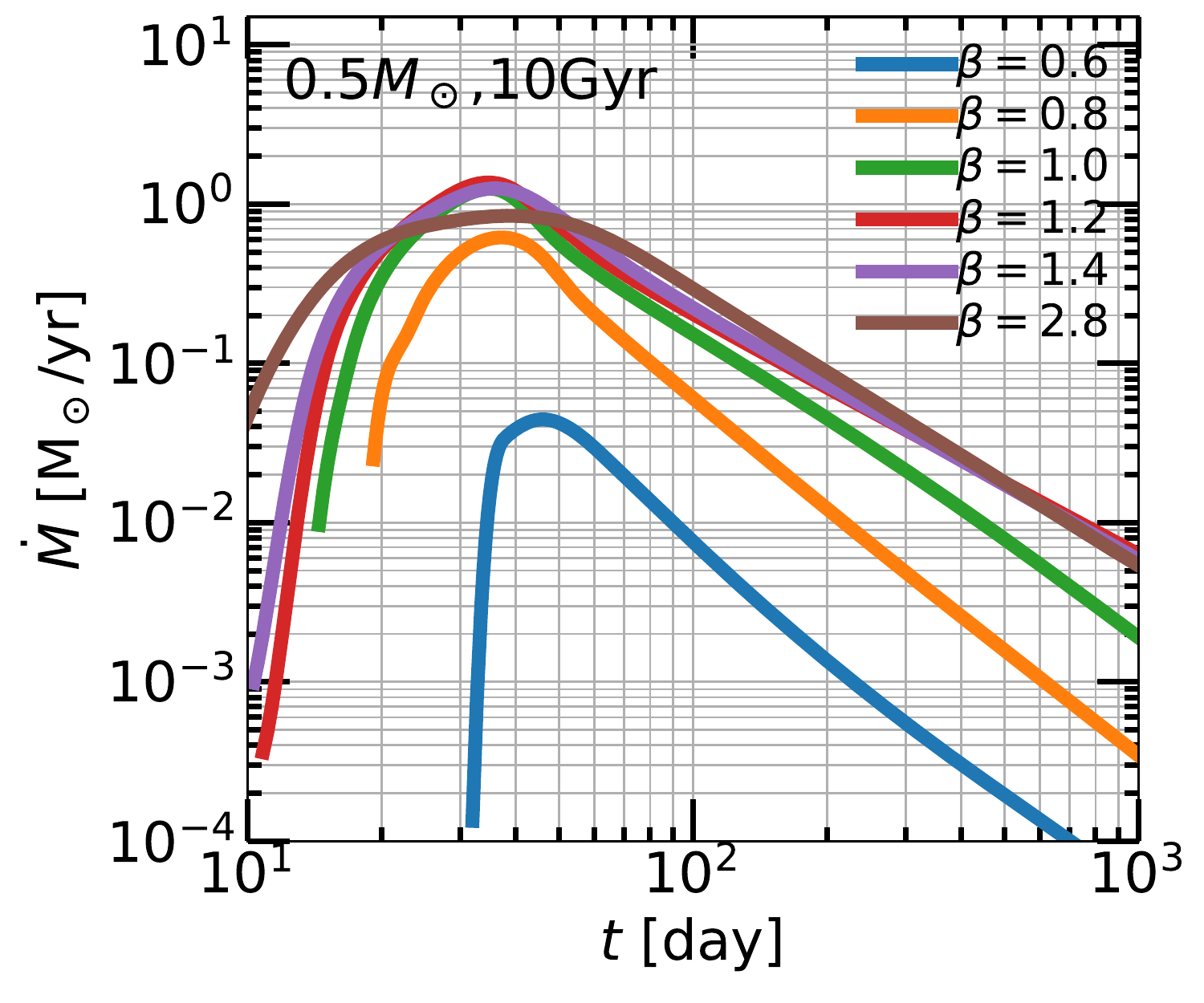}
\plotone{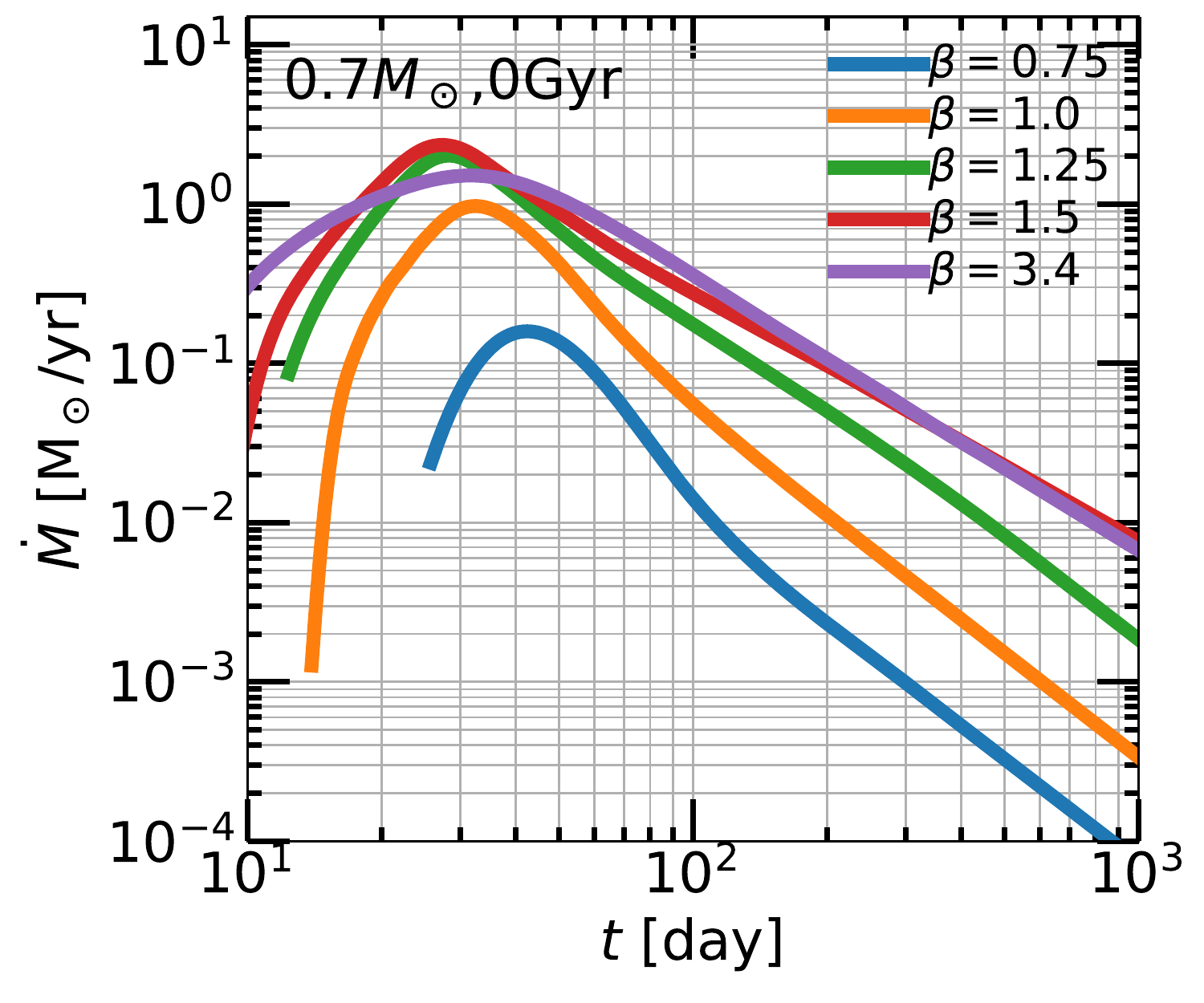}
\plotone{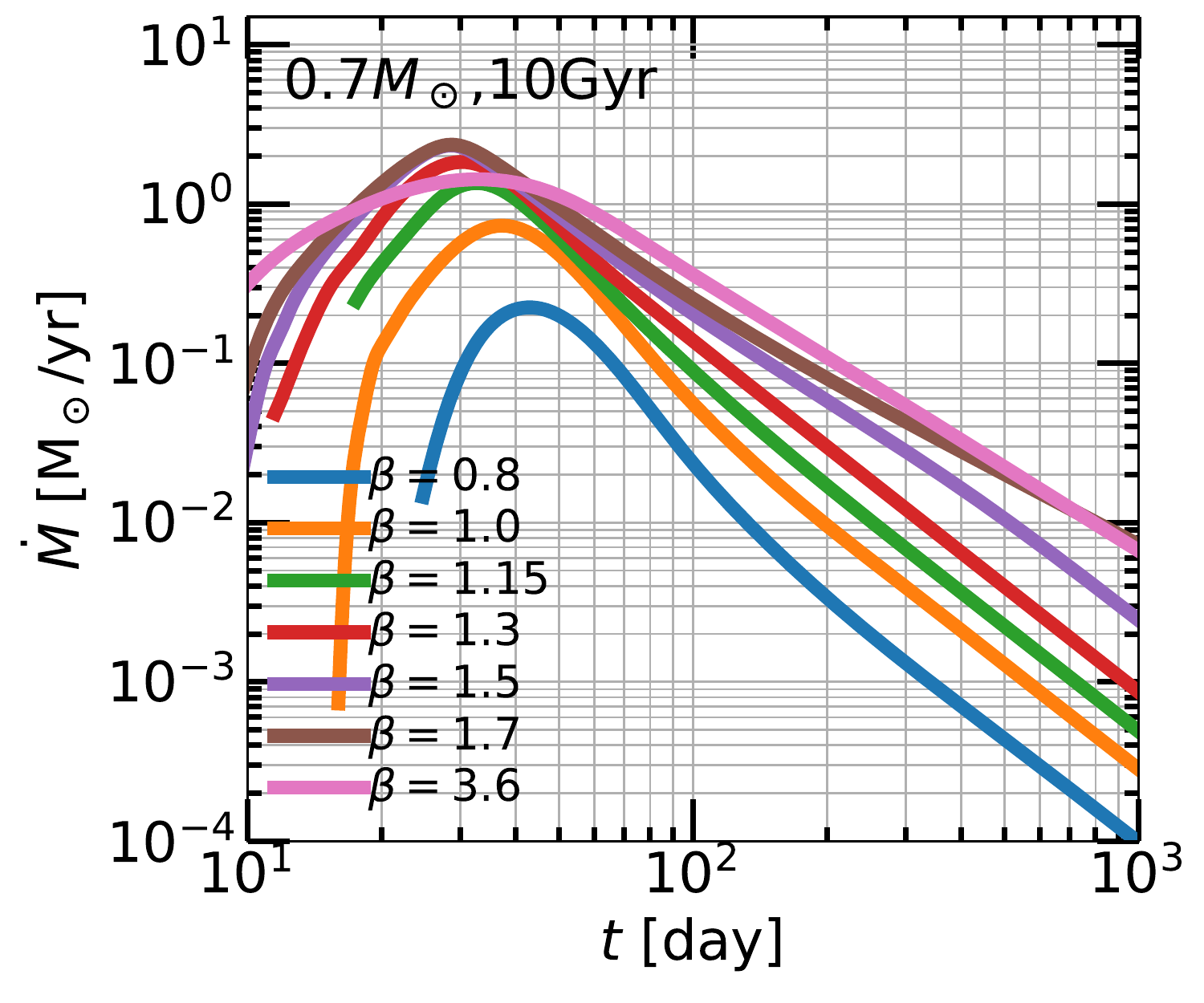}
\plotone{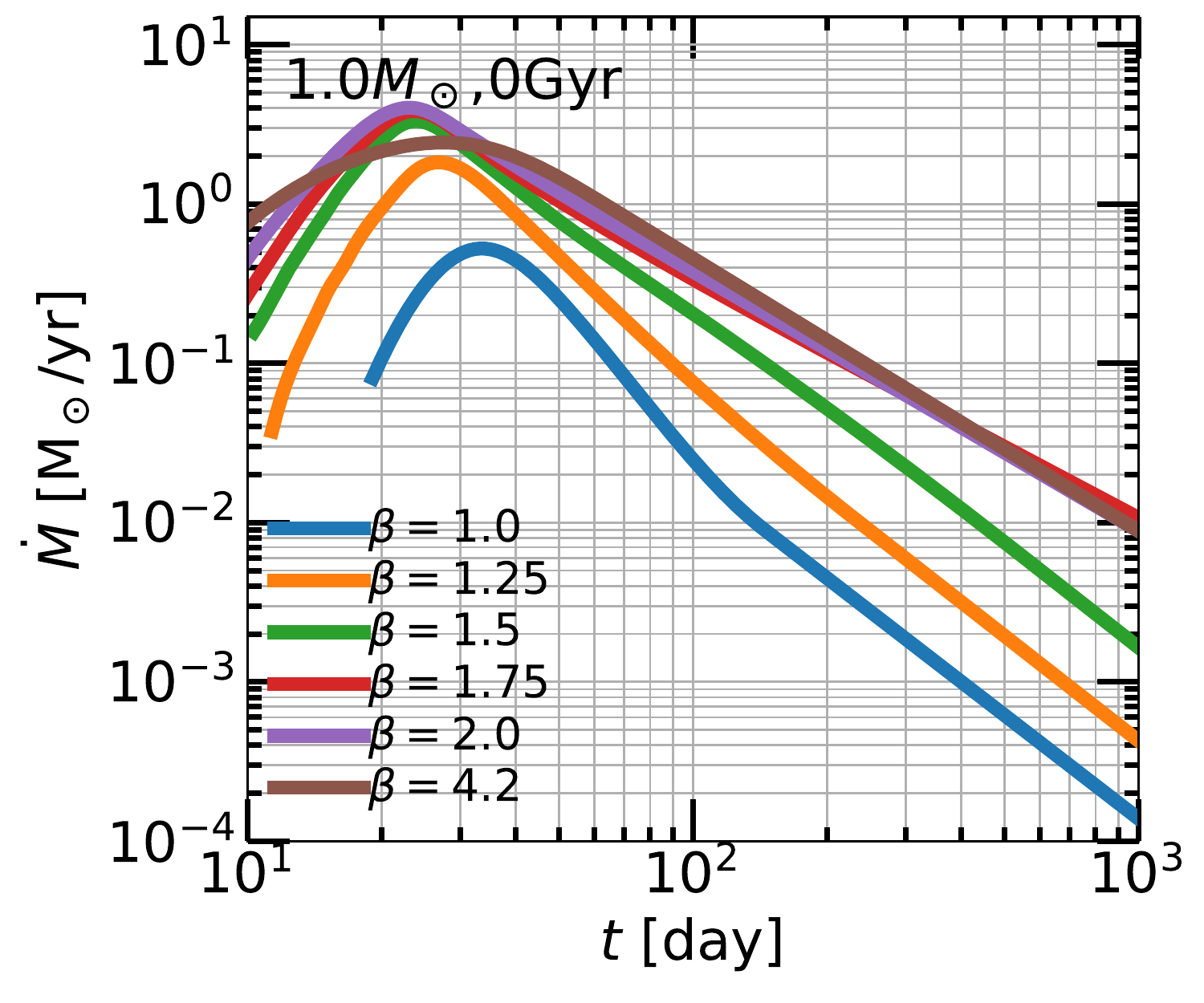}
\plotone{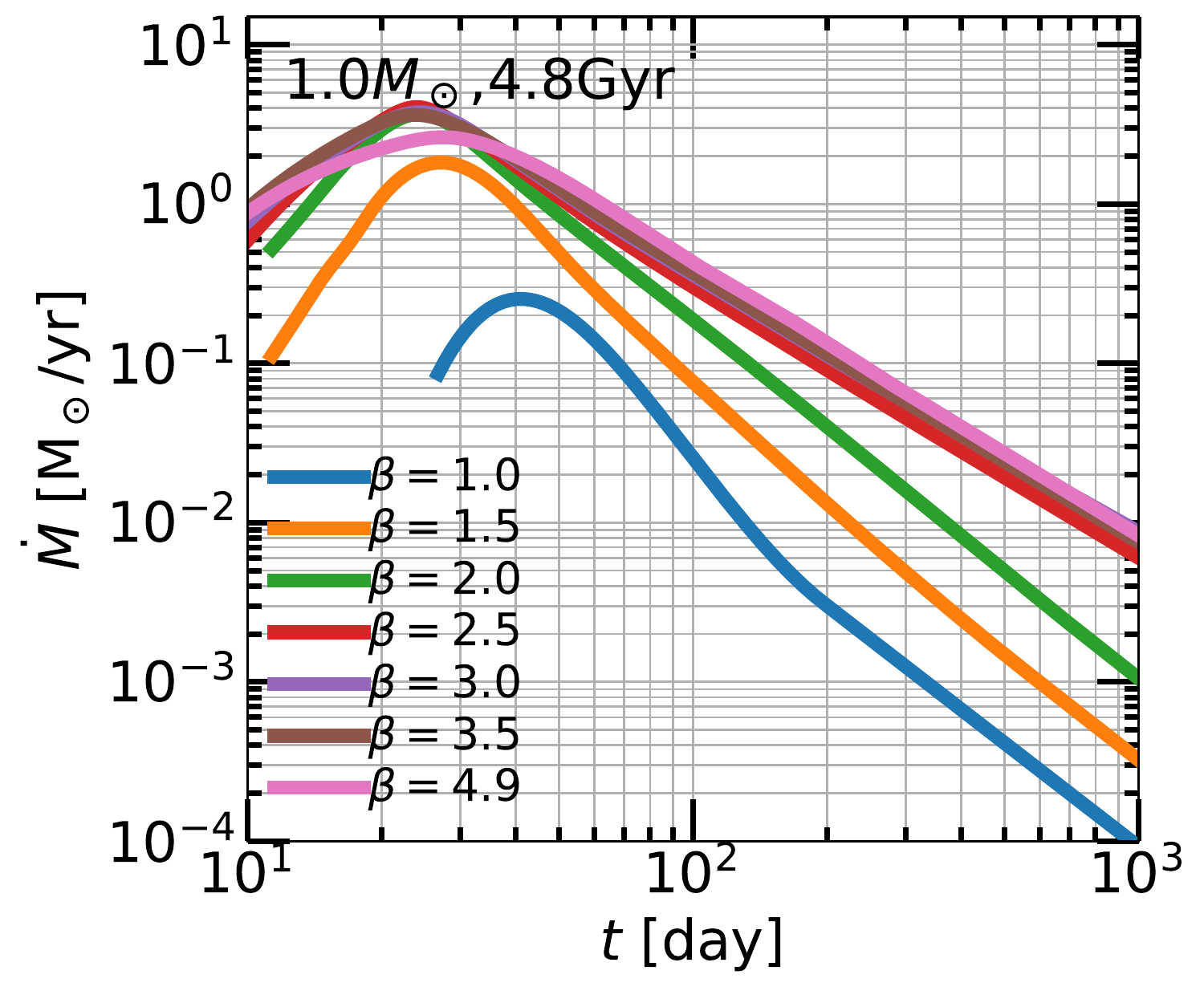}
\plotone{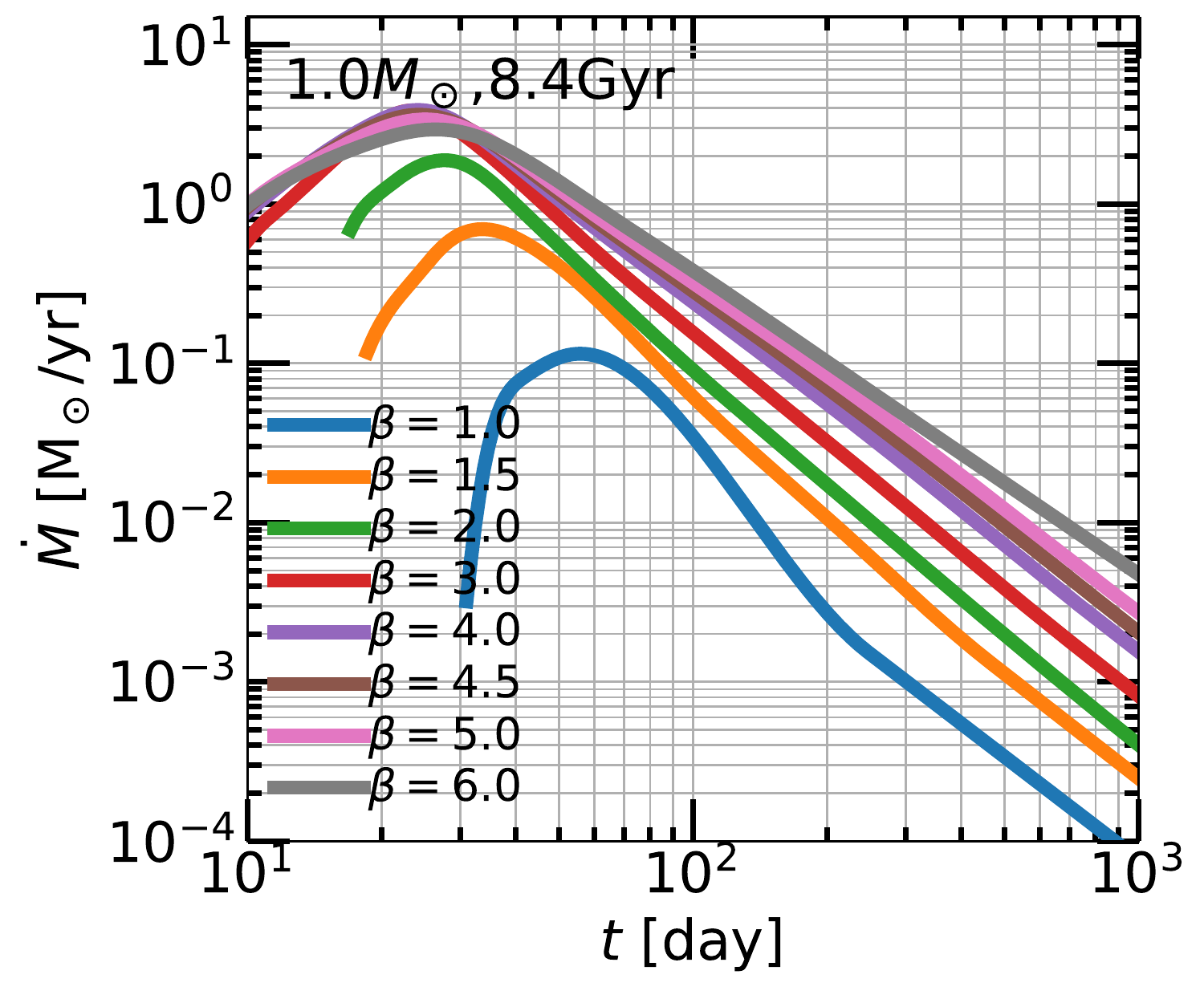}
\plotone{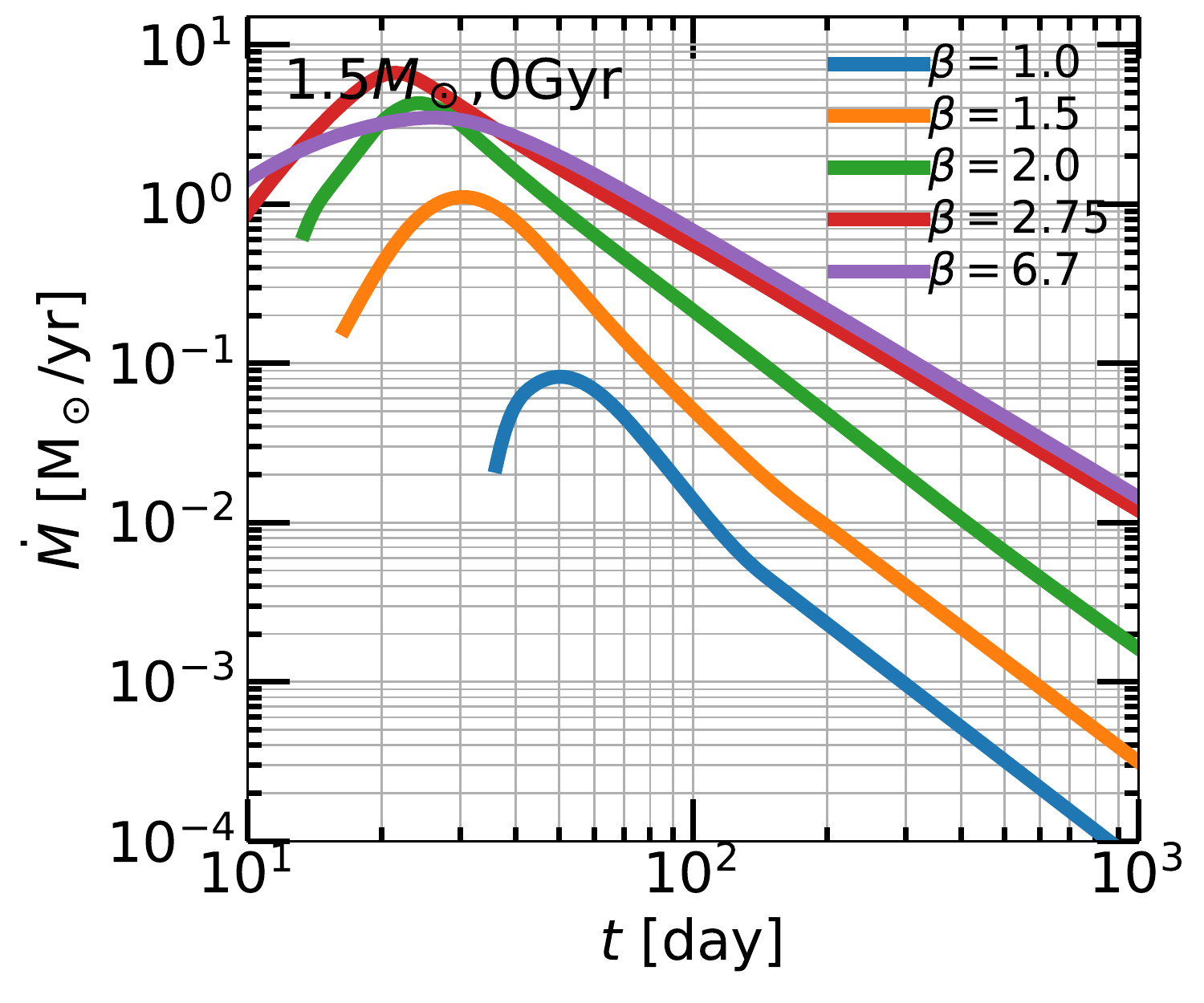}
\plotone{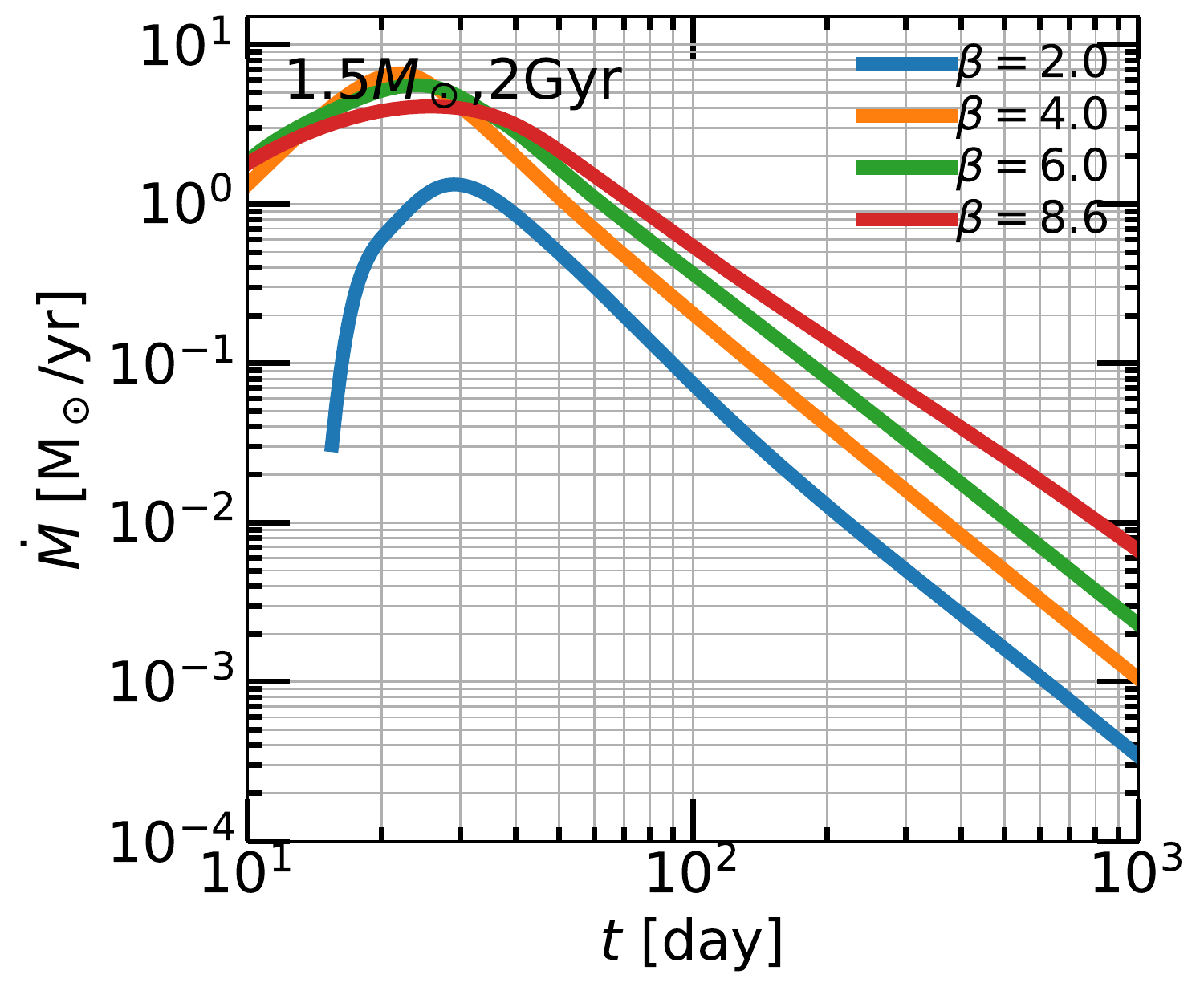}
\plotone{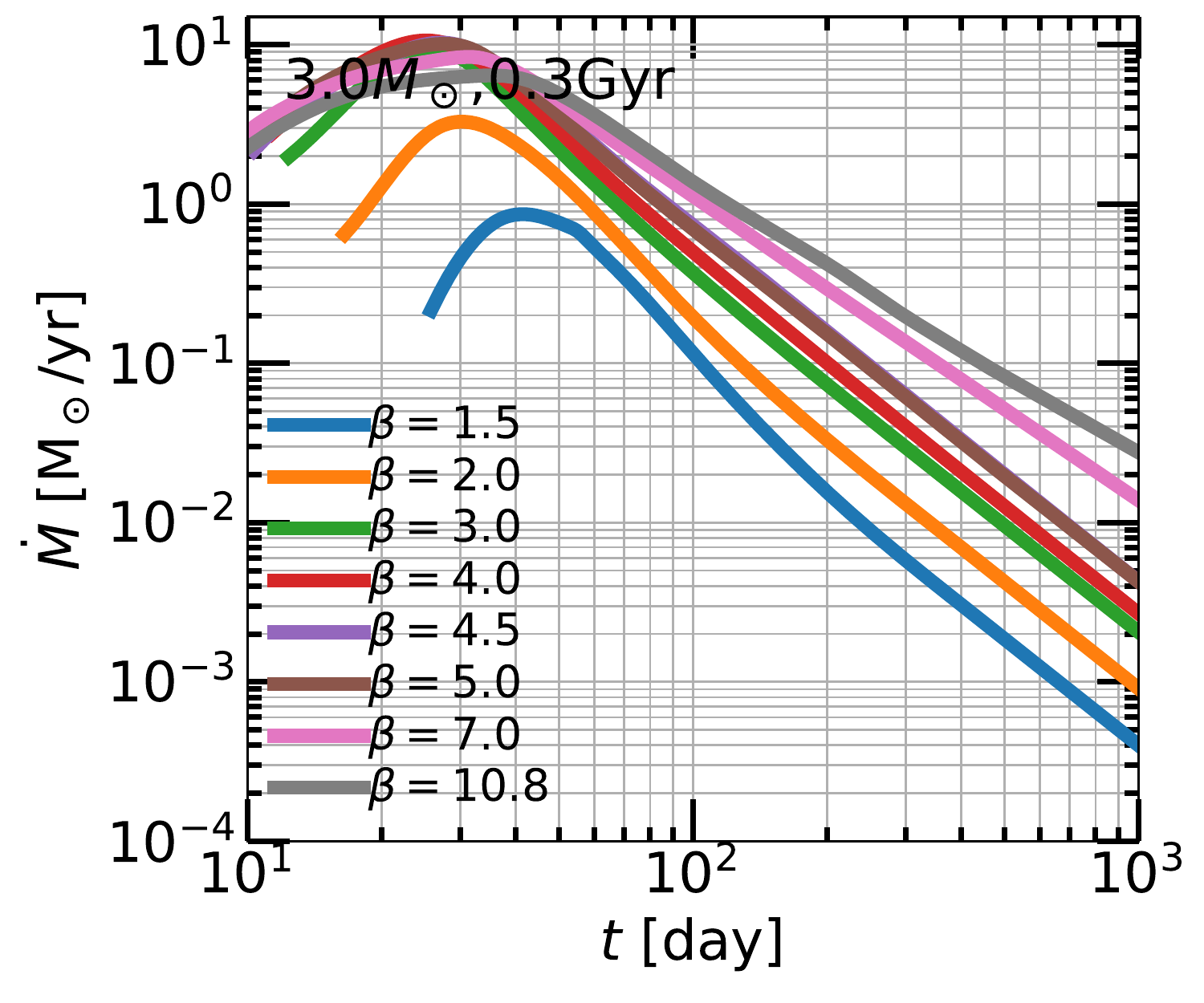} 
\caption{
Mass fallback rate $dM/dt$ to the BH as a function of time for all of our simulations. Each panel is a different star (stellar mass and stellar age, labeled in top left). Colors correspond to different impact parameters $\beta$.
\label{fig:mdots_allgrouped}
}
\end{figure*}

Figure~\ref{fig:mdots_allgrouped} shows mass fallback rates $dM/dt$ for all of our simulations, grouped by star. Each panel is a particular stellar mass and age, and shows all impact parameters $\beta$ for that object. One can perhaps pick out trends in the shape of the $dM/dt$ curves that we will explore in more detail below. Higher $\beta$ generally corresponds to shifting the $dM/dt$ curve upwards and to the left, giving a shorter peak timescale and higher peak fallback rate. After the critical $\beta$ for full disruption, the $dM/dt$ curve flattens, particularly near the peak. The rise and decay slopes also vary with stellar mass, stellar age, and impact parameter.

\begin{figure*}[htp!]
\epsscale{0.55}
\plotone{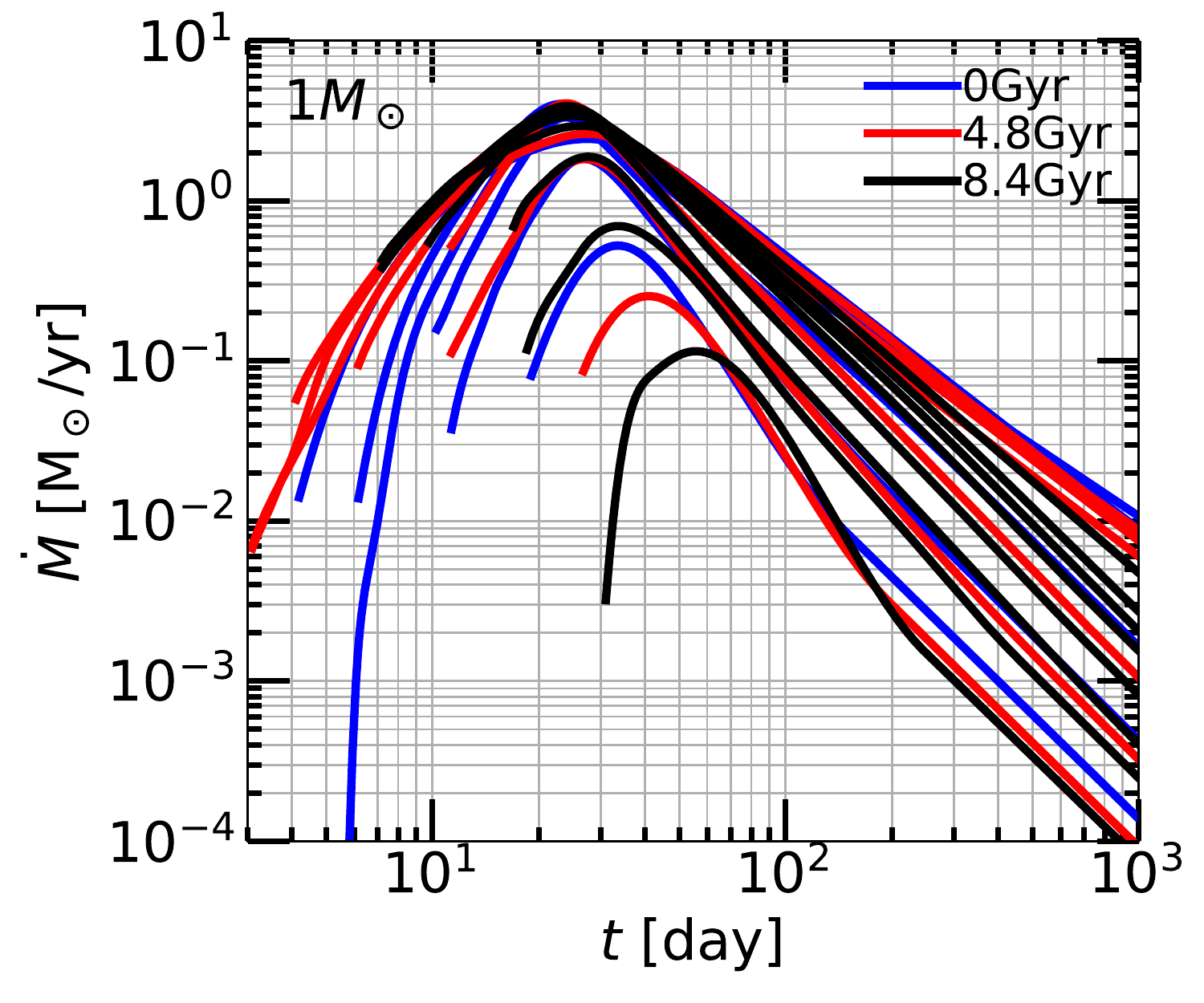}
\plotone{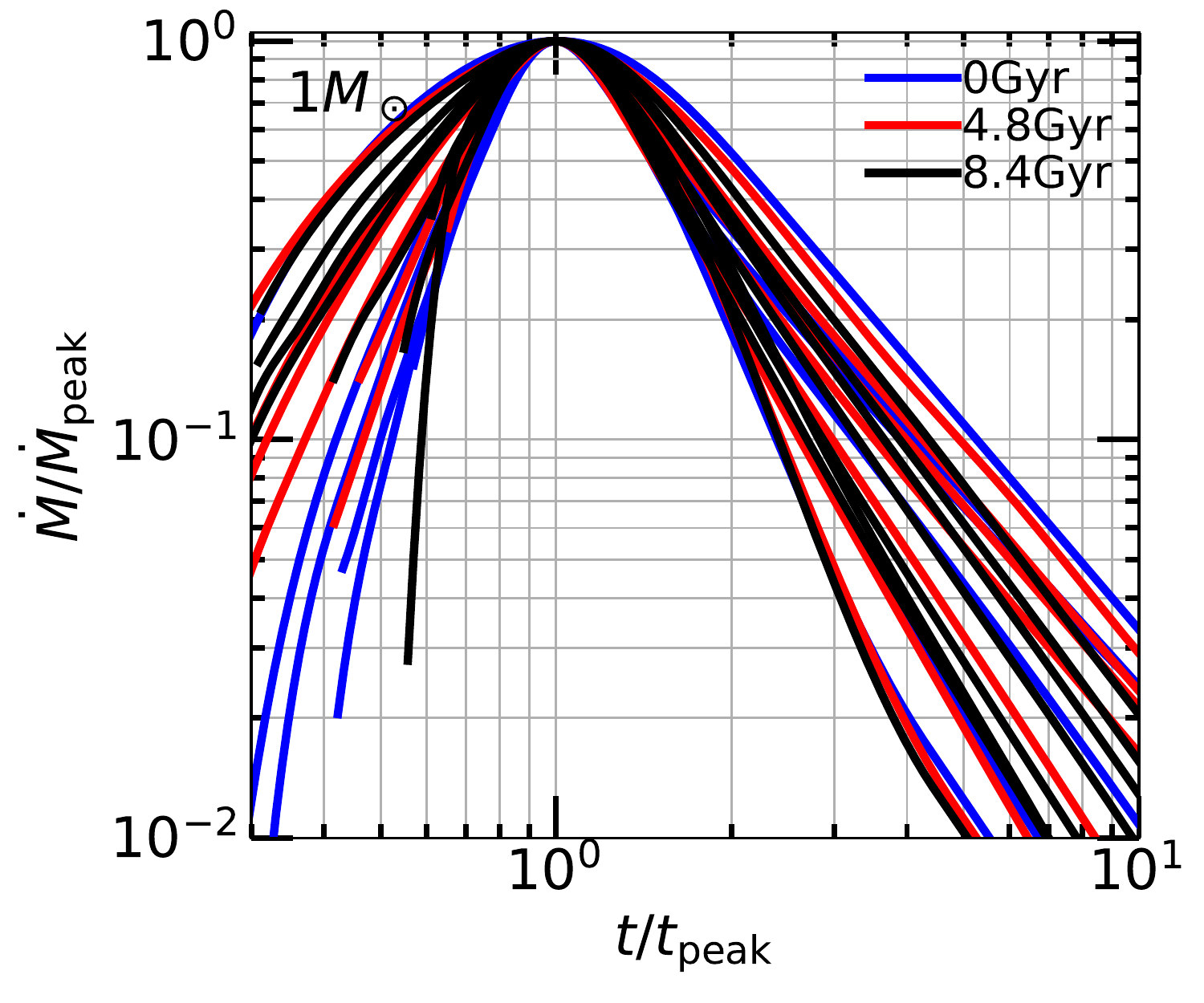}
\plotone{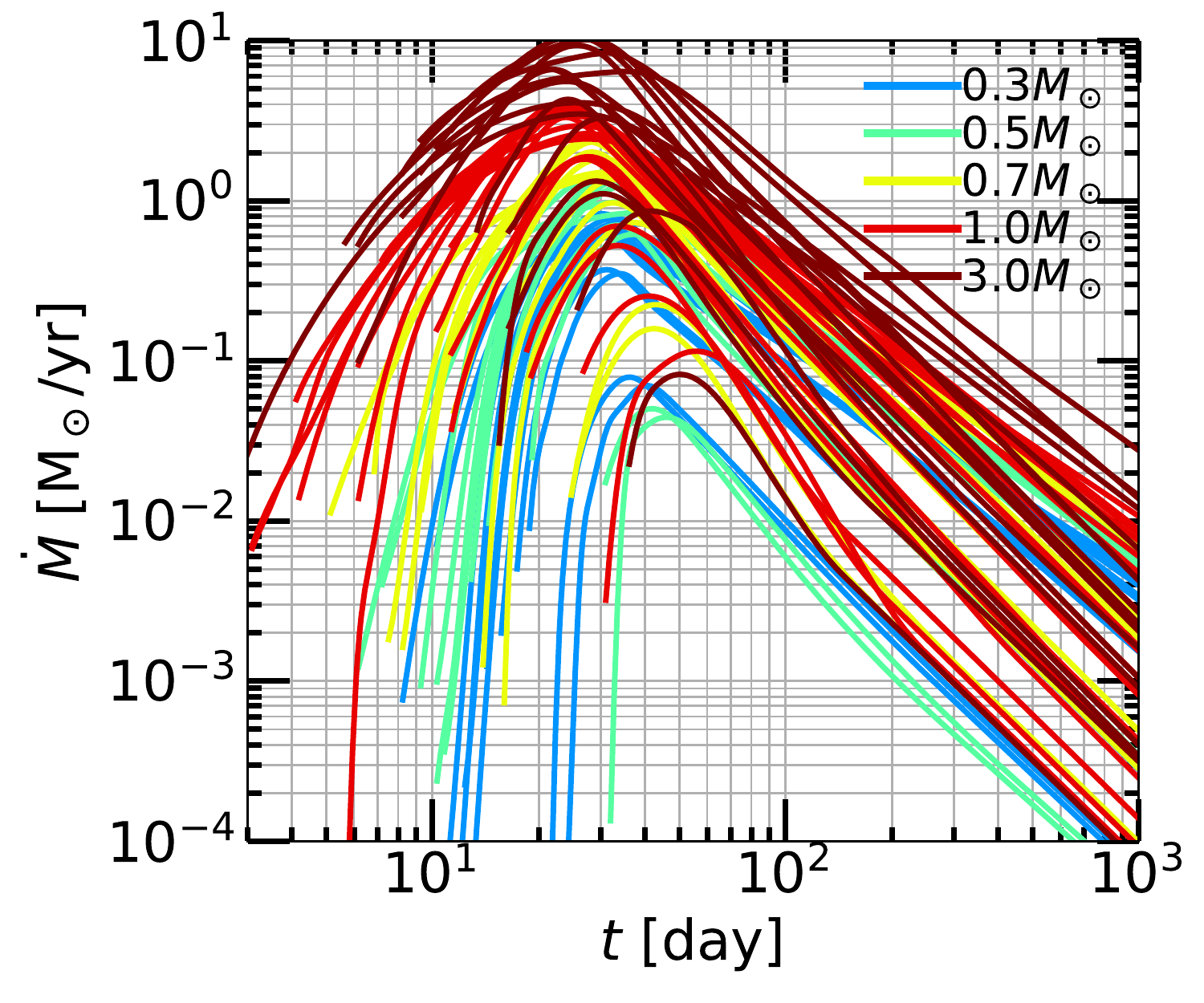}
\plotone{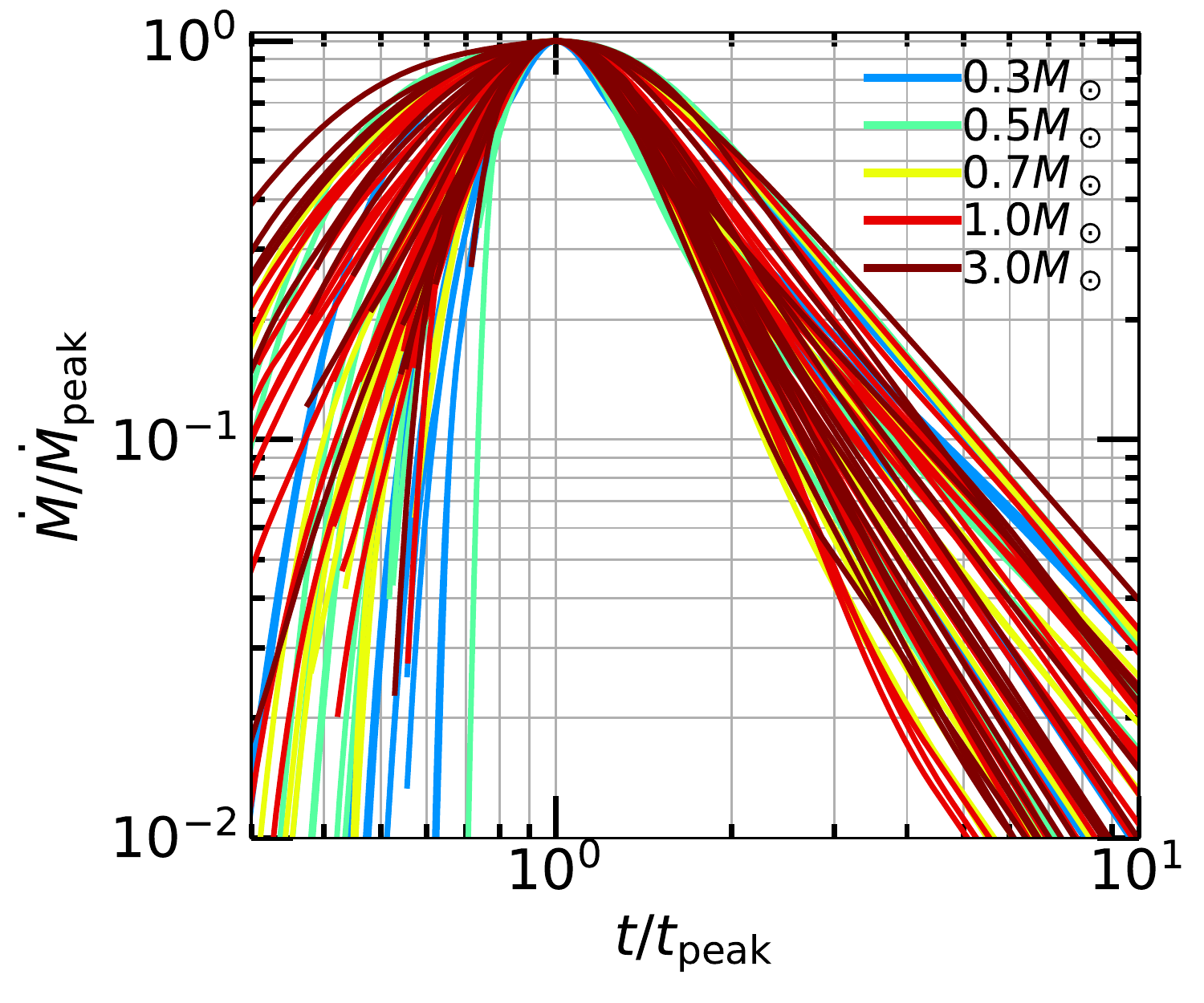}
\plotone{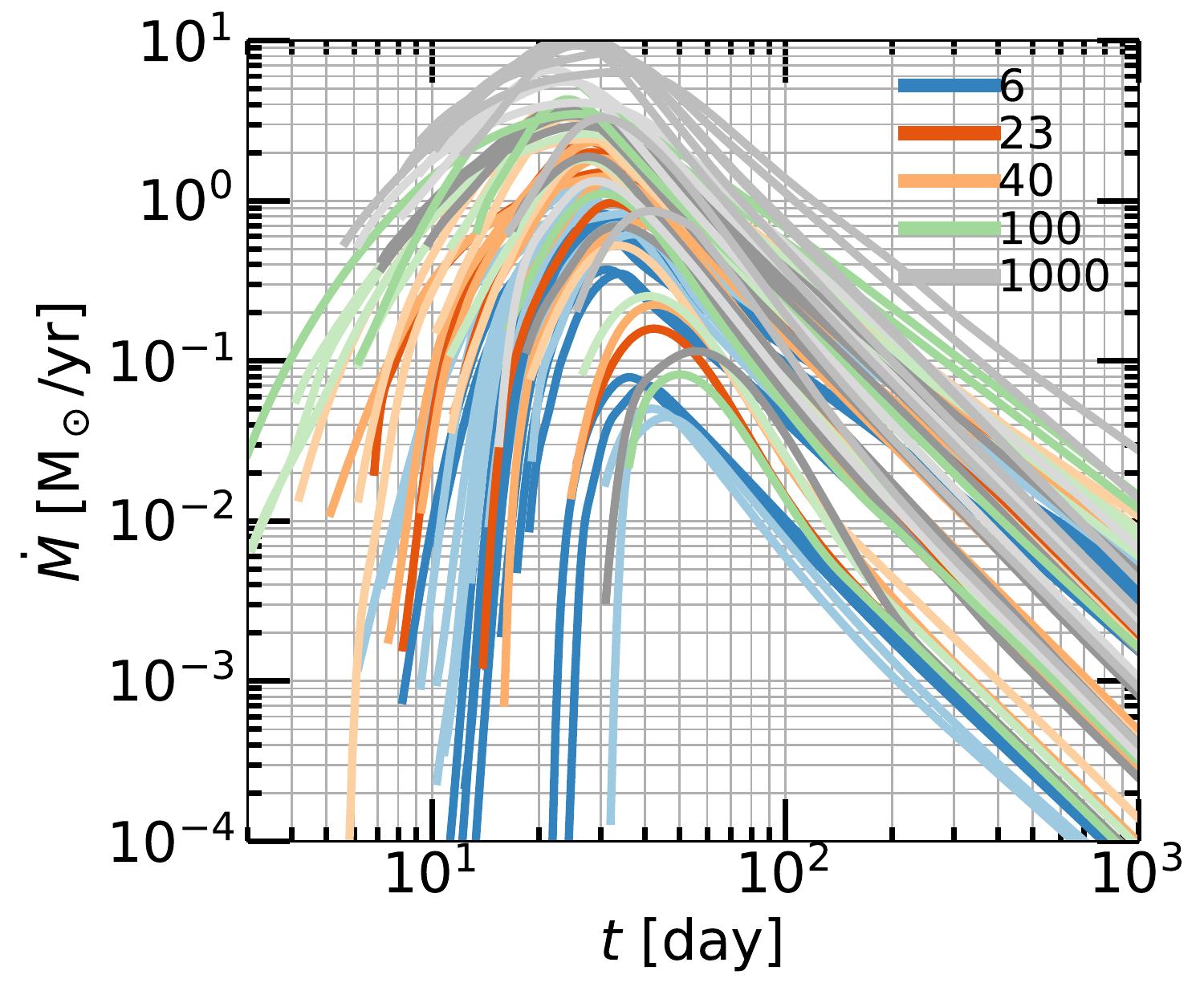}
\plotone{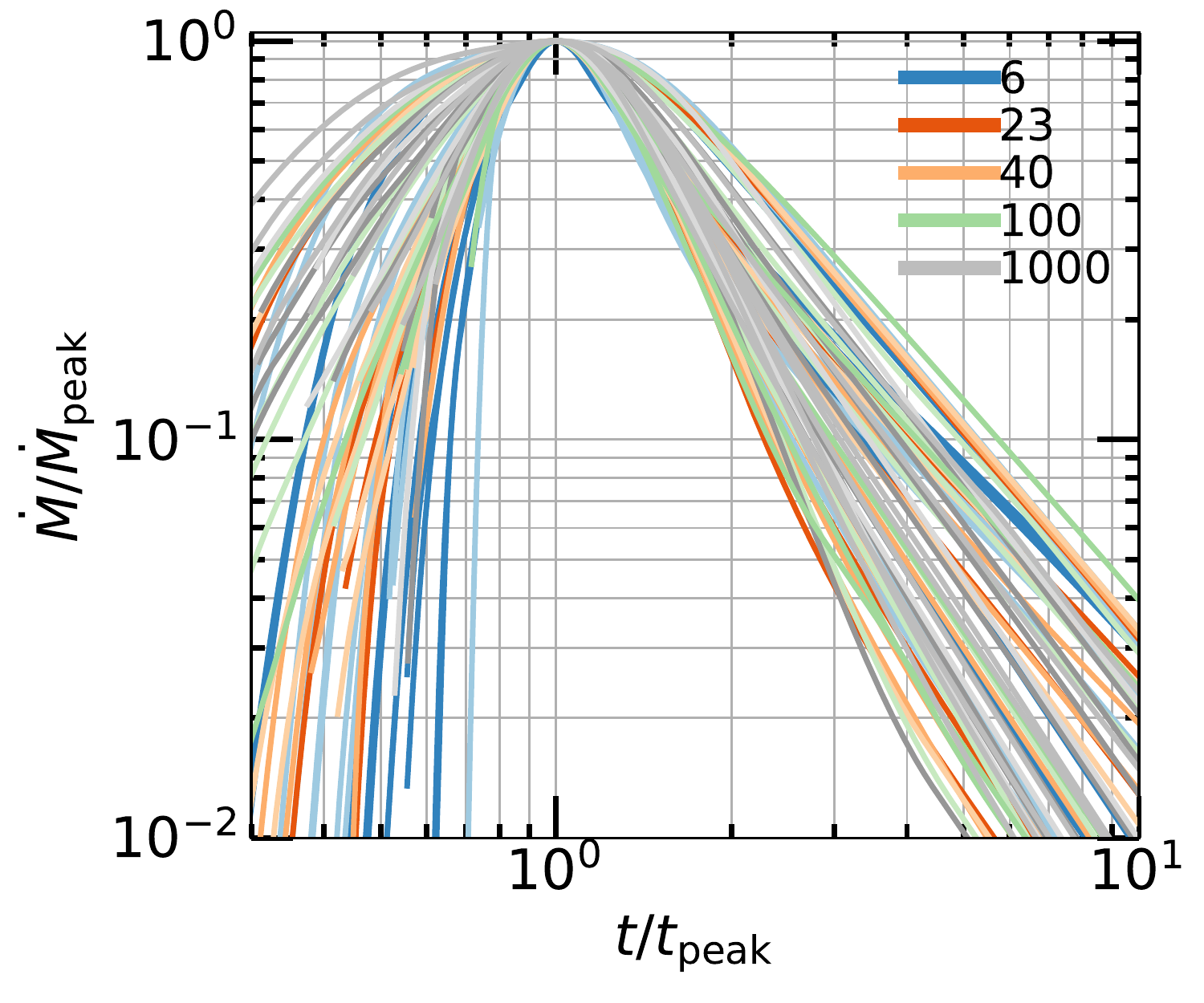}
\caption{
Mass fallback rate to the BH as a function of time for all of our simulations. 
Left panels: raw; right panels: normalized to peak.
Top: only $M_\star=1M_\sun$, all $\beta$'s, 3 stellar ages. \deleted{This demonstrates the degeneracy between stellar structure and $\beta$.}
Middle: color corresponds to $M_\star$.
Bottom: color corresponds to $\rho_c/\bar\rho$ for the star (with the same colors as in Figure~\ref{fig:structure}) and the legend indicates a few reference values.
\label{fig:mdots_allone}
}
\end{figure*}

The top panels of Figure~\ref{fig:mdots_allone} show $dM/dt$'s for a single stellar mass, 1$M_\sun$, for all $\beta$'s and stellar ages, in order to demonstrate the effect of stellar age. 
Note that the $\beta$'s are not the same for each stellar age (see Table~\ref{tab:grid} or Figure~\ref{fig:mdots_allgrouped}).
The left panel shows the raw curves and the right panel is normalized to the peak fallback rate ($\dot M_{\rm peak}$) and the associated peak fallback time ($t_{\rm peak}$) in order to focus on the rise and decay slopes.
In the left panel, the curves largely overlap, demonstrating the degeneracy between stellar structure and $\beta$.
This is similar to the shifting of $dM/dt$'s from $\gamma=5/3$ and $\gamma=4/3$ polytropes to lie on top of each other demonstrated in \citet{2013ApJ...767...25G}.
This is why there is a degeneracy in the \texttt{MOSFiT} TDE fitting between stellar mass and impact parameter (and so \added{radiative} efficiency) \citep{2019ApJ...872..151M, 2020arXiv200712198M}.
In the right panel, one sees that the rise and decay slopes also largely overlap for a single stellar mass at multiple ages. 

Thus, it is more challenging to identify the age of the star by the light curve alone, due to the degeneracy introduced by $\beta$. However, stars of different age have significantly different compositions, and thus the stellar debris will have different chemical abundances. This is a promising avenue to distinguish stellar age, and indeed also stellar mass, if one can tie the composition of the fallback debris to observed spectral features. See \citet{2019ApJ...882L..25L} for a first study of the chemical abundance of the debris.

The middle and bottom panels of Figure~\ref{fig:mdots_allone} show $dM/dt$'s for all of our simulations.
The middle panels are colored by stellar mass, with multiple stellar ages and $\beta$'s for each mass. From the middle left panel, evidently, more massive stars produce $dM/dt$ curves with higher normalizations than less massive stars, but the peak timescales are similar if BH mass is constant. 
This in principle would offer a way to determine stellar mass in an observed TDE, but there is a degeneracy between stellar mass and efficiency in the current fitting of TDE observations \citep[e.g.,][]{2019ApJ...872..151M, 2020arXiv200712198M}. 
One can have more efficient conversion of matter to radiation with a smaller total stellar debris mass, or a less efficient conversion of matter to radiation with a larger total stellar debris mass.

From the middle right panel, one sees that, generally, the rise slope is shallower with increasing stellar mass and (though this is more difficult to see) the decay slope is steeper with increasing stellar mass. The rise slope is a stronger trend, whereas the decay slope appears to mainly be a function of the fact that for more massive stars, a higher fraction of encounters studied in this work are partial disruptions. Indeed, with lower $\beta$ encounters being more likely, it is more likely to have a steeper decay slope for a more massive star. We also tabulate the asymptotic power law indices $n_\infty$ later on in this section.

The bottom panels are colored by $\rho_c/\bar\rho$. In the bottom right panel, one can see that more centrally concentrated stars have shallower rise slopes and (though it is difficult to see because of the density of lines) steeper decay slopes. 
The fact that more centrally concentrated stars have slower rises can be understood by thinking about (1) the density profile of the star and (2) the hydrodynamics of disruption. 
First, stars with higher values of $\rho_c/\bar\rho$ have relatively more extended, lower density envelopes. Just considering the undisturbed spherically-symmetric profile (see Figure~\ref{fig:MESA_profiles_vs_polytropes_norm}), the material that will be most bound to the BH has a slower transition in density, leading to a shallower rise slope. Stars with lower values of $\rho_c/\bar\rho$, on the other hand, have more uniform density profiles, and their outer layers exhibit a sharper transition in density, leading to a steeper rise slope.
Second, there is a hydrodynamical effect that, for a more centrally concentrated star, the outer layers are more vulnerable to tidal disruption and this material is thus further stretched out post-disruption (compare the density contours in Figure~\ref{fig:butterfly}). This results in a greater difference in rise slopes than might be predicted from the undisturbed density profiles alone. 
This can be seen more clearly in Figure~\ref{fig:mdots_full}, which shows only full disruptions and has fewer profiles plotted.
The fact that more centrally concentrated stars have faster decays relates to the survivability of the core. While full disruptions decay closer to $\propto t^{-5/3}$, the presence of a surviving core changes the binding energies of the material bound to the BH \citep{2013ApJ...767...25G}.
This results in a steeper decay slope when there is a surviving core \citep[][discuss this effect in more detail]{2019ApJ...883L..17C}.
\added{See also \citet{2020arXiv200103503R} for further discussion.}
The transition between partial disruptions and full disruptions occurs at higher $\beta$'s with increasing $\rho_c/\bar\rho$; thus, more centrally concentrated stars are more likely to exhibit steeper decays.

The structure of the star is imprinted on the shape and slope of the fallback curve. 
It is thus easier to determine the stellar structure of the disrupted star in an observed TDE than more degenerate properties such as stellar mass and stellar age.
This also hints that in principle, there may be a mapping from a single structural parameter such as $\rho_c/\bar\rho$, in combination with $\beta$, to the shape of the $dM/dt$. One can imagine a mapping in which $M_\star$ gives the normalization of the $dM/dt$ while $\rho_c/\bar\rho$ gives the ``tilt'' corresponding to the rise and decay slopes.

\subsubsection{Full disruptions}

\begin{figure*}[htp!]
\epsscale{0.55}
\plotone{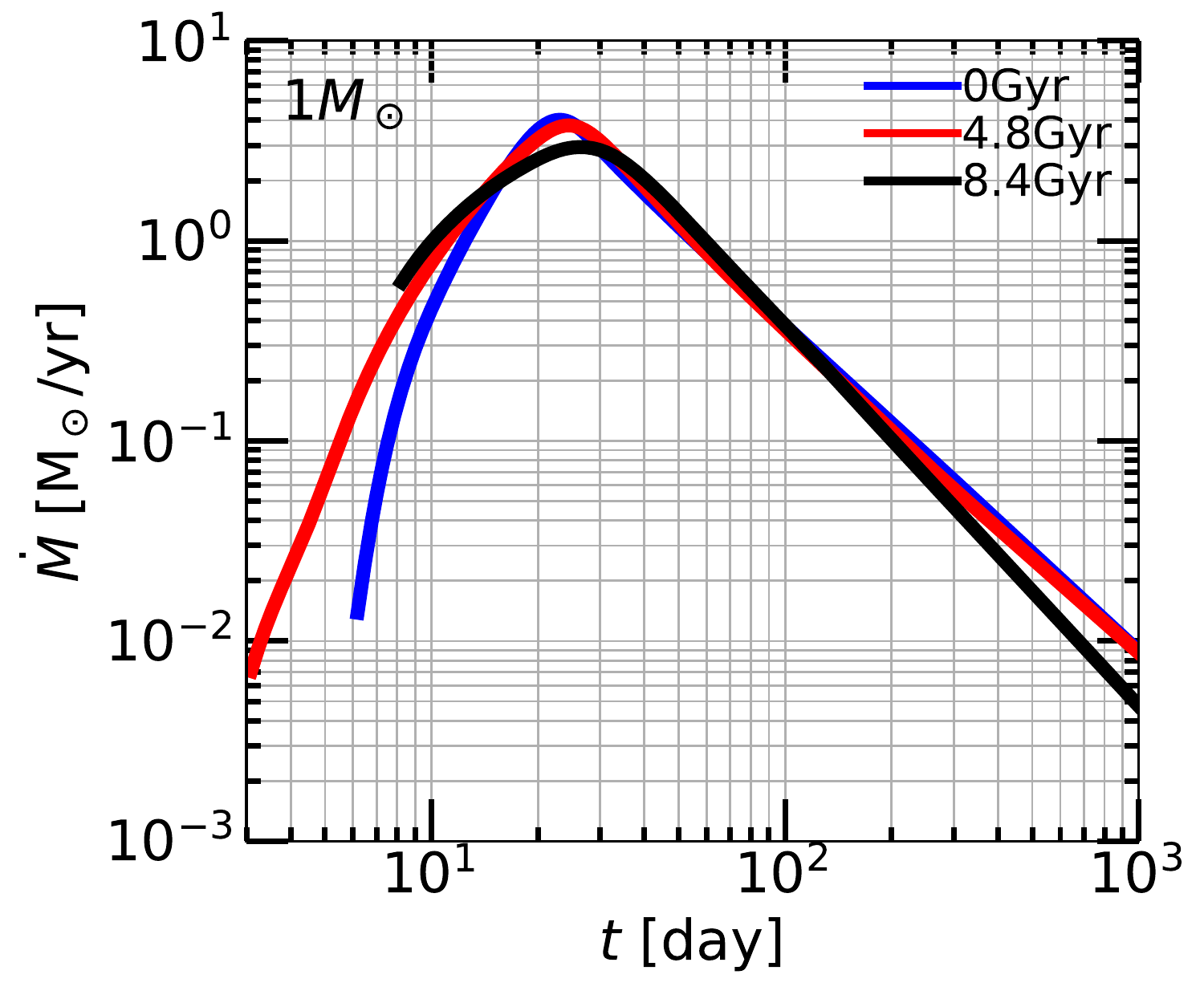}
\plotone{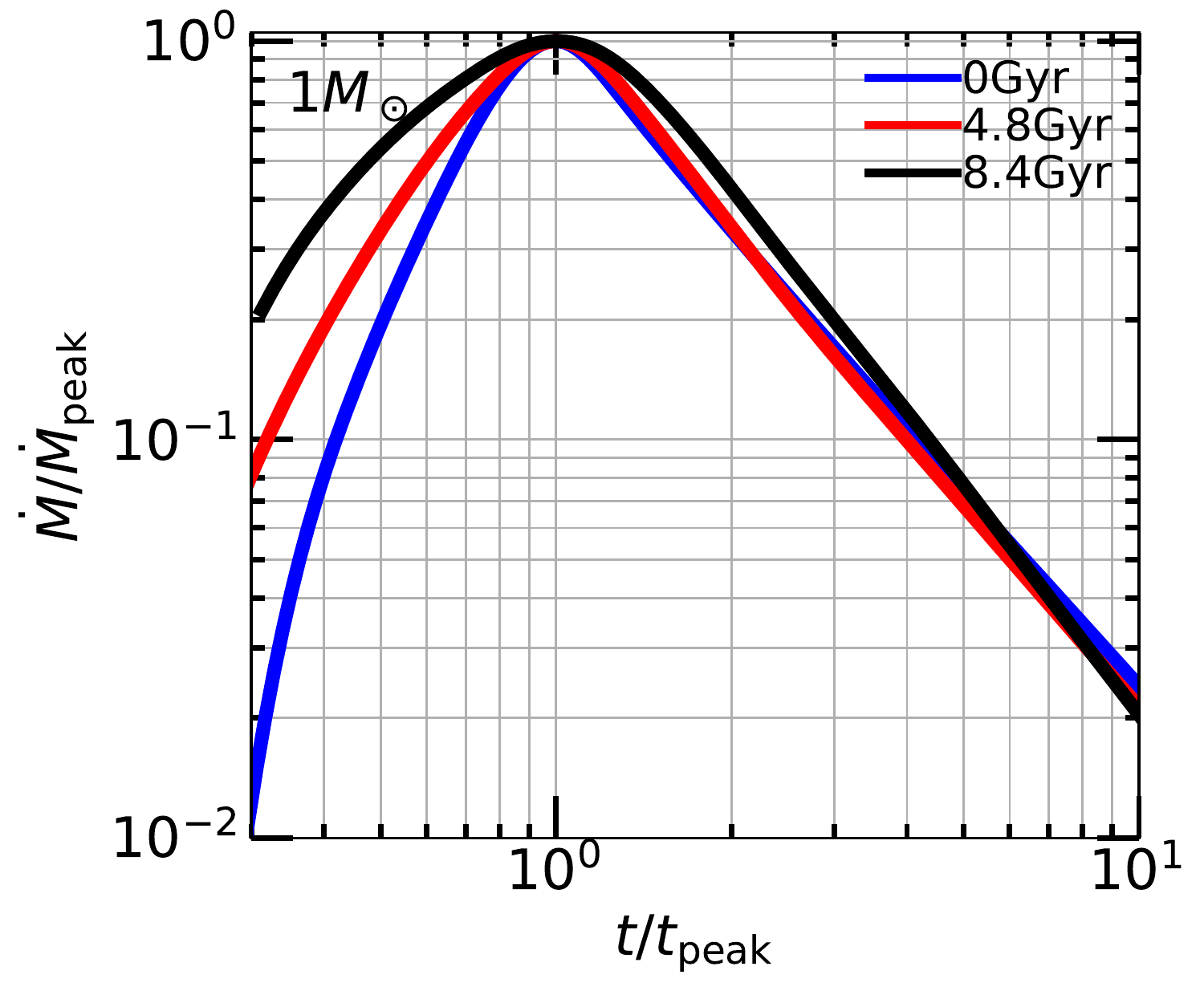}\\
\plotone{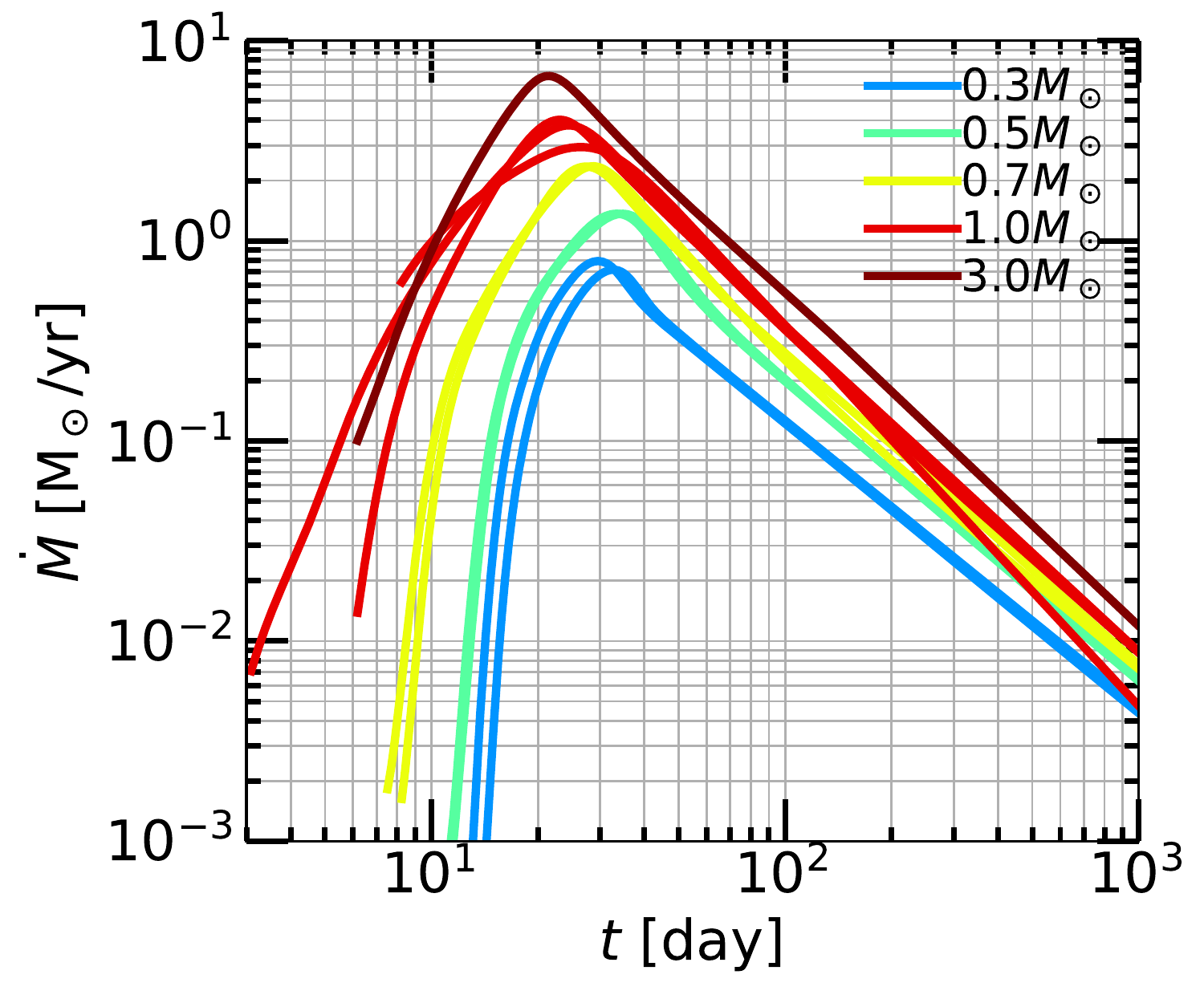}
\plotone{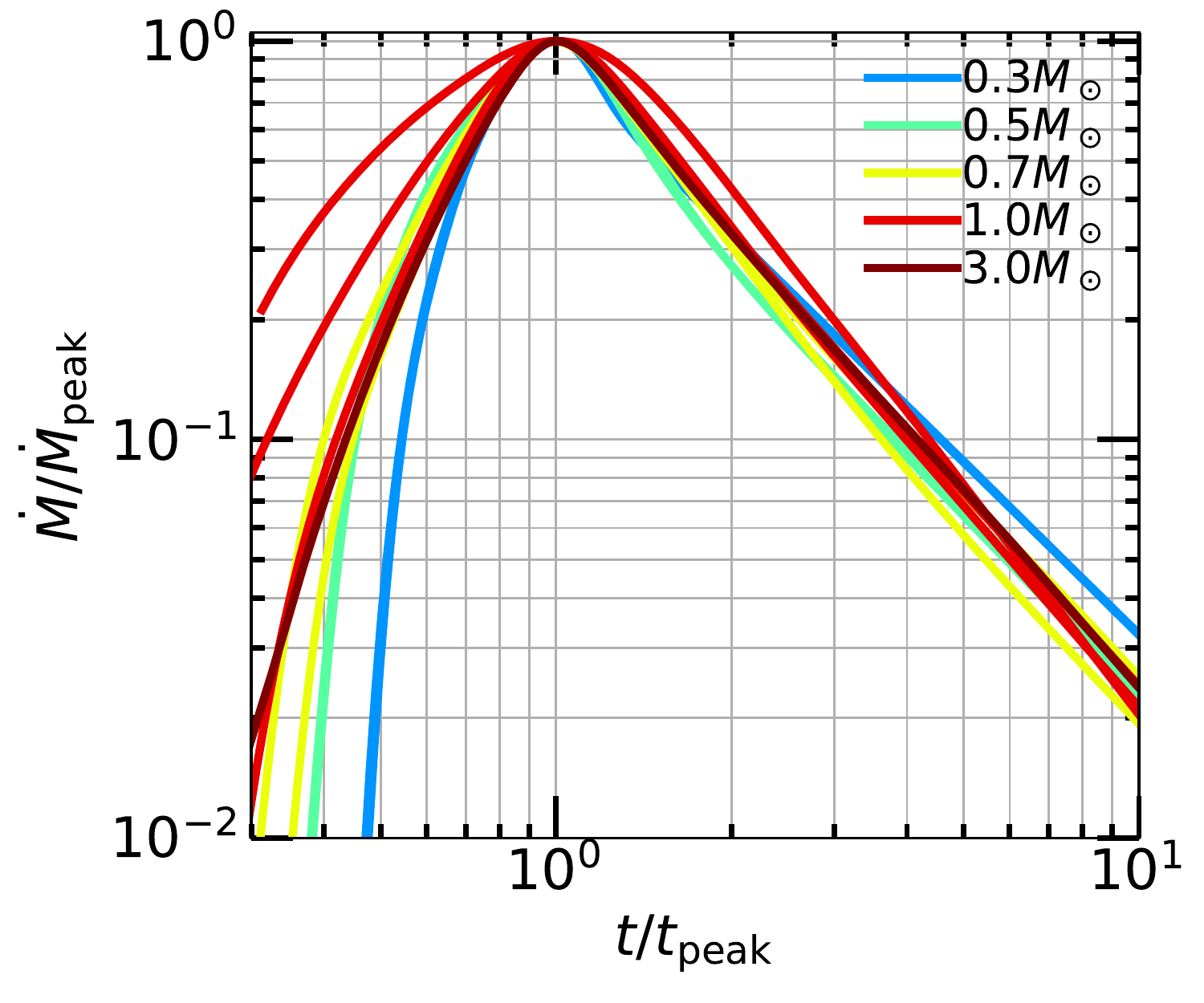}\\
\plotone{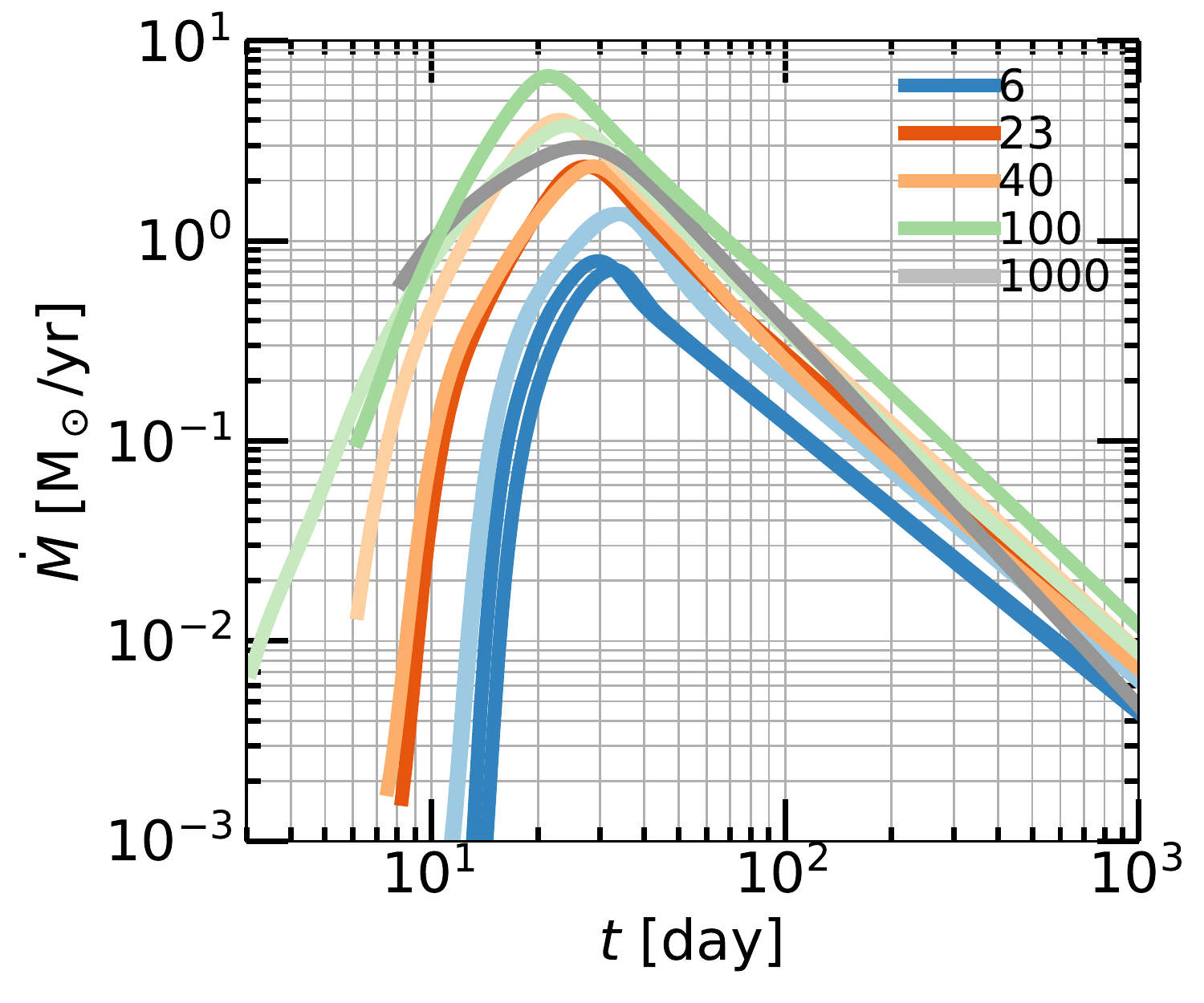}
\plotone{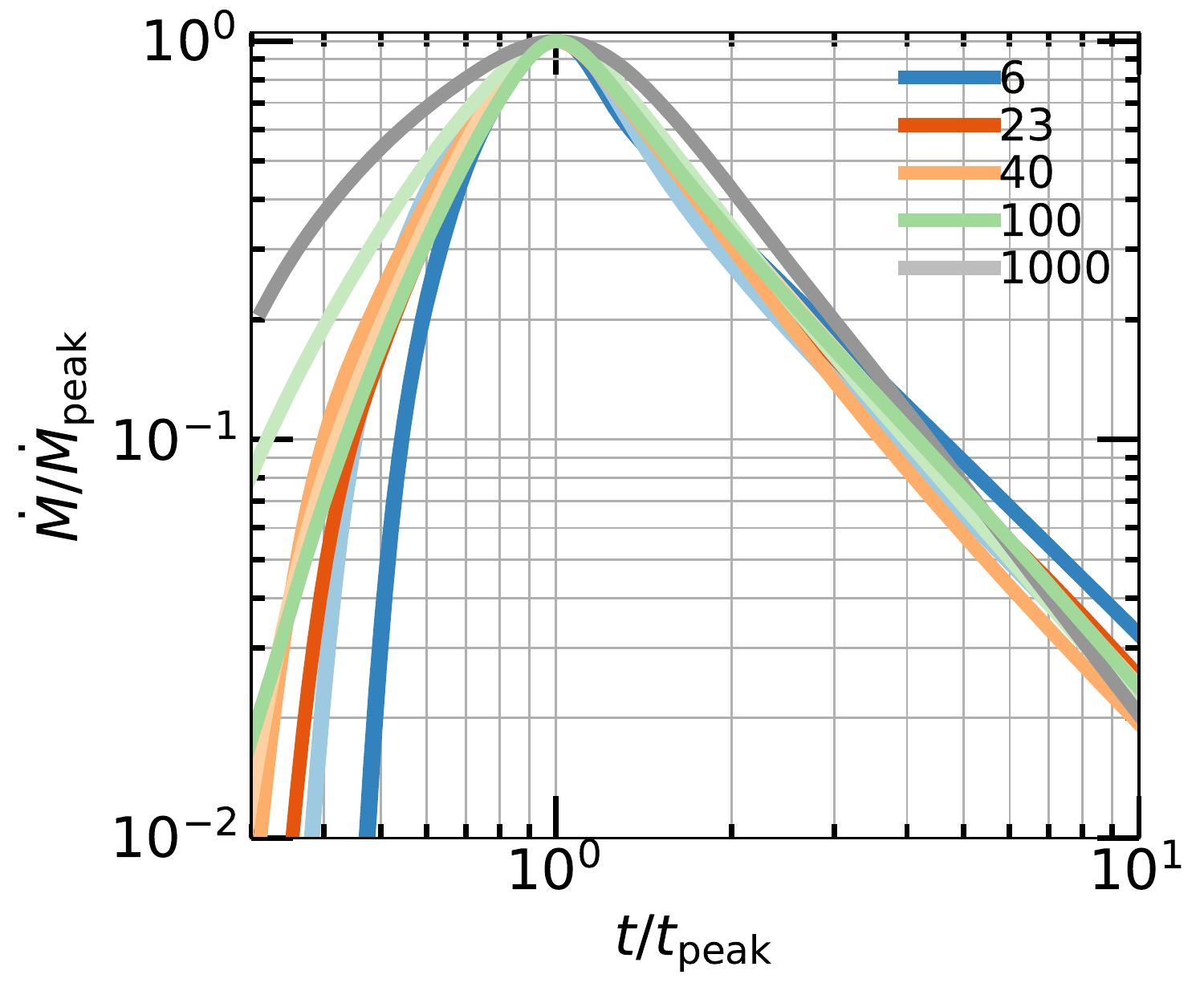}\\
\caption{
Same as Figure~\ref{fig:mdots_allone}, but only critical (full) disruptions.
Top: only $M_\star=1M_\odot$, colored by stellar age.
Middle: colored by stellar mass.
Bottom: colored by $\rho_c/\bar\rho$.
\label{fig:mdots_full}
}
\end{figure*}

Figure~\ref{fig:mdots_full} shows only critical (full) disruptions for all stellar masses and ages. These encounters thus all have the same fractional mass lost $\Delta M/M_\star \approx 1$. Here the trends are more evident than in Figure~\ref{fig:mdots_allone}, though the decay slopes are relatively similar, as all full disruptions give $n_\infty \approx -5/3$ (see below in the section). 
The top panels show a single stellar mass $M_\star=1M_\sun$ in order to highlight the stellar age dependence. Here the actual amount of mass unbound from the star $\Delta M$ and thus the mass eventually fed to the BH, $\approx \Delta M/2$, is identical between the stars. 
The rise slopes are shallower with increasing stellar age.

The middle panels are colored by stellar mass. 
In the left panel one primarily sees the normalization to $dM/dt$ given by $M_\star$.
With increasing $M_\star$, the $dM/dt$ curve is shifted upwards and slightly to to the left.
In the right panel, one sees that the rise slopes are shallower with increasing stellar mass.
There is a weak trend towards steeper decay slopes with increasing stellar mass.

The bottom panels are colored by $\rho_c/\bar\rho$. Here one sees perhaps most clearly the trends discussed above, as we are coloring by the key physical parameter. 
It appears that, for the same $\Delta M/M_\star$, $\rho_c/\bar\rho$ nearly uniquely determines the rise slope of $dM/dt$ (this is also true for the decay slope, but is more clear when we study $n_\infty$ vs. $\beta$ below). 
Increasing central concentration leads to shallower rise slopes, and, for full disruptions, similar decay slopes. 
See discussion above for the physical intuition behind this effect.

\subsubsection{At fixed mass lost}

\begin{figure*}[htp!]
\epsscale{0.38}
\plotone{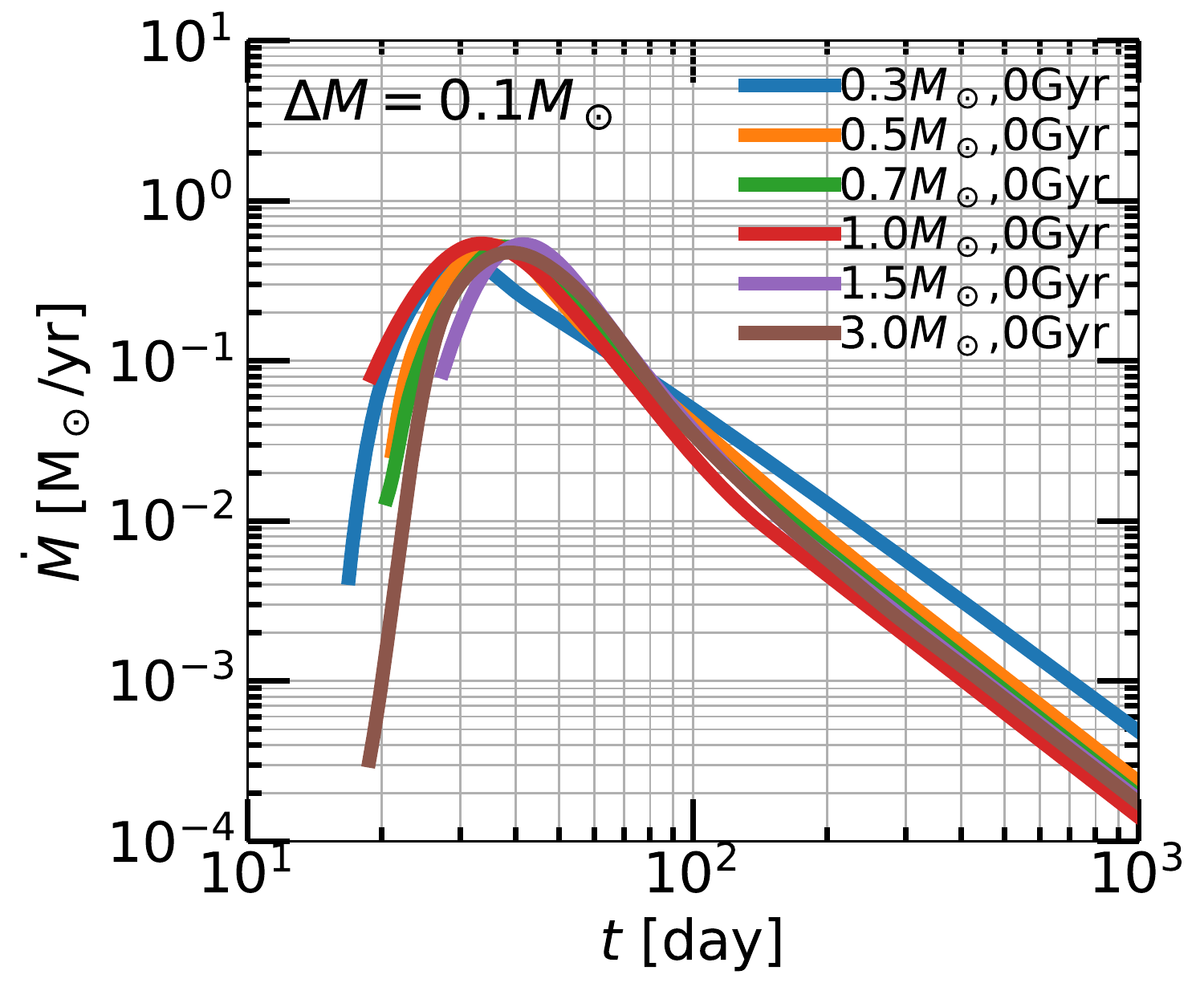}
\plotone{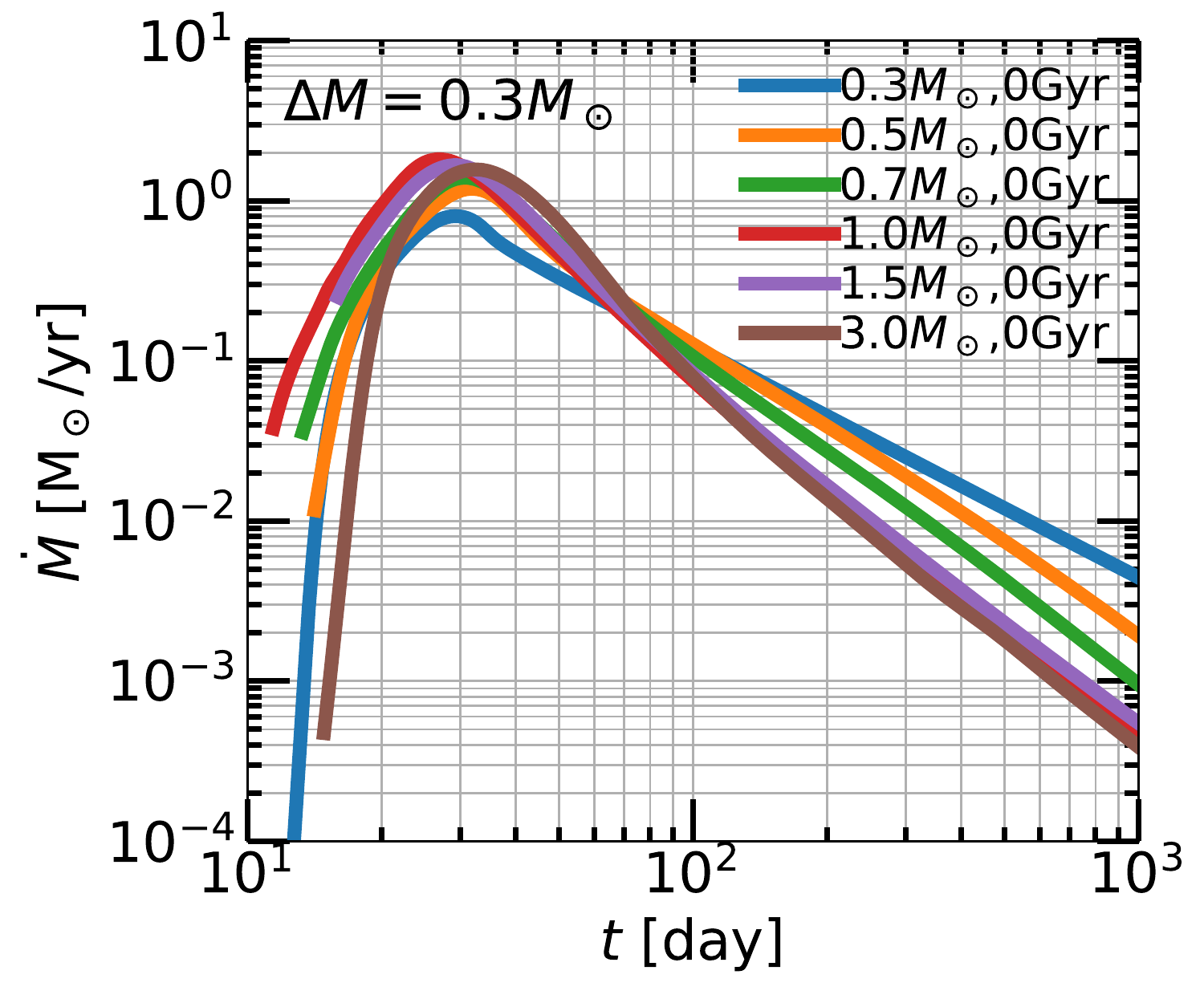}
\plotone{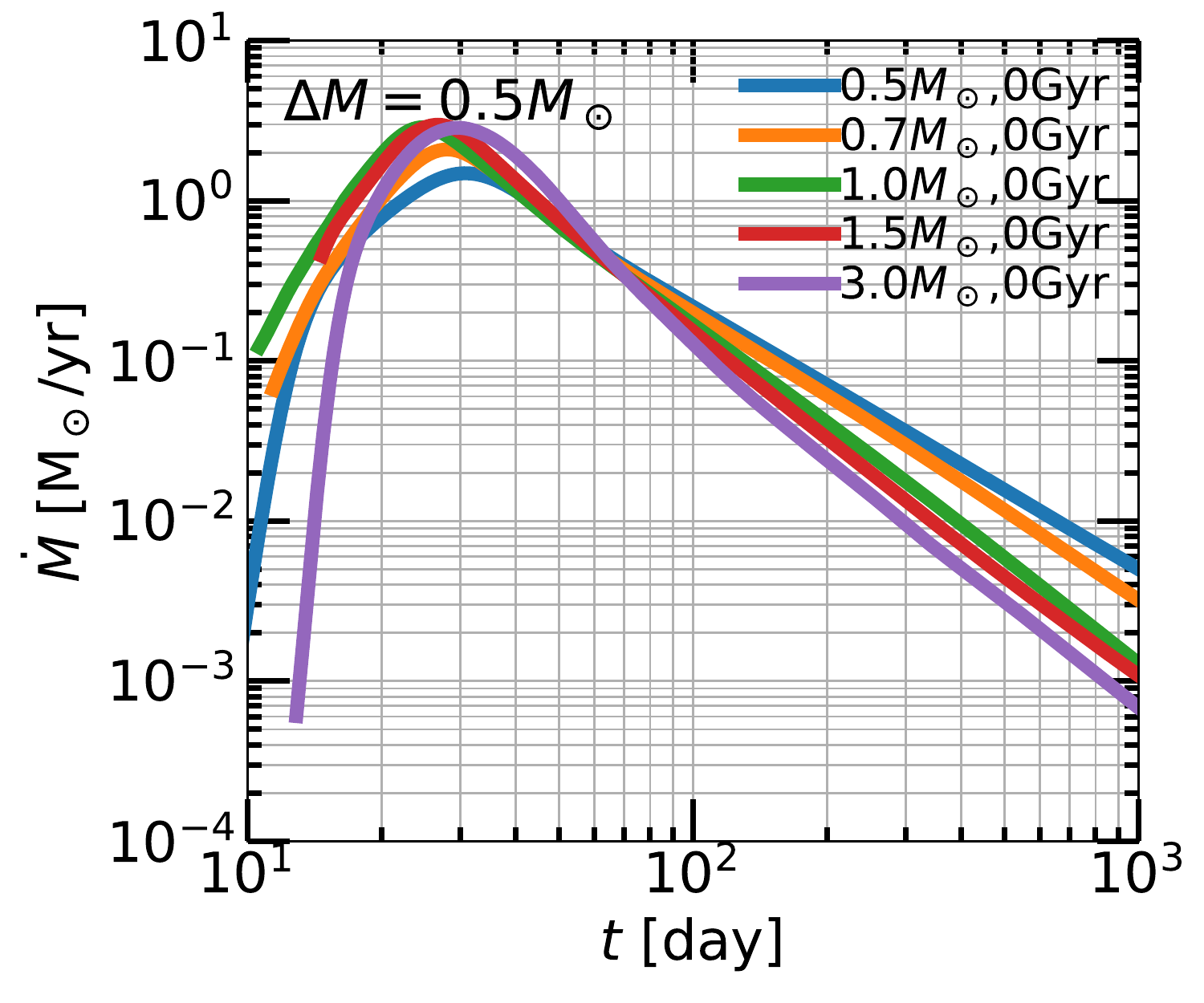}
\plotone{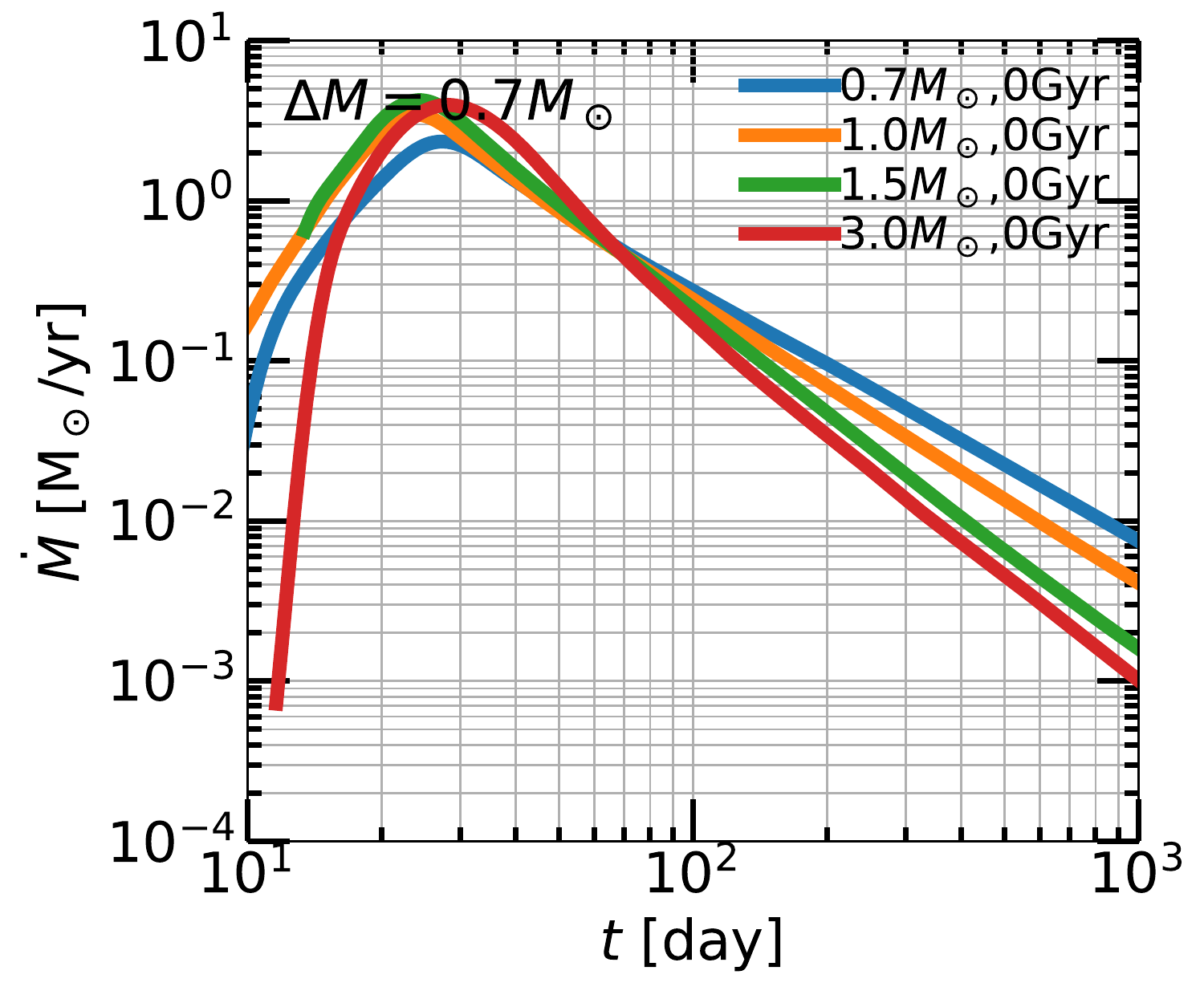}
\plotone{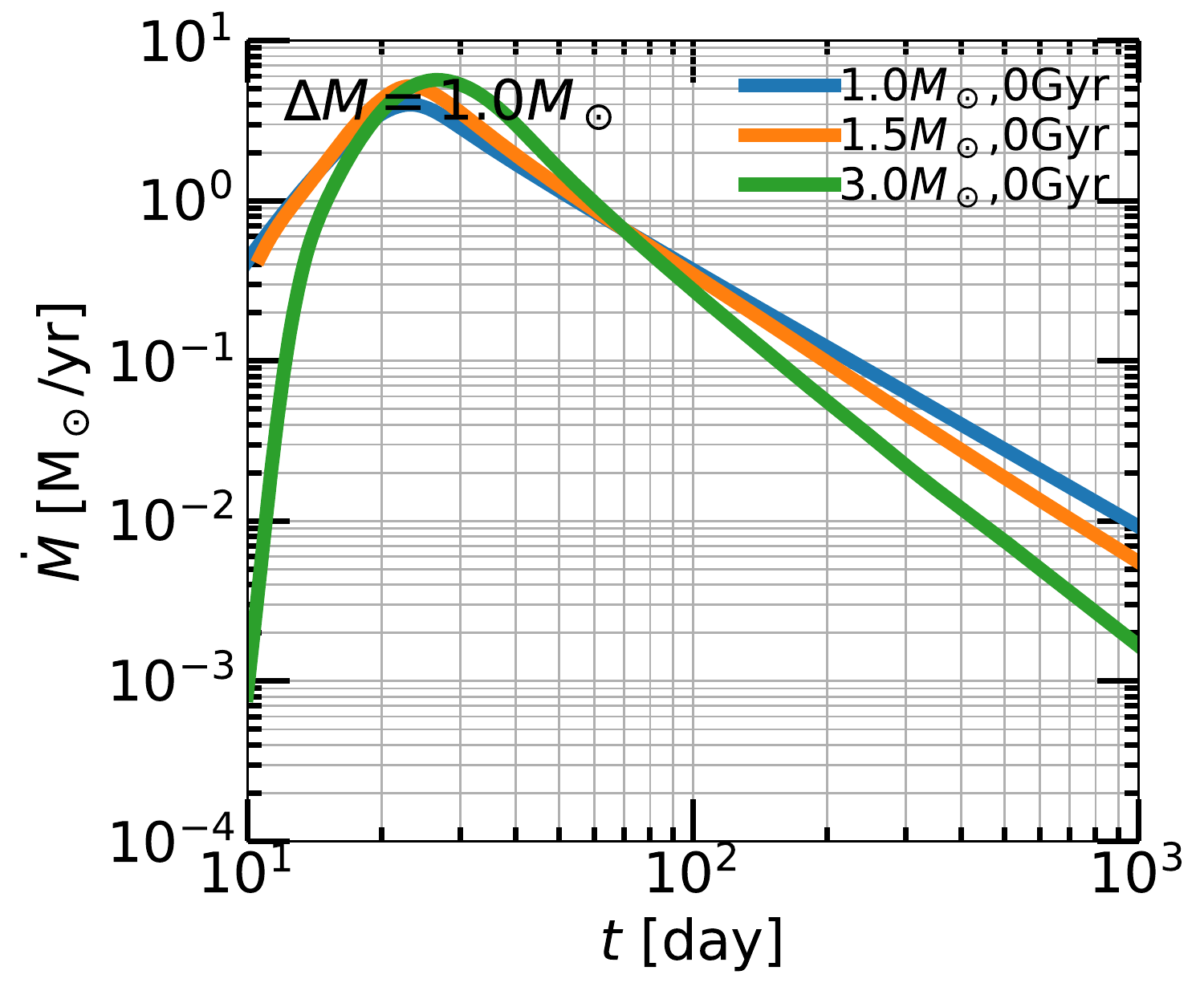}\\
\plotone{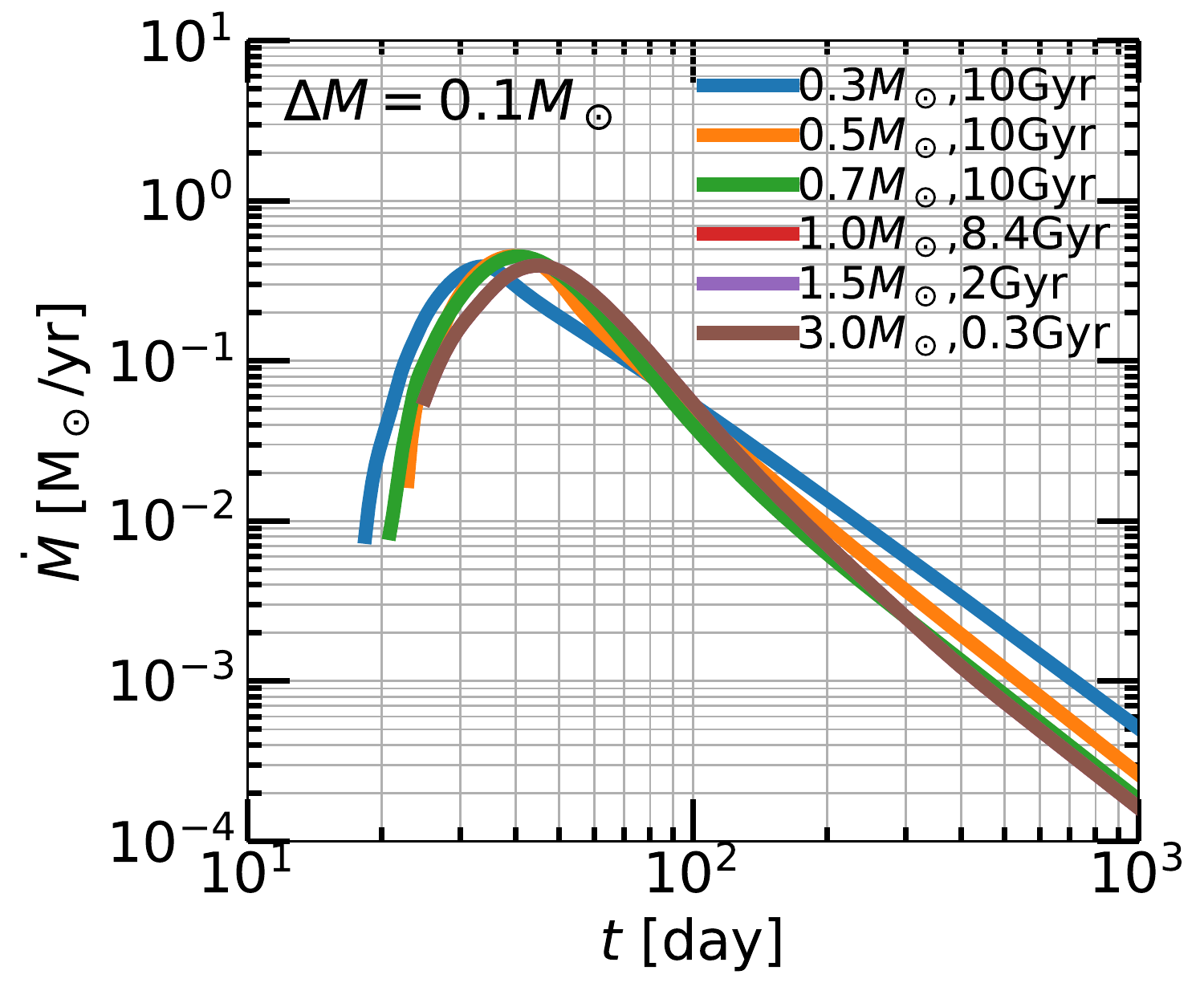}
\plotone{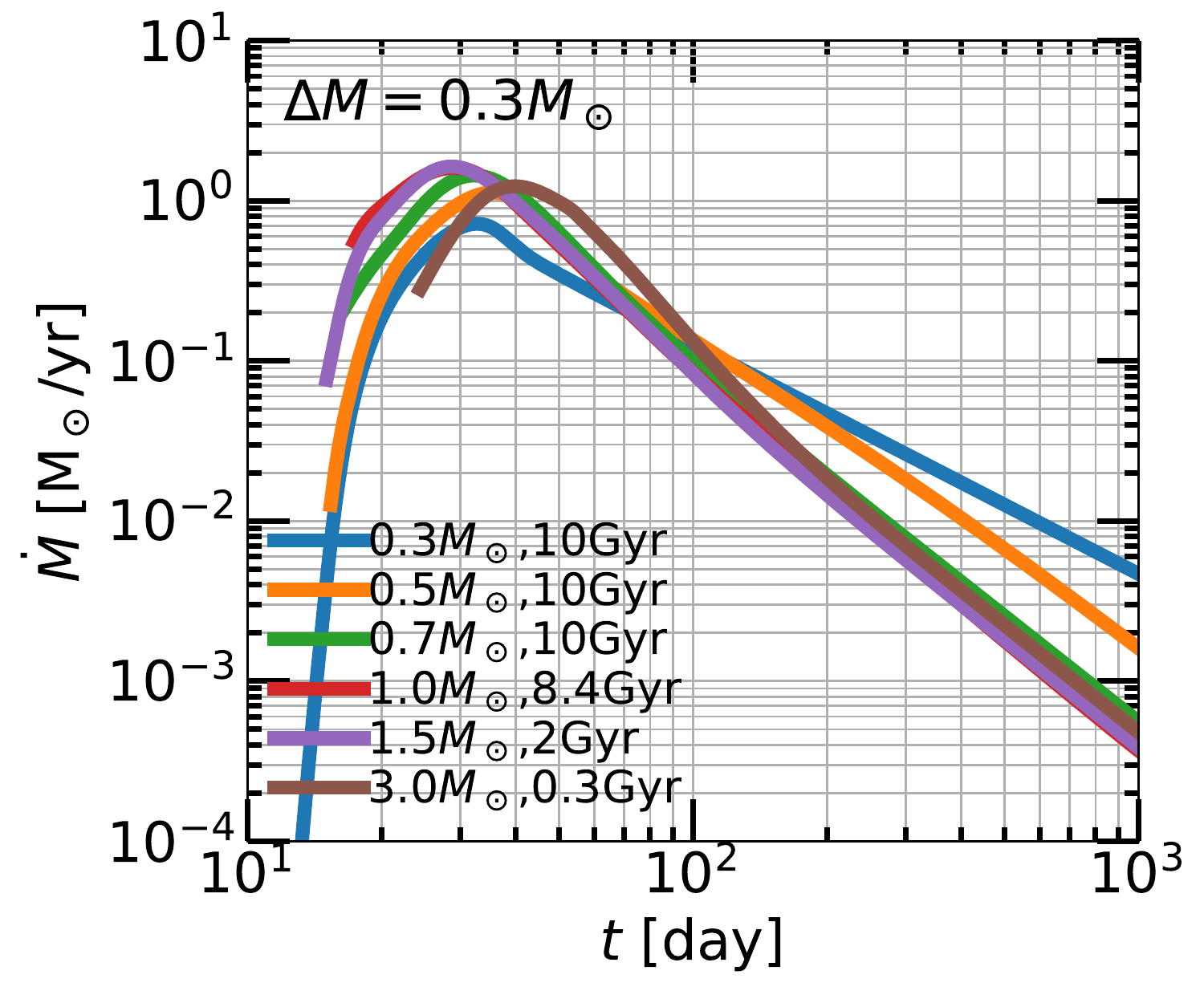}
\plotone{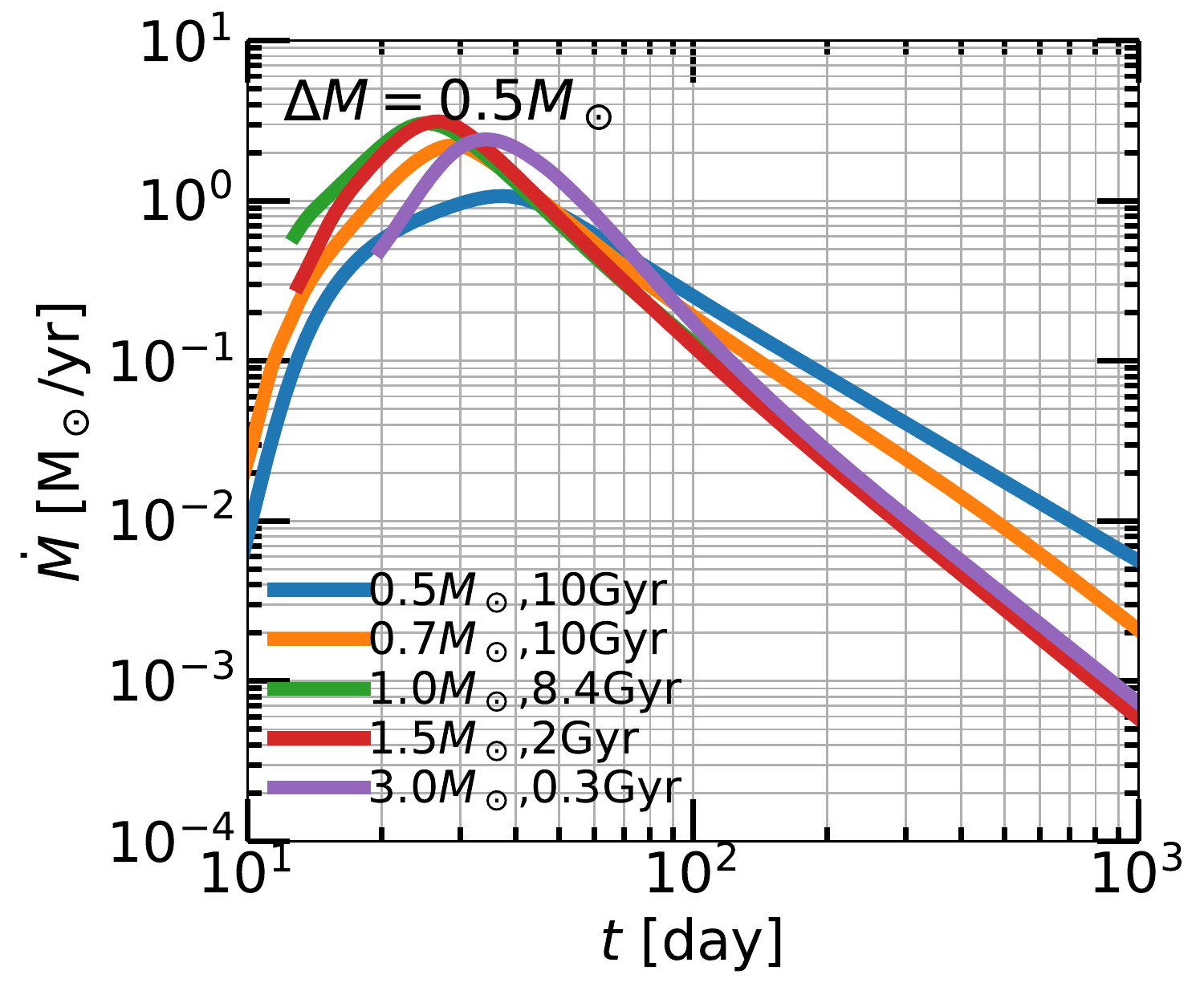}
\plotone{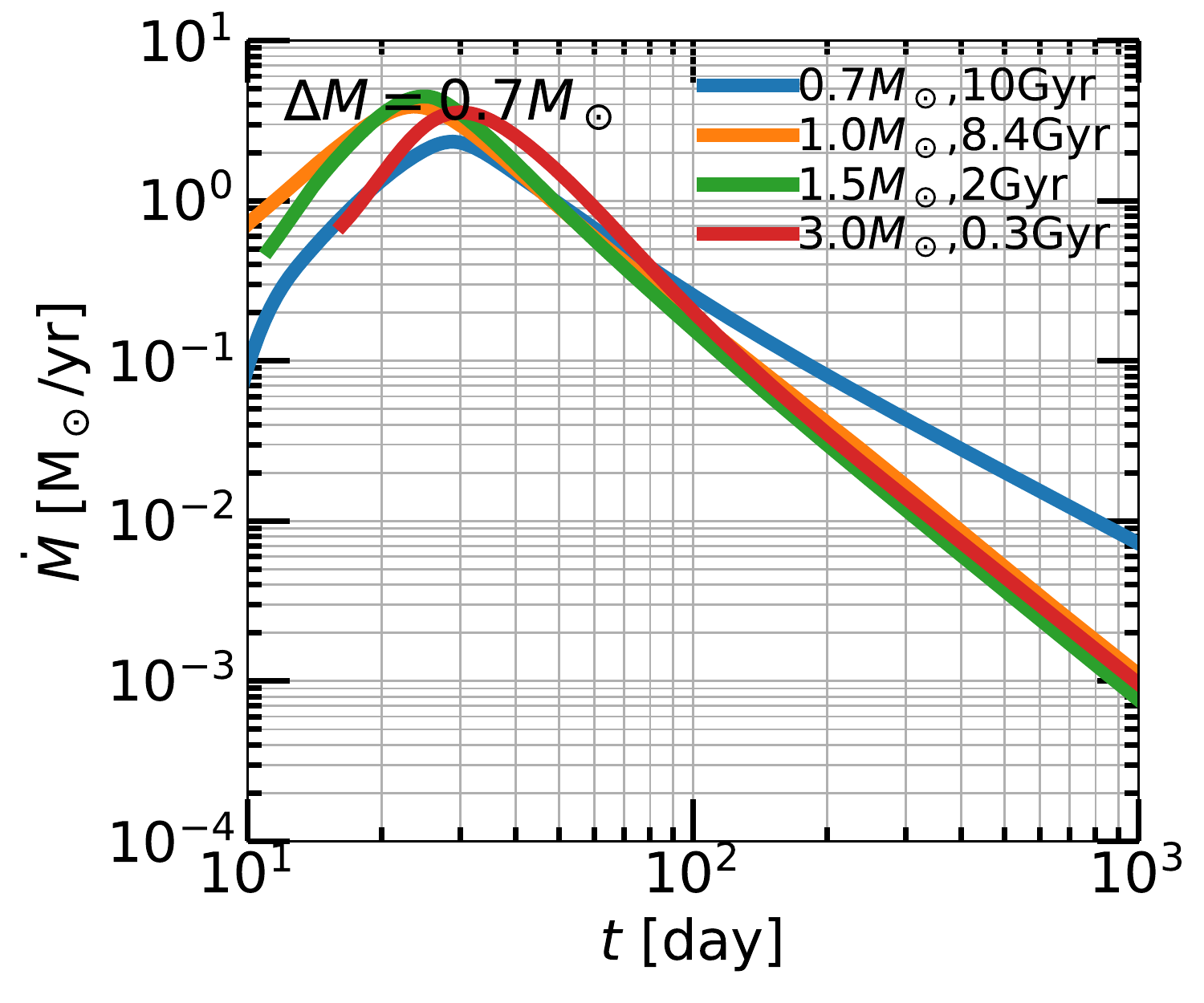}
\plotone{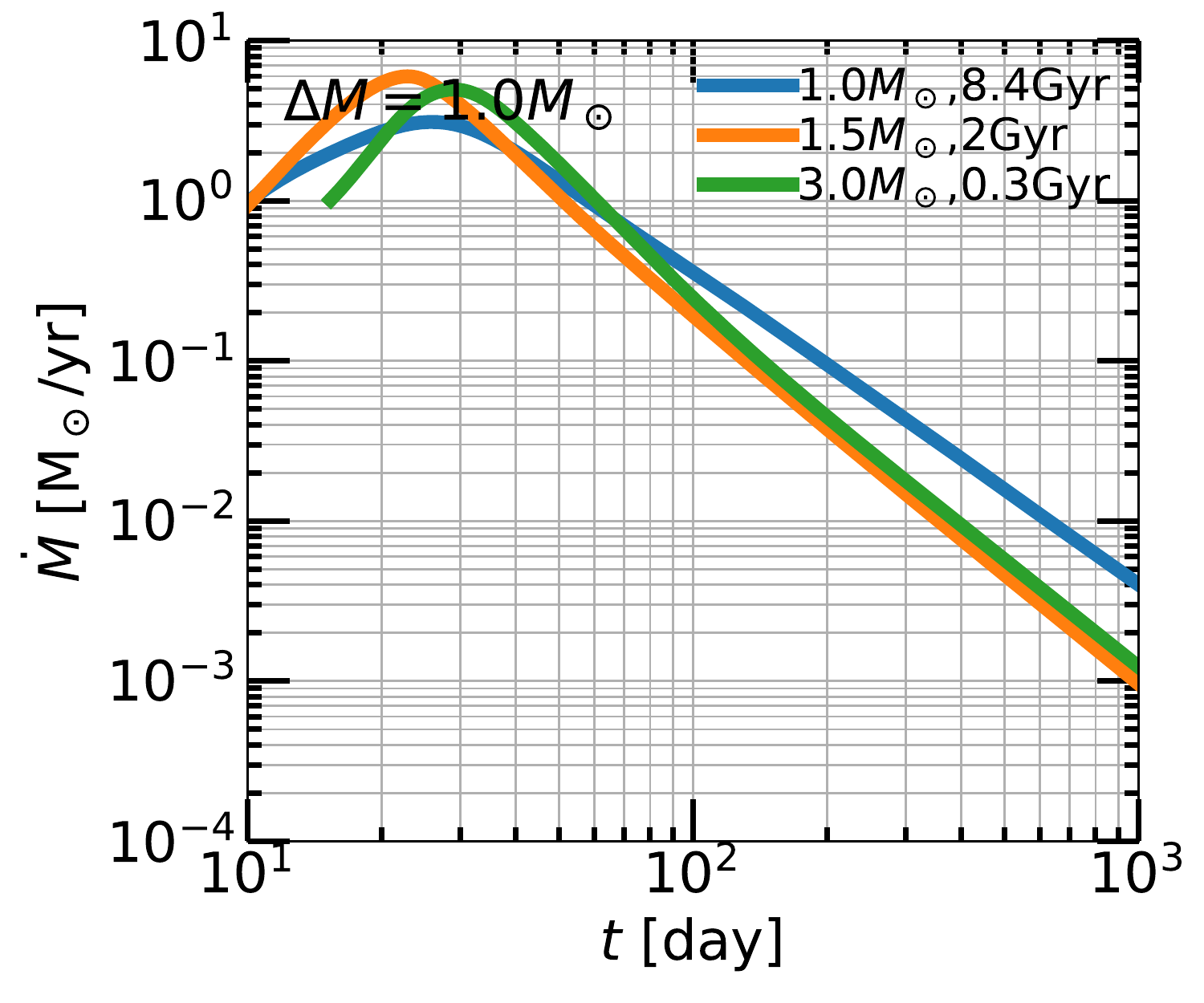}
\caption{
Comparison of $dM/dt$ curves at a fixed mass lost $\Delta M$. Top 5 panels are stellar ages of 0 Gyr and bottom 5 panels are stellar ages of 10 Gyr (or TAMS for $M_\star \geq M_\sun$). 
These are interpolated $dM/dt$ curves at the particular $\beta$ for each star that corresponds to the $\Delta M$ in the top left of each panel.
}
\label{fig:mdots_fixed_deltam}
\end{figure*}

Figure~\ref{fig:mdots_fixed_deltam} shows a comparison of $dM/dt$ curves at fixed mass lost $\Delta M$. 
This allows one to compare fallback rates for the same amount of total material supplied to the BH.
We compare at $\Delta M = (0.1, 0.3, 0.5, 0.7, 1.0)M_\sun$; half of this unbound material returns to the BH. Of course, some values of $\Delta M$ are inaccessible to some stellar masses.
To do this, we construct a very finely interpolated grid (see Appendix) of impact parameters for each star, and select the $\beta$ for which the $dM/dt$ integrates to the given $\Delta M$.

The top five panels show stars at 0 Gyr (ZAMS) and the bottom five panels show stars at 10 Gyr (or TAMS for $M_\star \geq M_\sun$). 
If the efficiency of conversion of matter to light is relatively constant, then the total energy radiated in a TDE is determined by $\Delta M$. 
Thus, this figure shows $dM/dt$'s from a stellar population at a single age, and (assuming constant efficiency) for a given radiated energy.

At fixed $\Delta M$, the shape of the $dM/dt$'s vary significantly with stellar mass. 
The peak fallback rate, the peak timescale, and the rise and decay slopes all vary significantly.
Most strikingly, for both stellar age populations, the decay slopes are steeper with increasing stellar mass. 
This is because for the same $\Delta M$, the encounter is more of a partial disruption for higher stellar masses.
Higher stellar masses also correspond to higher $\dot M_{\rm peak}$ values (by up to a factor of $\approx$5) at fixed $\Delta M$.
The differences in the shape of the $dM/dt$'s are not particularly more or less significant for different values of $\Delta M$.
These differences suggest that fitting TDEs with this library of fallback rates will allow for better determination of stellar properties, and perhaps that the nature of the disrupted star can be determined from the light curve alone.

\subsection{Derived quantities}

\subsubsection{Peak timescale and fallback rate}

\begin{figure*}[htp!]
\epsscale{0.55}
\plotone{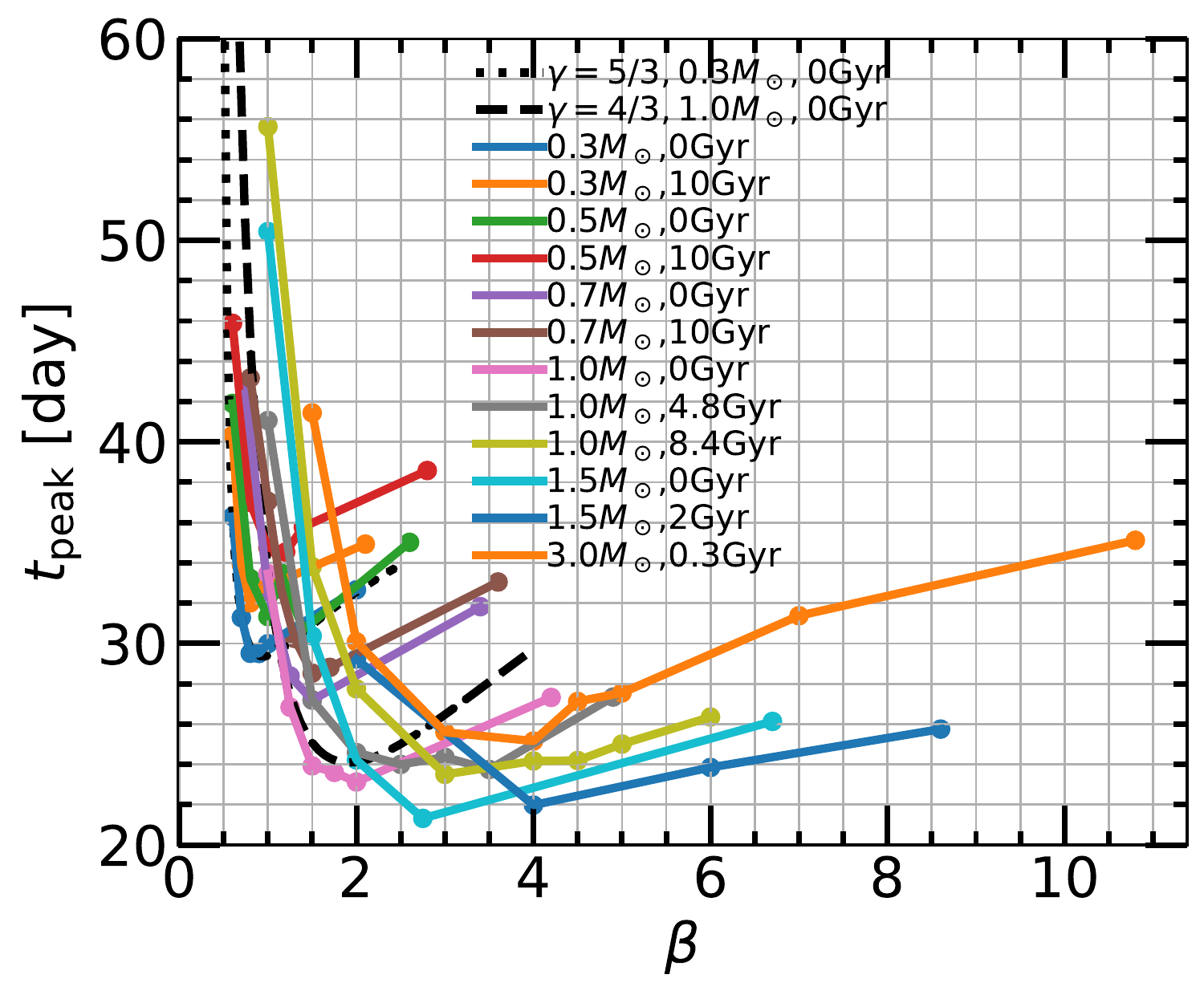}
\plotone{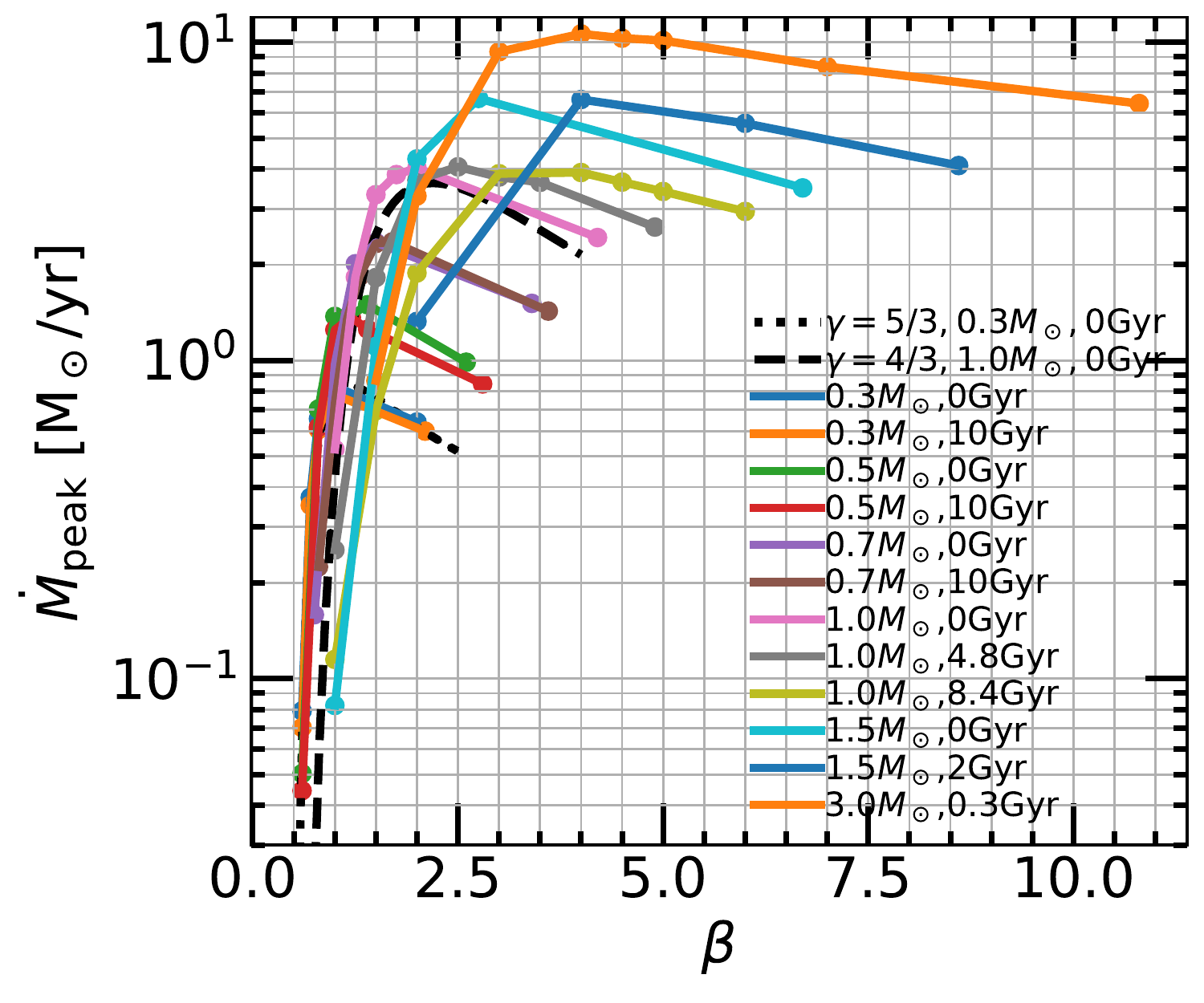} \\
\plotone{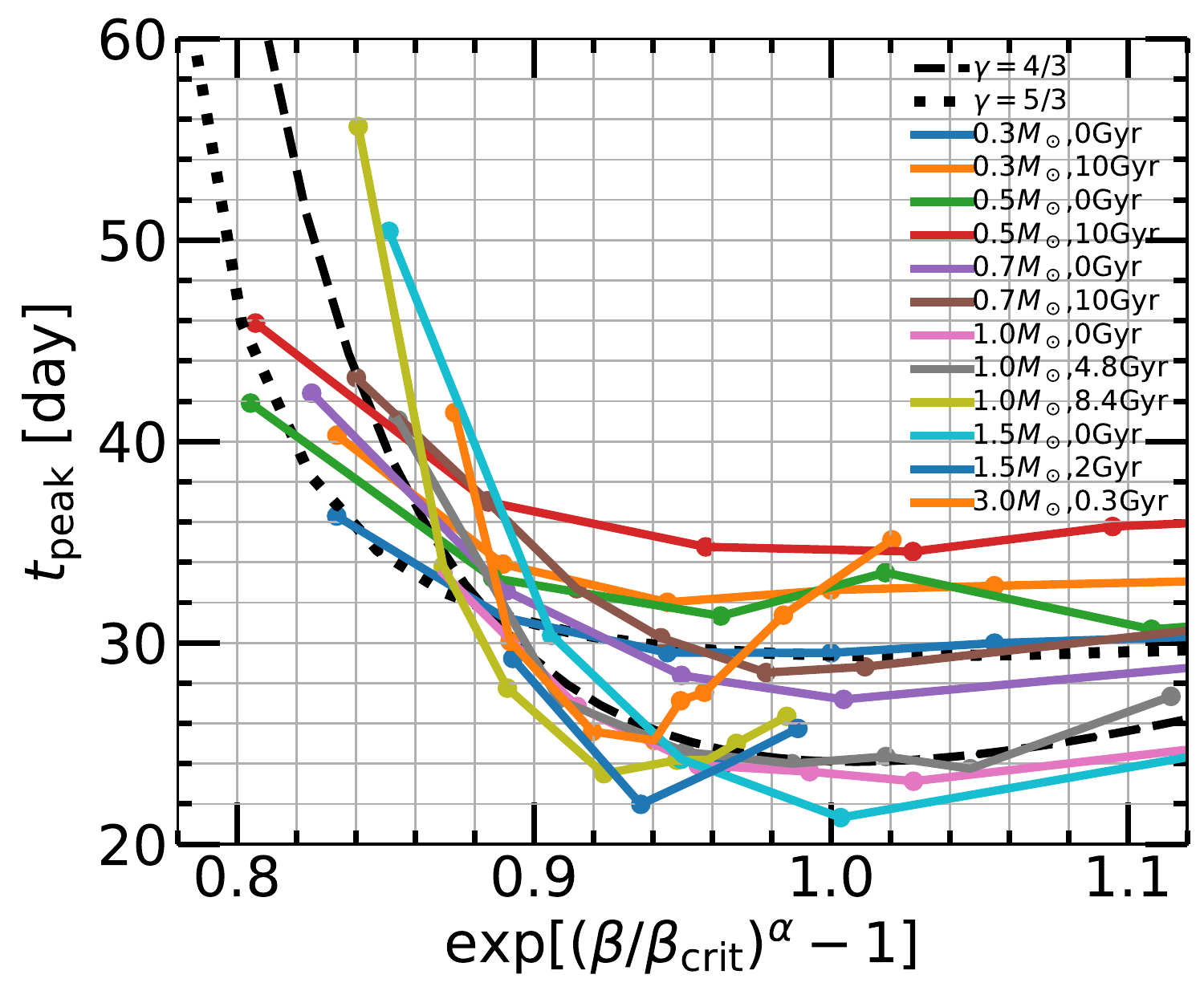}
\plotone{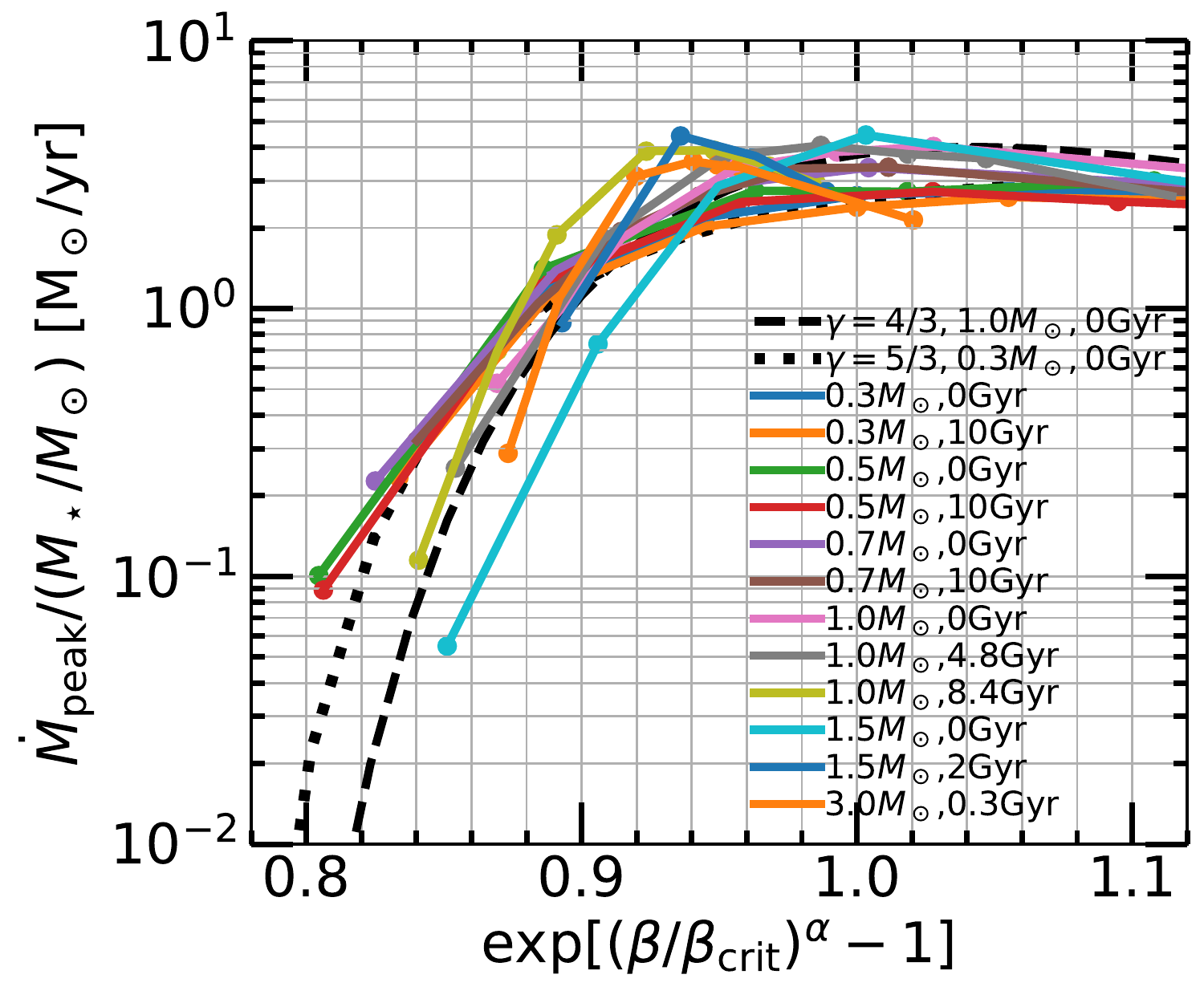} \\
\caption{
Time of peak (left panels) and peak mass fallback rate (right panels) as a function of impact parameter. Results from \citet{2013ApJ...767...25G} for a $\gamma=5/3$ and $\gamma=4/3$ polytrope are shown by the dotted and dashed lines respectively.
Bottom panels are where $x$-axis is scaled with a structural parameter, $\alpha = (\rho_c/\bar\rho)^{-1/3}$.
In the bottom right panel, $\dot M_{\rm peak}$ is normalized by the stellar mass.
\label{fig:mdotpeaktpeak}
}
\end{figure*}

Next we discuss quantities derived from the $\dot M_{\rm fb}$ results. 
The top panels of Figure~\ref{fig:mdotpeaktpeak} show the peak mass fallback rate $\dot M_{\rm peak}$ and the associated peak time $t_{\rm peak}$ as a function of impact parameter $\beta$. 
The $t_{\rm peak}$'s are plotted with a linear $y$-axis and so may appear to vary significantly, but nearly all $t_{\rm peak}$ values are $\approx$20--40 days.
This is because the mass-radius relationship of main sequence stars nearly cancels out the $M_\star$ and $R_\star$ dependence ($M_\star^{-1} R_\star^{3/2}$) in Equation~(\ref{eq:mdotpeaktpeakscalings}). 
The general trend is that $t_{\rm peak}$ decreases sharply with $\beta$ until full disruption and then increases more slowly post critical-$\beta$. 

The $\dot M_{\rm peak}$ values show a marked $M_\star$ dependence, and span several orders of magnitude. 
We can understand this because of the $M_\star$ and $R_\star$ dependence ($M_\star^{2} R_\star^{-3/2}$) in Equation~(\ref{eq:mdotpeaktpeakscalings}) has a higher power on $M_\star$ than in the $t_{\rm peak}$ scaling. 
The general trend is that $\dot M_{\rm peak}$ increases sharply with $\beta$ until full disruption and then decreases more slowly post critical-$\beta$. 
Trends for both $t_{\rm peak}$ and $\dot M_{\rm peak}$ are qualitatively similar to the \citet{2013ApJ...767...25G} simulations, but our study spans a wider range due to the many different stellar structures. 
Focusing on the 1$M_\sun$ star at 3 ages for comparison, at the same $\beta$, the more centrally concentrated (older) star peaks later and has a lower peak fallback rate.

In the bottom panels of Figure~\ref{fig:mdotpeaktpeak}, the $x$-axis is scaled with the structural parameter $\alpha = (\rho_c/\bar\rho)^{-1/3}$ in the same way as in Figure~\ref{fig:deltaM_vs_beta}. 
Here $\dot M_{\rm peak}$ is also normalized by the stellar mass.
Note that unlike for $\Delta M/M_\star$ in Figure~\ref{fig:deltaM_vs_beta}, the $y$-axes of these two plots are not normalized to be unitless, and so have some spread (note that $t_{\rm peak}$ appears to have more spread, but it is on a linear scale).
Despite this, nearly all of the simulations collapse into one coherent relationship.
Fitting formulae for $t_{\rm peak}$ and $\dot M_{\rm peak}$ are provided in the Appendix, such that one can obtain the $t_{\rm peak}$ and $\dot M_{\rm peak}$ for any stellar mass, stellar age, and impact parameter.

\begin{figure}[htp!]
\epsscale{1.1}
\plotone{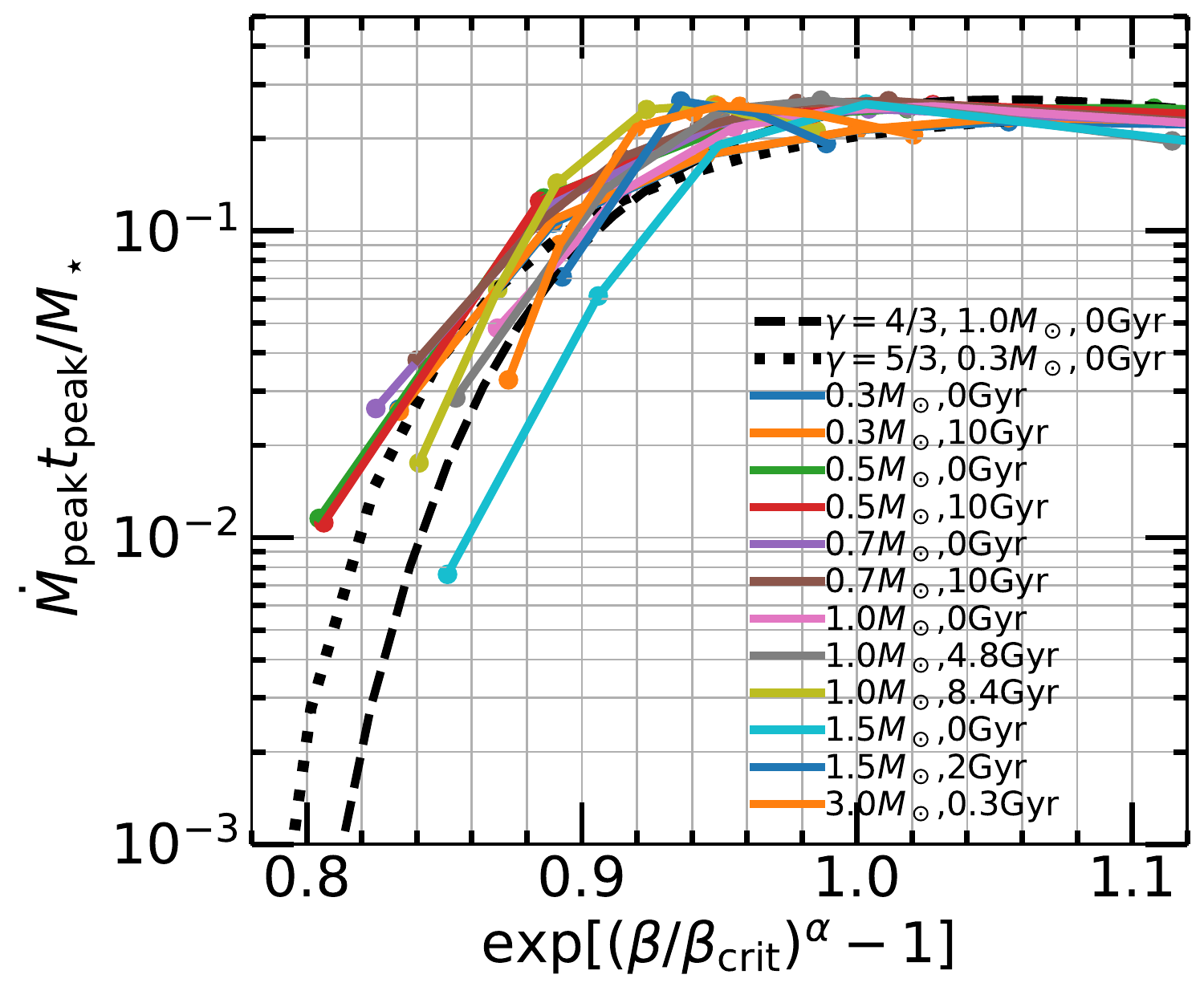}
\caption{
$\dot M_{\rm peak} \times t_{\rm peak}/M_\star$ as a function of $\beta$ scaled with structural parameter $\alpha = (\rho_c/\bar\rho)^{-1/3}$.
\label{fig:mdotpeak_times_tpeak}
}
\end{figure}

Figure~\ref{fig:mdotpeak_times_tpeak} shows the quantity $\dot M_{\rm peak} \times t_{\rm peak}/M_\star$ vs. $\beta$ scaled with structural parameter as above. 
To order-of-magnitude, $\dot M_{\rm peak} \times t_{\rm peak}/M_\star \approx \Delta M_{\rm bound}/M_\star \approx \Delta M/2M_\star$, so this relation is similar to the bottom left panel of Figure~\ref{fig:deltaM_vs_beta}. 
This is a good internal consistency check.
Note that the integrated $dM/dt$ is only approximately equal to the product of its peak values (e.g., for full disruptions $\Delta M/2M_\star = 0.5$, while $\dot M_{\rm peak} \times t_{\rm peak}/M_\star$ asymptotes to $\approx$0.25).

\subsubsection{Power law indices}

\begin{figure*}[htp!]
\epsscale{0.55}
\plotone{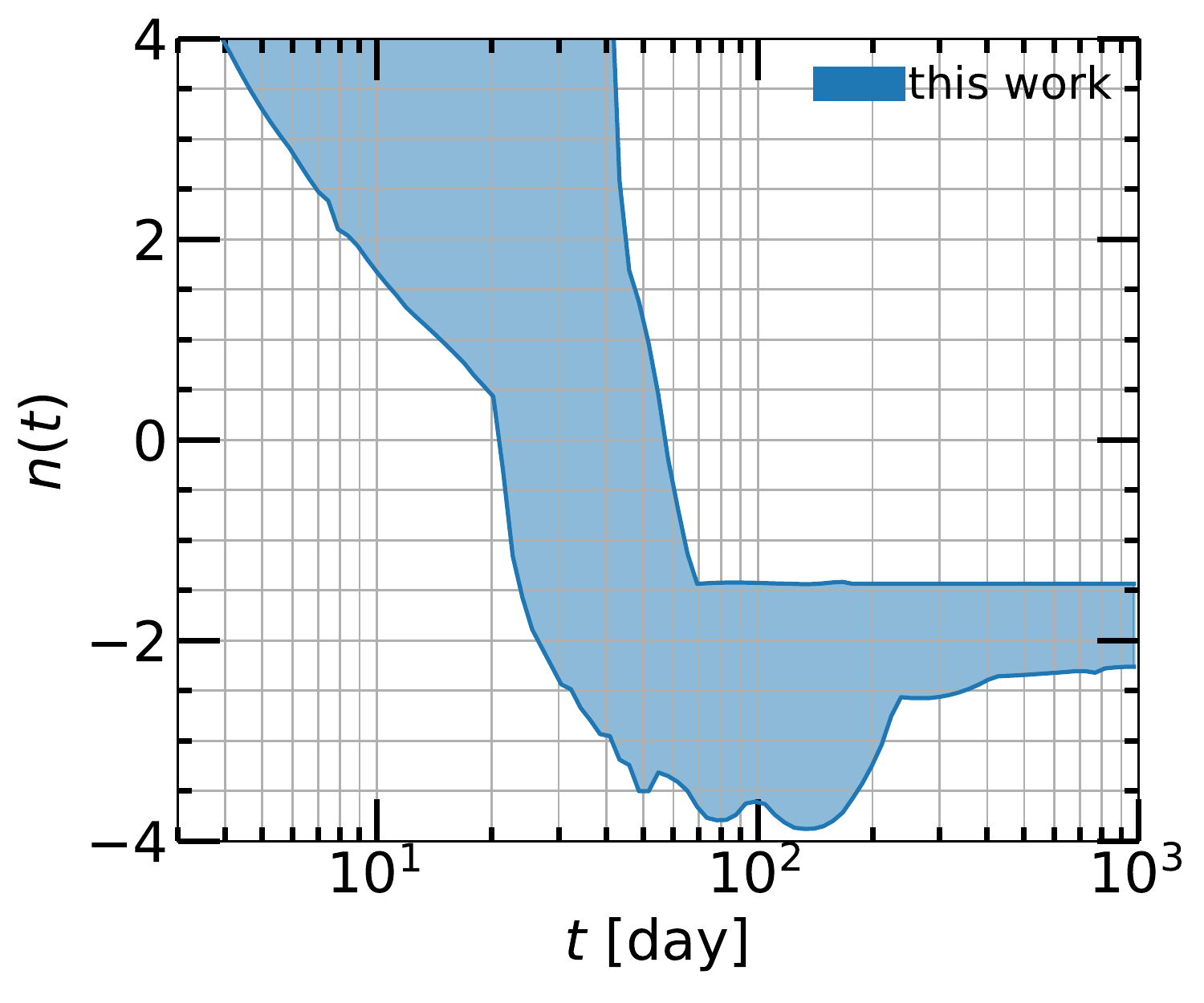}
\plotone{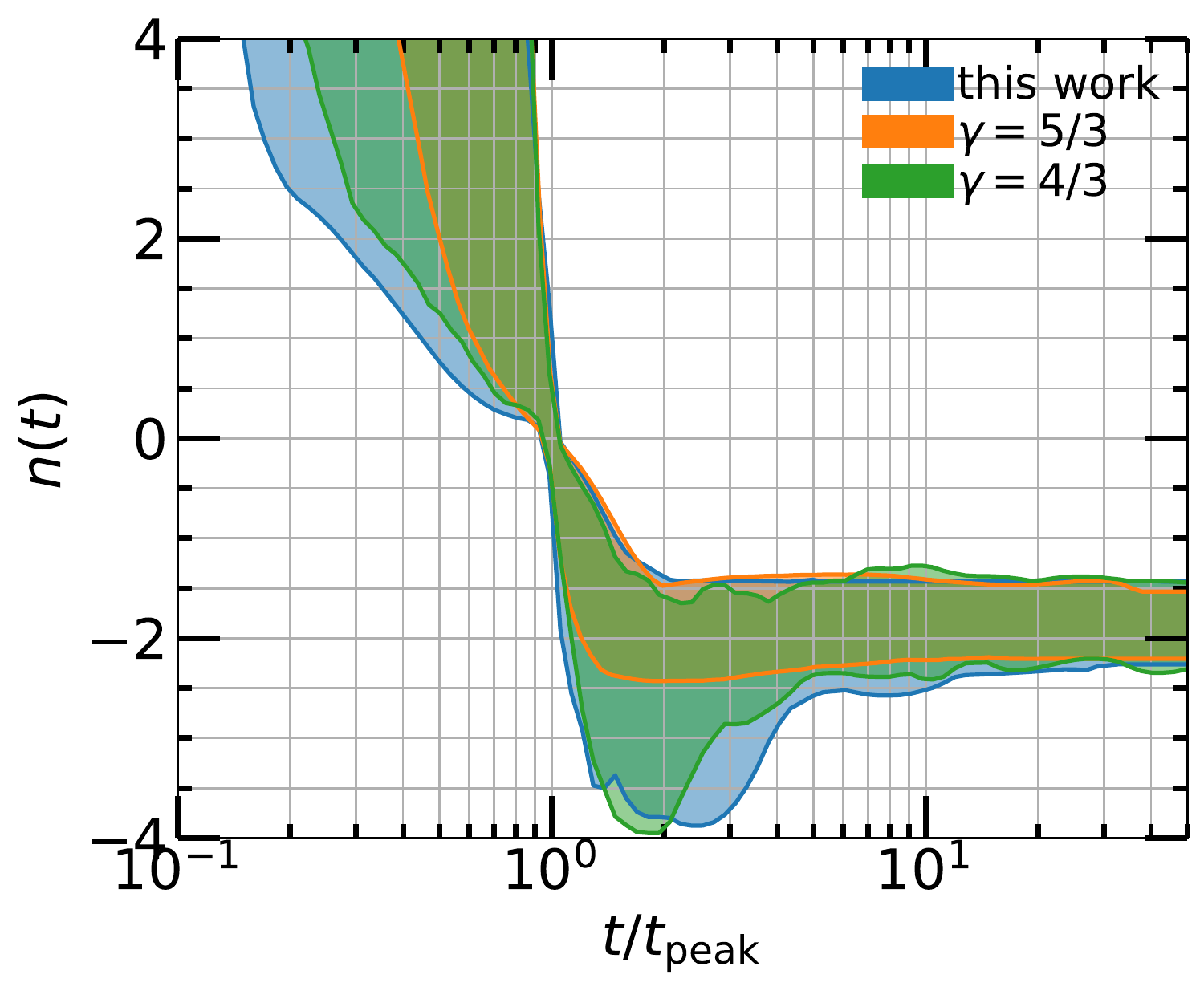}
\plotone{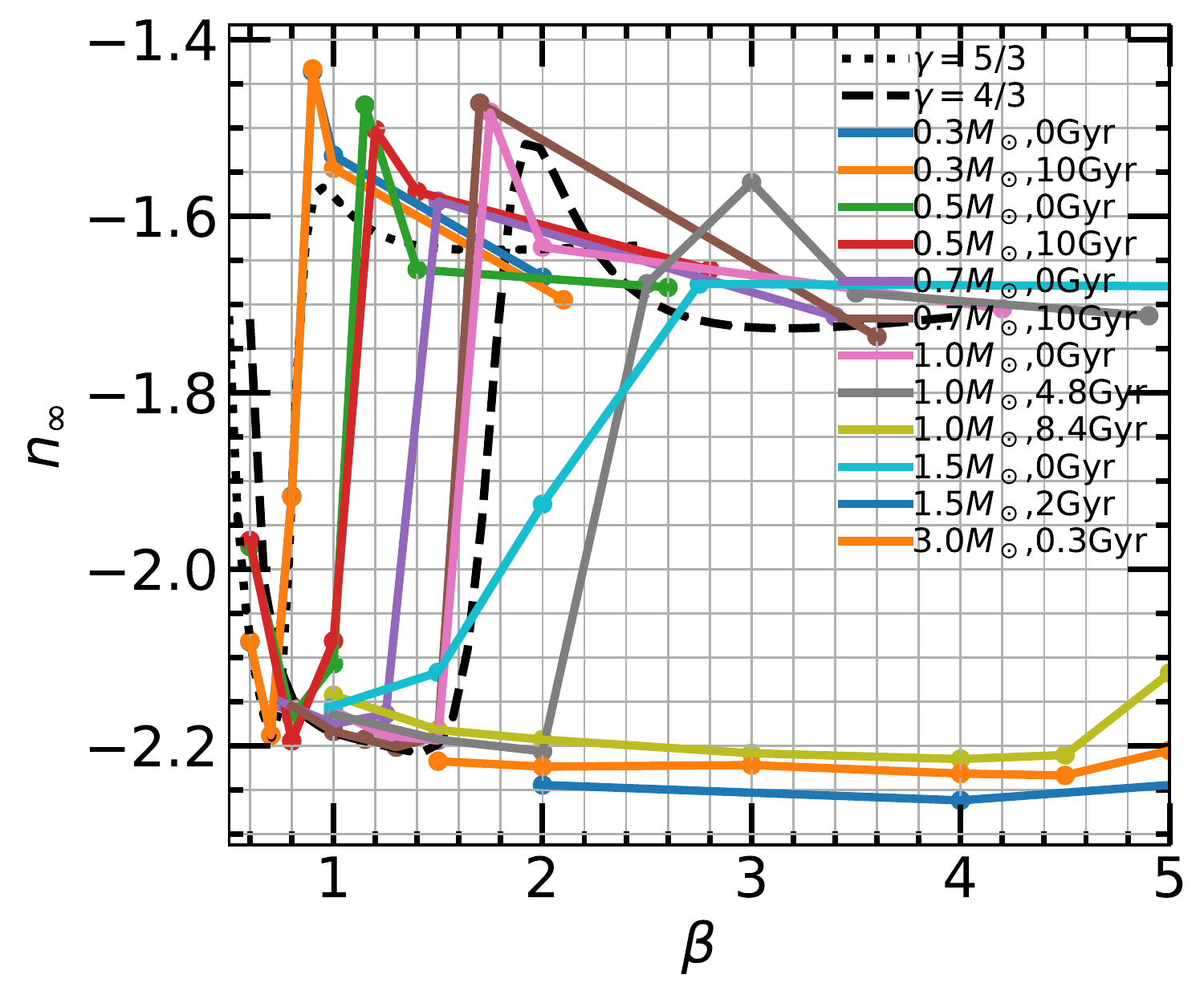}
\plotone{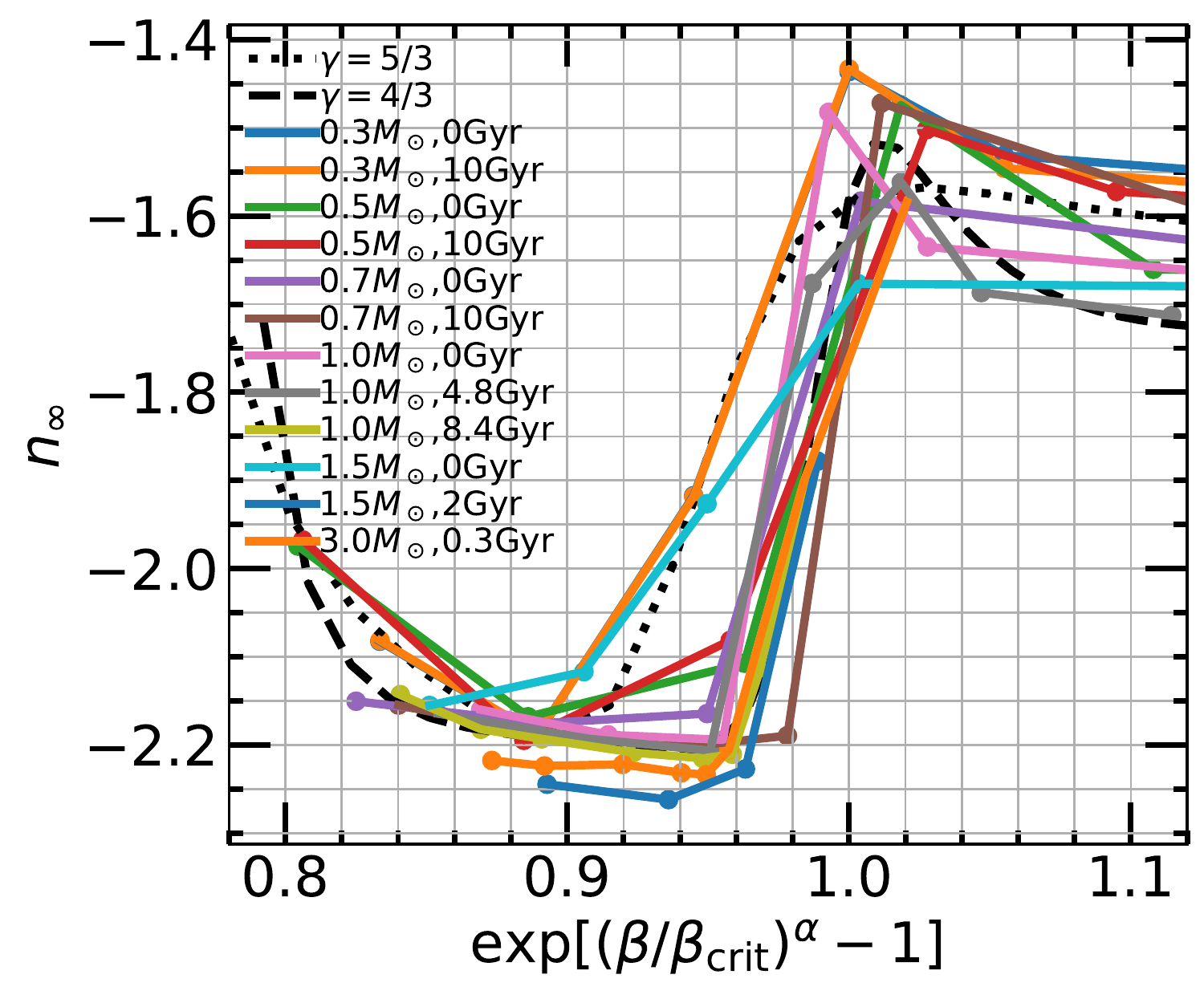}
\caption{
Top panels: instantaneous power law index $n(t)$ of the mass fallback rate as a function of time. 
Top left: absolute units; top right: $x$-axis normalized by the peak time.
Blue regions correspond to this work, while orange and green regions correspond to results from \citet{2013ApJ...767...25G} for $\gamma=5/3$ and $\gamma=4/3$ polytropes. 
Bottom panels: asymptotic decay power law index $n_\infty$ as a function of impact parameter. (In case of confusion due to the repeated colors, the 0.3$M_\sun$ stars follow the $\gamma=5/3$ polytrope.)
The bottom right panel has $x$-axis scaled with the critical impact parameter and a structural parameter $\alpha = (\rho_c/\bar\rho)^{-1/3}$. 
\label{fig:indices}
}
\end{figure*}

The top panels of Figure~\ref{fig:indices} show the instantaneous power law index $n(t)$ of the mass fallback rate as a function of time. 
The top left panel shows the parameter space occupied by tidal disruptions of MS stars for a $10^6 M_\sun$ BH in absolute units.
The $x$-axis in the top right panel is normalized by the peak time, and we also compare to the results of \citet{2013ApJ...767...25G} for $\gamma=5/3$ and $\gamma=4/3$ polytropes. The regions are comparable, but this work covers more parameter space, as expected by the larger range of stellar structures studied.
The different range of $\beta$'s sampled may also account for some differences.

The bottom panels of Figure~\ref{fig:indices} show the asymptotic decay power law index $n_\infty$ for individual stars as a function of $\beta$, and as a function of the scaled $x$-axis incorporating stellar structure familiar from earlier figures. 
Partial disruptions generally have $n_\infty \approx -2.2$ \citep[$\approx 9/4$;][]{2019ApJ...883L..17C}, while full and post-critical disruptions generally have $n_\infty \approx -5/3$, though the $\beta$-dependence is more nuanced than this.
For example, $n_\infty$ peaks near the critical $\beta$ for full disruption, then falls slightly for post-critical disruptions.
There is also a small region at small $\beta$ where there is a sharp transition from higher $n_\infty$ to $n_\infty \approx -2.2$---this was better captured in \citet{2013ApJ...767...25G}, who sampled $\beta$ more finely in this region.

In the bottom left panel, one can see that the more centrally concentrated stars (which correspond roughly to the more massive stars) have $n_\infty \approx -2.2$ for progressively higher $\beta$'s. 
Aside from fitting individual TDEs, one can use this information with a statistical population of TDEs:
given that the TDE rate depends inversely on $\beta$ to some power (see Section~\ref{sec:conclusion} for more detailed discussion), one can use $n_\infty$ as a probe of the stellar structure and more indirectly the stellar mass and age. 
More specifically, the lower-mass stars spend only a small portion of $\beta$ parameter space at $n_\infty \approx -2.2$, while the higher-mass stars spend much more of $\beta$ parameter space here, especially when weighted inversely by $\beta$ to some power. 
If the stellar mass function of TDEs is roughly flat for $M_\star \lesssim M_\sun$ \citep{2016MNRAS.461..371K}, and especially if mass segregation in galactic centers is important (see Section~\ref{sec:conclusion}) then if an $n_\infty \approx -2.2$ is observed, it is more likely due to the partial disruption of a more massive star. Less massive stars are more likely to be full disruptions, and thus if $n_\infty \approx -5/3$ is observed, it is more likely due to the full disruption of a lower-mass star.

In the bottom right panel, after rescaling the $x$-axis with $\rho_c/\bar\rho$, the simulations reduce into a single relationship describing $n_\infty$ for any star and impact parameter. We provide a B-spline fit to this relation in the Appendix, so that one can obtain the $n_\infty$ for any stellar mass, stellar age, and impact parameter.

%%%%%%%%%%%%%%%%%%%%%%%%%%%%%%%%%%%%%%%%%%%%%%%%%%%%%%%%%%%%%%%%%%%%%%%%%%%%%%%%%%%%%%%%%%%%%%%%%%%%%%%%%%%%%%%%%%%%%%%%%%%%% 
\section{Conclusion}\label{sec:conclusion}

\subsection{Comparison to other work}

We compare with \citet{2013ApJ...767...25G} throughout, but perhaps the most salient point is that, as we stress in this paper, every tidal disruption quantity that we have calculated depends primarily on the stellar structure, parameterized by $\rho_c/\bar\rho$, and impact parameter $\beta$. 
Thus, the study of \citet{2013ApJ...767...25G} is only directly applicable to stars with exactly $\gamma=5/3$ or $\gamma=4/3$ stellar structures, which do not represent the range of stellar masses and stellar ages involved in TDEs.
This work offers a significant improvement upon the \citet{2013ApJ...767...25G} grid of simulations, and applies to any main-sequence stellar mass and stellar age.
The fitting formulae (see Appendix) we find for key disruption quantities can be used to predict the disruption quantities for any star, provided one knows the $\rho_c/\bar\rho$.
Additionally, the \texttt{STARS\_library} tool \citep[][see Appendix]{jamie_a_p_law_smith_2020_4062018} allows one to retrieve the $dM/dt$ for any stellar mass and age.

Our results are consistent with the hydrodynamical simulations of \citet{2019MNRAS.487..981G}, \citet{2019MNRAS.487.4790G} in the non-relativistic limit, and \citet{2019ApJ...882L..26G} for the impact parameters and stars they consider.
The asymptotic decay power law indices we find are broadly consistent with the analytic predictions of \citet{2019ApJ...883L..17C} ($n_\infty=-9/4$ for partial disruptions, $n_\infty=-5/3$ for full disruptions), though our results indicate a more nuanced $\beta$ dependence (see Figure~\ref{fig:indices} and associated discussion).
We note that we find $n_\infty \approx -2.2$ for partial disruptions, slightly larger than the $n_\infty=-2.25$ found by \citet{2019ApJ...883L..17C}.

Our results are broadly consistent with those of \citet{2020arXiv200103501R,2020arXiv200103502R,2020arXiv200103503R,2020arXiv200103504R}, though we note some differences.
Rather than comparing $dM/dt$'s in detail, we note that the critical impact parameters we find for full disruption differ slightly. 
\added{This is partly due to the fact that \citet{2020arXiv200103501R,2020arXiv200103502R,2020arXiv200103503R,2020arXiv200103504R} study a single stellar age (MAMS), whereas we study 2 or 3 stellar ages for each star, so a direct comparison of results for a given stellar mass is misleading, as the stellar structures differ.}
See Figure~\ref{fig:beta_crit_vs_Mstar2} in the Appendix for a comparison.
One disagreement is at the 15\% level in our $\beta_{\rm crit}=0.9$ result for the 0.3 $M_\sun$ star (which has an identical profile to a $\gamma=5/3$ polytrope) for $M_{\rm BH}=10^6 M_\odot$; \citet{2020arXiv200103501R,2020arXiv200103502R,2020arXiv200103503R,2020arXiv200103504R} argue that this discrepancy is due to GR effects at $r_{\rm p} = 22 r_{\rm g}$. The difference is modest, but we note that our $\beta_{\rm crit}=0.9$ result has been independently found by other authors for polytropic stellar models, and the non-relativistic regime is one in which our grid-based adaptive-mesh FLASH framework has also been independently verified with two other numerical methods (SPH and moving-mesh) and resolutions by other authors \citep{2013ApJ...767...25G, 2017A&A...600A.124M, 2019MNRAS.487..981G, 2019ApJ...882L..26G, 2019MNRAS.487.4790G}.

\subsection{Caveats/extensions}

Our calculations are Newtonian, and thus do not capture the minority of encounters in which relativistic effects significantly alter the $dM/dt$. 
\citet{2014PhRvD..90f4020C}, \citet{2017MNRAS.469.4483T}, and \citet{2019MNRAS.487.4790G} \citep[for a review of TDEs in GR see also][]{2019GReGr..51...30S} studied differences between the fallback rates from relativistic versus Newtonian tidal disruption simulations in detail, and found that for $r_{\rm p}>10 r_{\rm g}$, the difference is $\lesssim10\%$\added{, and that the most significant corrections from relativistic effects are at pericenter distances smaller than this}.\added{\footnote{Though we note that \citet{2020arXiv200103504R} argue for slightly larger effects due to the inclusion of GR.}}
\added{They found that for $r_{\rm p}>10 r_{\rm g}$} the error in a non-relativistic tidal disruption simulation \added{scales approximately as $v_{\rm p}^2 \sim r_{\rm g}/r_{\rm p}$ (though we emphasize that GR is a non-linear theory, and order unity effects are possible for the rare extreme encounters in which GR dominates the dynamics)}.
For encounters in which relativistic effects are significant, the decay power law index remains approximately the same, and the most important differences are that the peak time is shifted to the right and the peak fallback rate is shifted down compared to the Newtonian case.
The fraction of disruptions that require relativistic hydrodynamics in order to accurately model the mass fallback rate is \added{$\sim \rm{max}(1, 10 r_{\rm g} / r_{\rm t})$ in the pinhole regime of loss cone repopulation, and a step function in the diffusive regime: $\sim$0 if $10 r_{\rm g} < r_{\rm t}$, and $\sim$1 if $10 r_{\rm g} > r_{\rm t}$}.
For a 1$M_\sun$, 1$R_\sun$ star, this is $\approx$5\% for a $10^6 M_\sun$ BH and $\approx$20\% for a $10^7 M_\sun$ BH.
See Figure~\ref{fig:mbh_beta_phase_space} in the Appendix for the range of applicability of Newtonian hydrodynamics simulations.
Note that while relativistic effects may cause significant apsidal precession, the effect on the mass fallback rate to the BH is relatively small---and it is this quantity that tracks the luminosity evolution of TDEs so closely.
While the library presented in this paper applies to the majority of TDEs, we plan to extend our setup to include relativistic encounters in future work.

Other extensions include studying orbits with different eccentricities, incorporating stellar spin \citep[rather than performing new simulations with spinning stars, we could scale our $dM/dt$ results with a parameterized stellar spin dependence based on the results of][]{2019ApJ...872..163G}, studying magnetic fields \citep[we already have the framework for this, first studied in][]{2017ApJ...834L..19G}, adding more stellar masses and ages to the grid, sampling more finely in $\beta$ (though this will not make a significant difference because of our $dM/dt$ interpolation), and various extensions and upgraded functionality of the \texttt{STARS\_library} tool \citep{jamie_a_p_law_smith_2020_4062018}. 
Finally, we plan to use the interpolated fallback-rate library as the new backbone for the publicly available TDE fitting software \texttt{MOSFiT} \citep{2018ApJS..236....6G,2019ApJ...872..151M}.

We do not perform an analysis of TDE rates as a function of stellar mass and stellar age, but below is a brief discussion.
In terms of broad demographics, \citet{2016MNRAS.461..371K} calculated that for $M_{\rm BH} \lesssim 10^7 M_\sun$, the typical TDE is due to a $0.3 M_\sun$ star, but that the mass function is relatively flat for $M_\star \lesssim M_\sun$.
This study only considered effects due to the initial mass function (IMF) and did not include any dynamical interactions between stars, such as mass segregation in galactic centers, which segregates more massive stars closer to the BH and ejects less massive stars \citep[e.g.,][]{2004ApJ...613.1143B,2016ApJ...819...70M}---this would cause the stellar mass function for TDEs to peak at higher masses than suggested by a convolution of the IMF and the luminosity function of TDEs \citep{2012ApJ...760..103D,2013ApJ...777..133M}.
See also Figure 15 in \citet{2012ApJ...757..134M} for the fraction of stars at different evolutionary stages contributing to TDE flares. For $M_{\rm BH} \lesssim 10^8 M_\sun$, MS stars make up $>$85\%. 
In more detail, \citet{2012ApJ...757..134M} find that the tidal disruption rate scales with the tidal radius of a given star as $\dot n \propto r_{\rm t}^\alpha$, where $\alpha \approx 0.2$--$0.5$. 
Ignoring other considerations, the physical cross-section increases with stellar age and leads to higher rates of tidal disruption for older MS stars.
However, the $\Delta M$ lost at a given $\beta$ decreases as the star becomes more centrally concentrated, leading to fainter flares.

The dynamical mechanisms operating in the relevant galactic center and the associated most likely age for a star to be disrupted are also important.
The two-body relaxation time, which is $\approx$2 Gyr for a $10^6 M_\sun$ BH \citep{2013ApJ...764...52B} (the MS lifetime of a 1.5$M_\sun$ star), gives a characteristic limit for the approximate mass and age of a star upon disruption.
However, there are many other mechanisms that disrupt stars earlier.
Stars can interact with one another coherently inside the sphere of influence of the BH, in contrast to two-body relaxation, leading to rapid angular momentum evolution \citep{1996NewA....1..149R}; for example, this occurs for eccentric nuclear disks \citep{2018ApJ...853..141M}. \deleted{galaxy mergers, or} \added{The presence of a SMBH binary can also dramatically increase the TDE rate, driven by a combination of strong scatterings and Kozai–Lidov cycles between individual stars and the SMBH binary \citep[e.g.,][]{2015MNRAS.451.1341L}}.
\added{Other mechanisms that may disrupt stars over timescales $\ll$~2~Gyr include strong (non-diffusive) two-body scatterings, which make up a fraction of the TDE rate in any galaxy \citep{2017MNRAS.468.1760W}, and radially anisotropic velocity fields or unusually dense star clusters, which may occur in post-starburst galaxies \citep{2018MNRAS.480.5060S}.}
\replaced{Finally}{Moreover}, the fact that TDEs appear to be observed preferentially in \added{rare} post-starburst galaxies \citep{2014ApJ...793...38A, 2016ApJ...818L..21F, 2017ApJ...850...22L, 2018ApJ...853...39G} means that many TDEs are sourced by unique stellar populations \added{with particular dynamical histories}.

\subsection{Summary}
We summarize the main points of this paper below.
\begin{enumerate}

\item We present a grid of tidal disruption simulations of stars with accurate stellar structures and chemical abundances, using MESA models as initial conditions to FLASH simulations with a Helmholtz EOS.

\item We interpolated this grid in 3D (stellar mass, stellar age, and impact parameter) to provide the \texttt{STARS\_library} tool (see Section~\ref{sec:stars_lib}), where one can request the $dM/dt$ for any stellar mass, stellar age, and impact parameter.

\item The quantities $\beta_{\rm crit}, \Delta M, t_{\rm peak}, \dot M_{\rm peak}$, and $n_\infty$ vary significantly with stellar mass and stellar age, but can be reduced to a single relationship that depends only on stellar structure, parameterized by $\rho_c/\bar\rho$, and impact parameter $\beta$ (see Figures~\ref{fig:deltaM_vs_beta}, \ref{fig:beta_crit_vs_Mstar}, \ref{fig:mdotpeaktpeak}, and \ref{fig:indices}).
We provide fitting formulae for these quantities applicable to any MS star (see Figure~\ref{fig:fitting_formulae} in Appendix).

\item For the same $\Delta M$, the shape of the $dM/dt$ curve varies significantly with stellar mass, promising the potential determination of stellar properties from the TDE light curve alone (see Figure~\ref{fig:mdots_fixed_deltam}).

\item The critical impact parameter for full disruption increases with increasing central concentration, and scales approximately as $(\rho_c/\bar\rho)^{1/3}$ for $\rho_c/\bar\rho \lesssim 500$ and $(\rho_c/\bar\rho)^{1/2.3}$ for $\rho_c/\bar\rho \gtrsim 500$ (see Figure~\ref{fig:beta_crit_vs_Mstar}).

\item In general, more centrally concentrated stars have steeper $dM/dt$ rise slopes and shallower decay slopes (see Figures~\ref{fig:mdots_allone} and \ref{fig:mdots_full}).

\item We show that the shape of $dM/dt$ depends only on the stellar density profile, and has little dependence on the internal EOS of the star (see Section~\ref{sec:stellar_structure_vs_EOS} where we compare a 1$M_\sun$ and 10$M_\sun$ star with nearly identical density profiles), thus extending the range of applicability of our interpolated library and fitting formulae to any MS star.

\end{enumerate}

We hope the community makes use of the \texttt{STARS\_library} tool (see Appendix), and we look forward to incorporating this library as the new backbone of the \texttt{MOSFiT} fitting software for TDEs, or any other TDE fitting framework, allowing for more accurate determinations of both the nature of the disrupted star and the BH.

\acknowledgements
We thank Jieun Choi, Josiah Schwab, Dongwook Lee, Kevin McKinnon, Katie Auchettl, Tiara Hung, Jane Dai, Roseanne Cheng, Emily Martin, Nick Leaf, \added{and the anonymous referee} for intellectual contributions.
We acknowledge use of the {\it lux} supercomputer at UCSC, funded by NSF MRI grant AST 1828315, and the HPC facility at the University of Copenhagen, funded by a grant from VILLUM FONDEN (project number 16599). 
We thank the Heising-Simons Foundation, the Danish National Research Foundation (DNRF132), and the NSF (AST-1911206 and AST-1852393) for support. D.C. acknowledges support from the NSF GRFP under grant DGE1339067.
We thank NVIDIA for helping with visualizations and volume renderings of the simulations---these were performed with the GPU-enabled NVIDIA IndeX plug-in for ParaView.
\software{MESA \citep{2011ApJS..192....3P}, 
FLASH \citep{2000ApJS..131..273F},
astropy \citep{2013A&A...558A..33A,2018AJ....156..123A}, 
yt \citep{2011ApJS..192....9T},
NVIDIA IndeX (\url{https://developer.nvidia.com/nvidia-index}),
\texttt{STARS\_library} \citep{jamie_a_p_law_smith_2020_4062018}.
}

%%%%%%%%%%%%%%%%%%%%%%%%%%%%%%%%%%%%%%%%%%%%%%%%%%%%%%%%%%%%%%%%%%%%%%%%%%%%%%%%%%%%%%%%%%%%%%%%%%%%%%%%%%%%%%%%%%%%%%%%%%%%% 
%%%%%%%%%%%%%%%%%%%%%%%%%%%%%%%%%%%%%%%%%%%%%%%%%%%%%%%%%%%%%%%%%%%%%%%%%%%%%%%%%%%%%%%%%%%%%%%%%%%%%%%%%%%%%%%%%%%%%%%%%%%%% 
\appendix
\section{Interpolated fallback-rate library}\label{sec:stars_lib}

The STARS library of interpolated fallback rates, as well as up-to-date instructions for loading and using the library, is available on GitHub\footnote{\texttt{STARS\_library}: \href{https://github.com/jamielaw-smith/STARS_library}{github.com/jamielaw-smith/STARS\_library}.} and version 1.0.0 is archived on Zenodo \citep{jamie_a_p_law_smith_2020_4062018}.
One can create custom interpolated grids and one can also query for any stellar mass, stellar age, and impact parameter.

The basic interpolation between $dM/dt$'s is the same as in the \texttt{MOSFiT} software \citep{2018ApJS..236....6G,2019ApJ...872..151M}.
We first interpolate in $\beta$ for a given stellar mass and age.
We then interpolate each of these $\beta$'s in stellar mass, for a given fractional main-sequence stellar age (e.g., 0.3$M_\sun$ ZAMS to 0.5$M_\sun$ ZAMS).
We then interpolate in stellar age for a given stellar mass.
In the query/retrieval mode, in order to retrieve a $dM/dt$ corresponding to a specific stellar mass, stellar age, and impact parameter that does not already exist in the interpolated library, we perform a quick series of interpolations on the nearest neighbor points in 3D space.

\begin{figure*}[htp!]
\epsscale{0.17}
\plotone{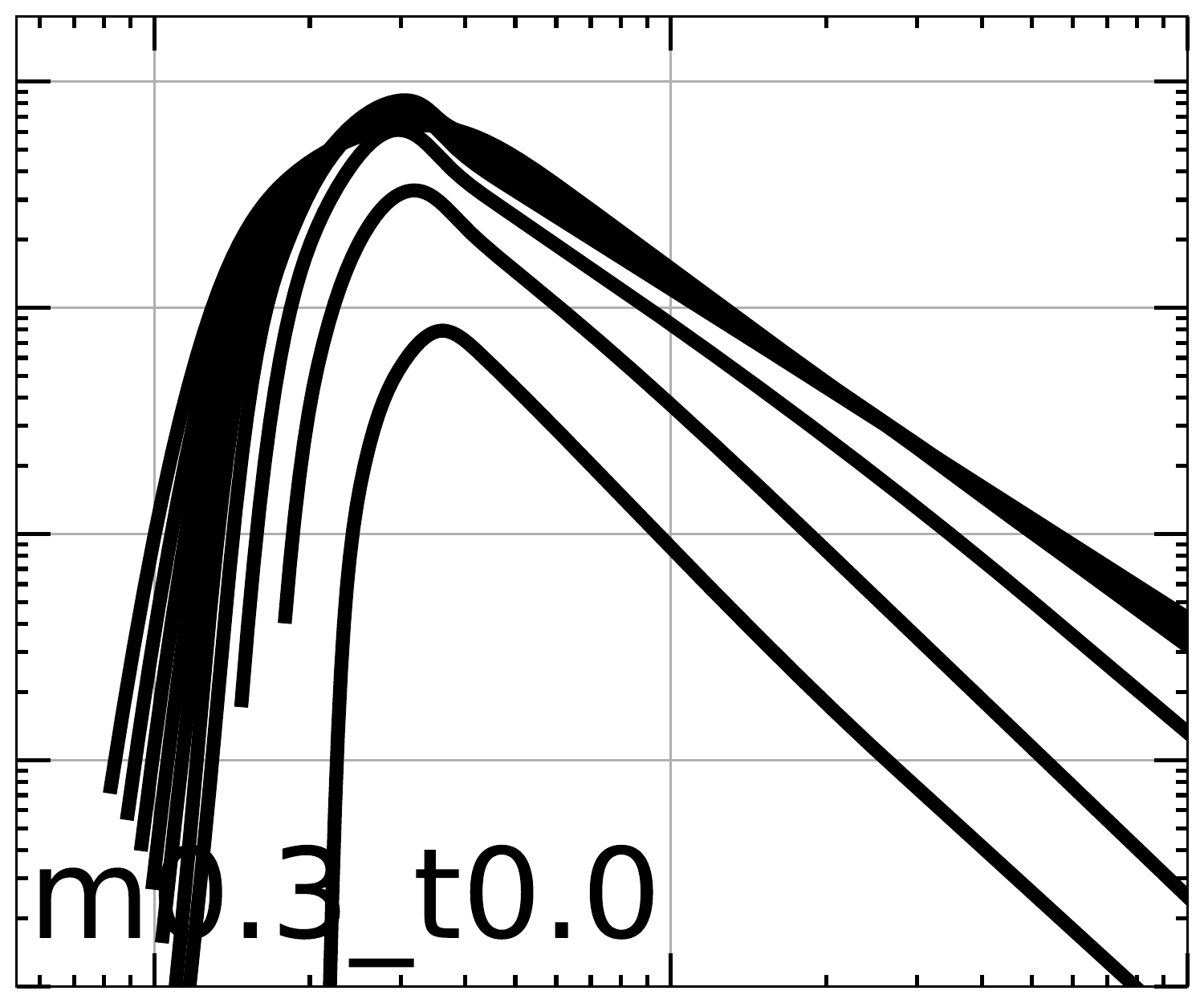}
\plotone{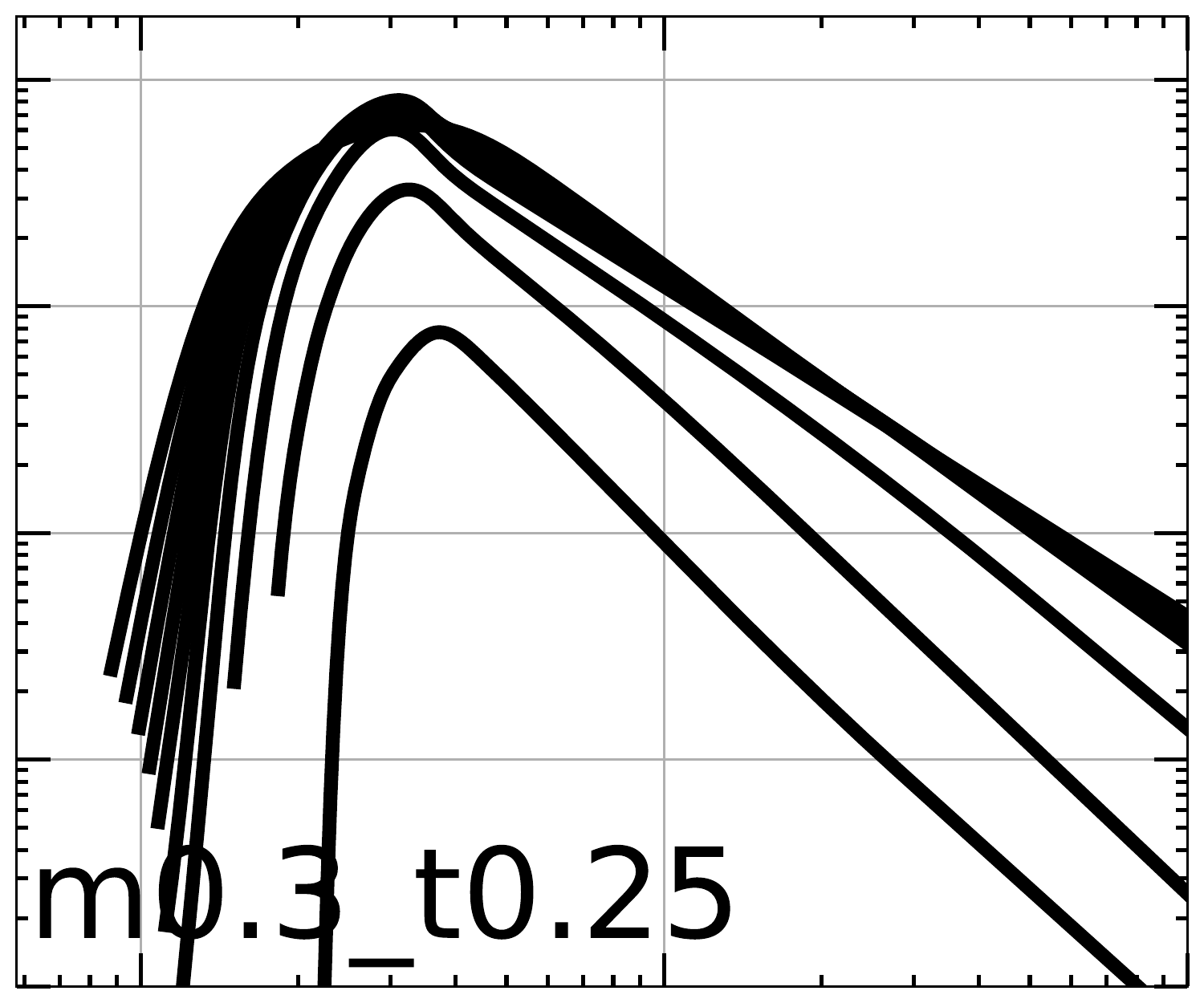}
\plotone{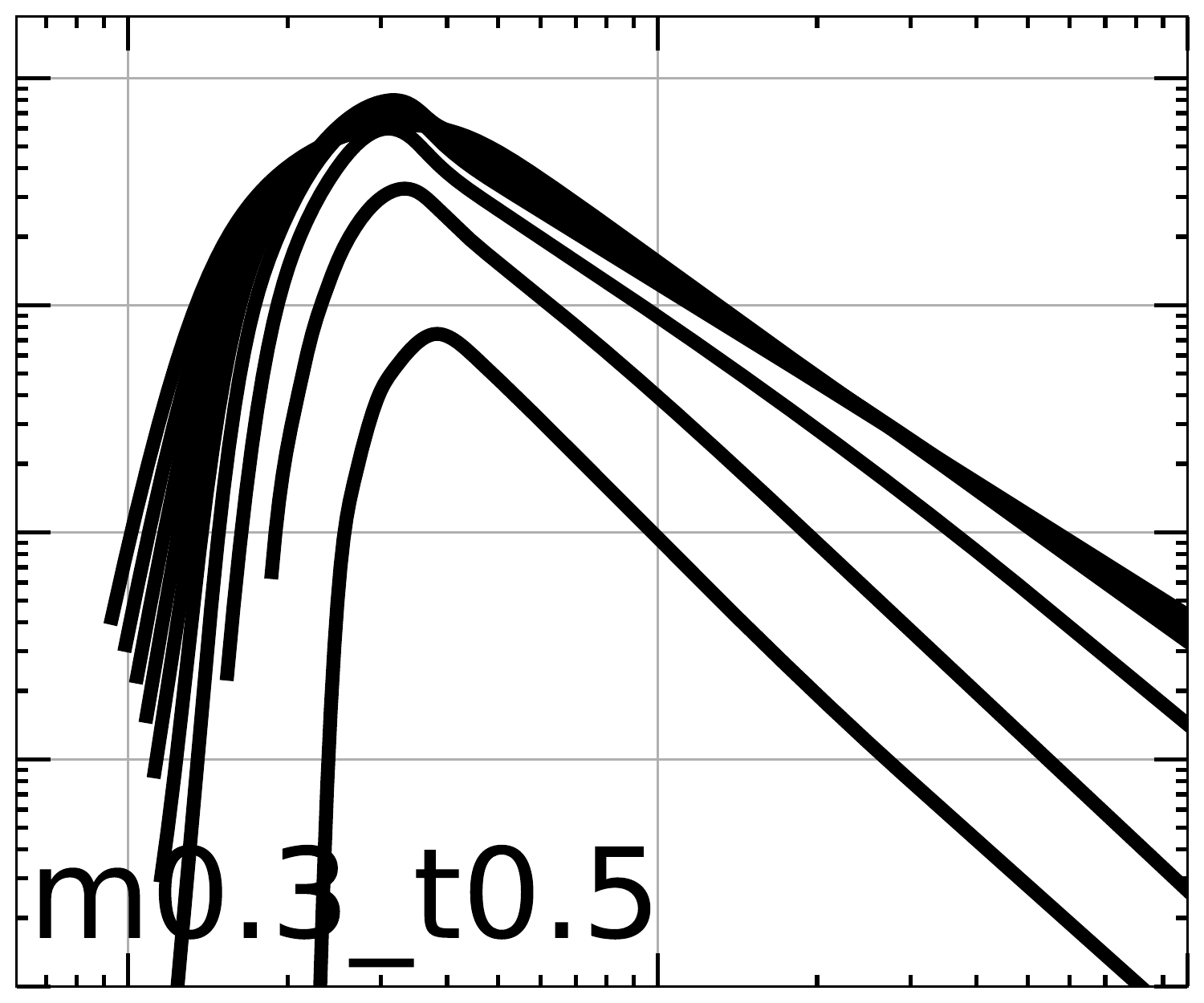}
\plotone{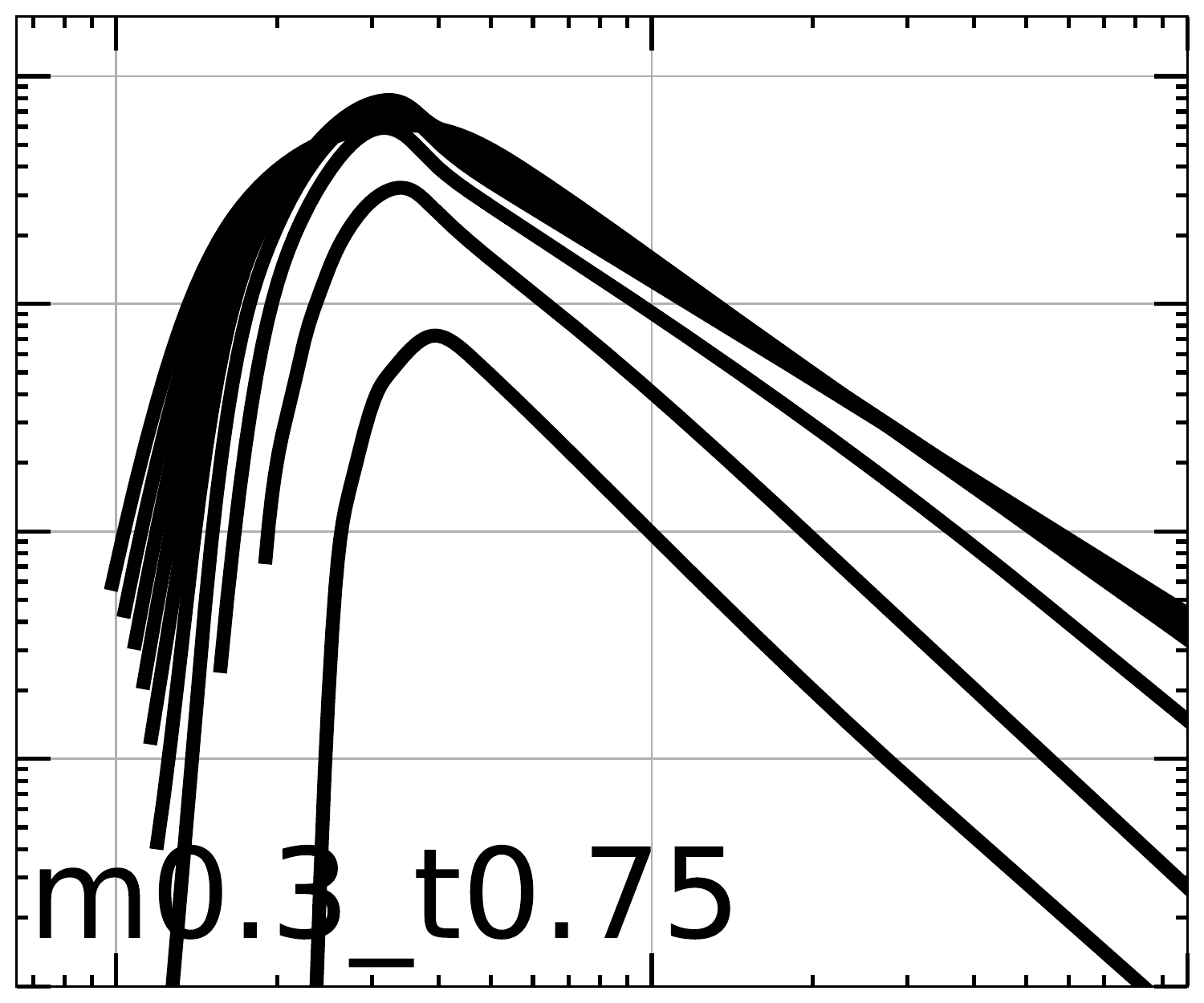}
\plotone{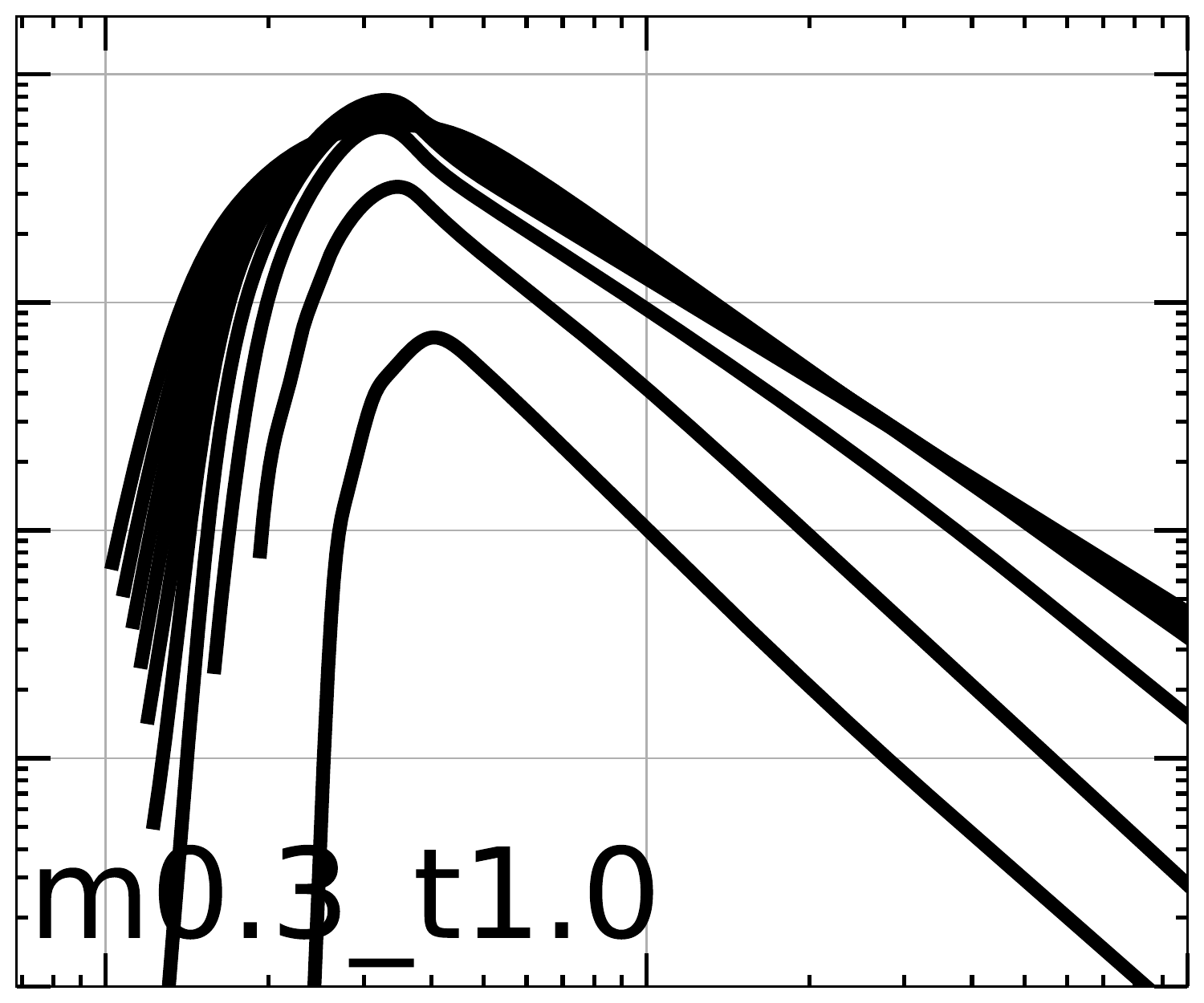}
\plotone{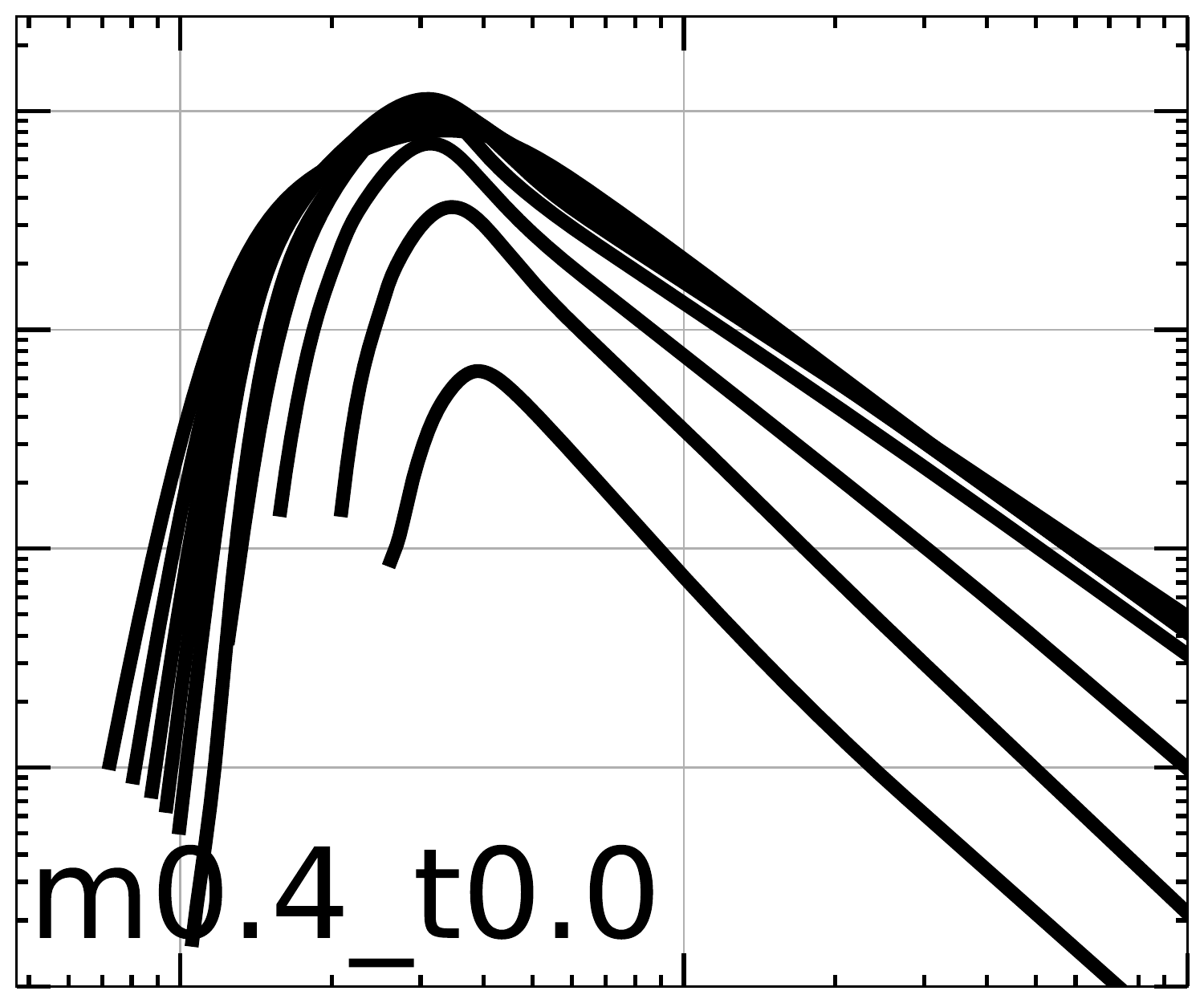}
\plotone{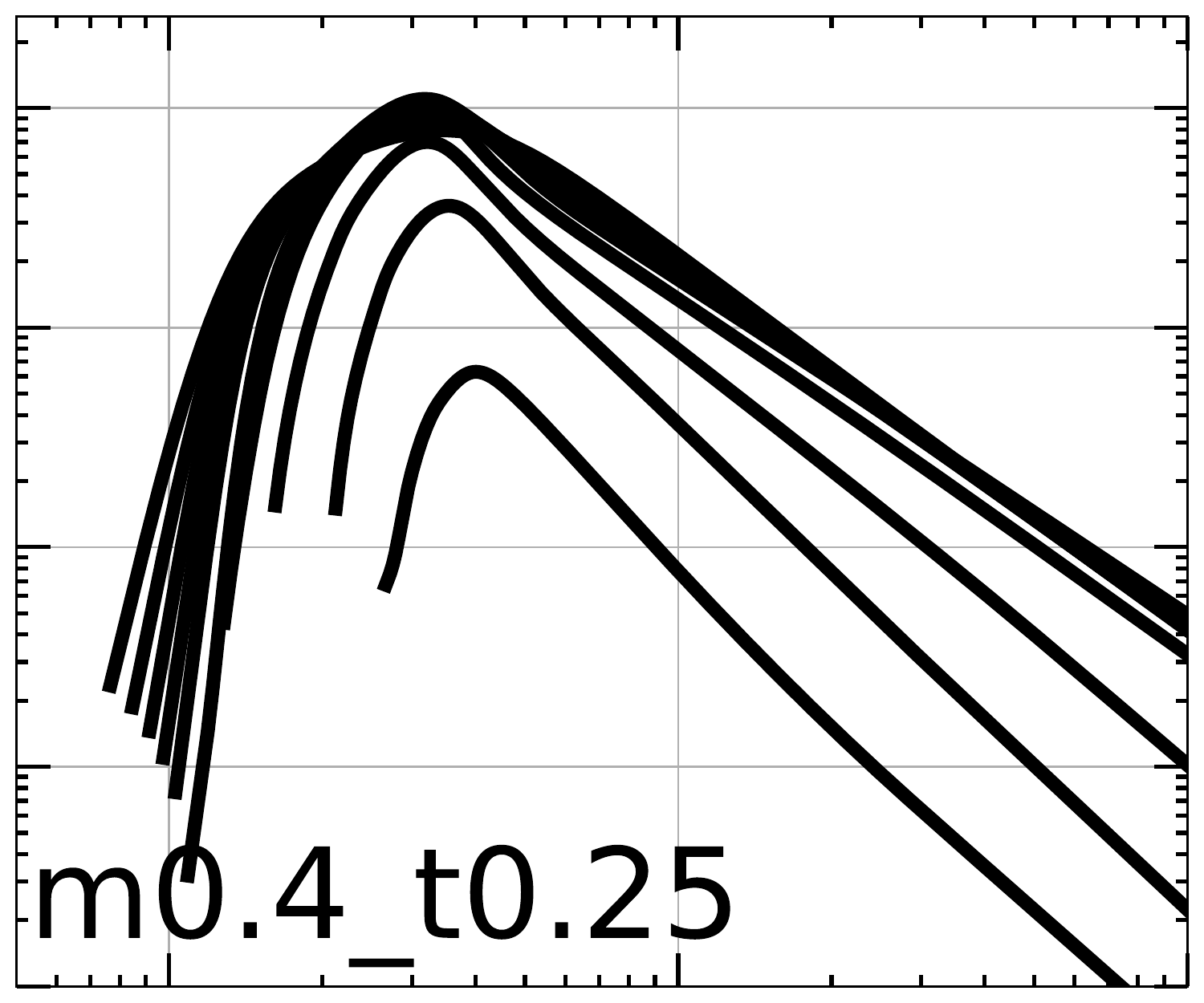}
\plotone{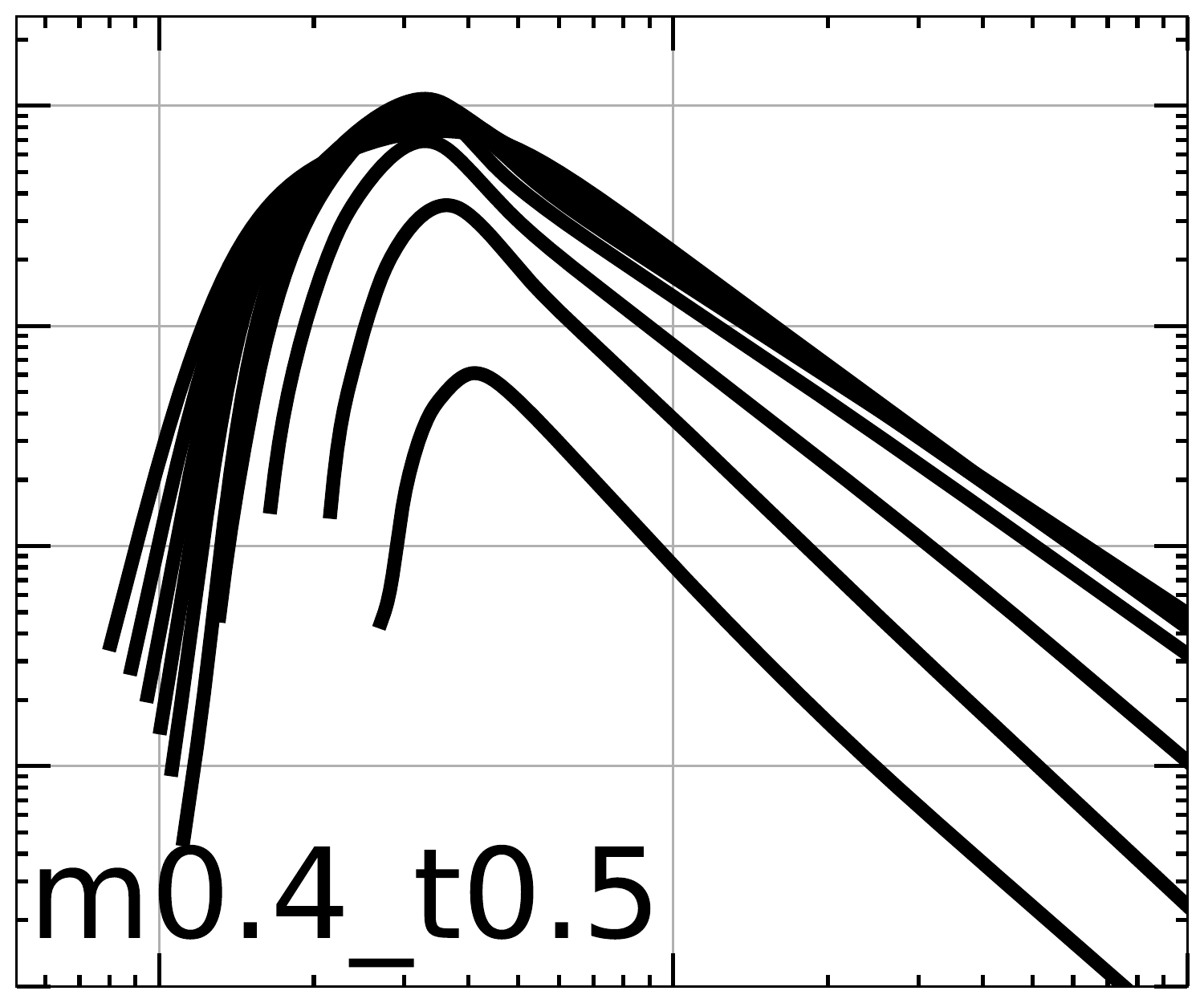}
\plotone{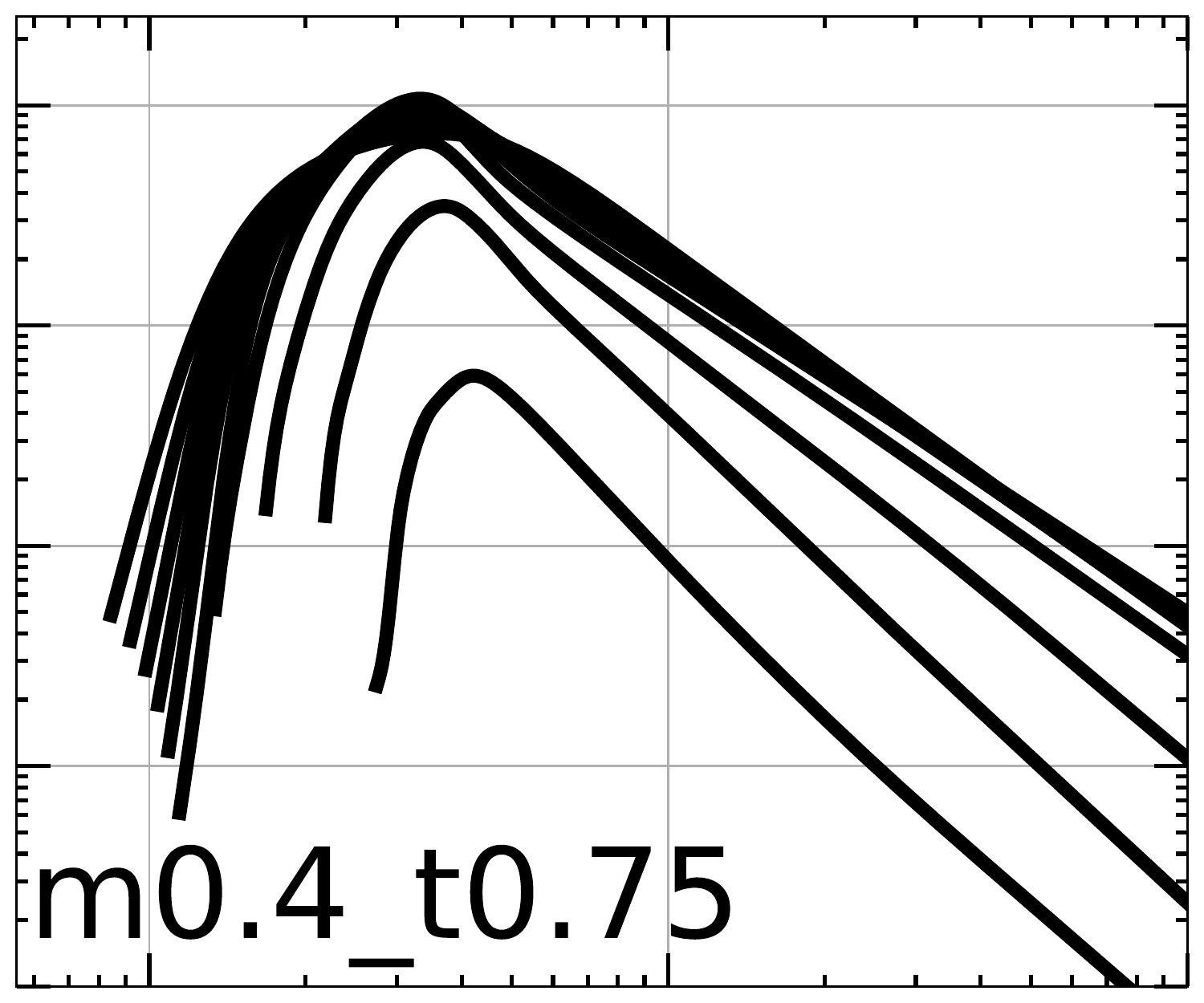}
\plotone{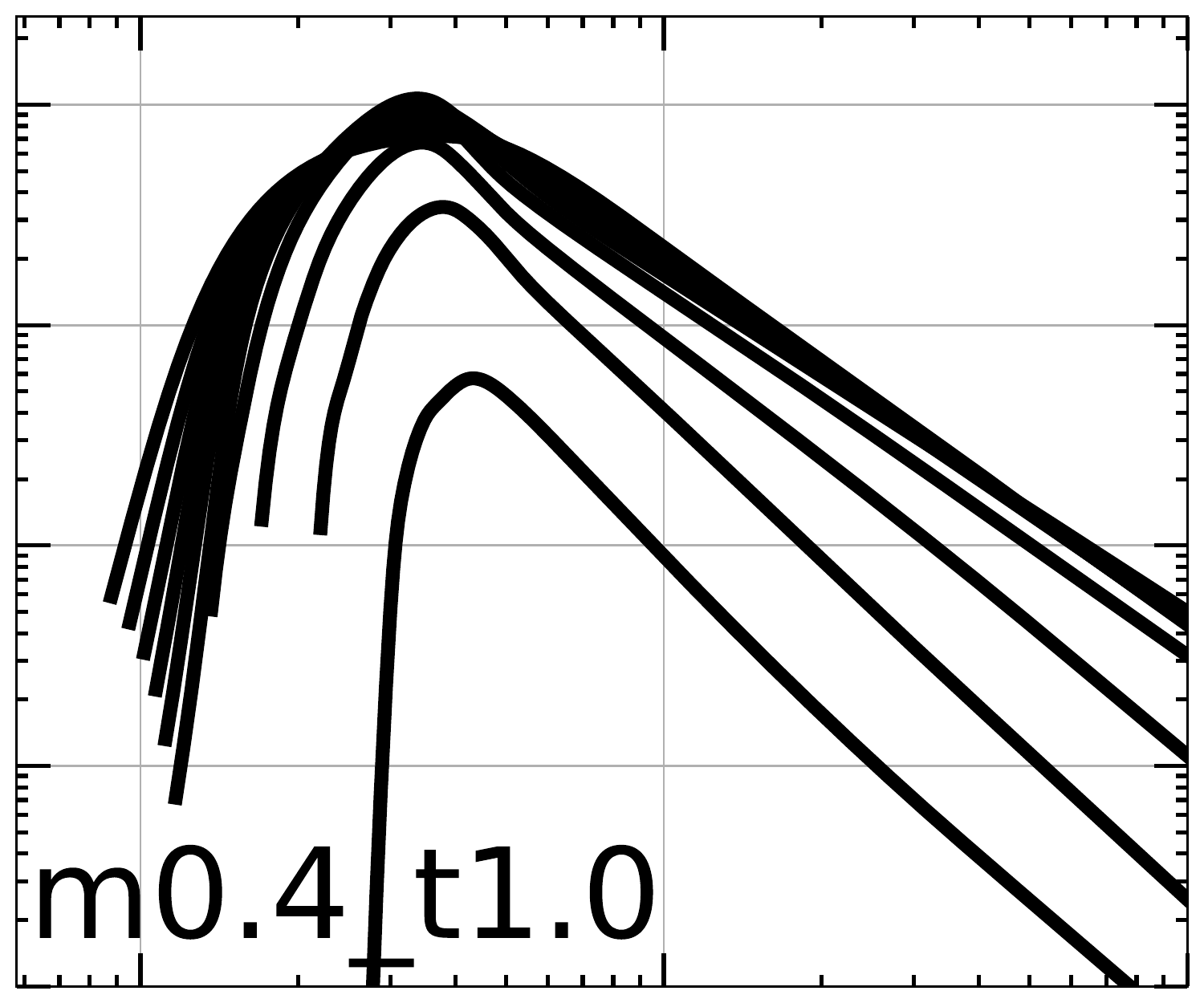}
\plotone{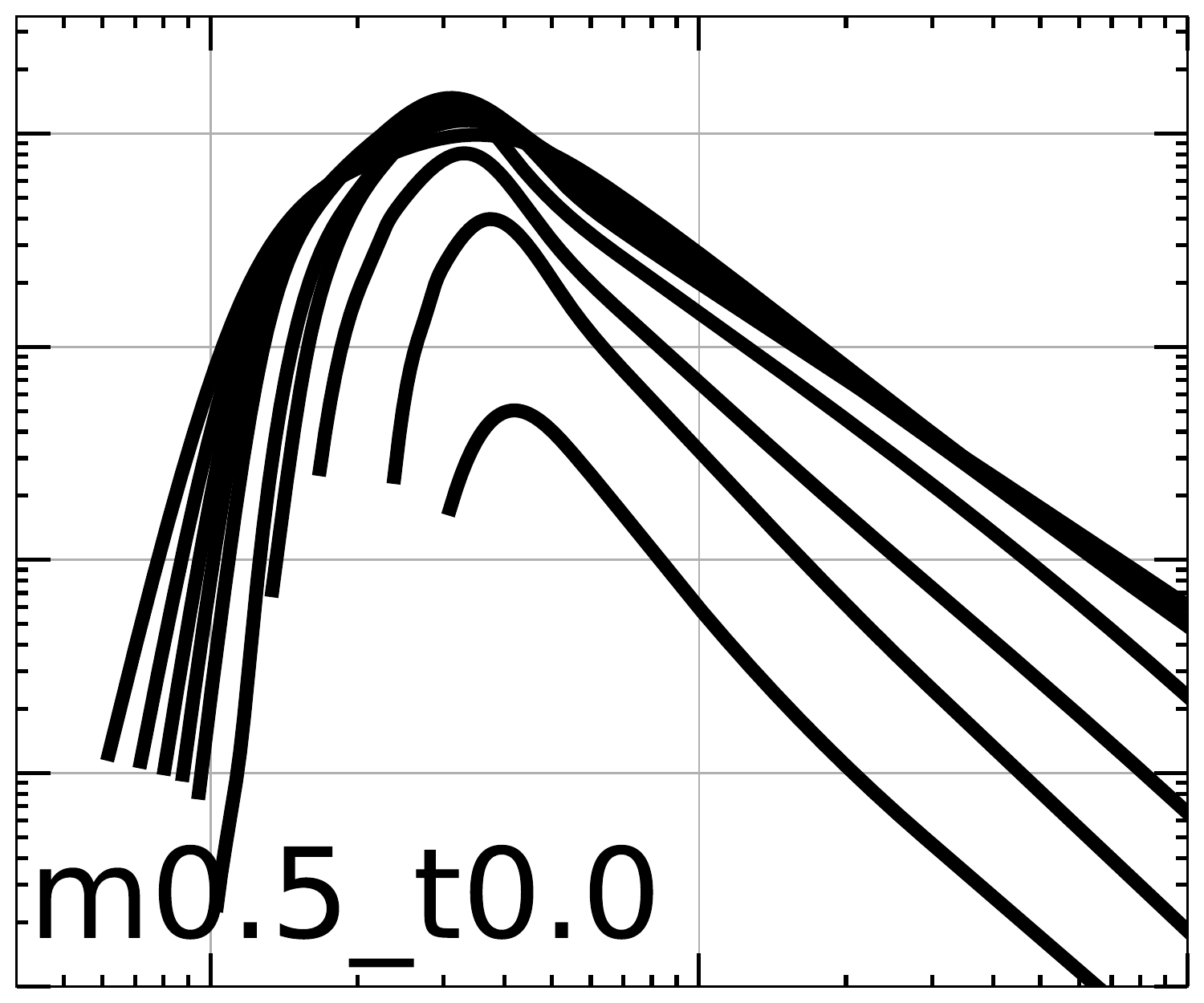}
\plotone{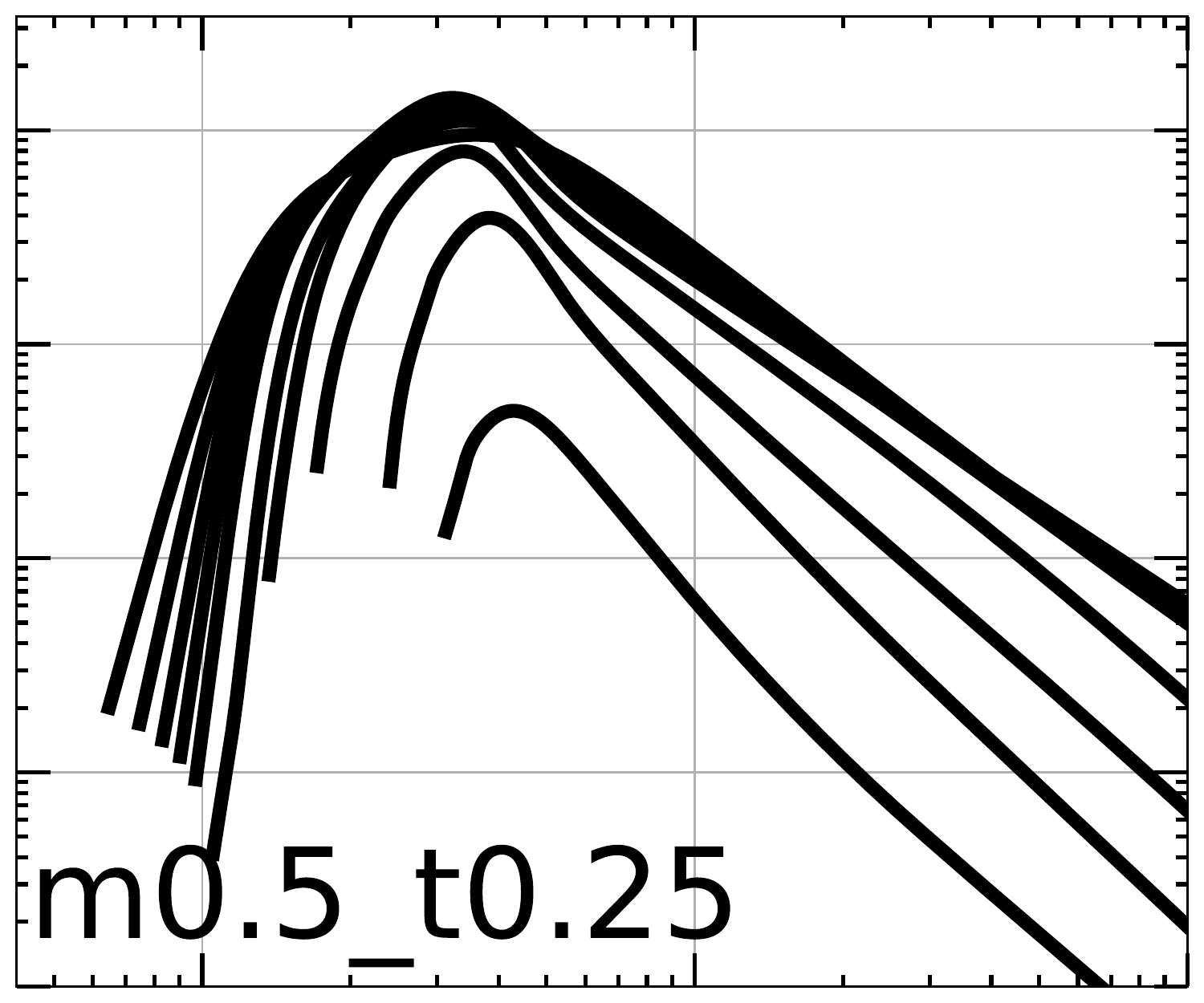}
\plotone{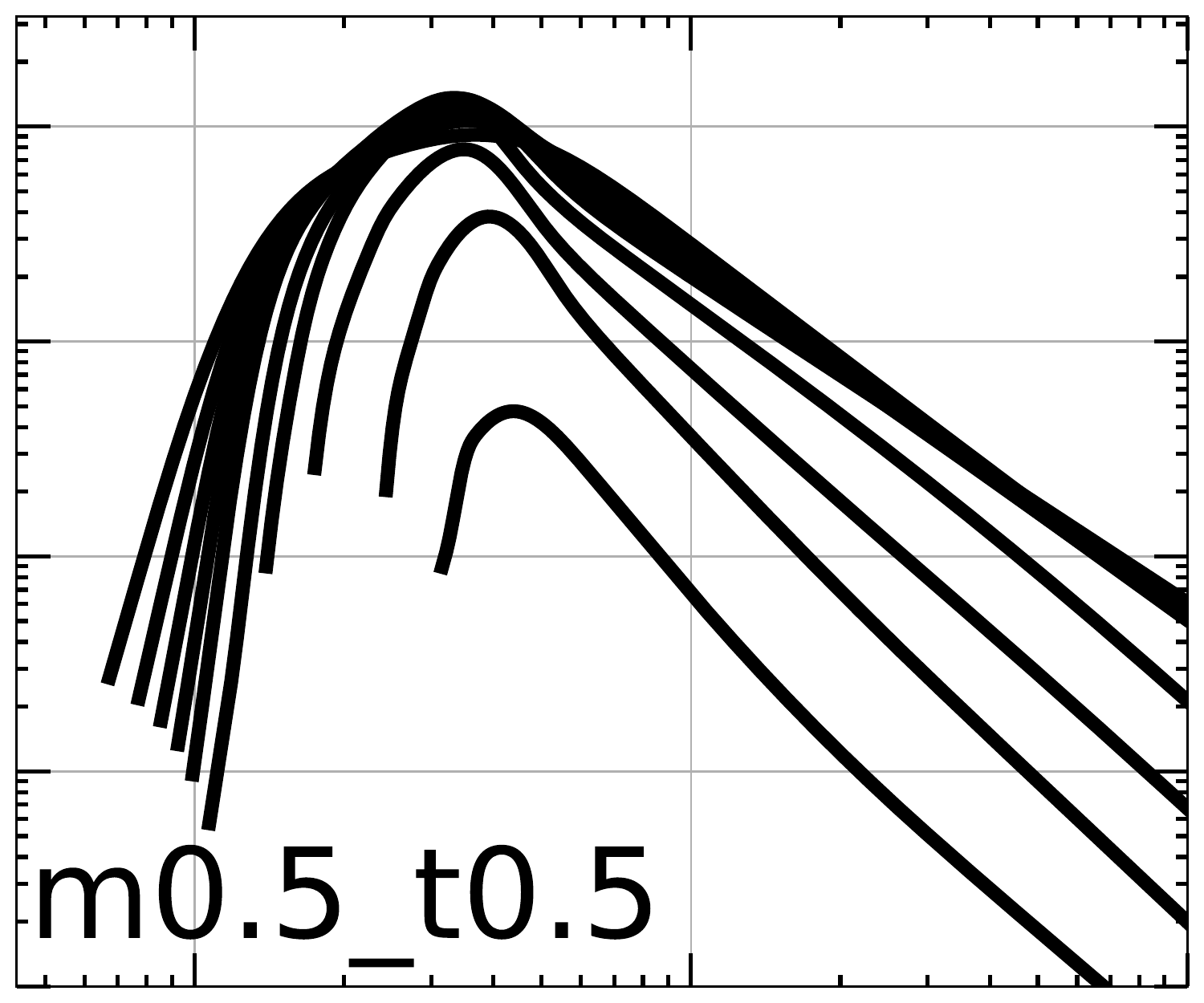}
\plotone{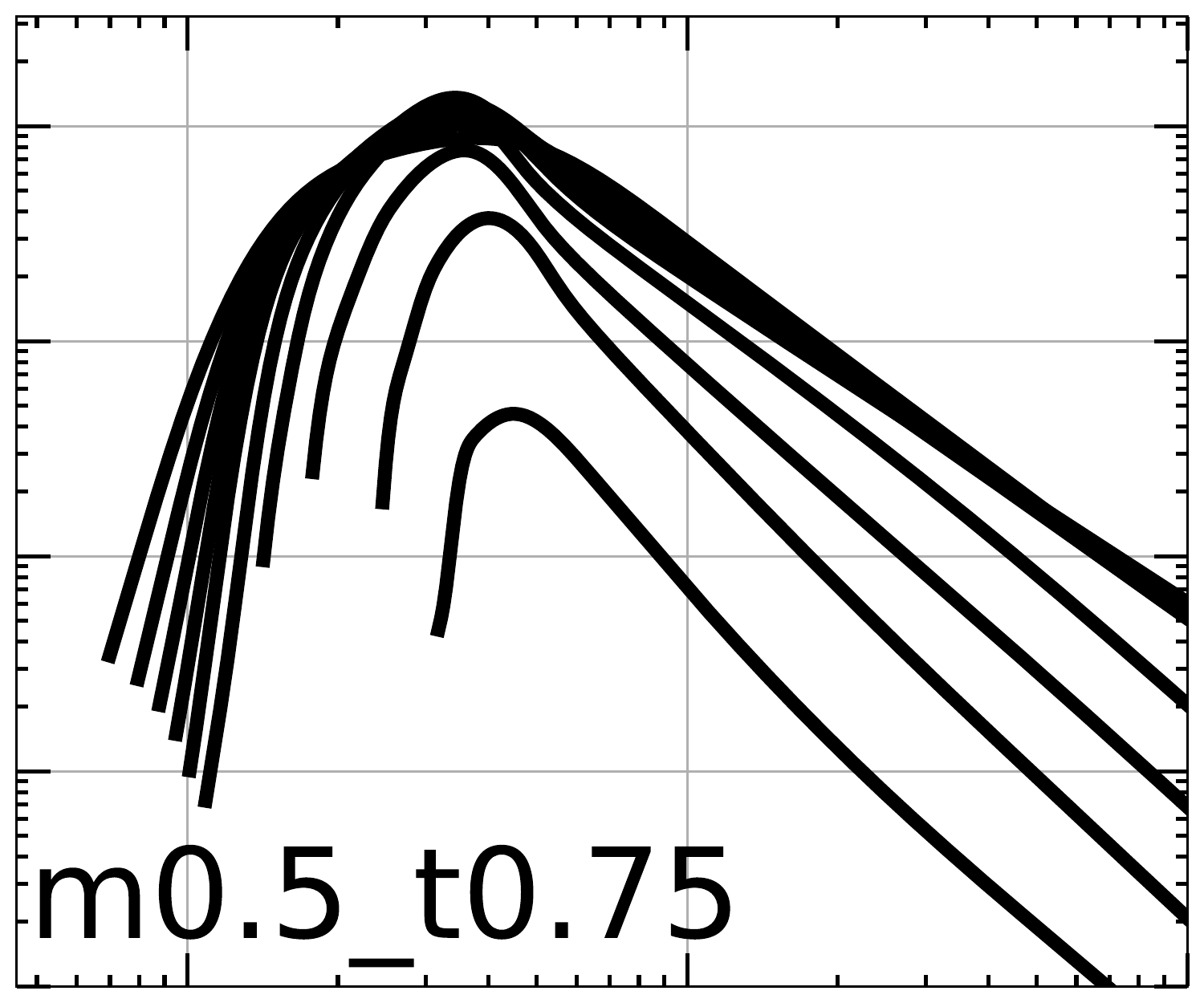}
\plotone{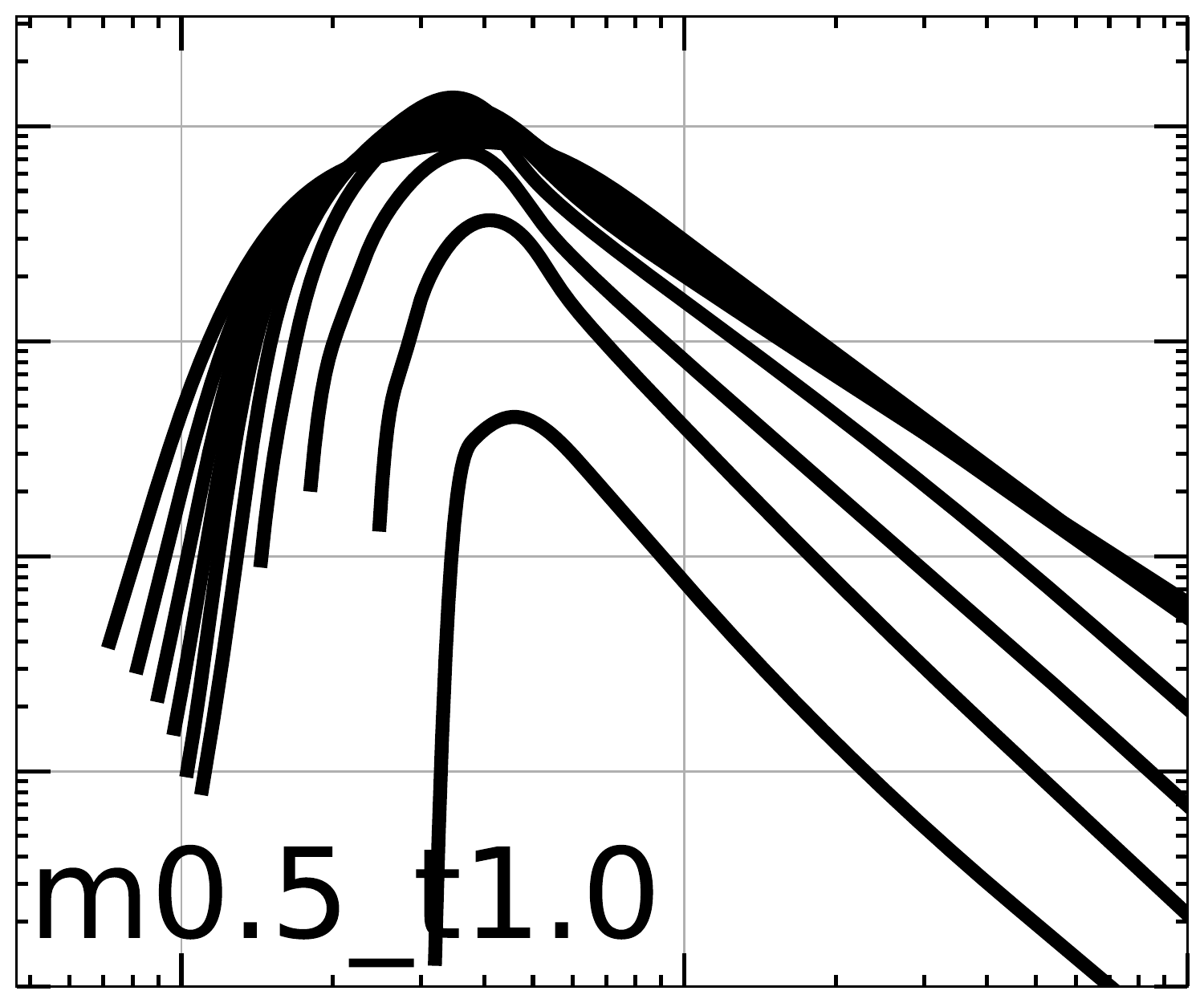}
\plotone{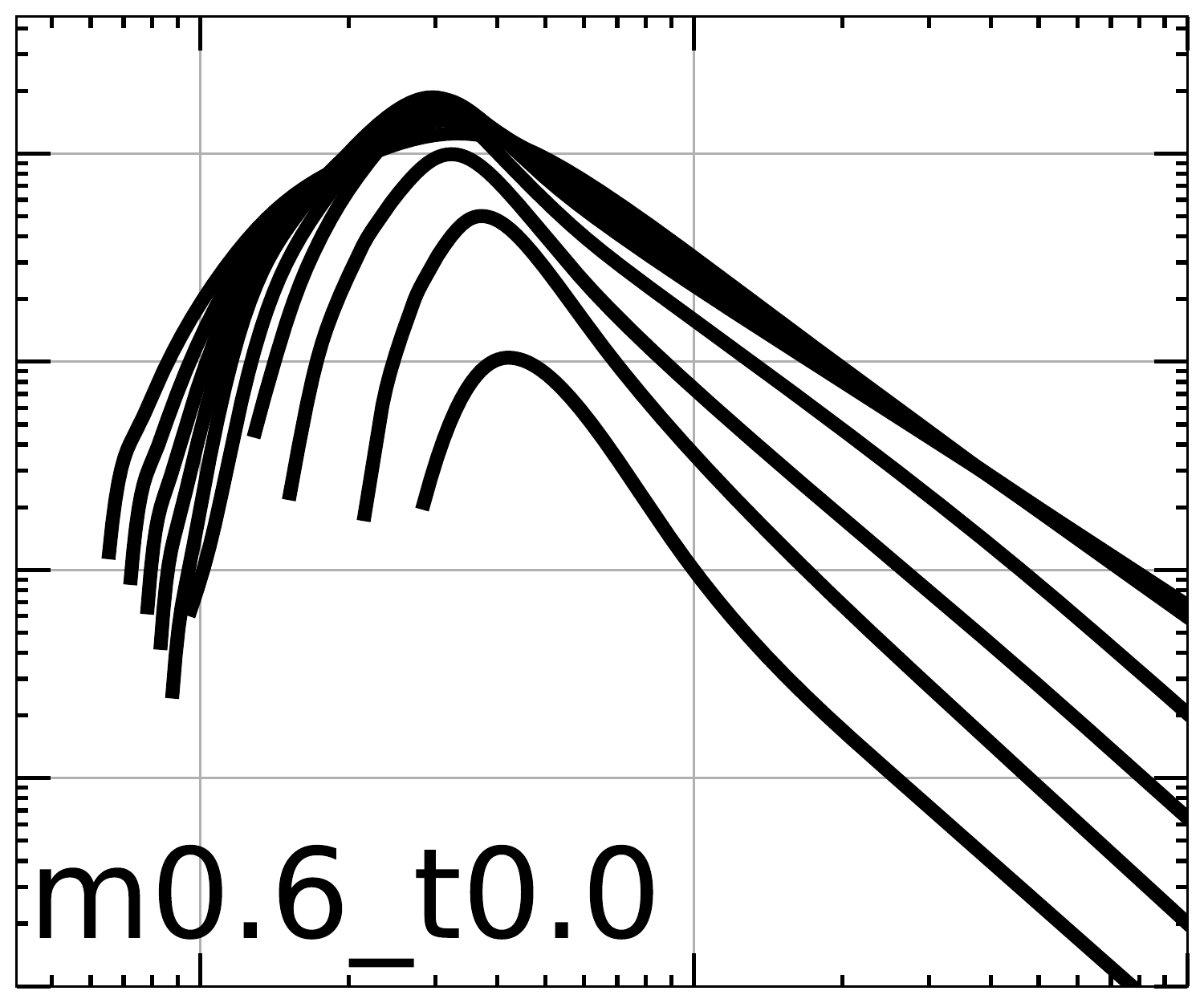}
\plotone{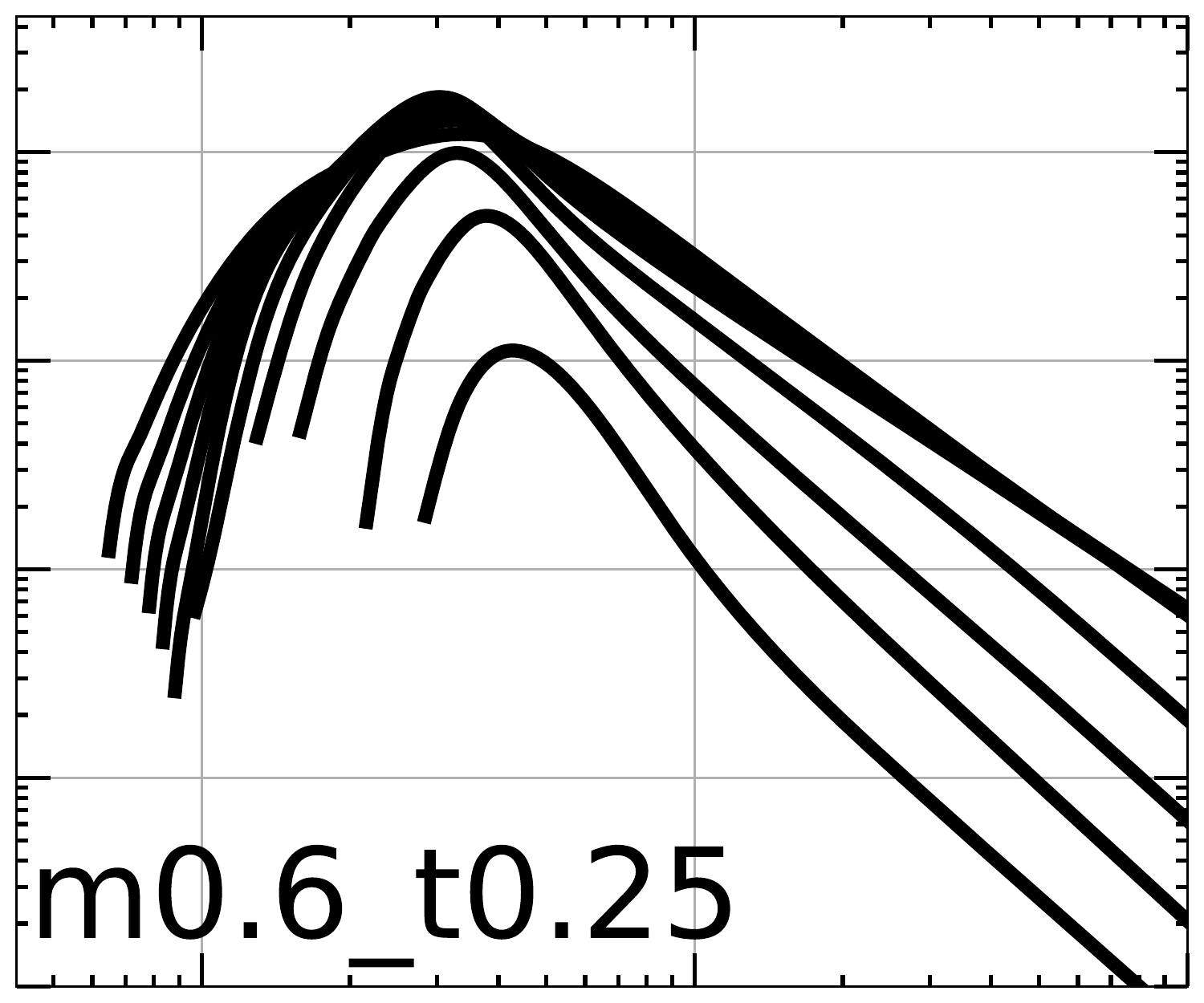}
\plotone{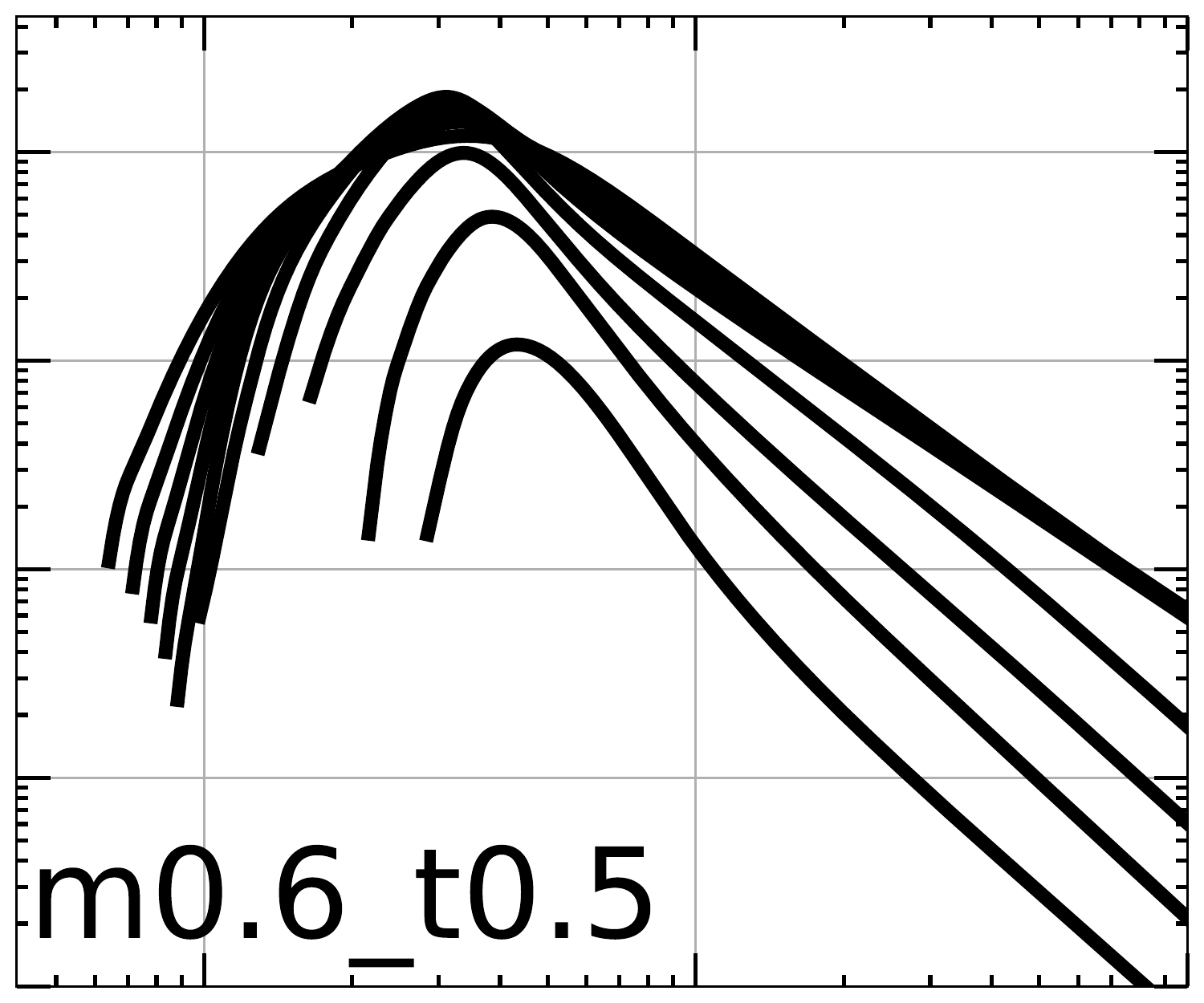}
\plotone{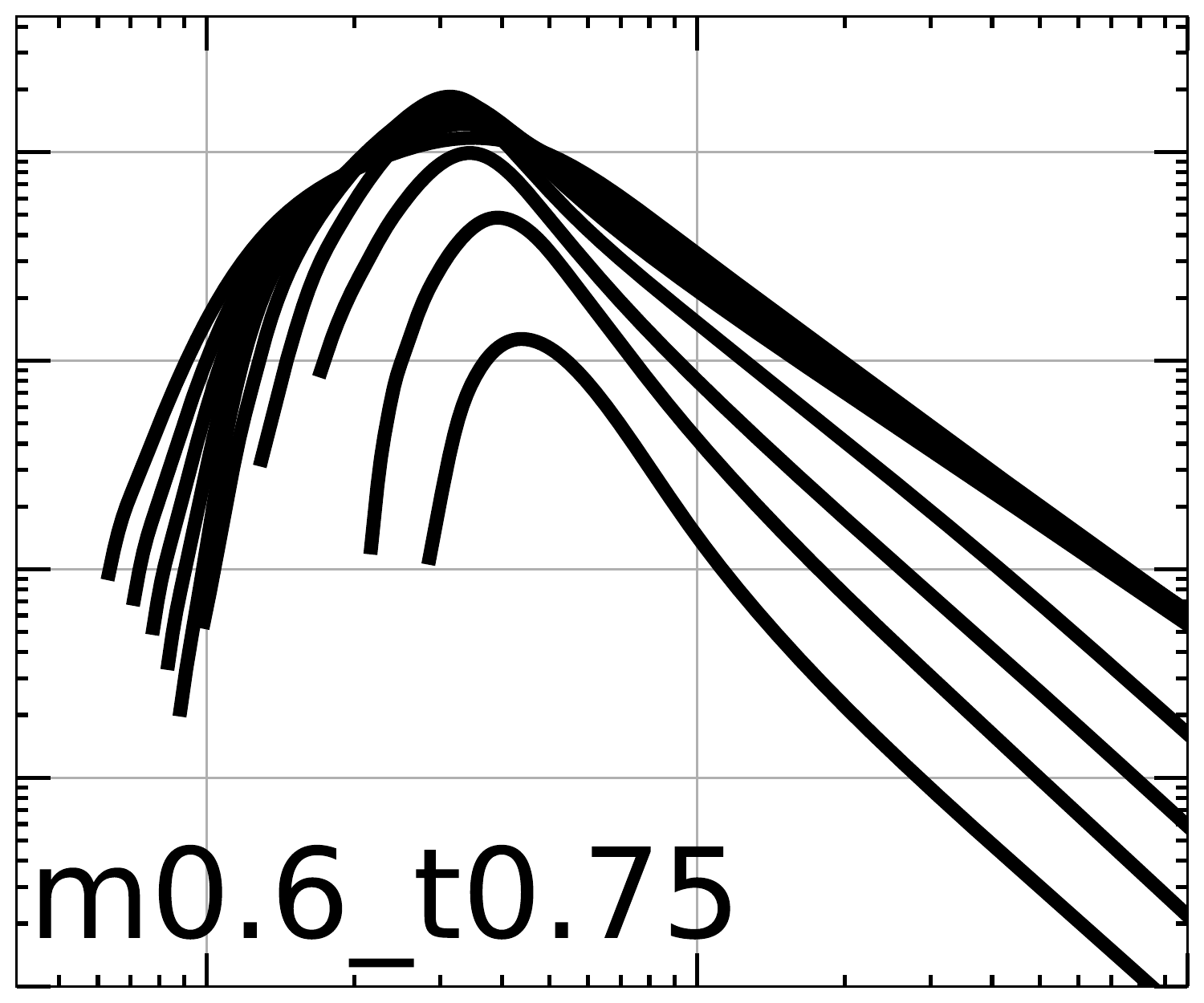}
\plotone{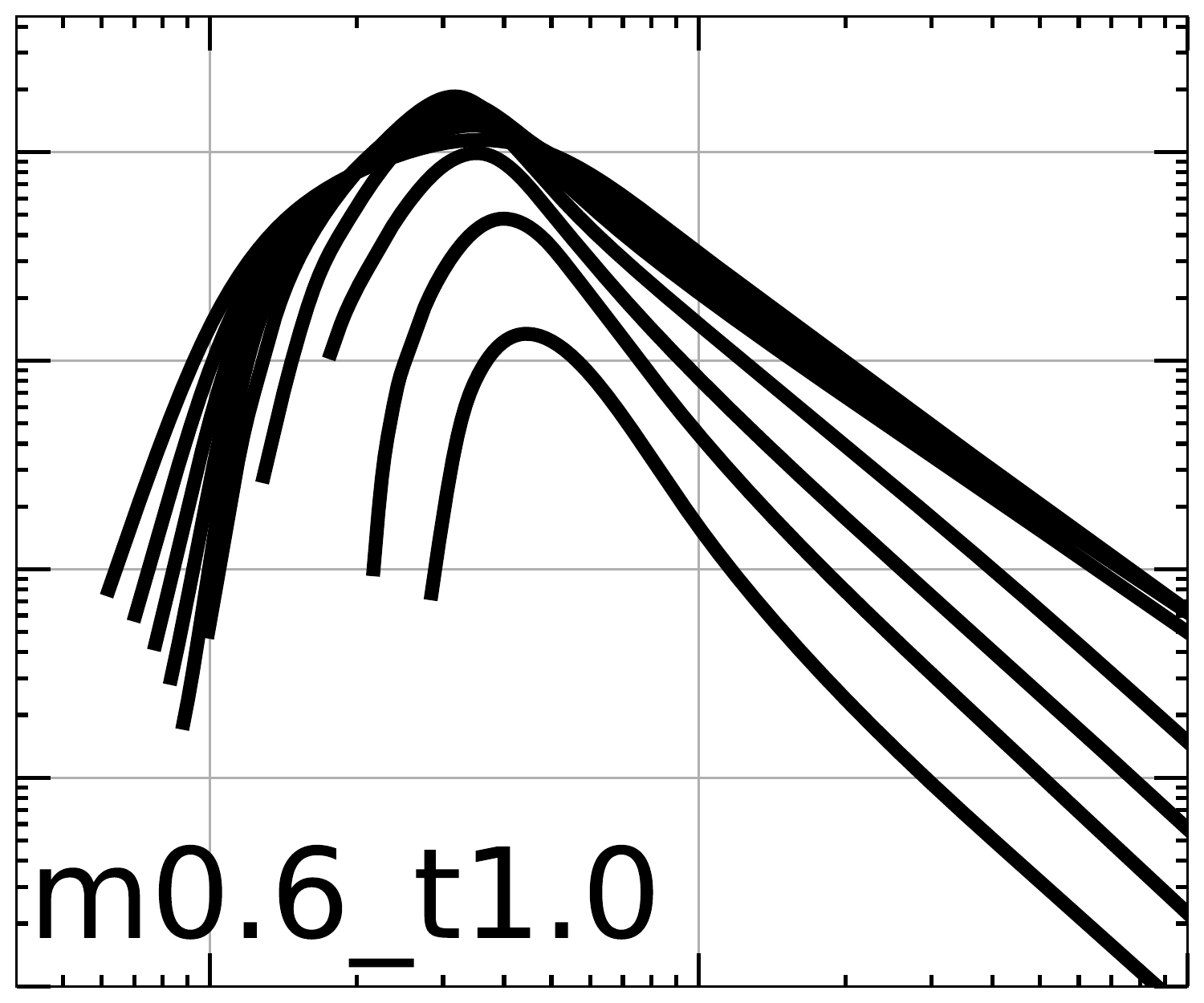}
\plotone{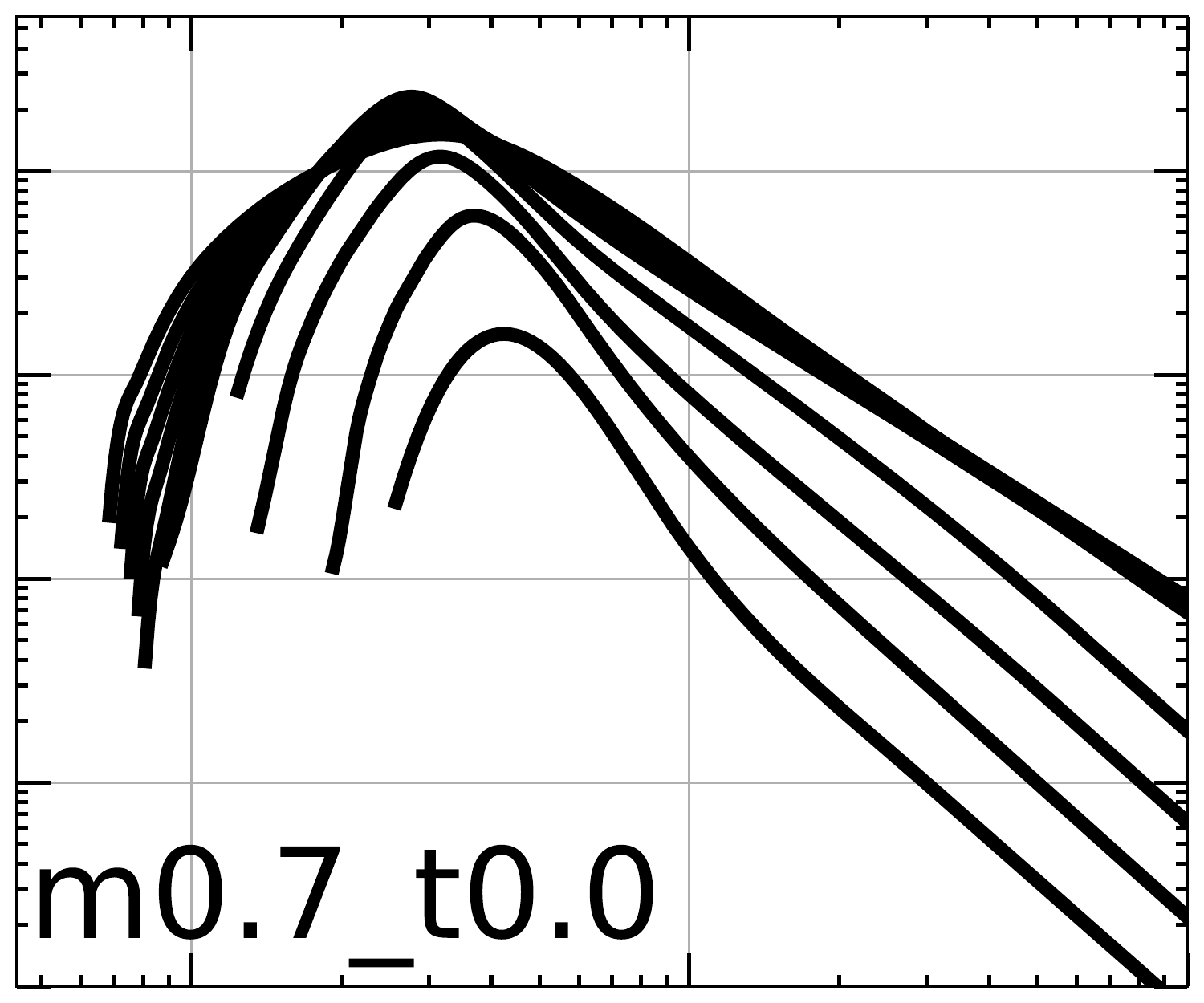}
\plotone{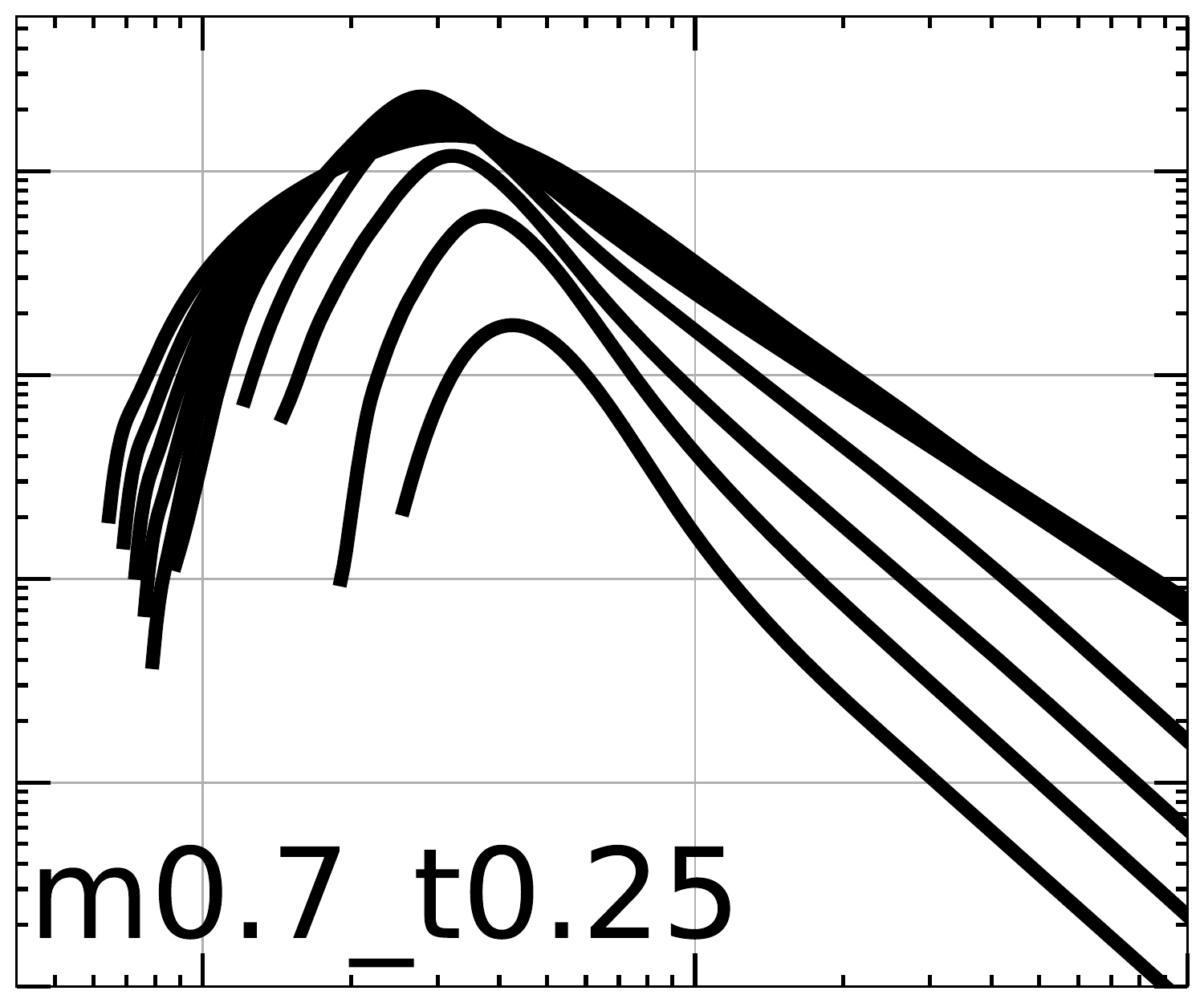}
\plotone{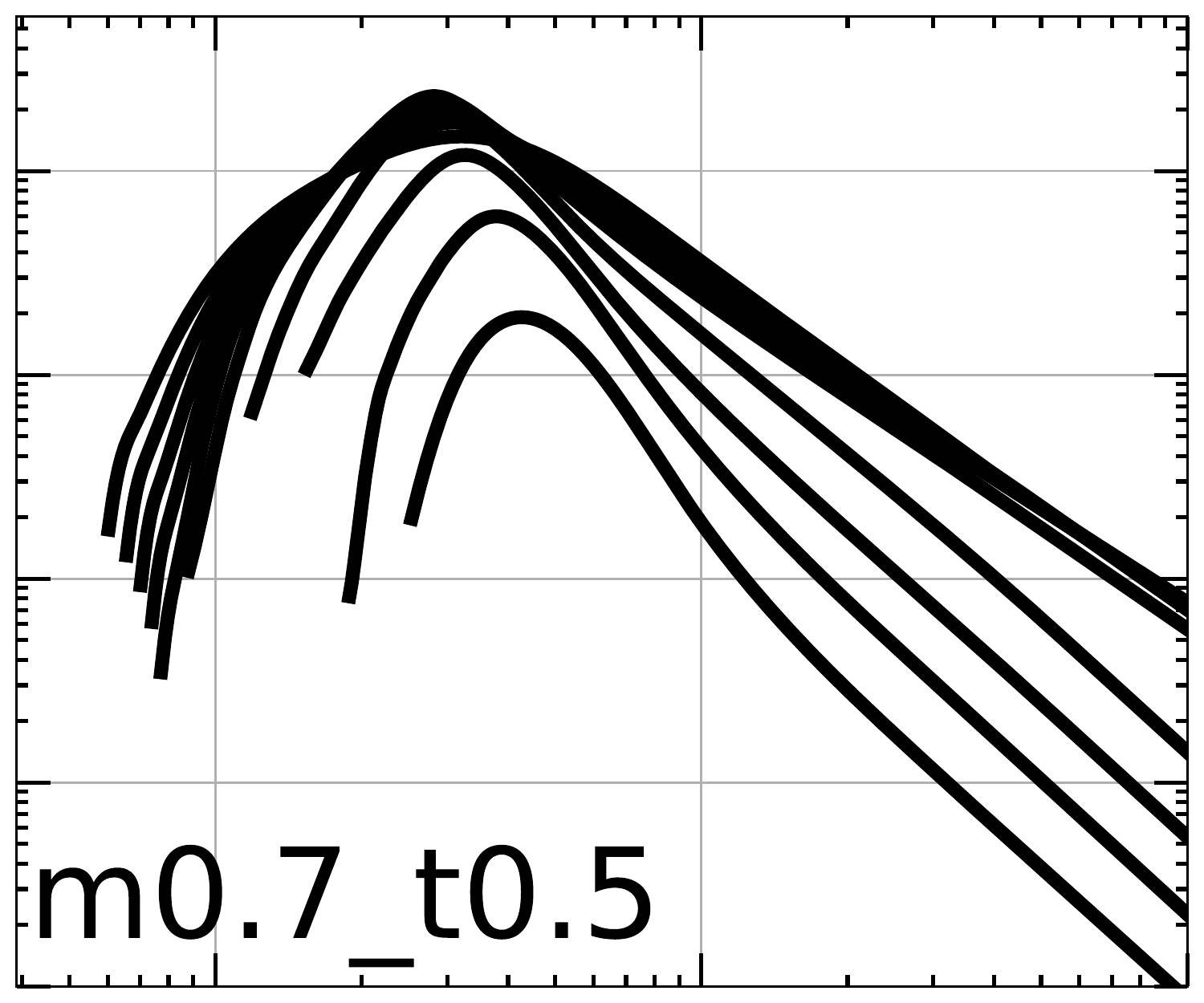}
\plotone{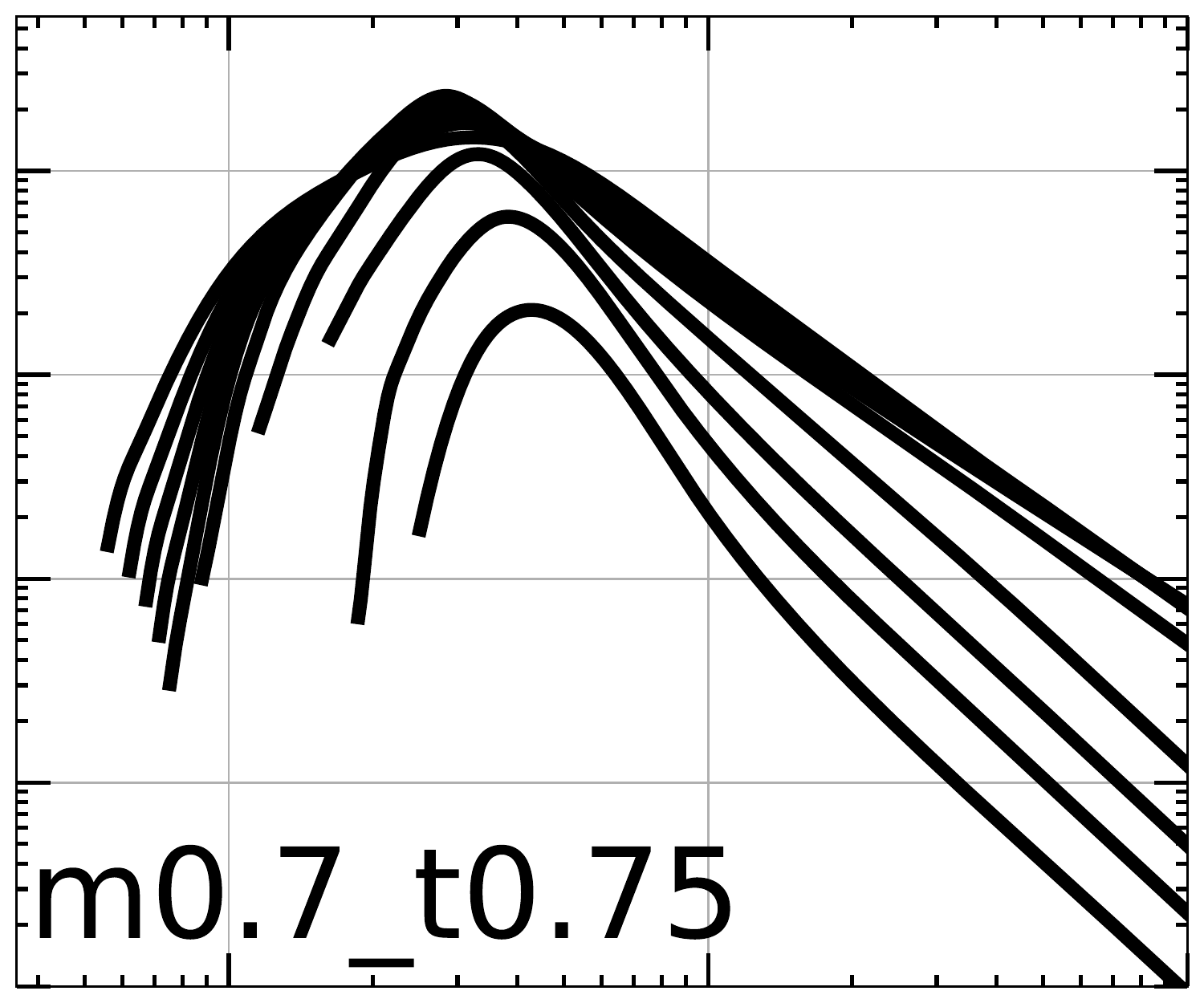}
\plotone{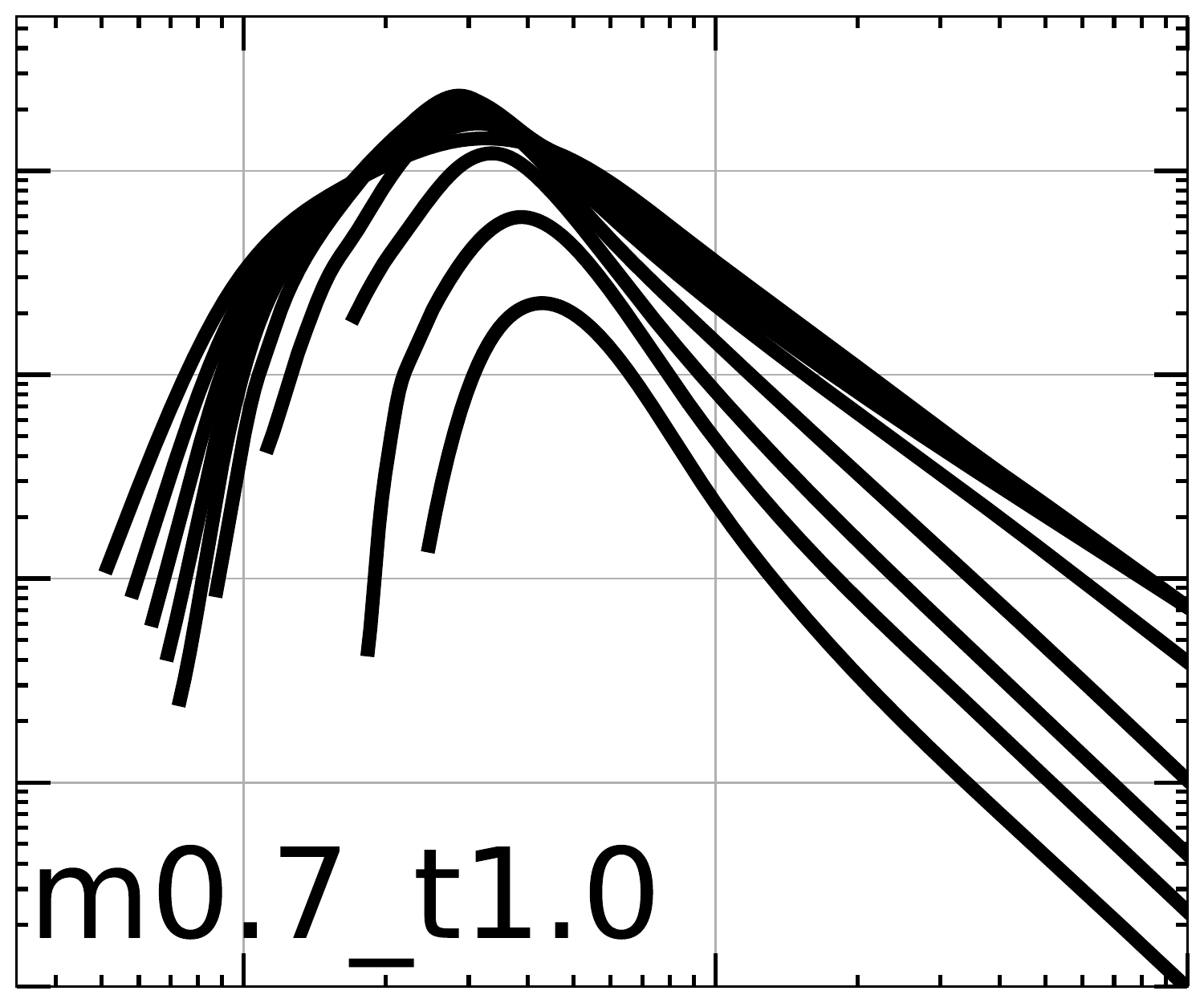}
\plotone{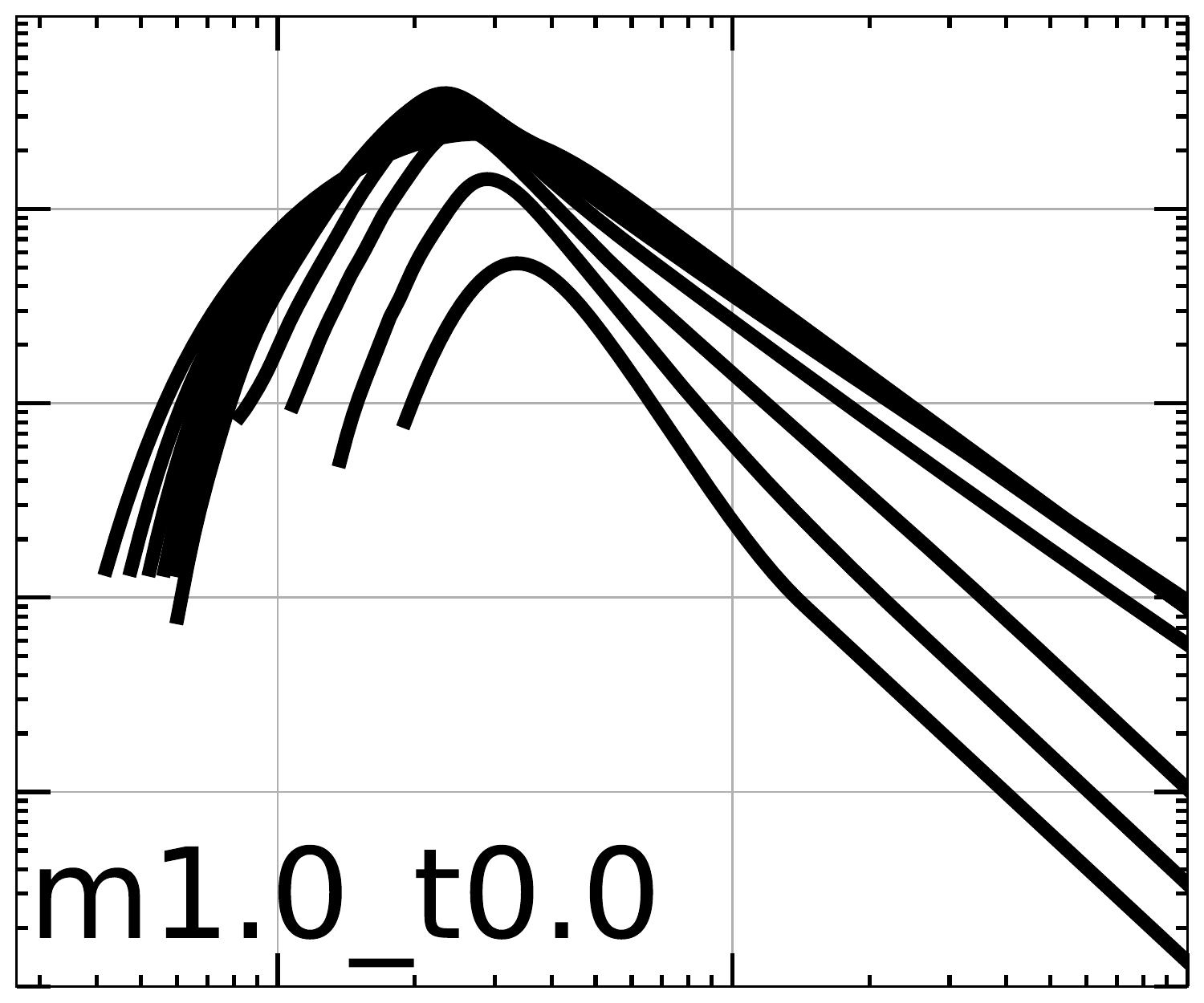}
\plotone{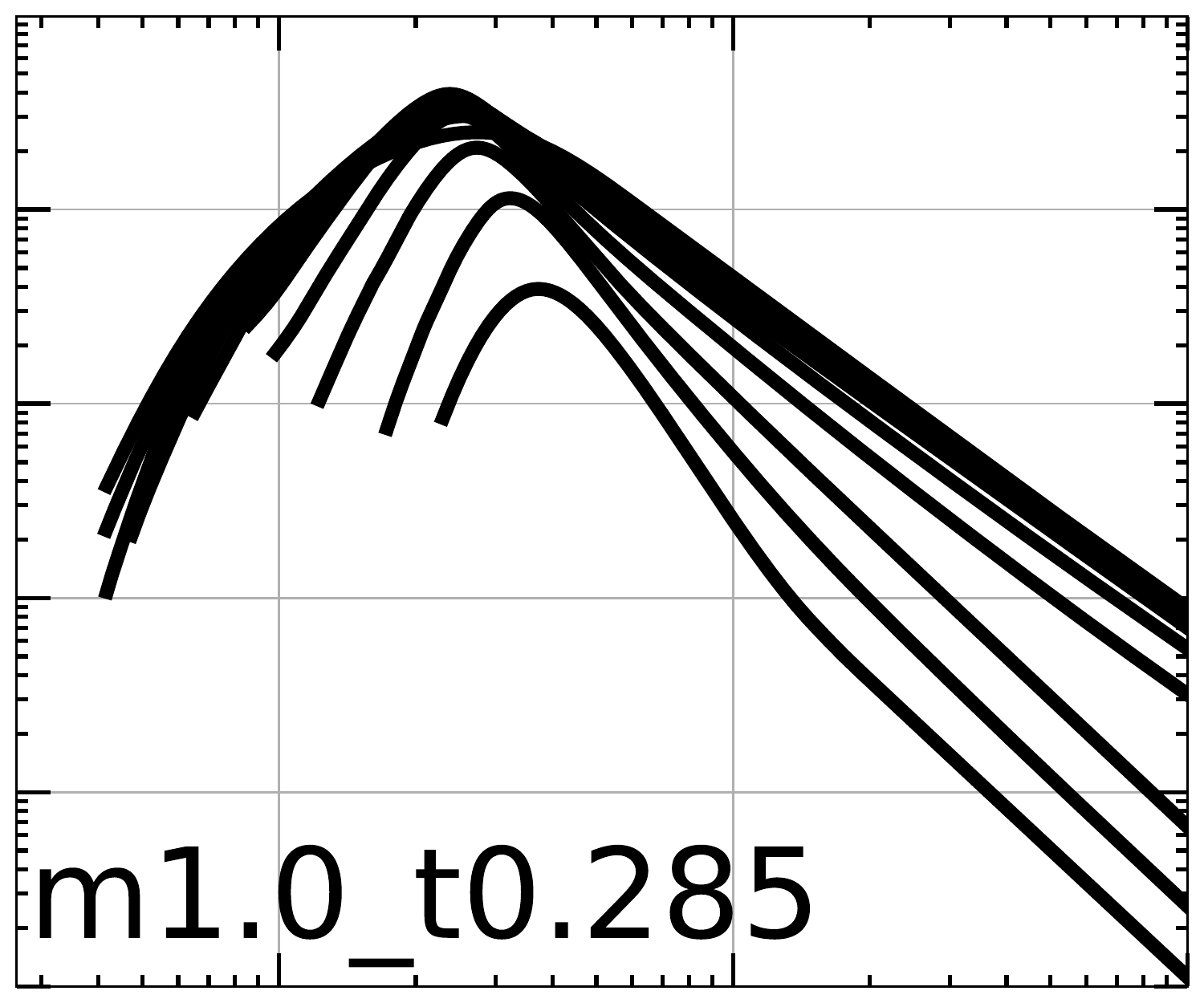}
\plotone{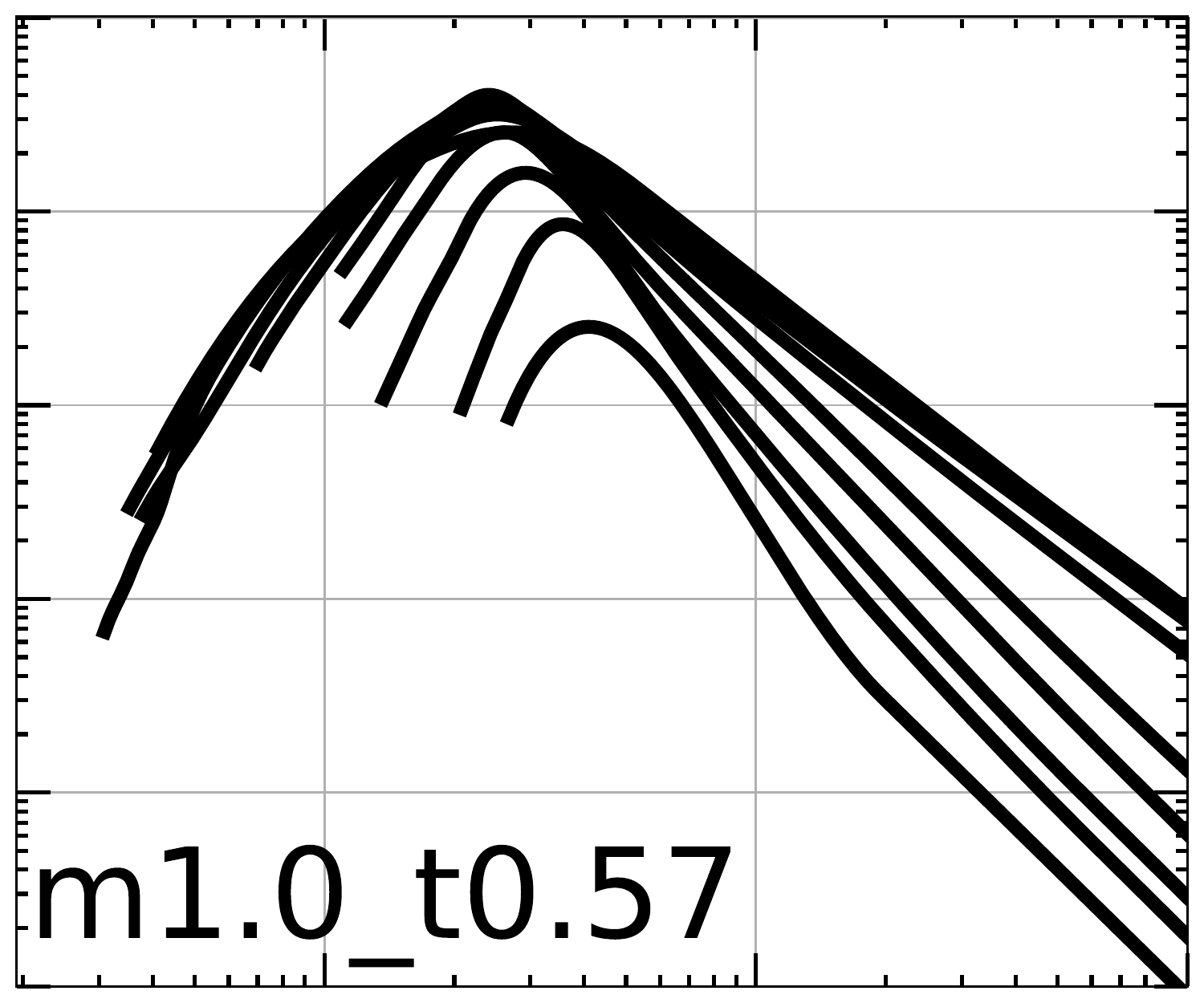}
\plotone{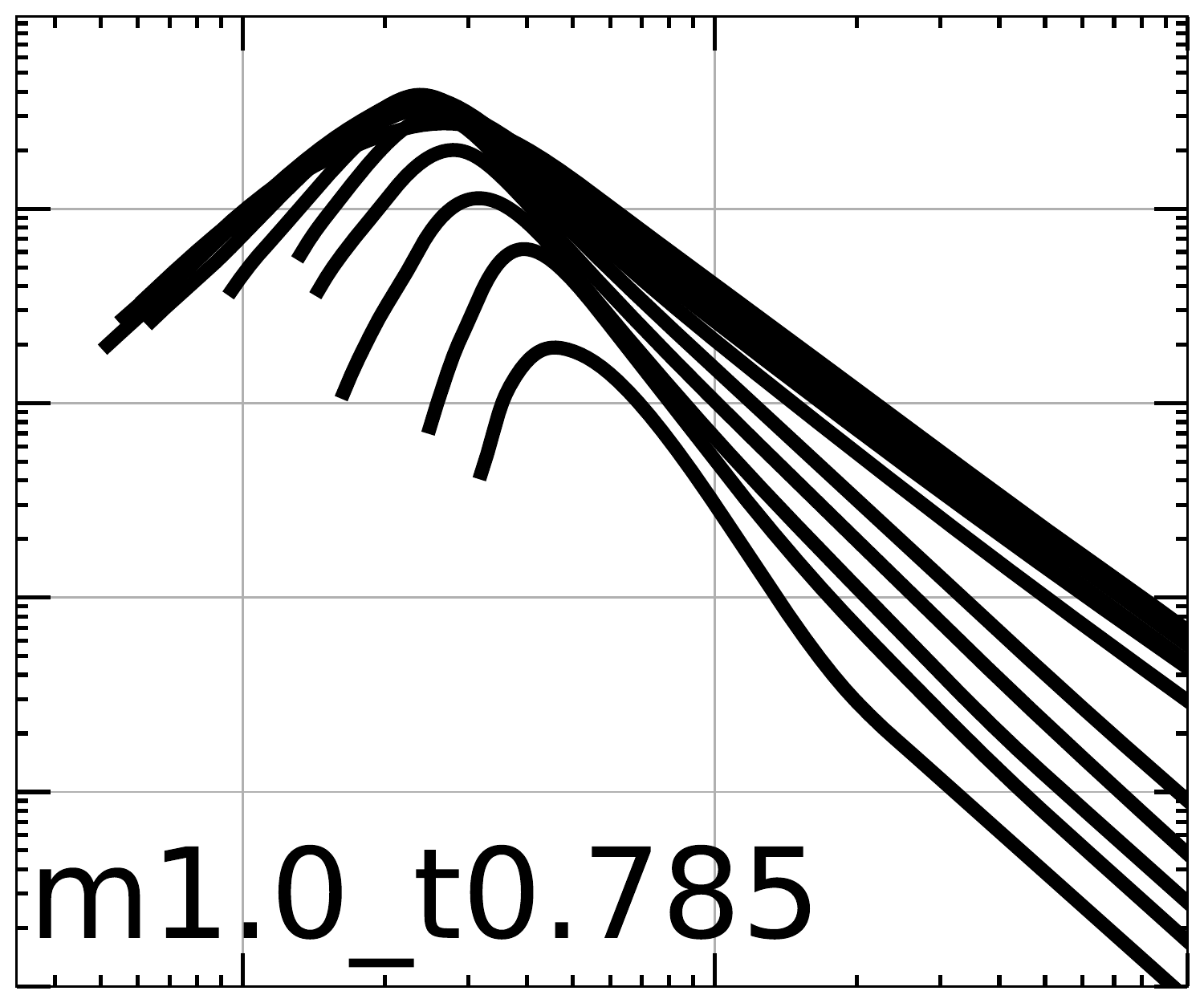}
\plotone{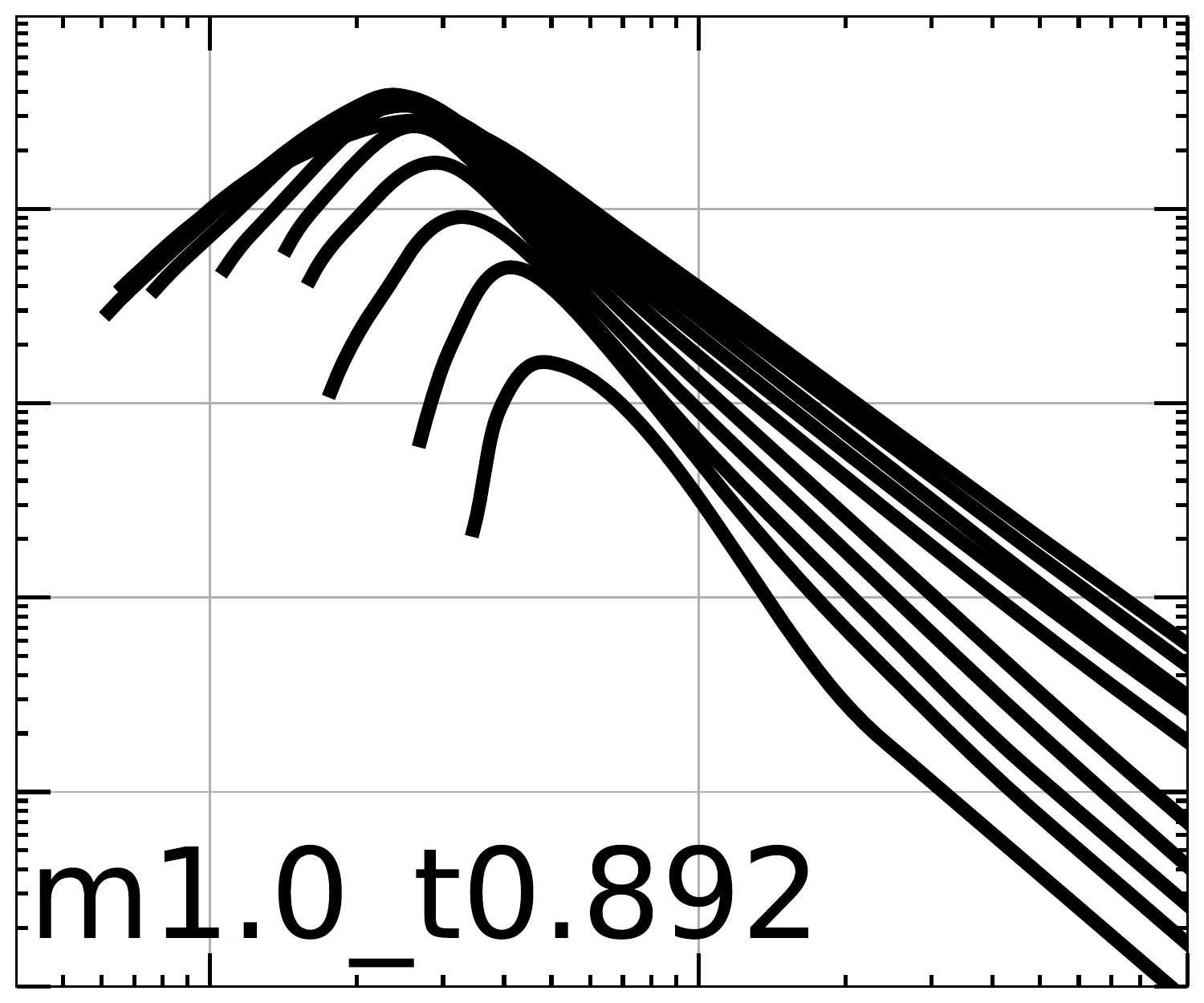}
\plotone{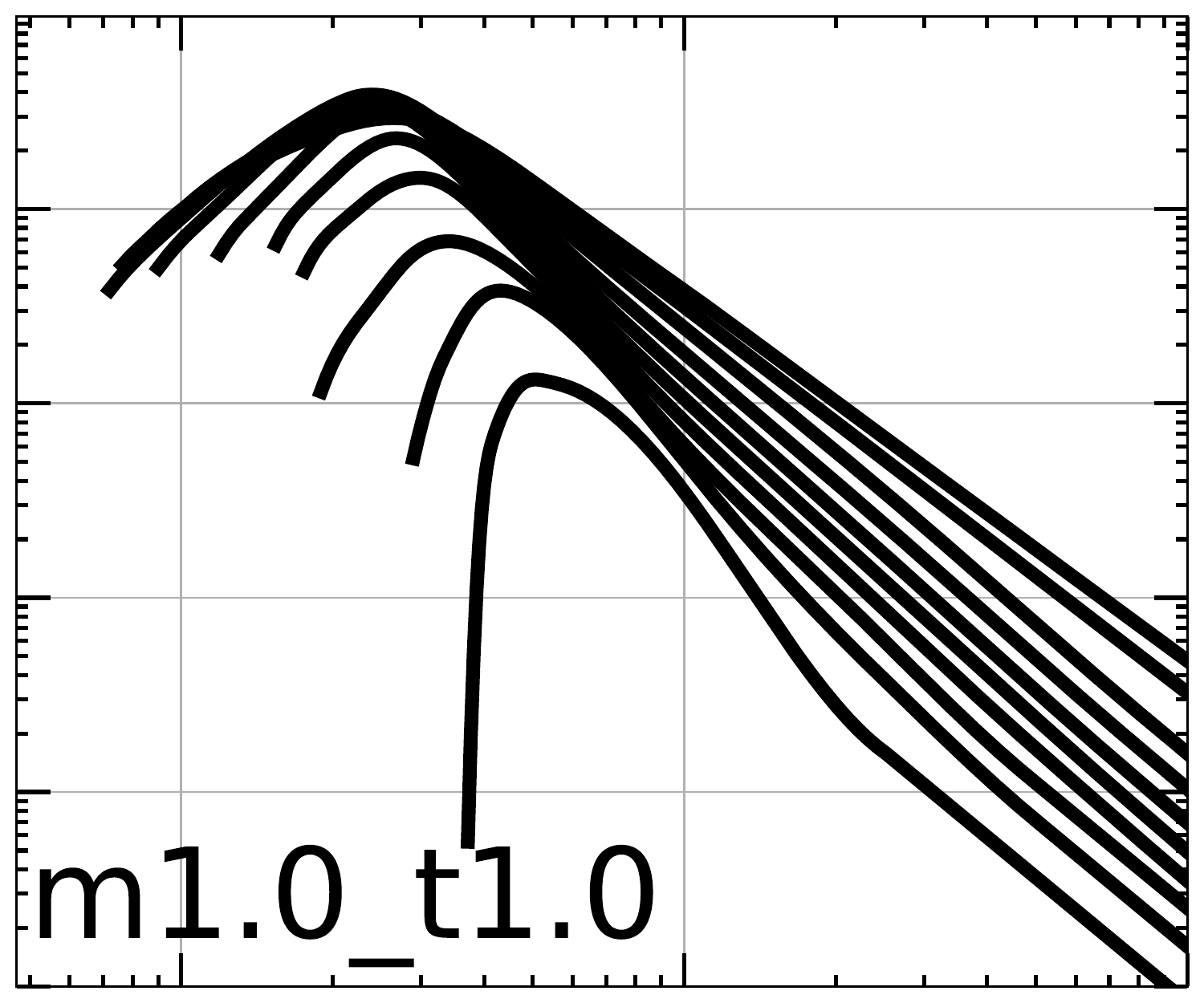}
\plotone{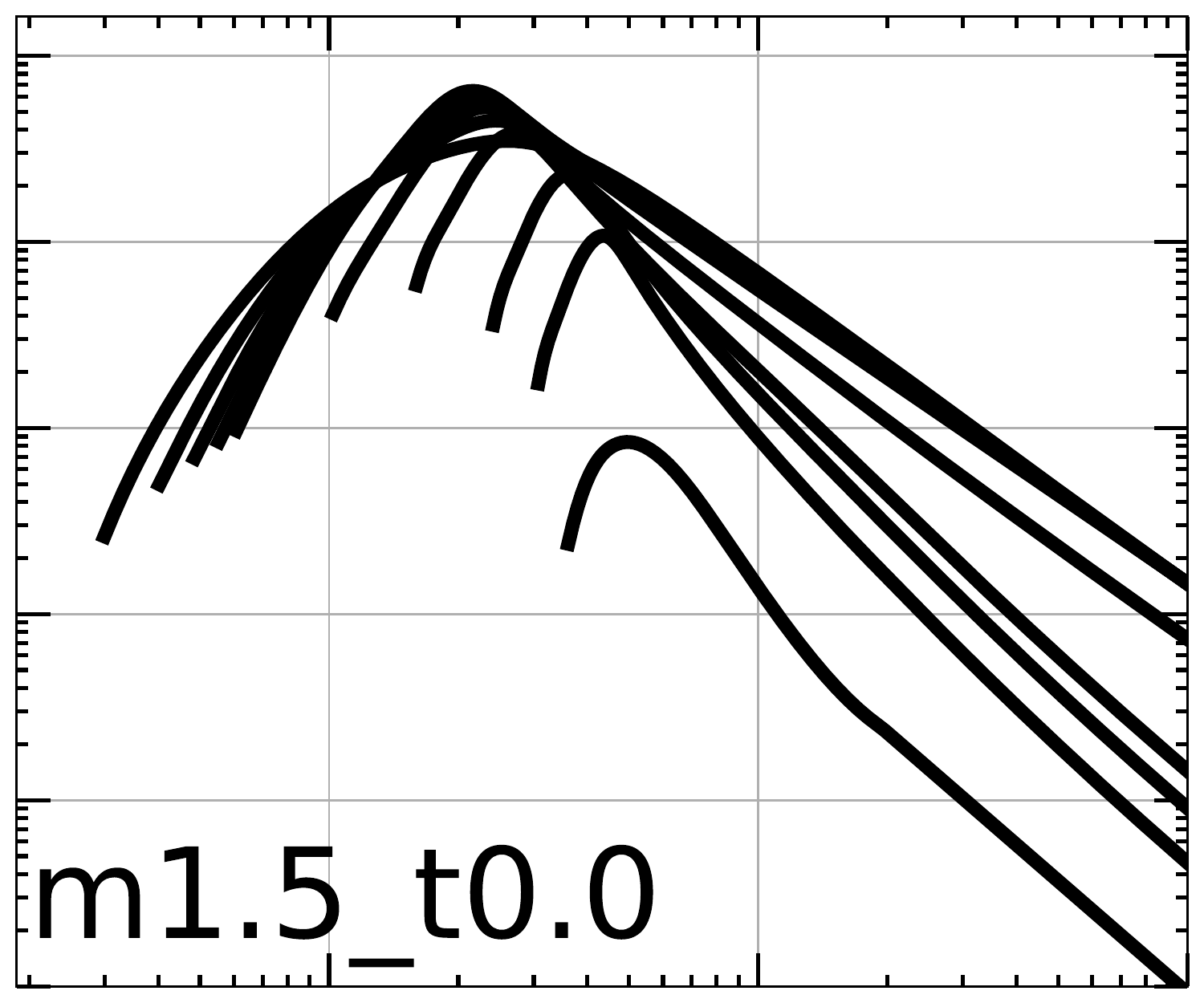}
\plotone{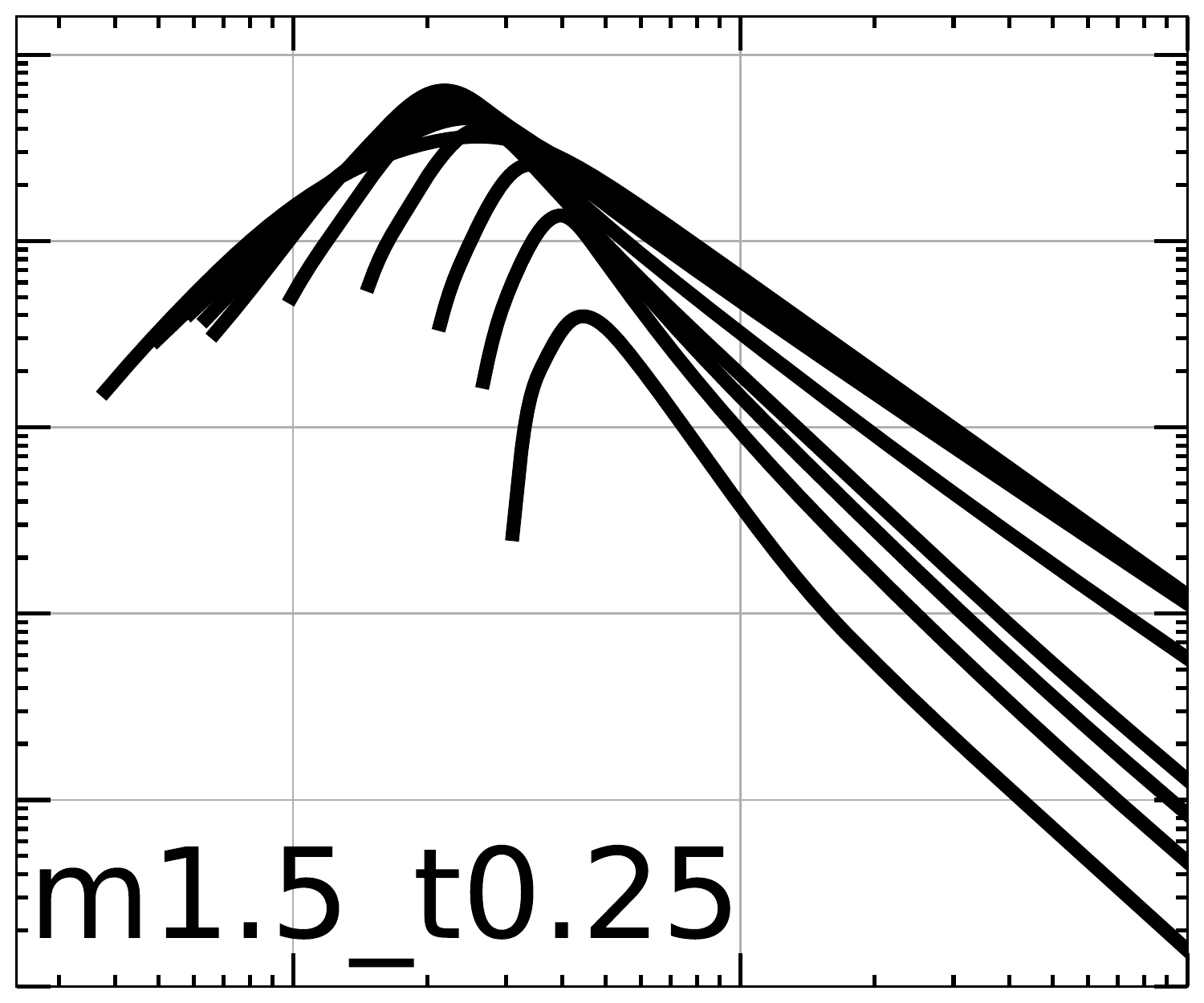}
\plotone{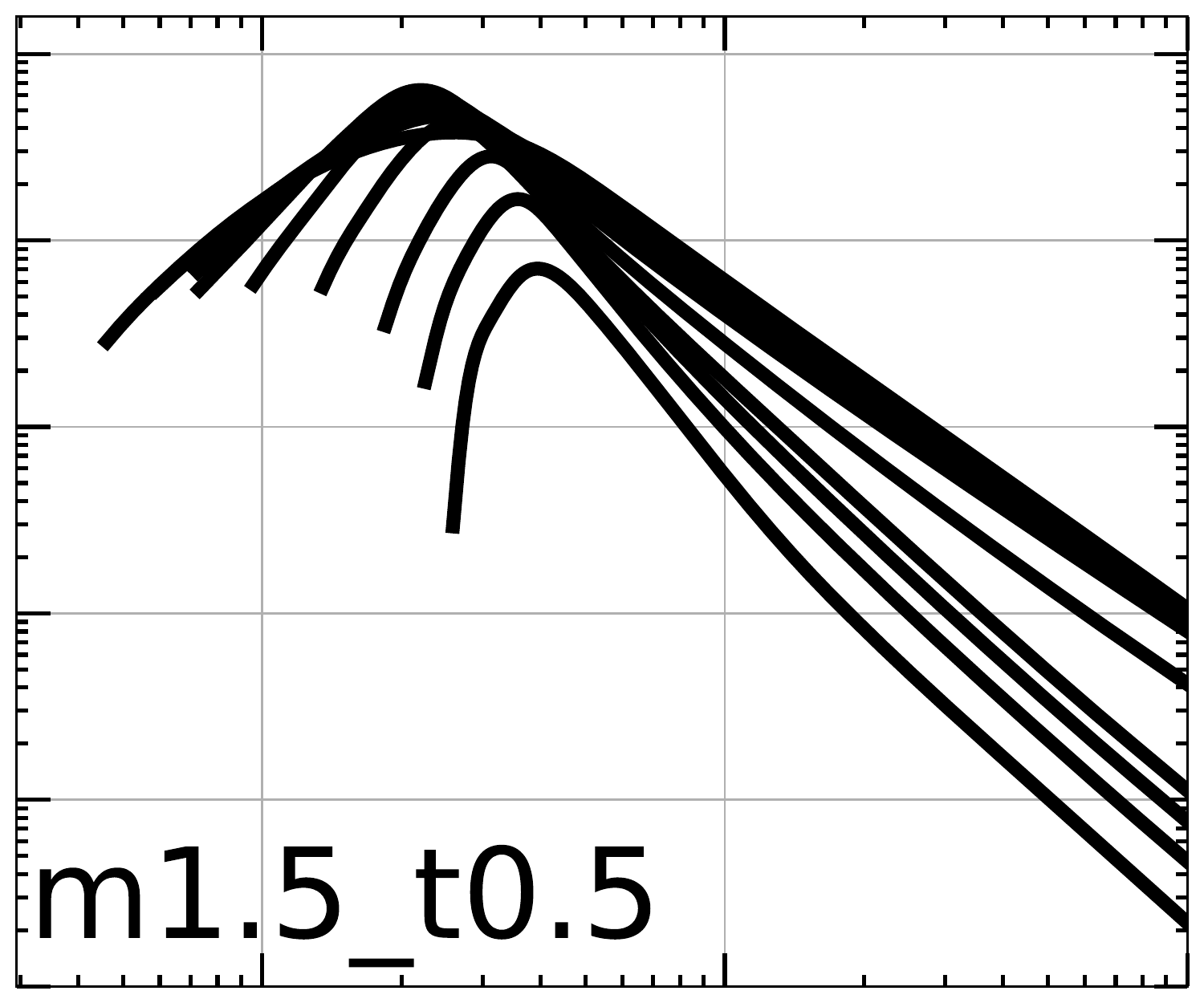}
\plotone{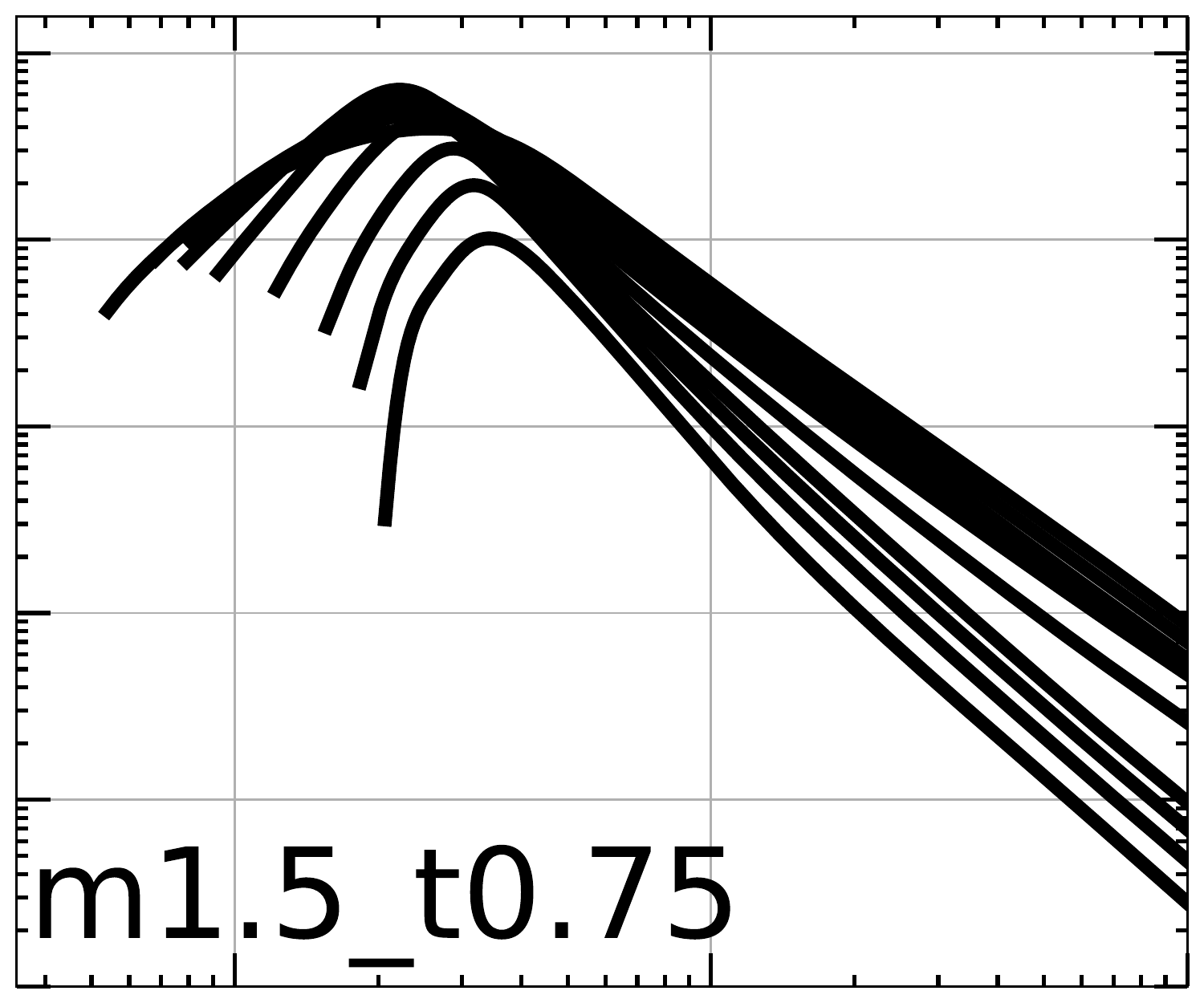}
\plotone{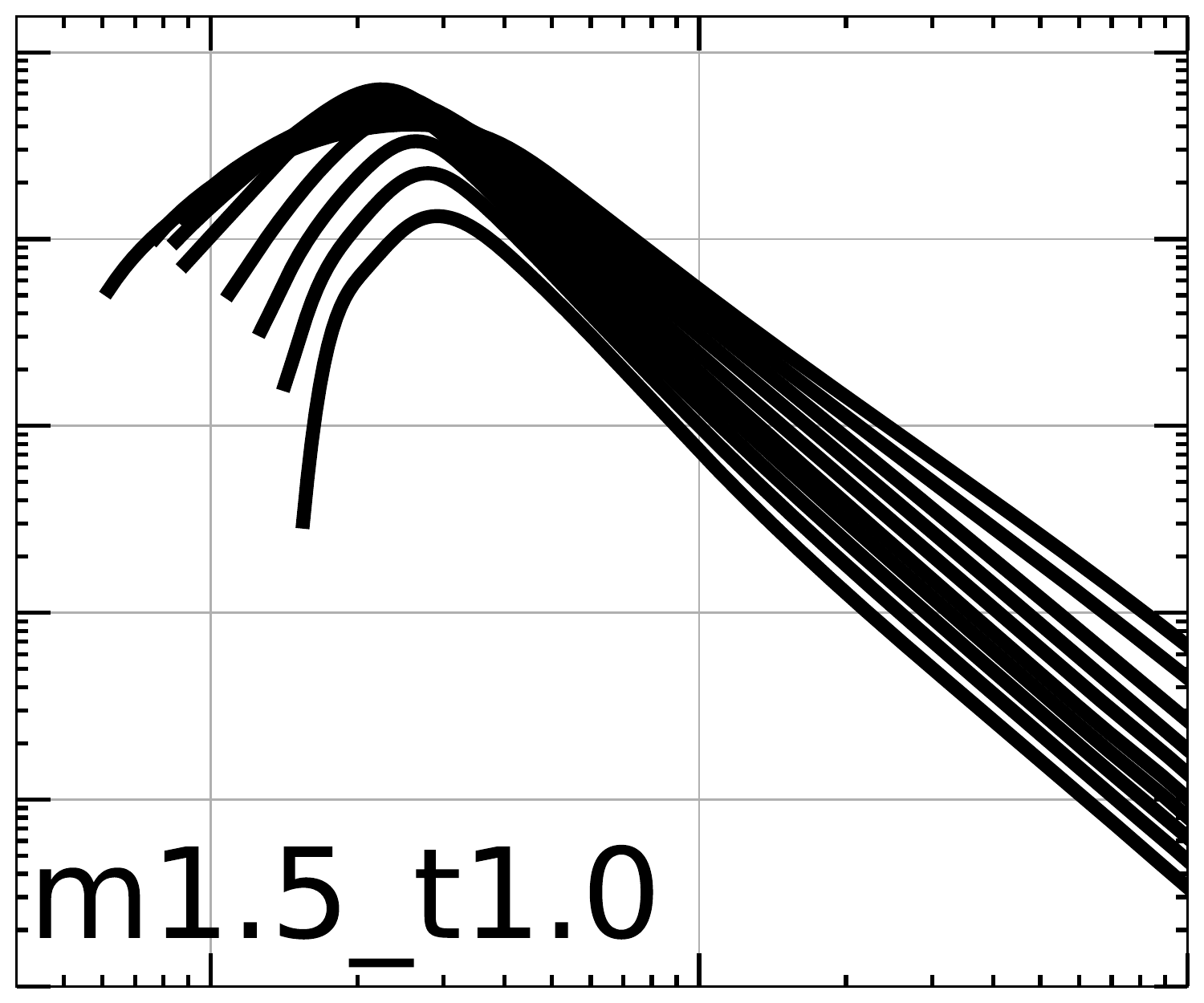}
\plotone{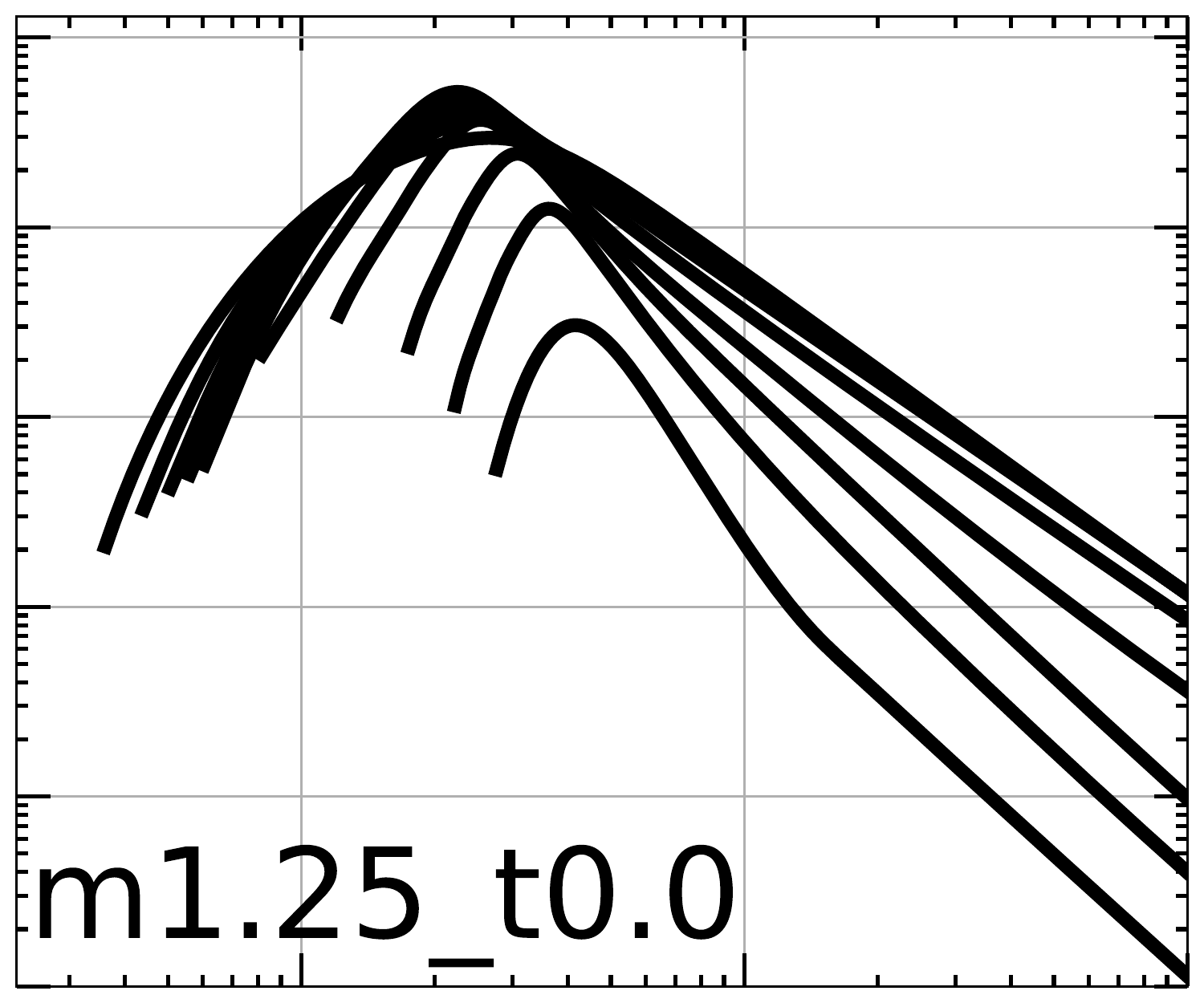}
\plotone{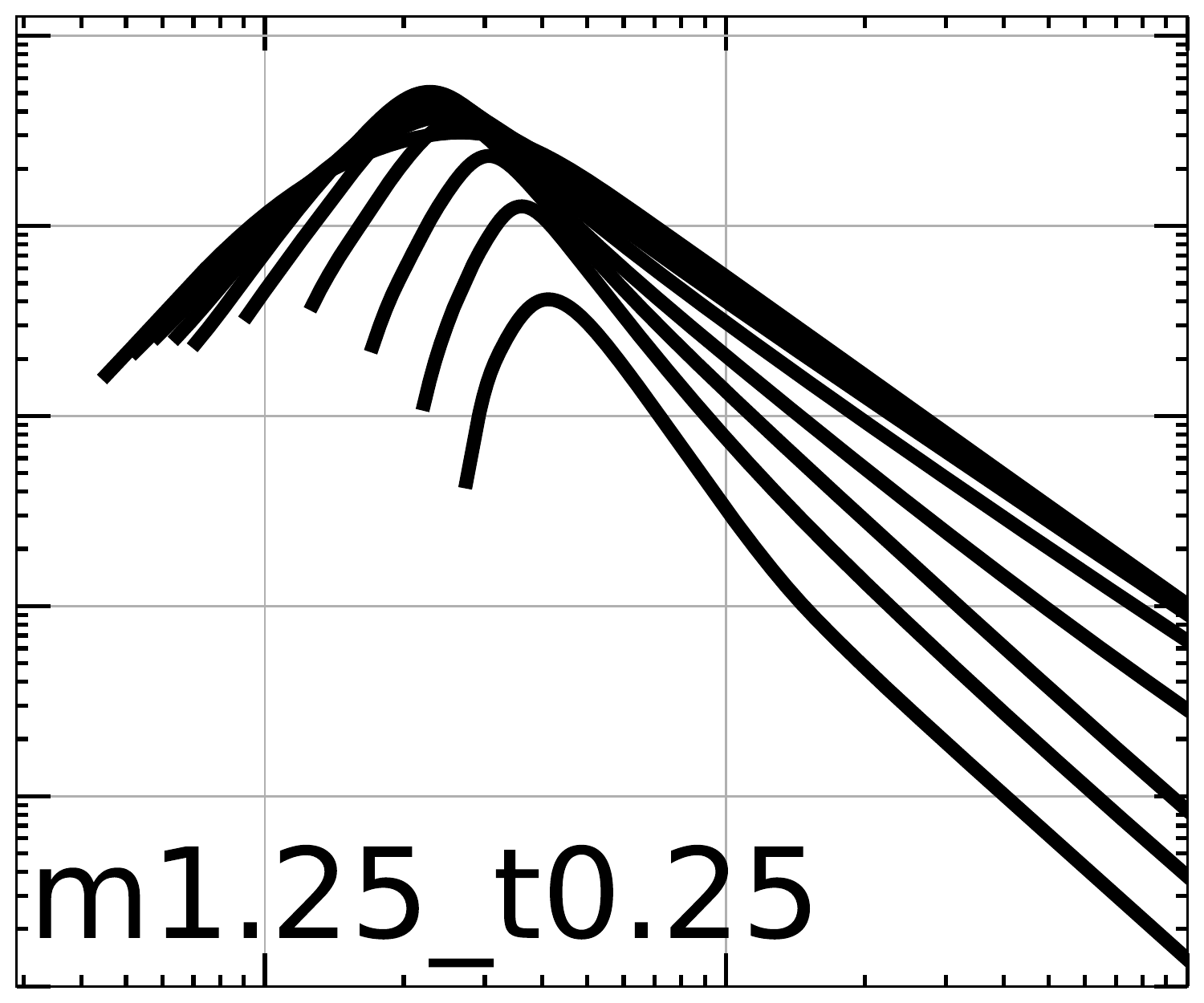}
\plotone{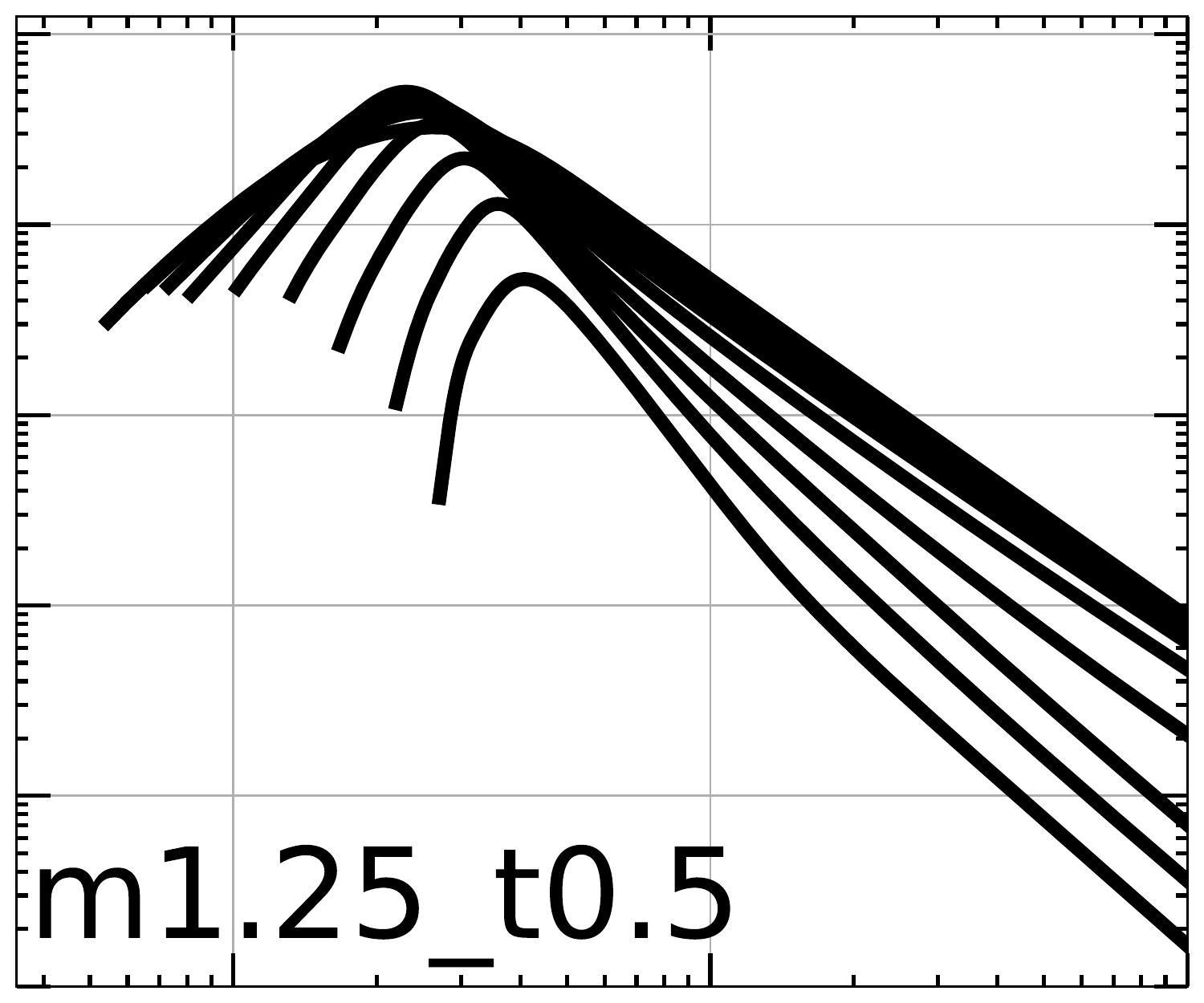}
\plotone{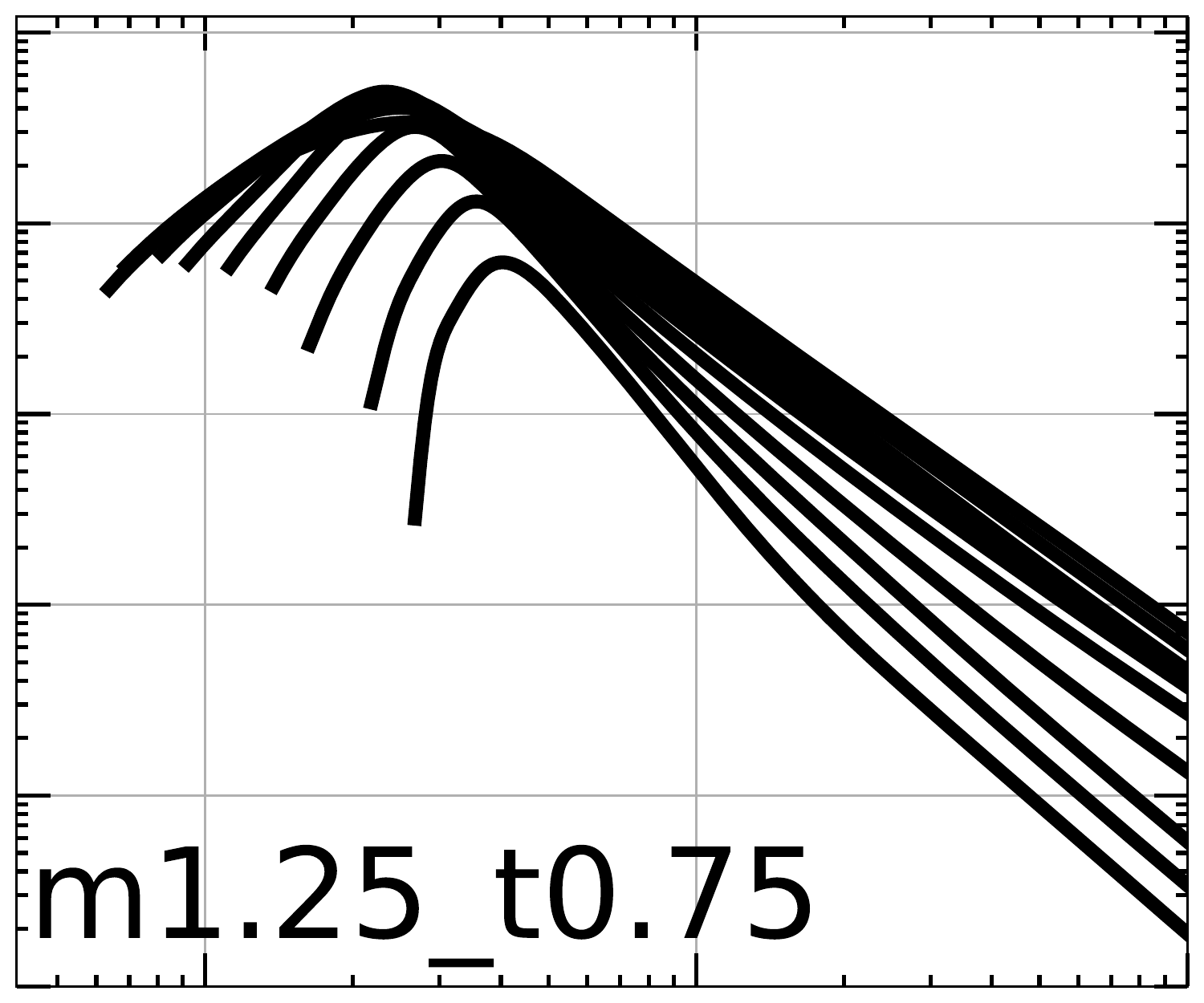}
\plotone{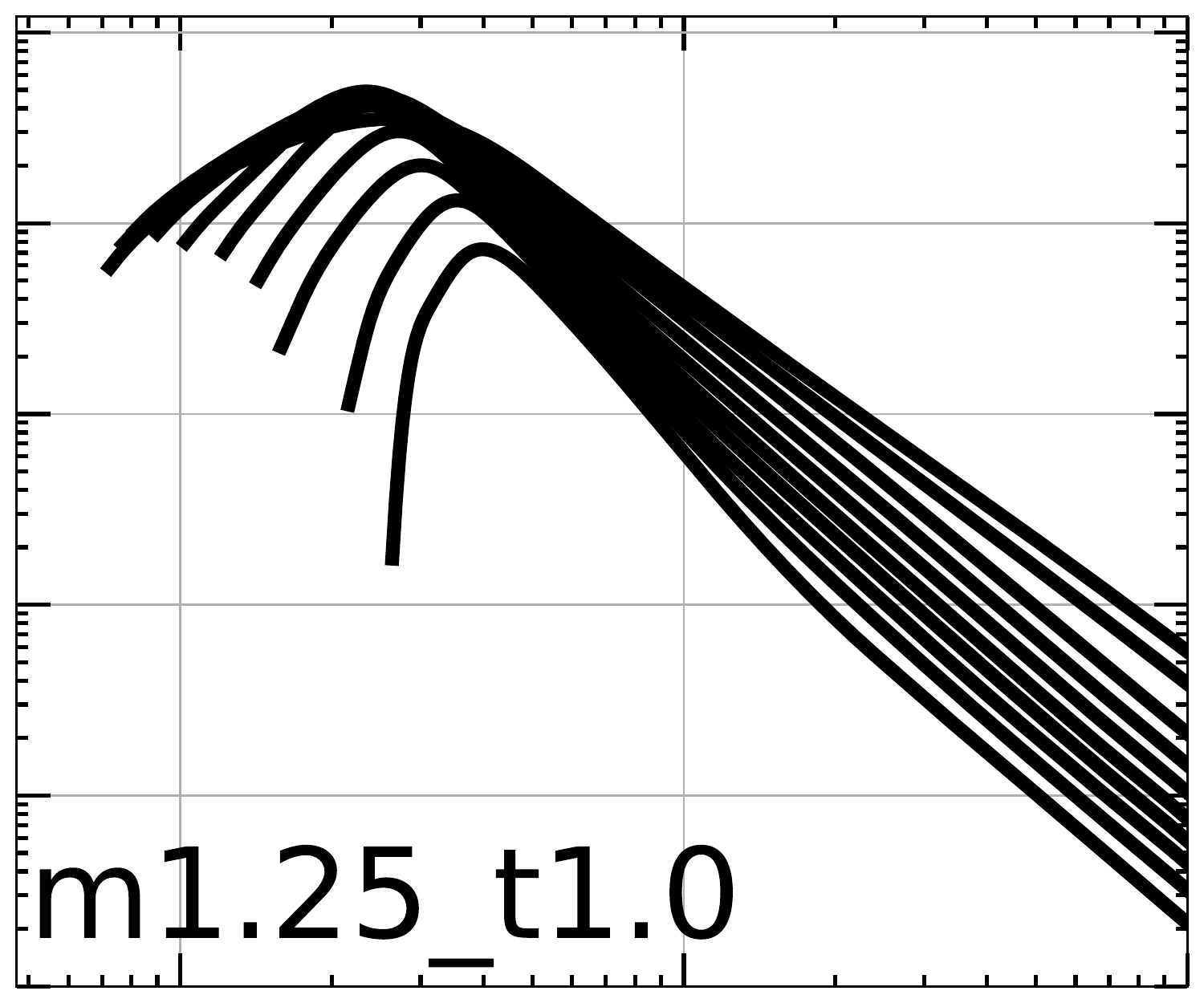}
\plotone{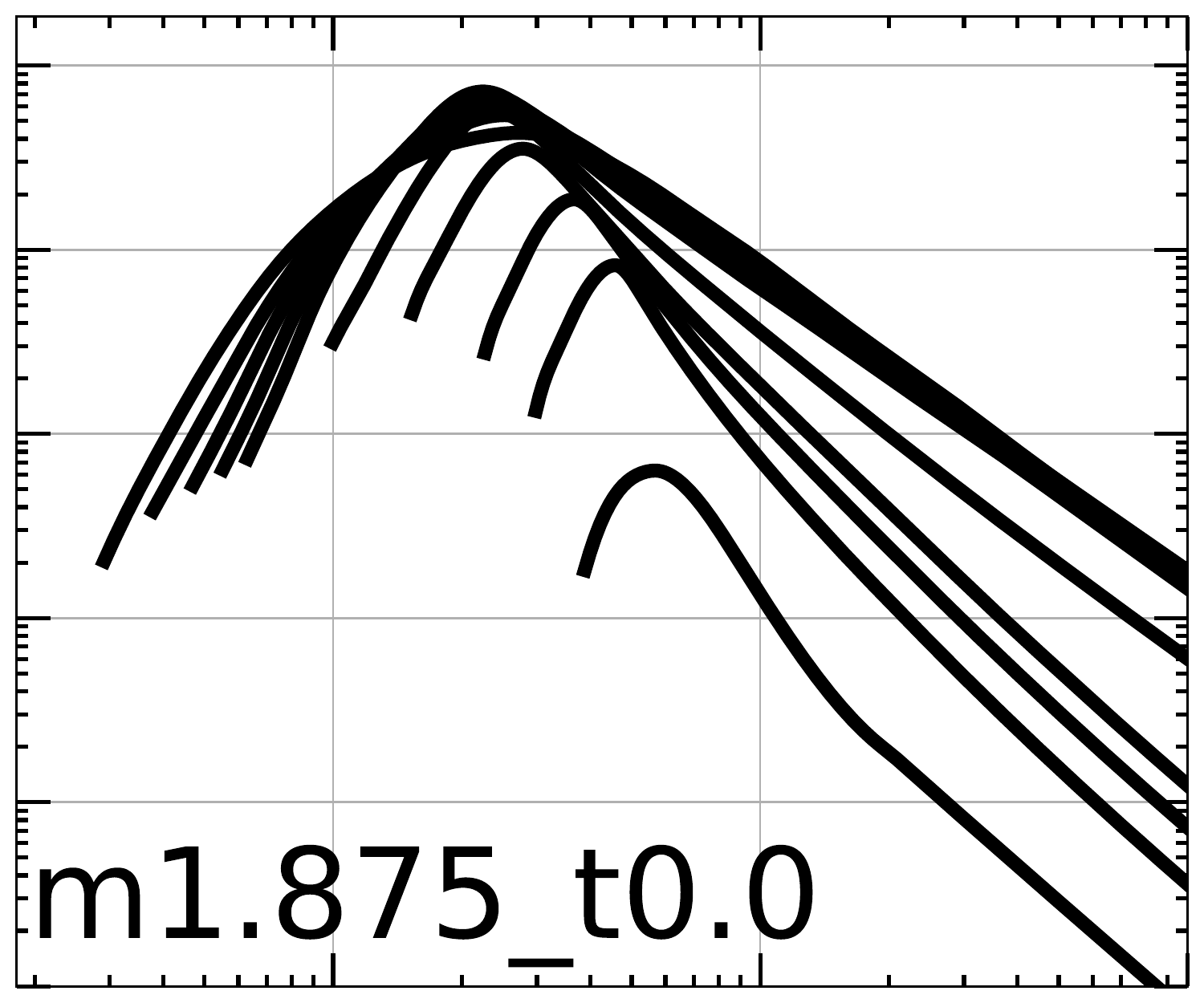}
\plotone{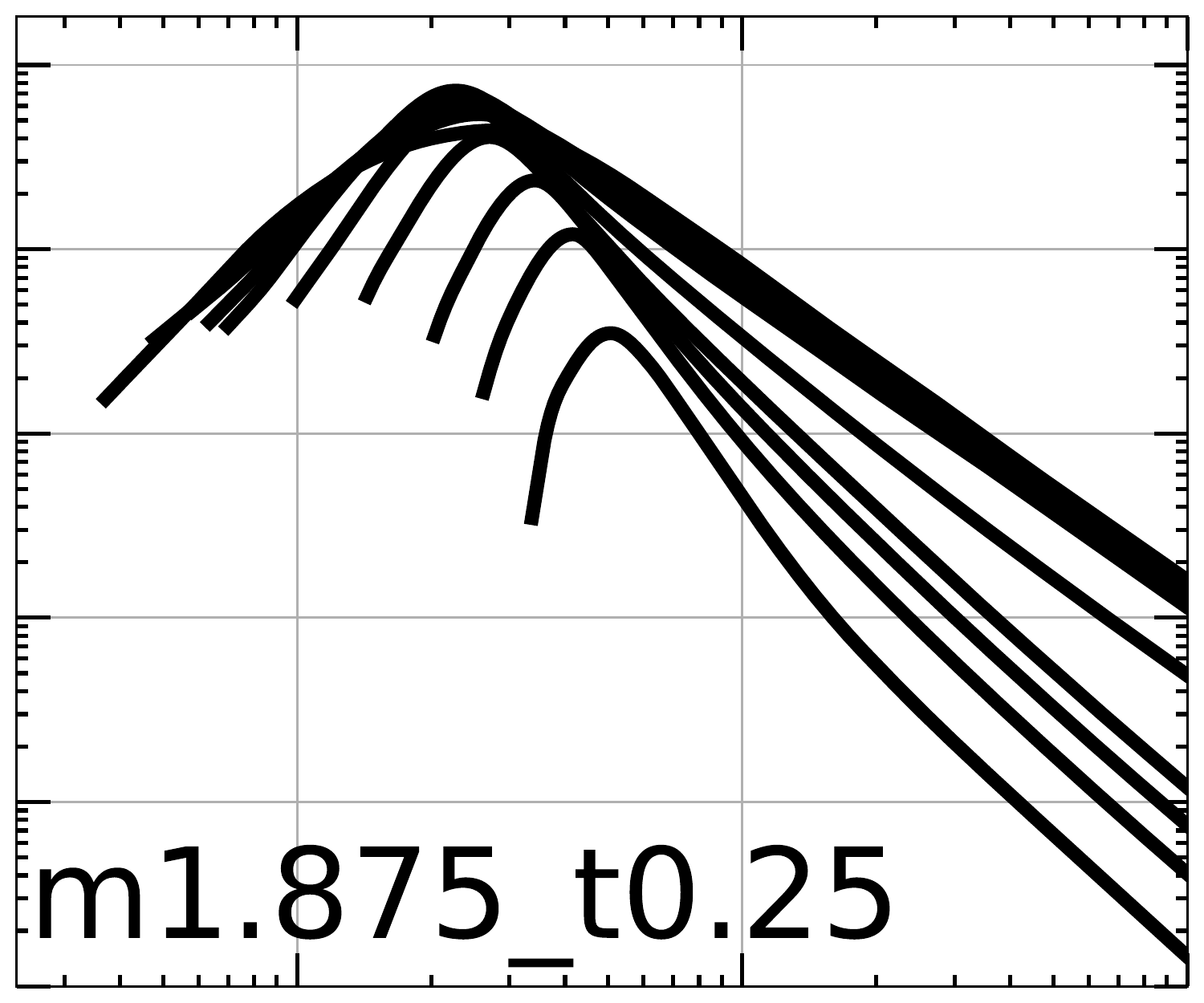}
\plotone{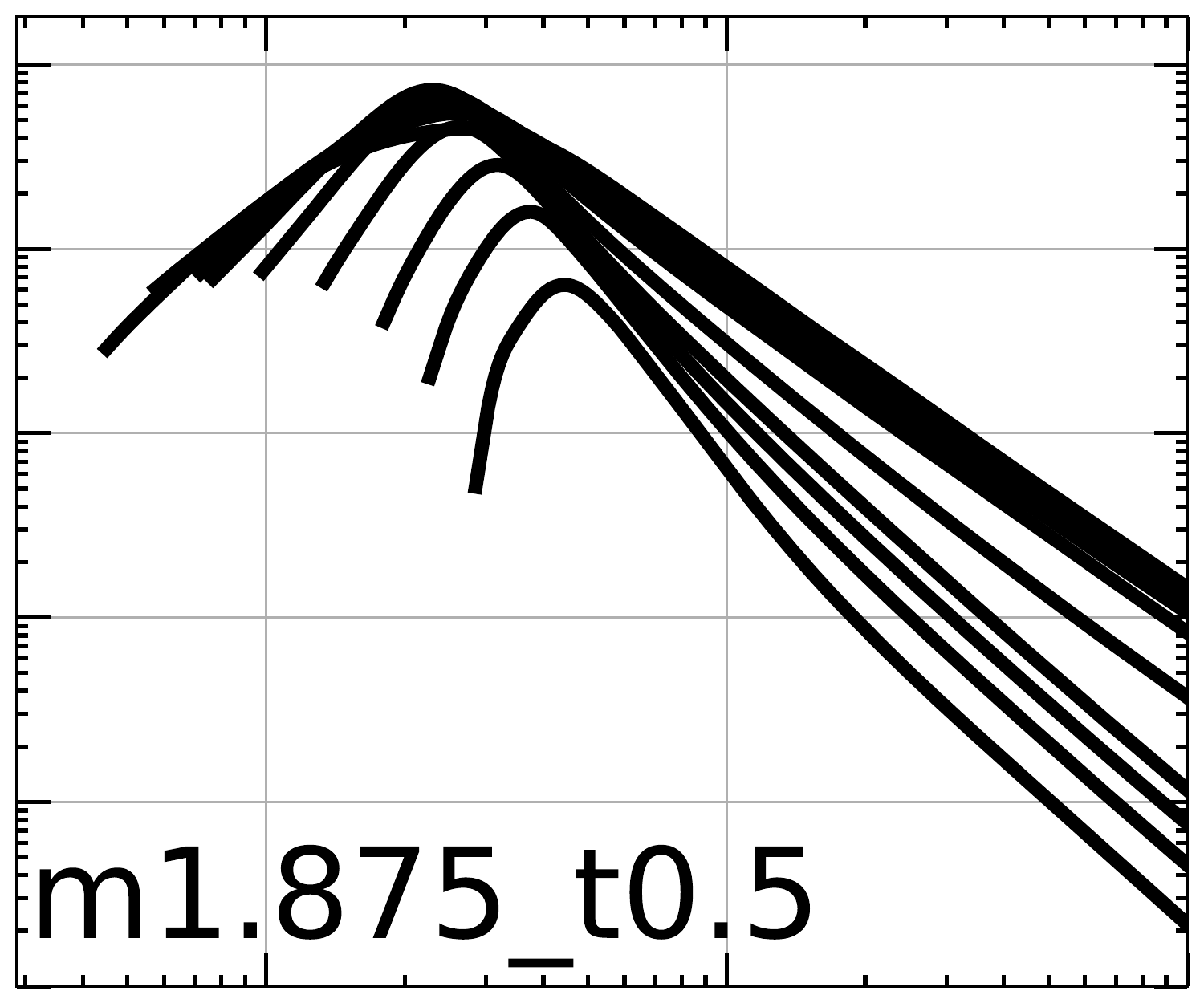}
\plotone{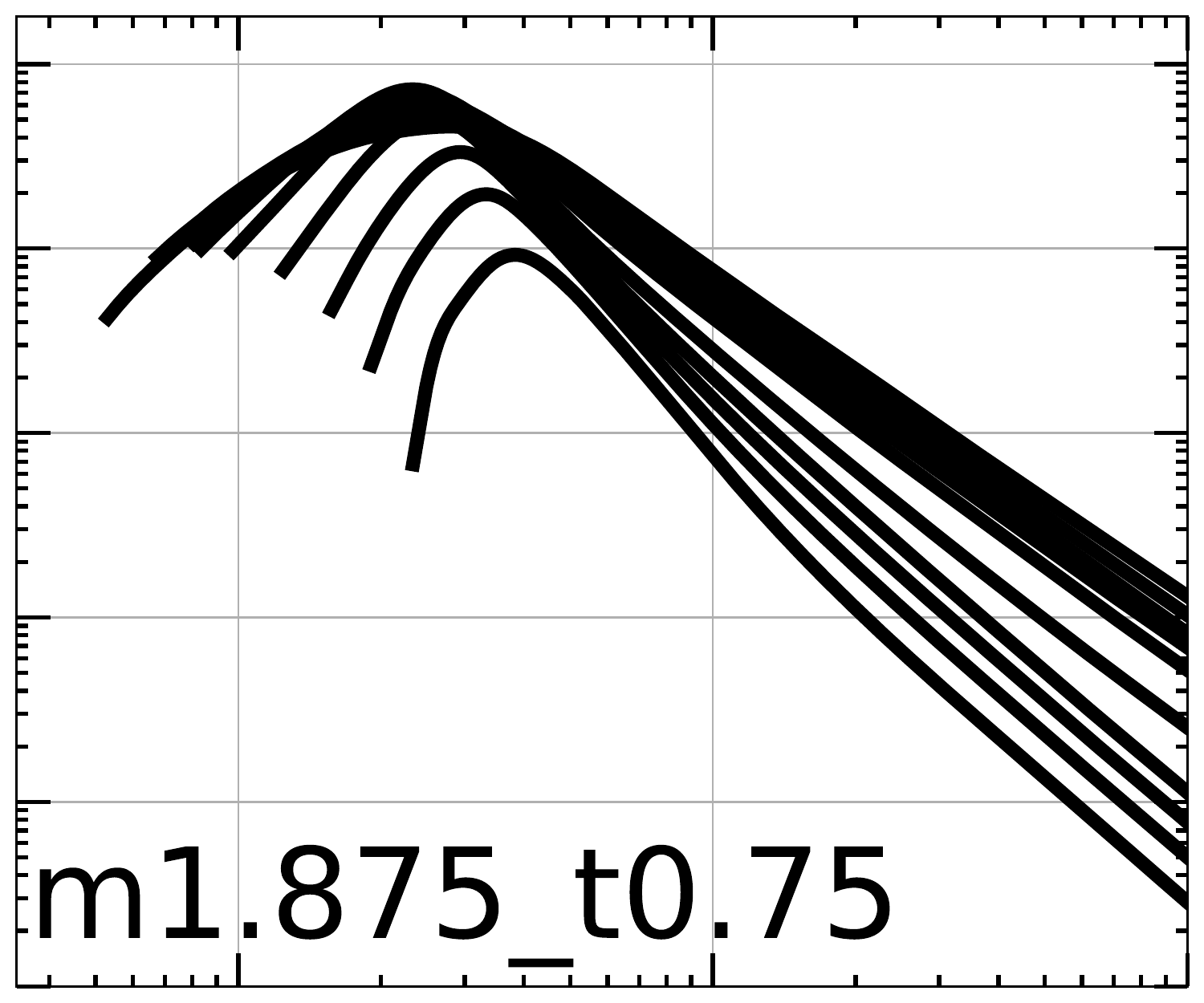}
\plotone{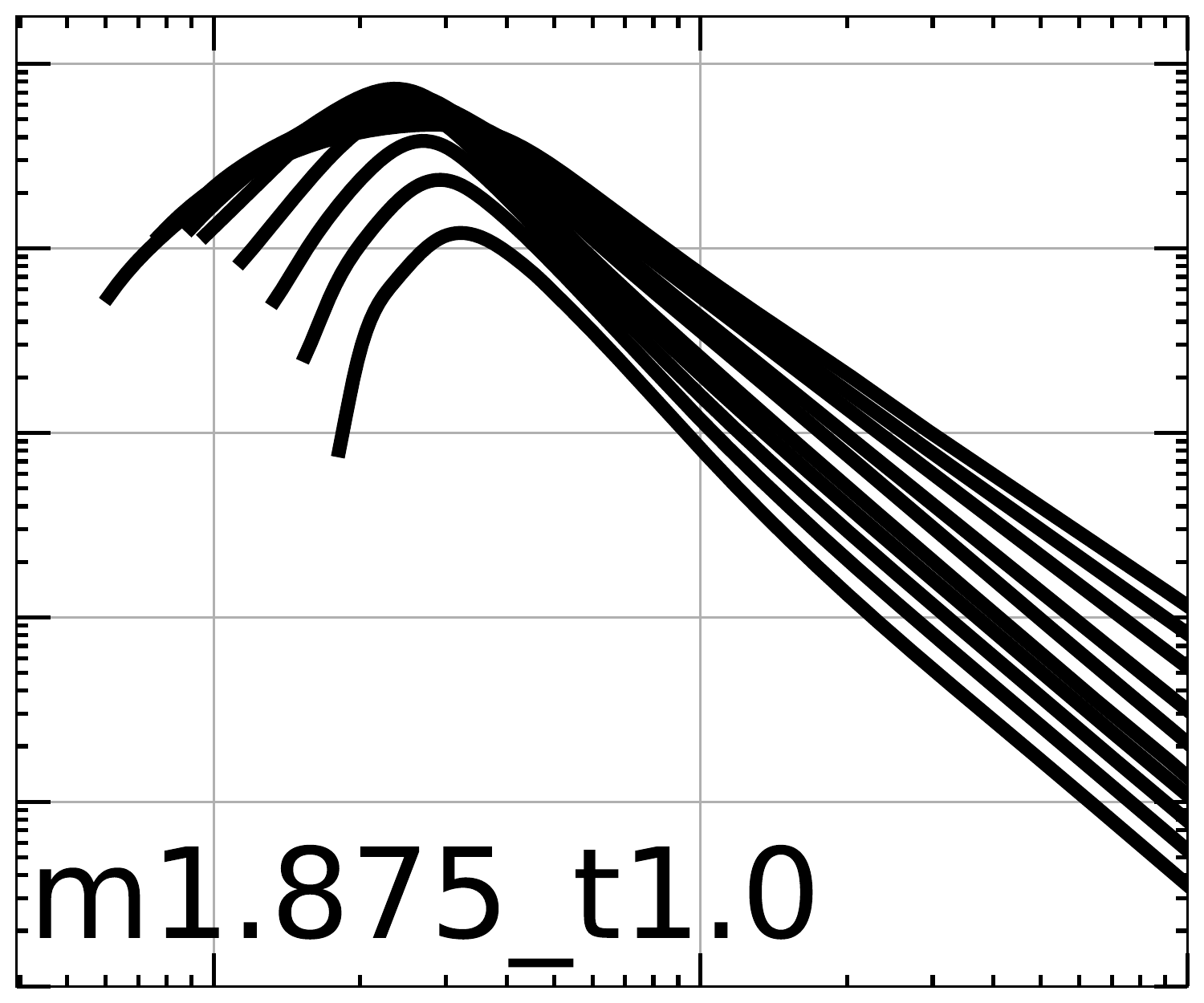}
\plotone{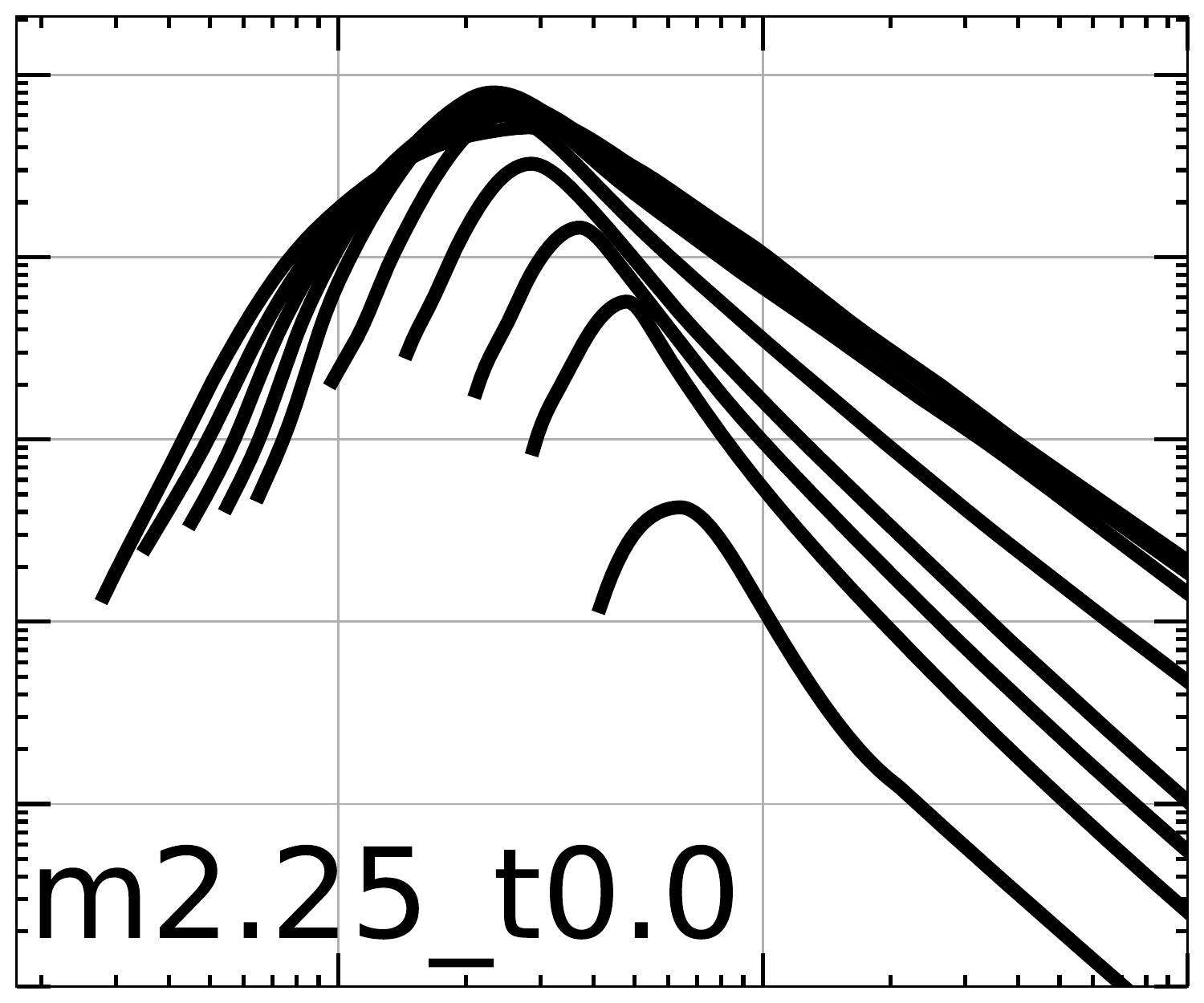}
\plotone{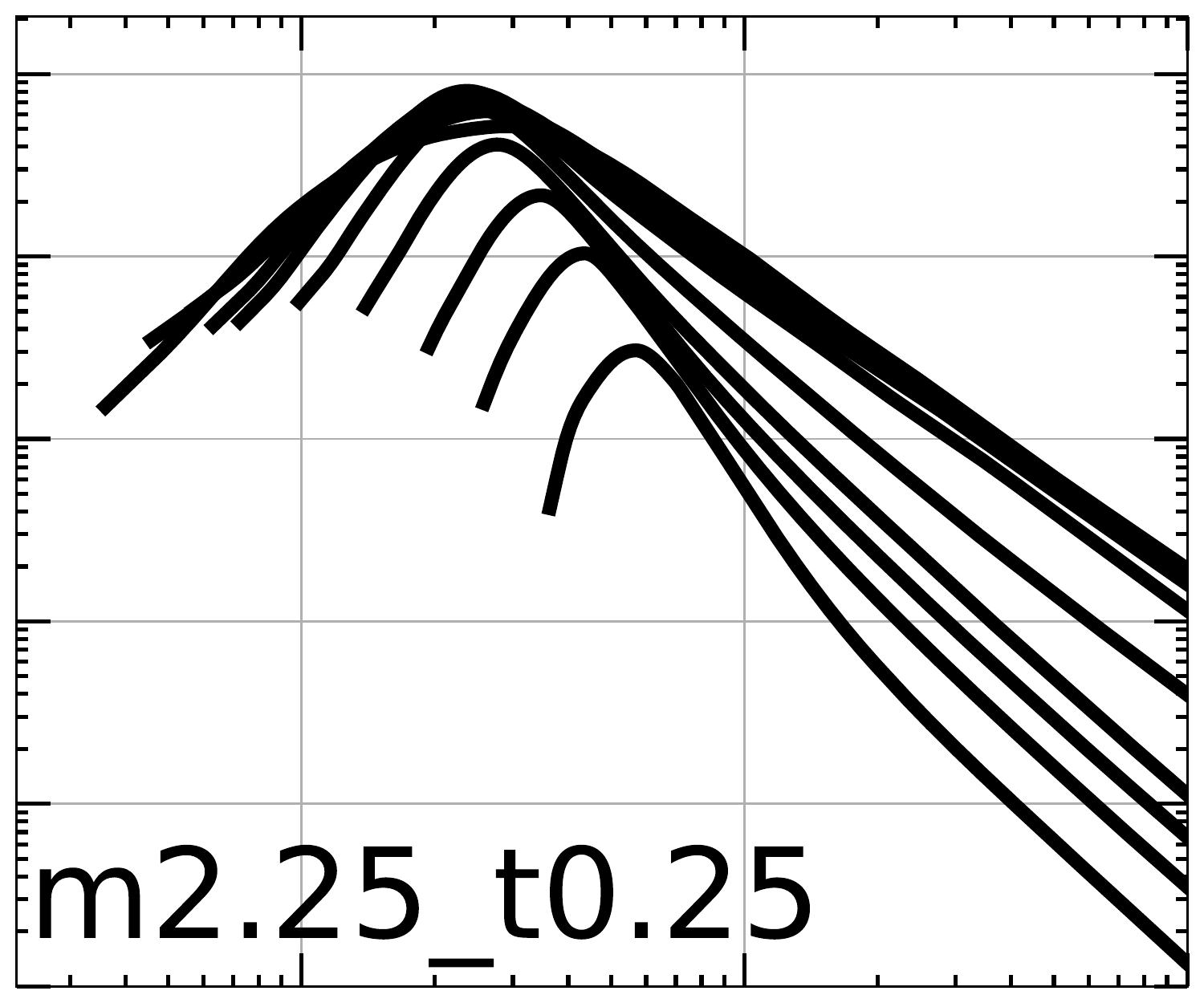}
\plotone{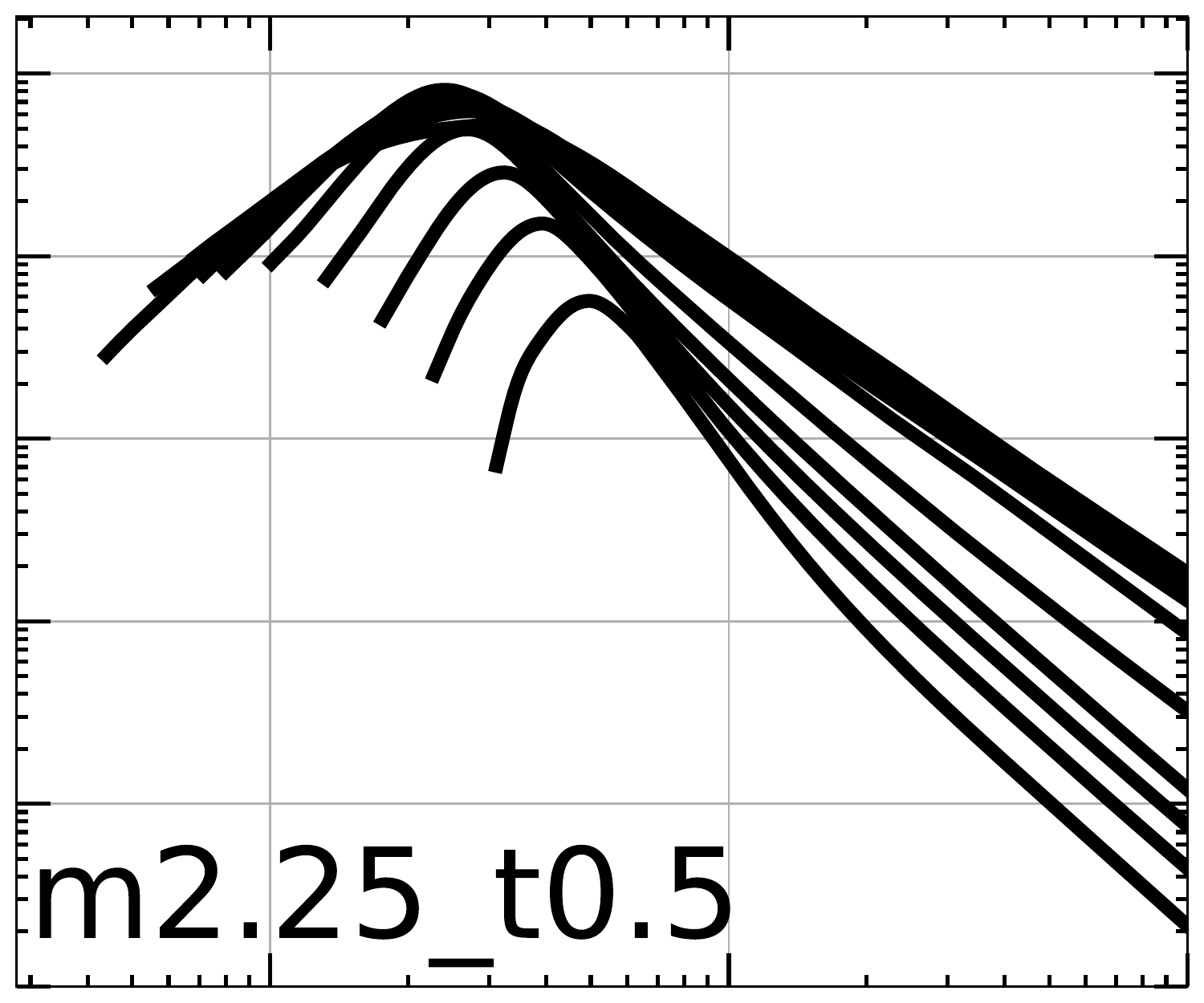}
\plotone{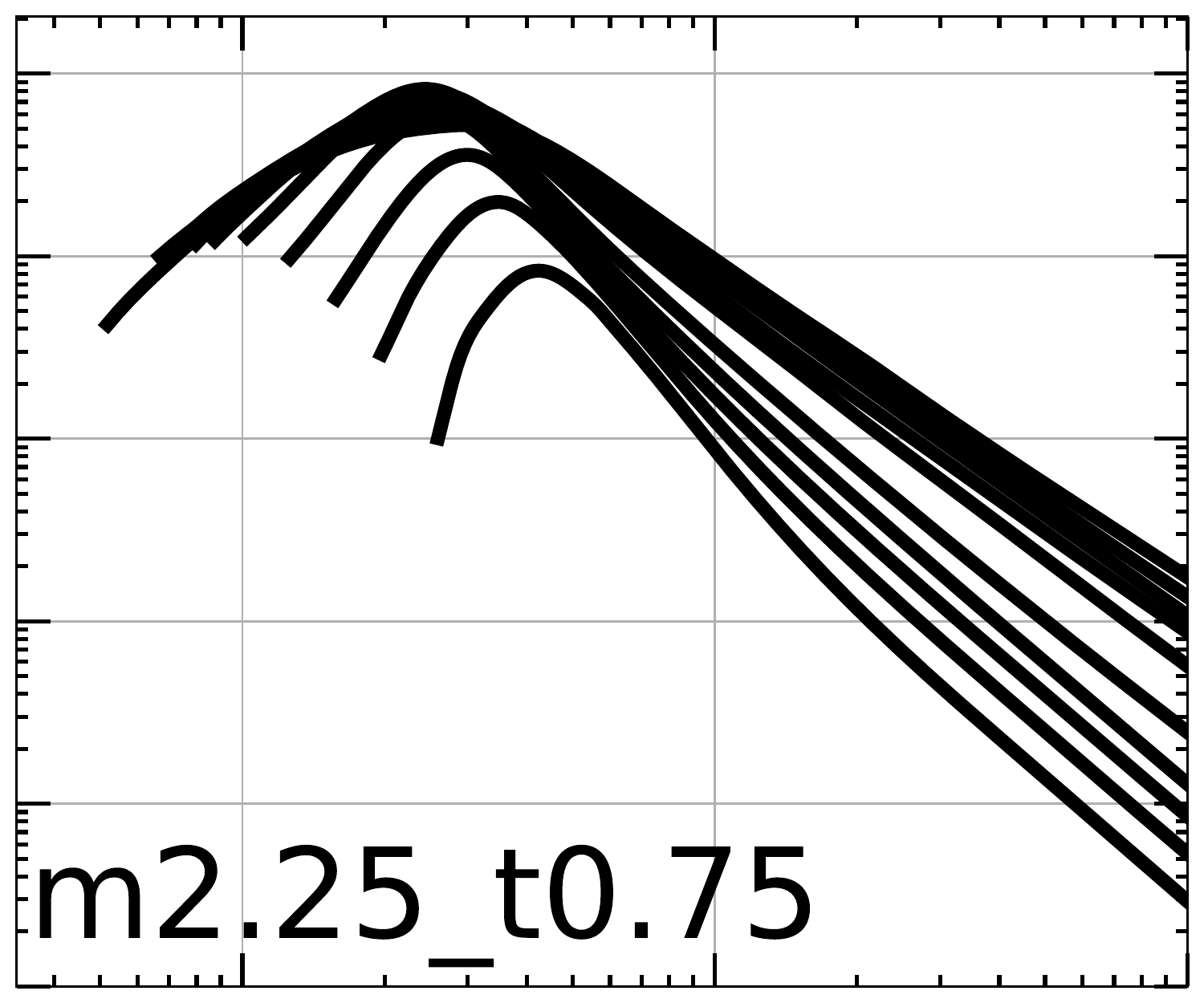}
\plotone{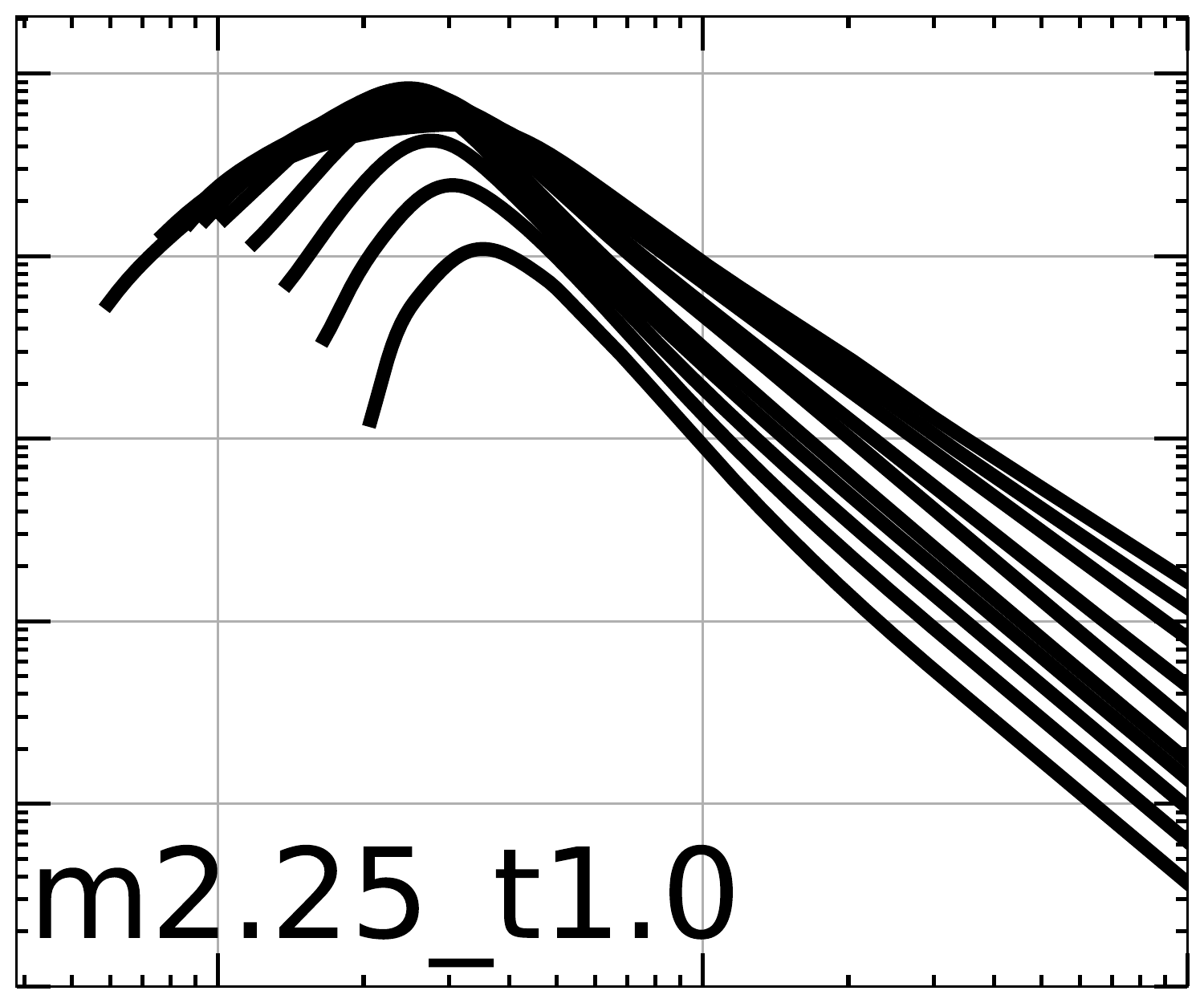}
\plotone{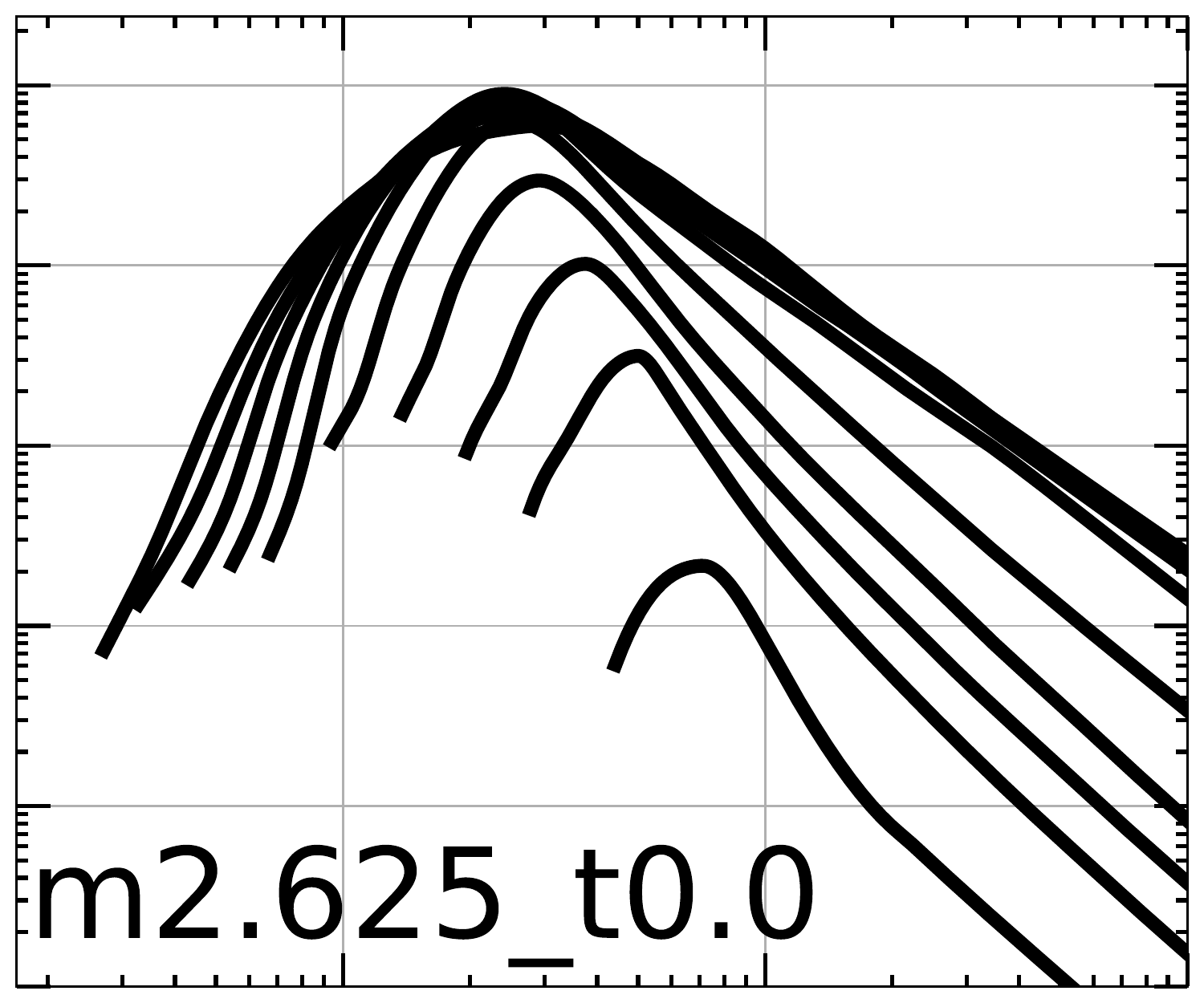}
\plotone{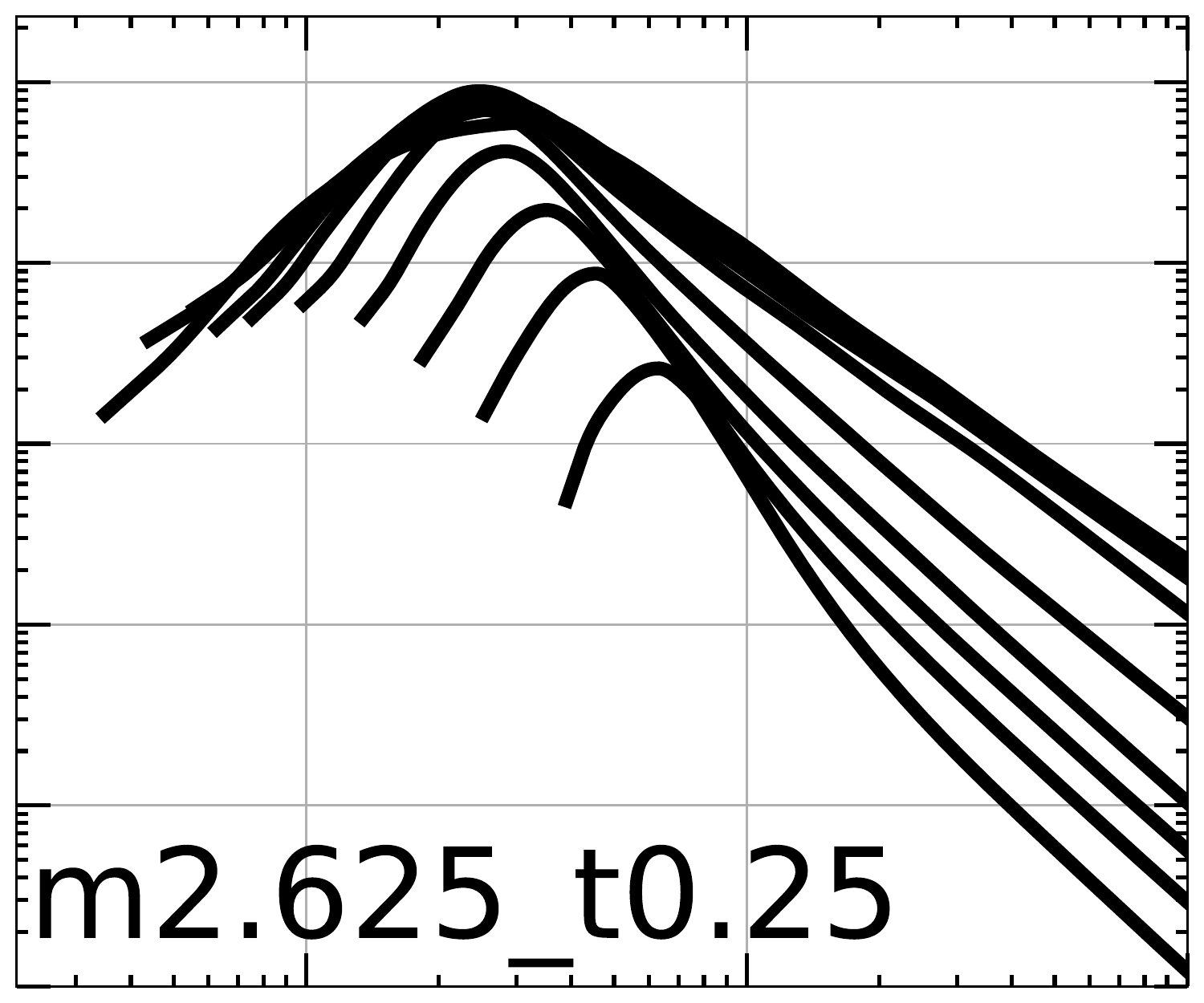}
\plotone{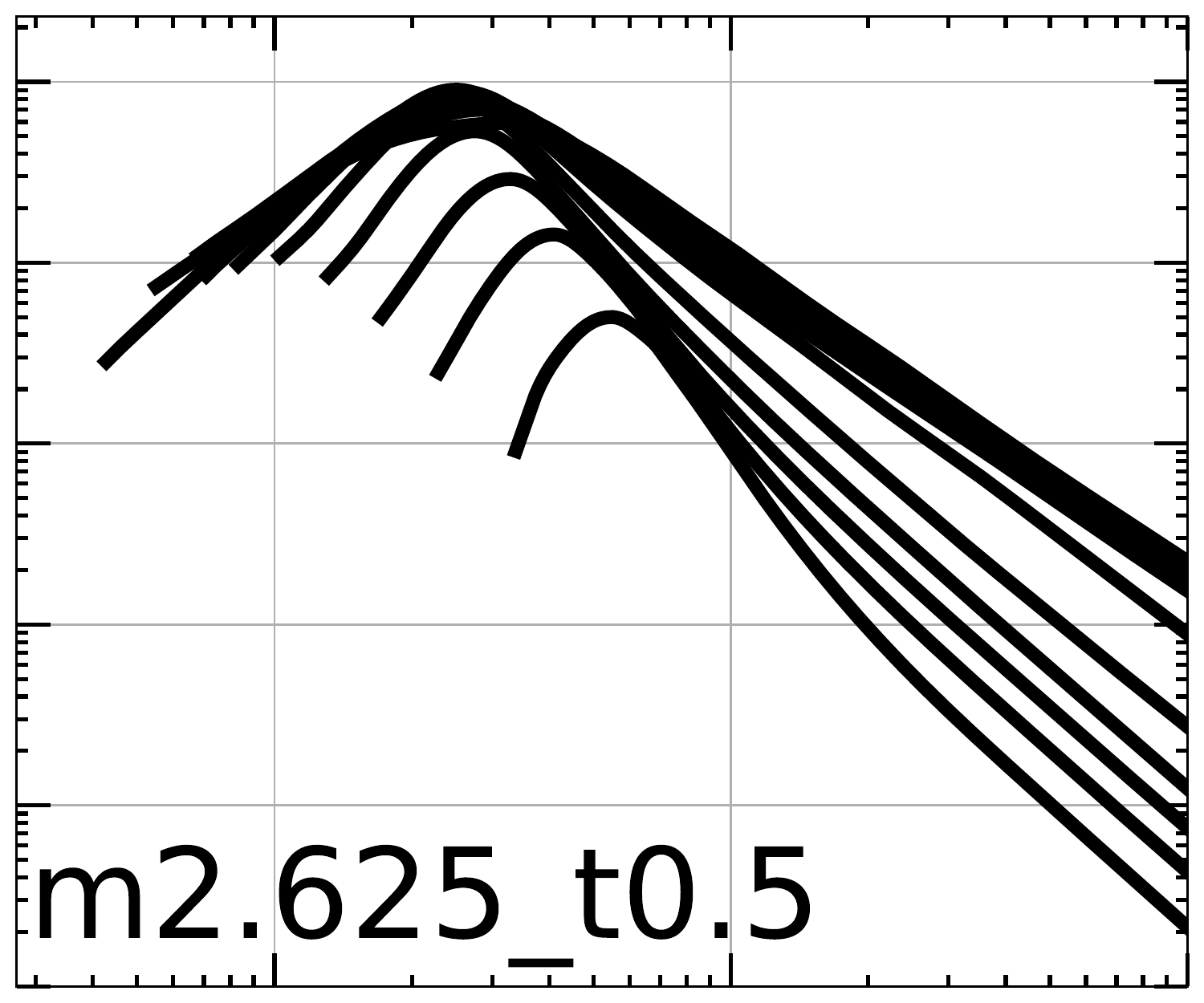}
\plotone{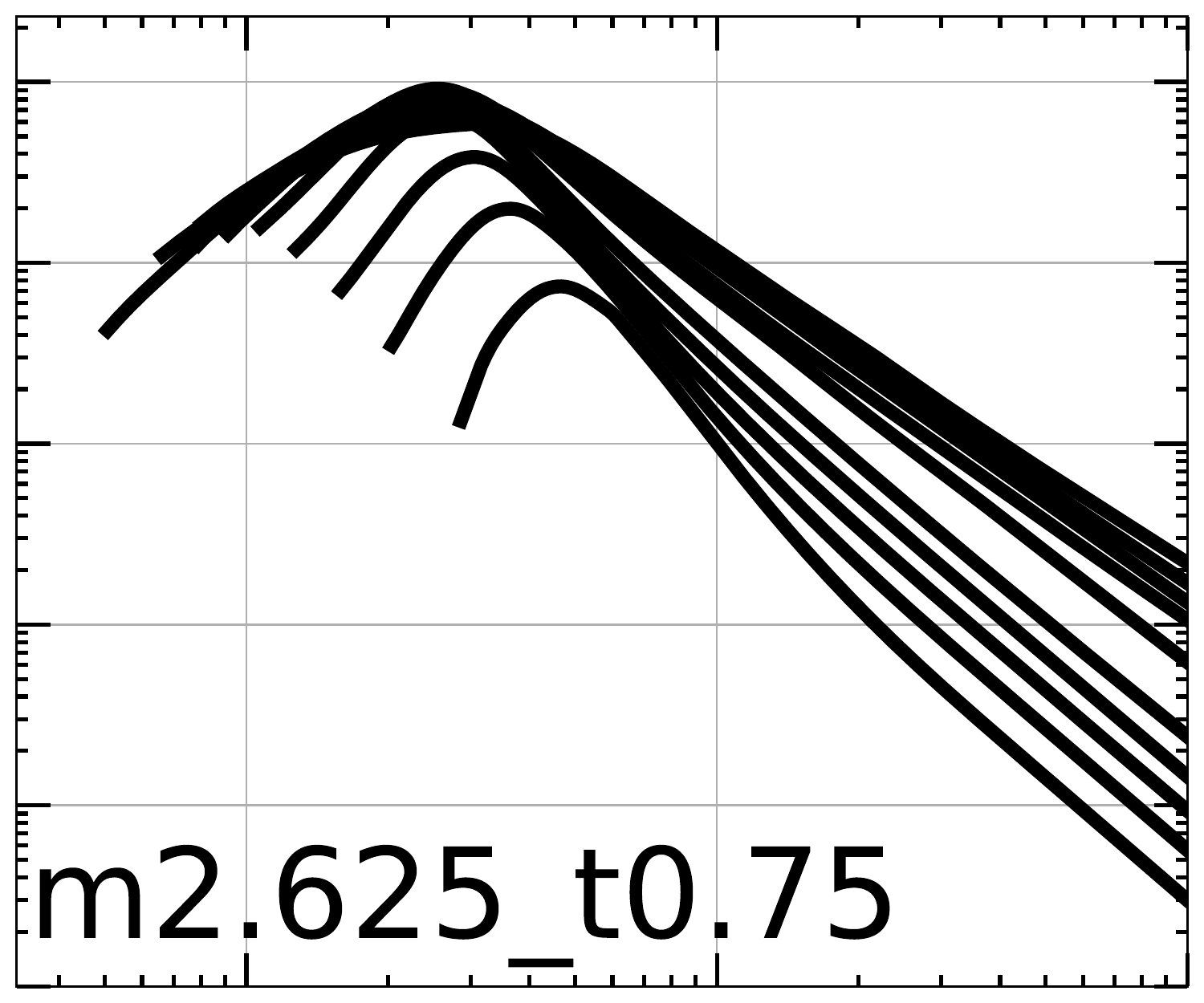}
\plotone{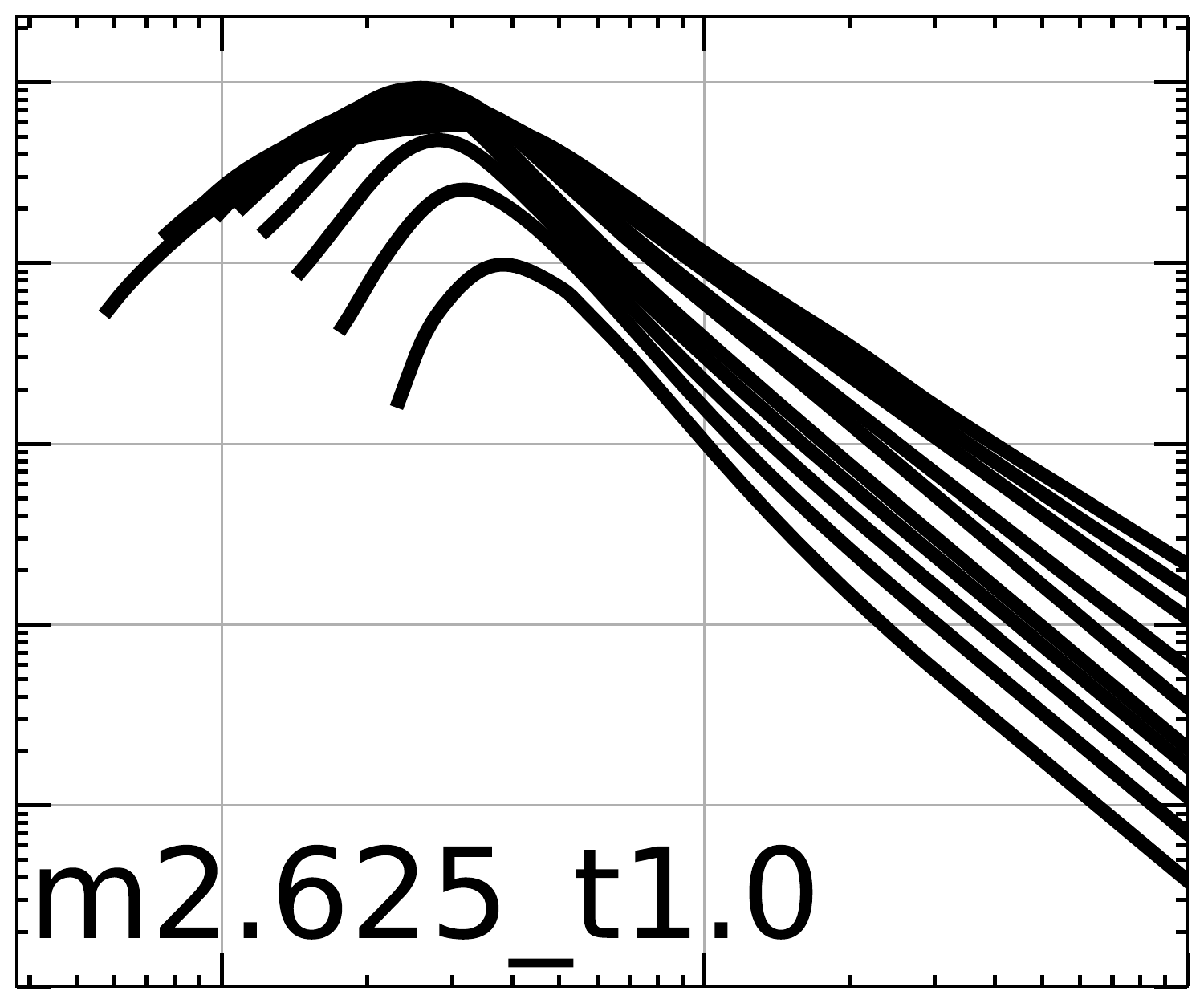}
\plotone{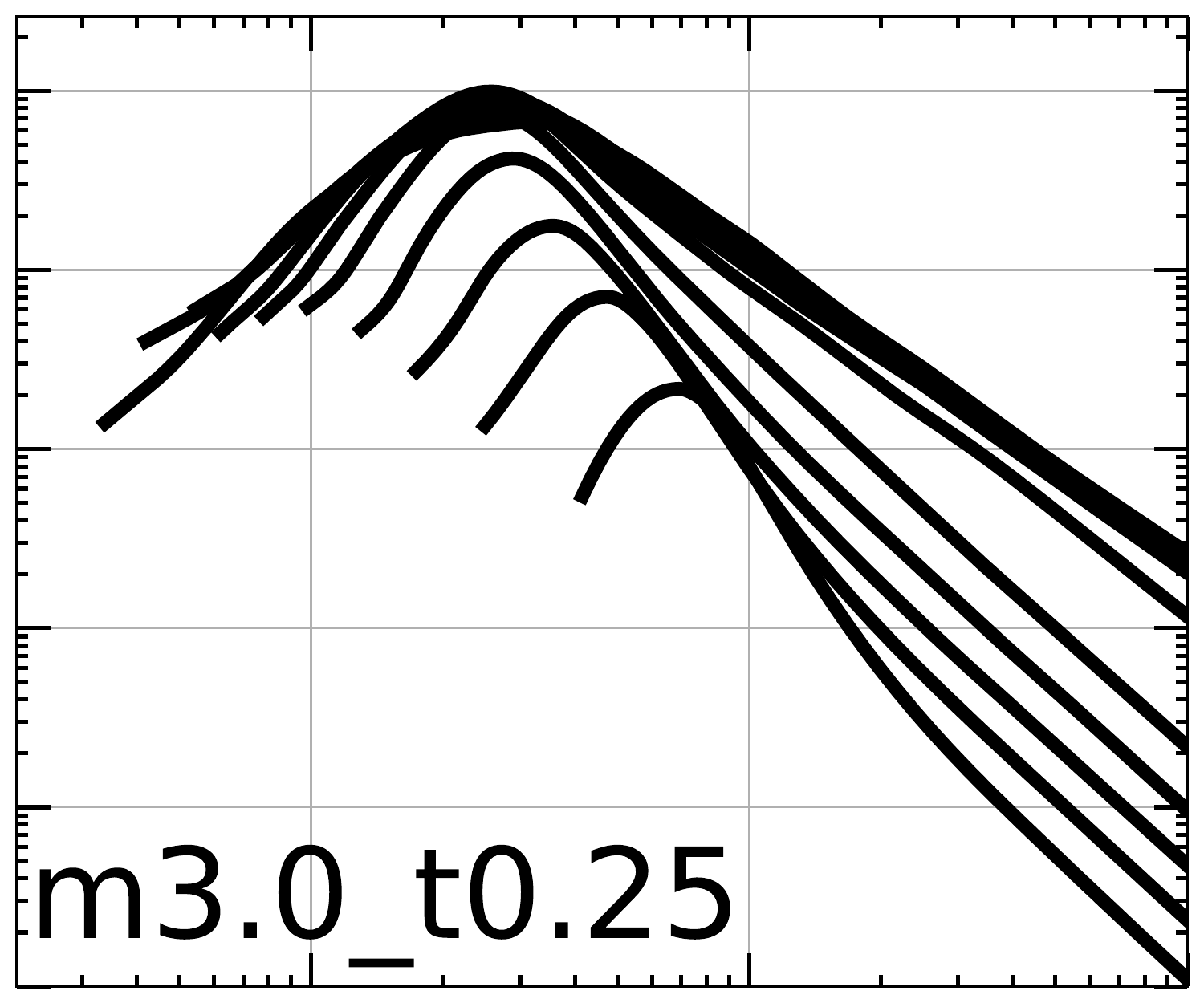}
\plotone{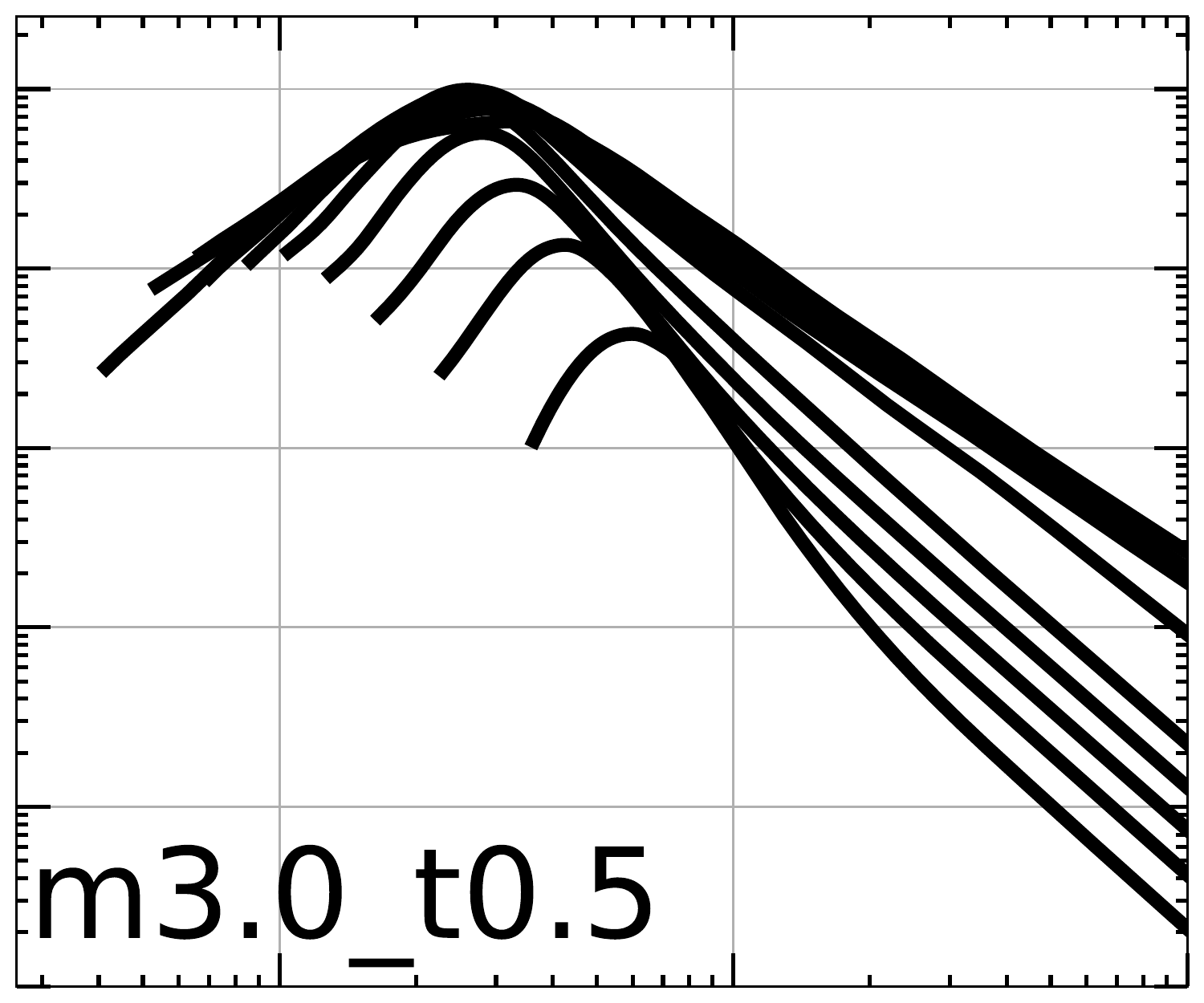}
\plotone{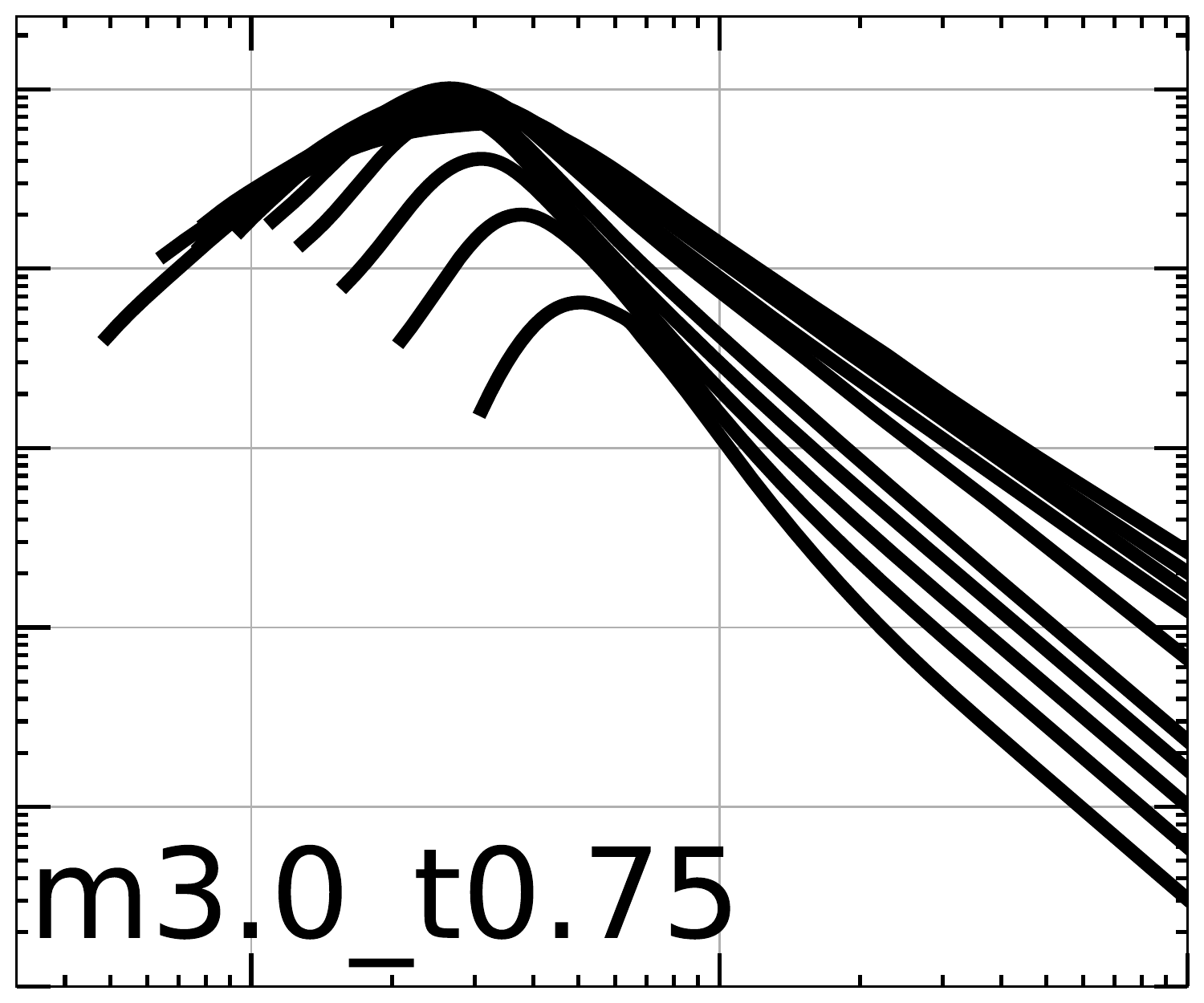}
\plotone{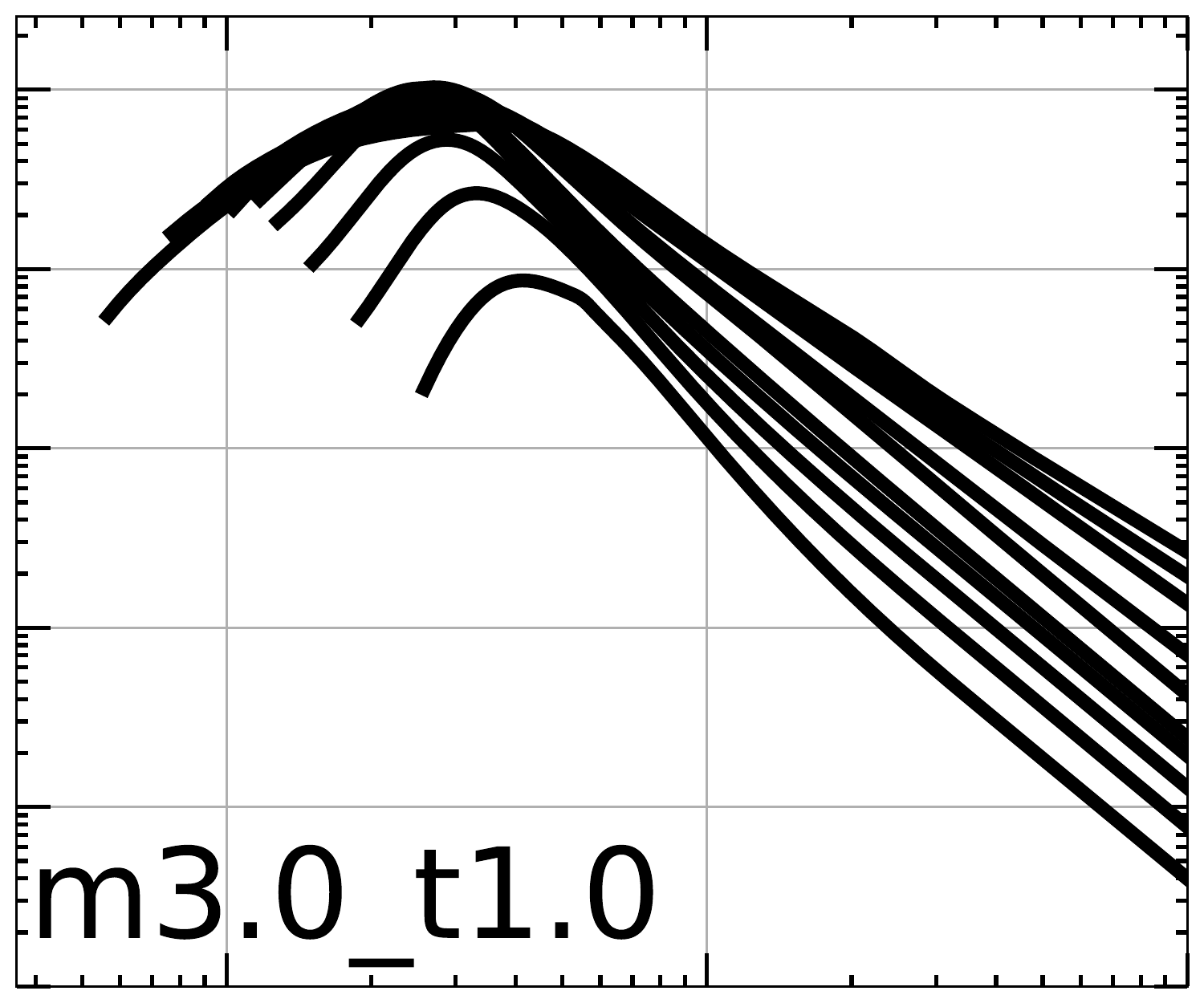}
\caption{
STARS library interpolated $dM/dt$'s for a small grid of stellar mass, stellar age, and impact parameter, and at a single BH mass $M_{\rm BH}=10^6M_\sun$. Axes labels are removed for clarity, but are the same as in other figures in this paper ($x$-axis is time, from 0 to 1000 days, $y-$axis is $\dot M$, from $10^{-4}$ to 10 $M_\sun$/yr, both log-scaled). Note that we provide more finely spaced interpolated grids for download (see text). The directory name, in format mXX\_tYY, where XX is the stellar mass in $M_\sun$ and YY is the fractional MS age, is in the lower left of each panel.
\label{fig:STARS_lib}
}
\end{figure*}

Figure~\ref{fig:STARS_lib} shows the STARS library interpolated $dM/dt$'s for a small grid in stellar mass, stellar age, and impact parameter, for a single BH mass $M_{\rm BH}=10^6M_\sun$. This grid has 10 interpolated $\beta$ points, 11 interpolated mass points, and 5 interpolated age points, with spacing in $\beta$ and stellar mass logarithmic, and spacing in stellar age linear.

\begin{figure*}[htp!]
\epsscale{0.55}
\plotone{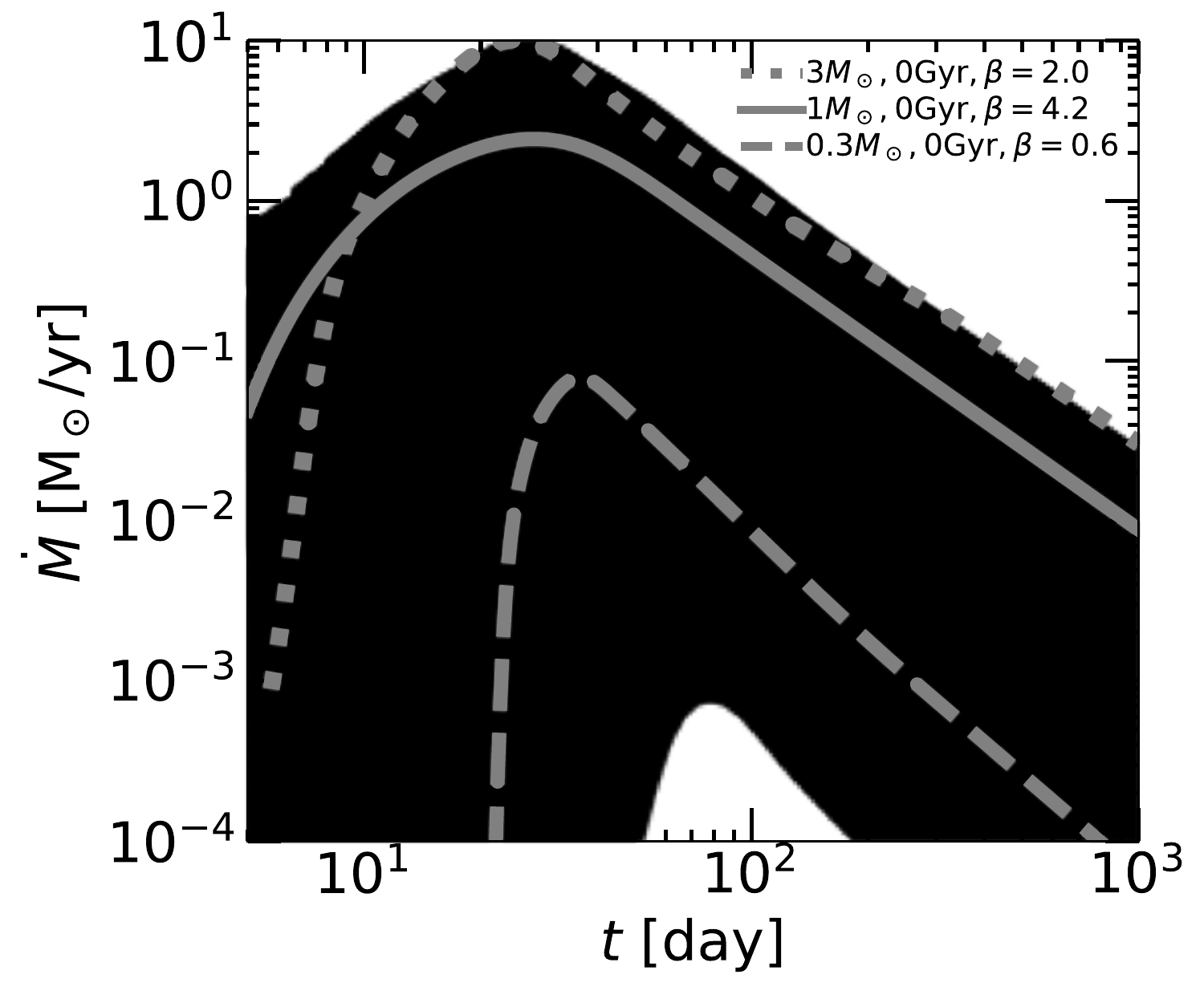}
\plotone{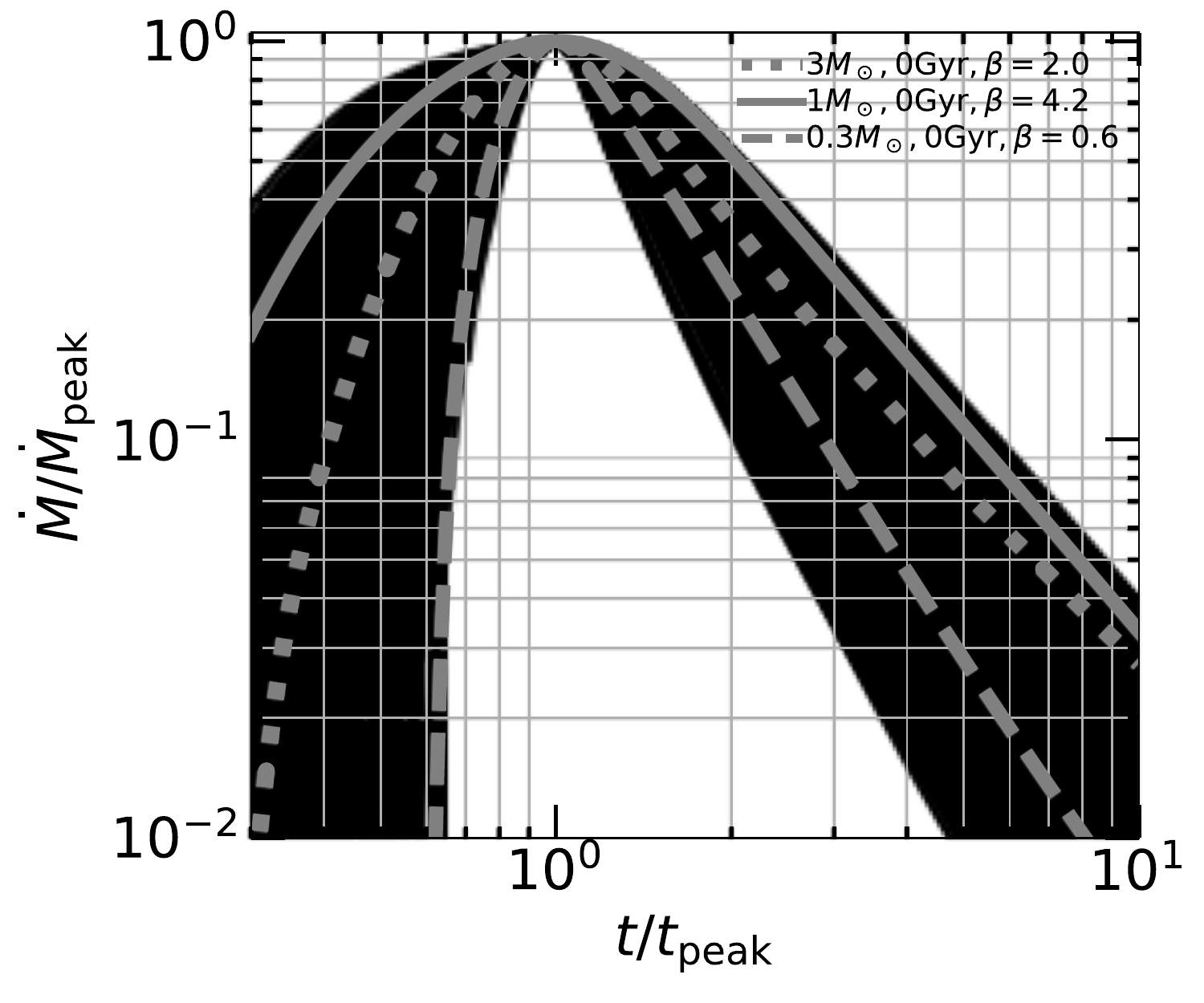}
\caption{
STARS library interpolated $dM/dt$'s, all in one plot, for a single BH mass $M_{\rm BH}=10^6M_\sun$. Left: absolute units. right: normalized to peak time and peak fallback rate. We applied a fill between $dM/dt$'s in order to emphasize the tune-able spacing of the 3D interpolation.
Three $dM/dt$'s are overplotted in gray: a ZAMS 0.3$M_\sun$ star with $\beta=0.6$ (dashed), a ZAMS 1$M_\sun$ star with $\beta=4.2$ (solid), and a ZAMS 3$M_\sun$ star with $\beta=2.0$ (dotted).
}
\label{fig:STARS_lib_allinone}
\end{figure*}

Figure~\ref{fig:STARS_lib_allinone} shows all of the STARS library interpolated $dM/dt$'s in the same plot, for a single BH mass $M_{\rm BH}=10^6M_\sun$. The left panel is in absolute units, in order to emphasize the several orders of magnitude in $\dot M$ and time covered by this grid even for a single BH mass---roughly 6 orders of magnitude in fallback rate and 4 orders of magnitude in time (we extend the $dM/dt$'s to later times than shown in this plot).
The right panel is normalized to peak time and peak fallback rate, in order to emphasize the range of rise and decay slopes exhibited by the library. See Section~\ref{sec:results} for more discussion of how these slopes depend on stellar structure. In general, more centrally concentrated stars have steeper rise slopes and shallower decay slopes. The more rounded behavior near peak is from post-critical-$\beta$ disruptions; see e.g. the high-$\beta$ behavior for a given star in Figure~\ref{fig:mdots_allgrouped}.
We also overplot three $dM/dt$'s in gray: a ZAMS 0.3$M_\sun$ star with $\beta=0.6$, a ZAMS 1$M_\sun$ star with $\beta=4.2$, and a ZAMS 3$M_\sun$ star with $\beta=2.0$.

%%%%%%%%%%%%%%%%%%%
\section{Fits to disruption quantities}

\begin{figure}[htp!]
\epsscale{0.55}
\plotone{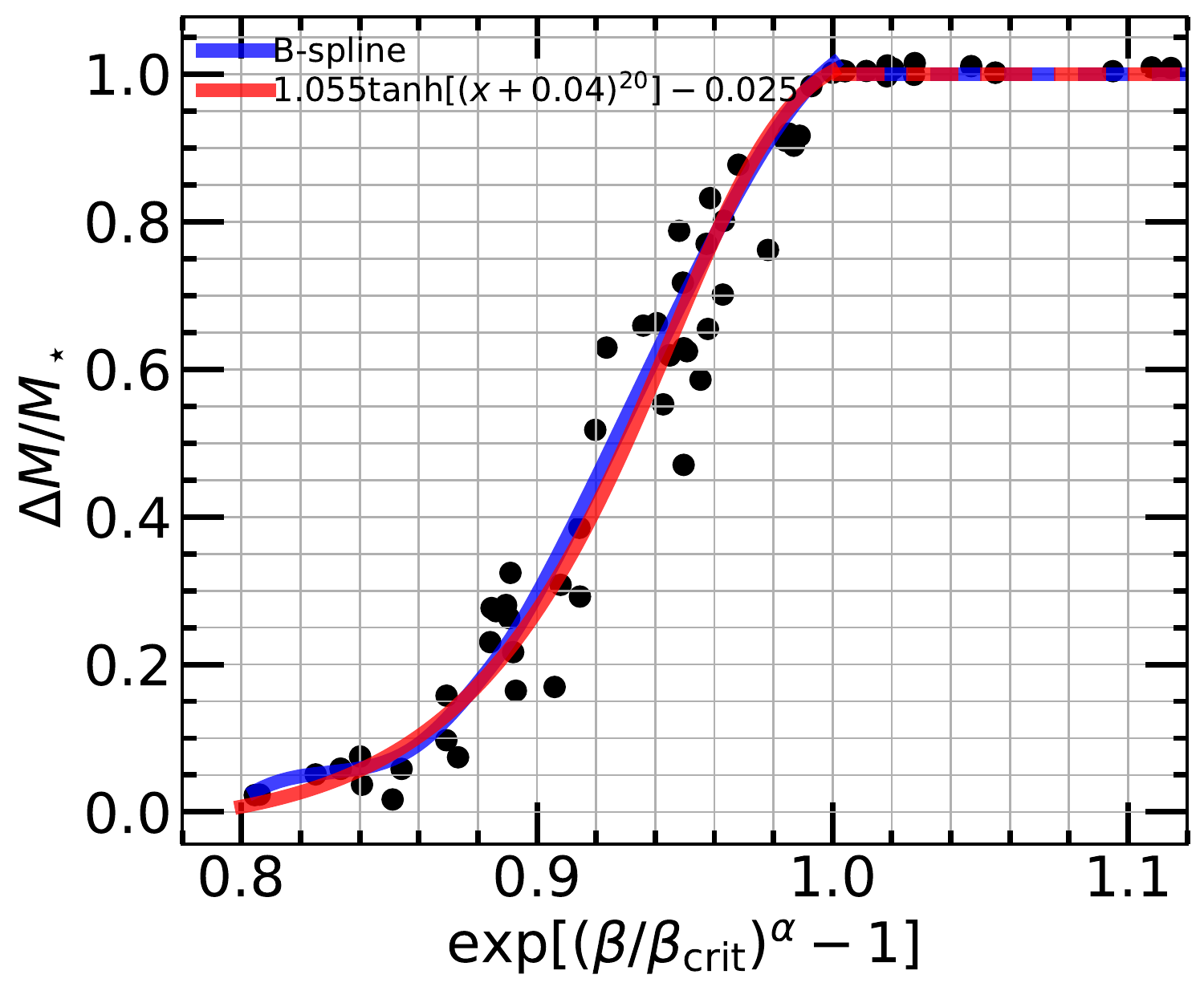}
\plotone{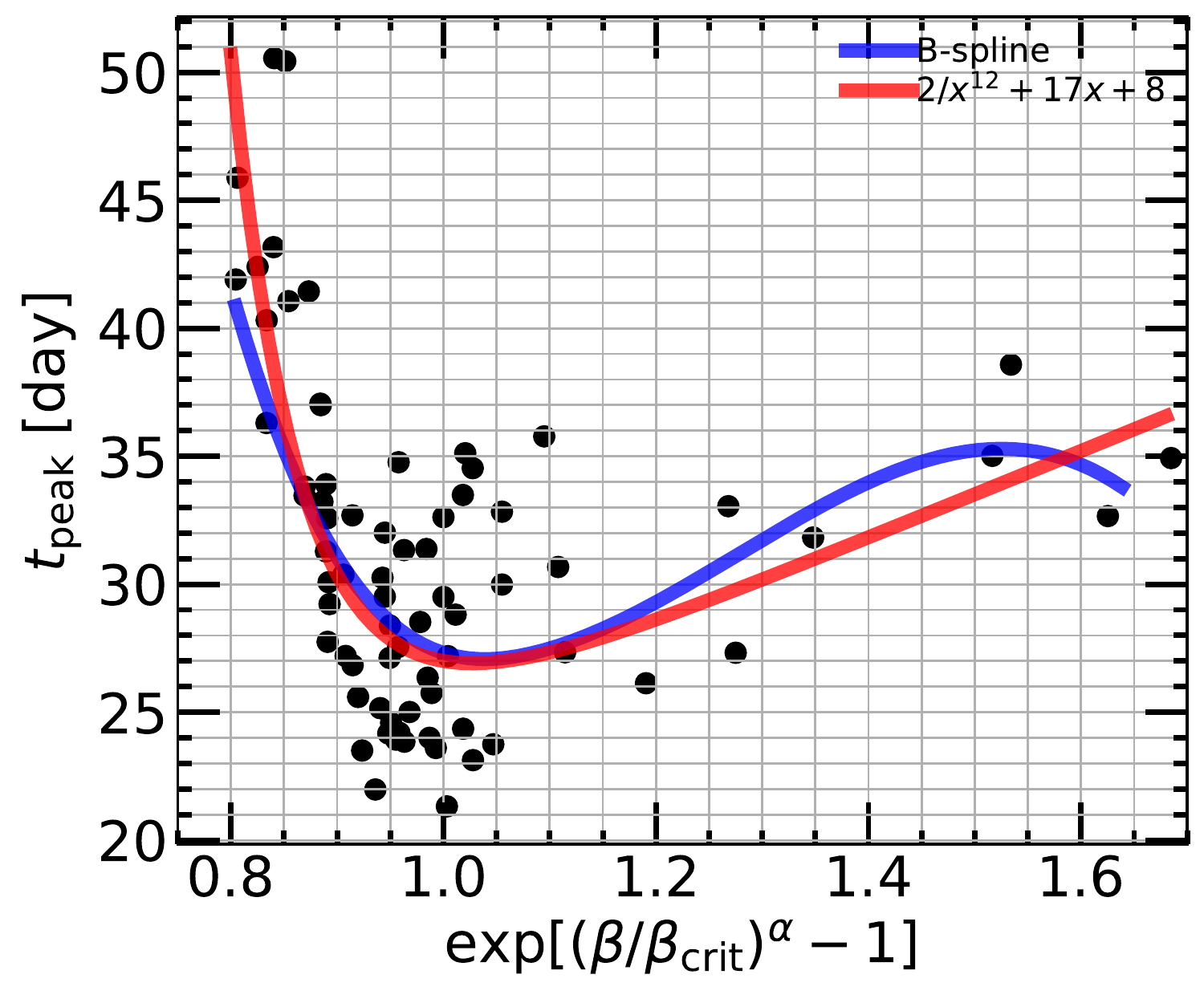}
\plotone{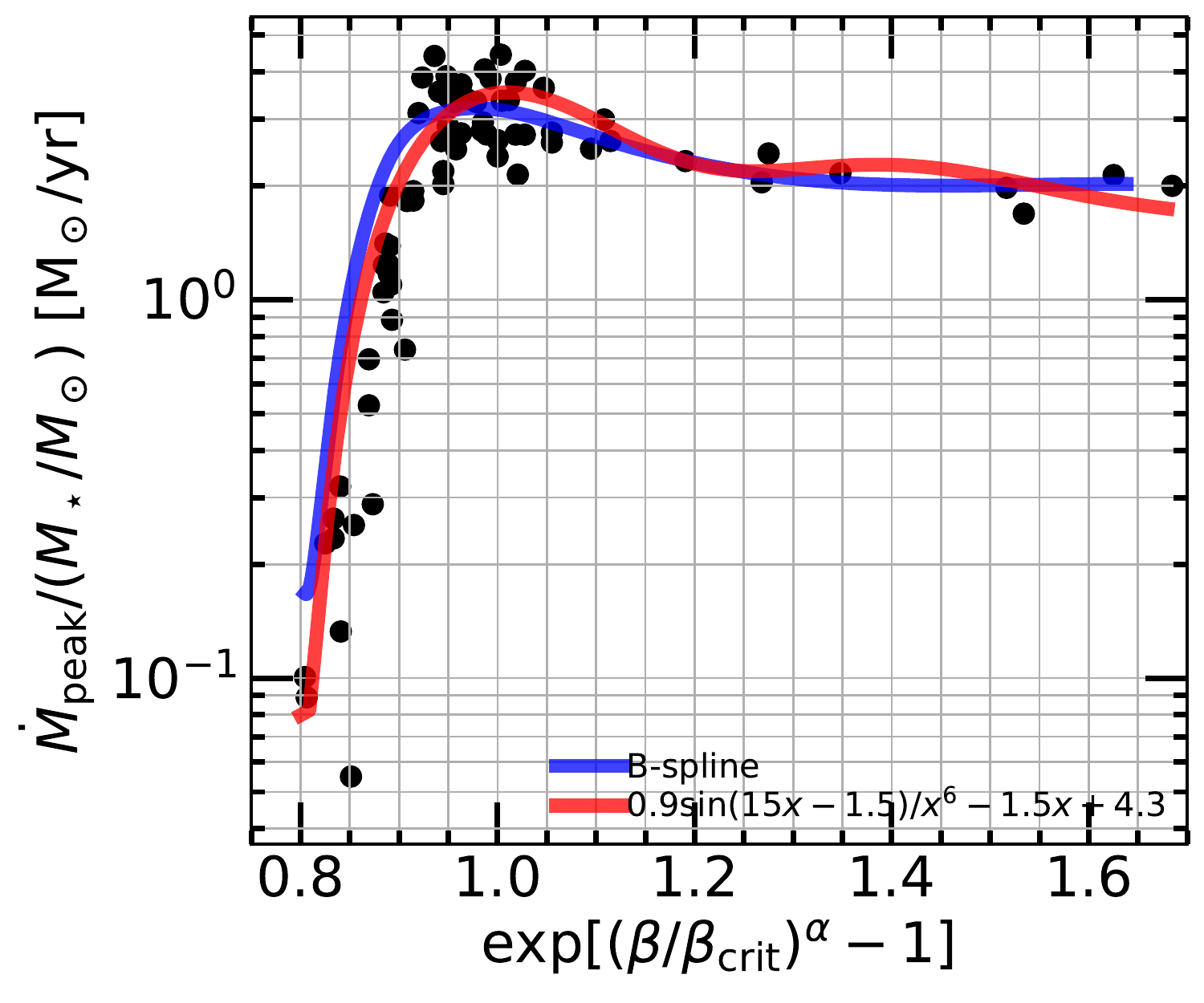}
\plotone{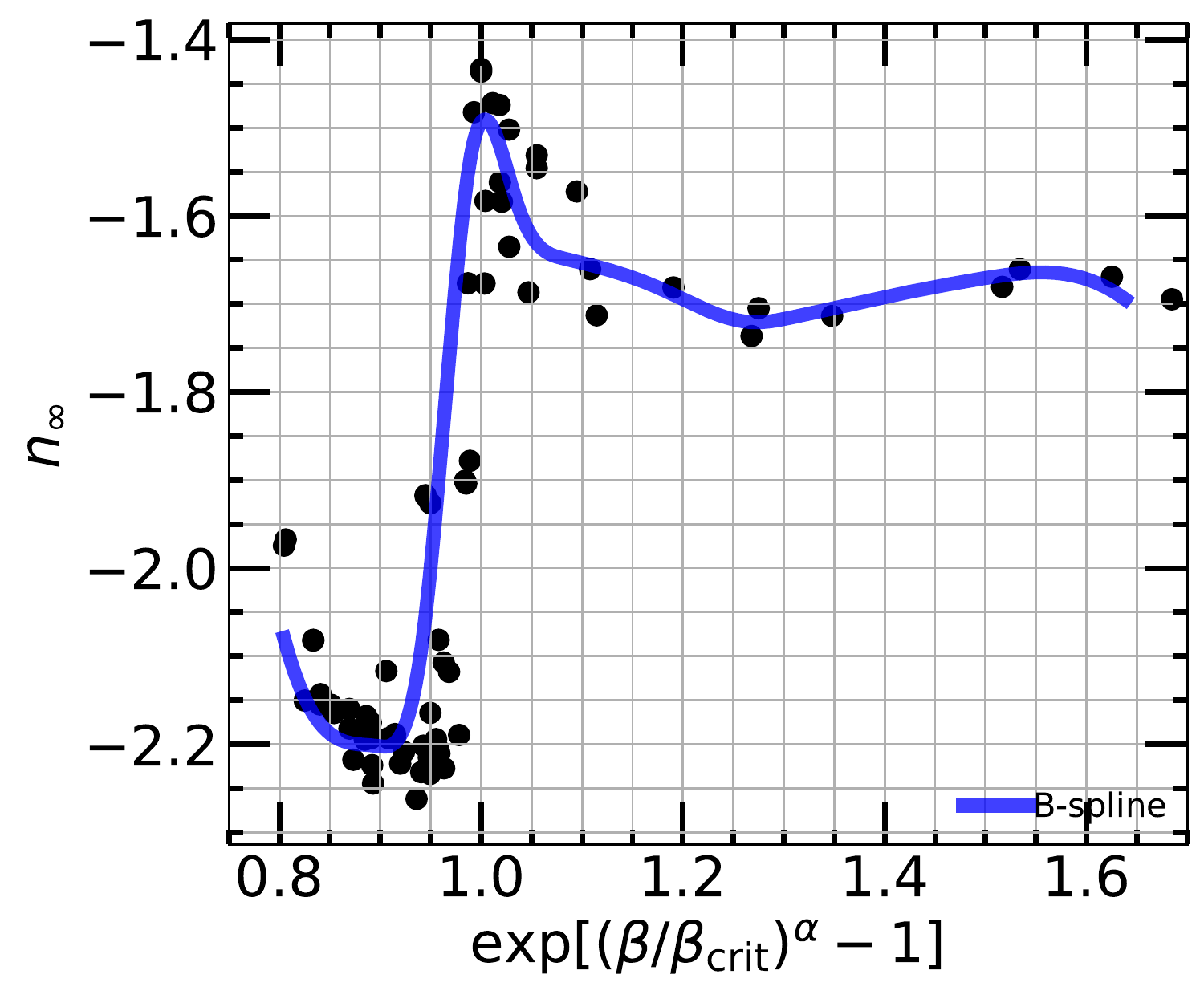}
\caption{
B-spline and analytic fits to disruption quantities $\Delta M/M_\star$, $t_{\rm peak}$, $\dot M_{\rm peak}$, and $n_\infty$.
Note for the top left panel, we extend $\Delta M/M_\star=1$ after full disruption ($x \geq 1$).
\label{fig:fitting_formulae}
}
\end{figure}

Figure~\ref{fig:fitting_formulae} shows fitted B-splines and analytic formulae for the mass lost from the star ($\Delta M/M_\star$), time of peak mass fallback rate ($t_{\rm peak}$), peak mass fallback rate ($\dot M_{\rm peak}$), and asymptotic decay power law index ($n_\infty$).
These fitting relations allow one to obtain the $\Delta M/M_\star$, $t_{\rm peak}$, $\dot M_{\rm peak}$, and $n_\infty$ for any stellar mass, stellar age, and impact parameter. One must first obtain the $\rho_c/\bar\rho$ value for this star, e.g. from Table~\ref{tab:grid}, a MESA model directly, or a pre-packaged grid of MESA models such as the MIST models \citep{2016ApJ...823..102C}.

The simple analytic fits are below. 
We were unable to fit the $n_\infty$ data with a simple formula, so the only option for this quantity is the B-spline.
\begin{eqnarray}
\Delta M/M_\star &=& 1.055 \tanh[(x+0.04)^{20}]-0.025\\
t_{\rm peak}\ {\rm [day]} &=& 2/x^{12} + 17x + 8\\
\dot M_{\rm peak}/(M_\star/M_\sun)\ {\rm [M_\sun/yr]} &=& 0.9 \sin(15x-1.5)/x^6 -1.5x+4.3\\
n_\infty &=& {\rm (unable\ to\ find\ simple\ analytic\ fit;\ see\ B\ spline\ below)}\\
x&=&\exp[(\beta/\beta_{\rm crit})^\alpha -1],\ \ \ \alpha=(\rho_c/\bar\rho)^{-1/3}\\
\beta_{\rm crit} &\approx& 
\begin{cases}
0.5 (\rho_c/\bar\rho)^{1/3},\ \rho_c/\bar\rho \lesssim 500\\
0.39 (\rho_c/\bar\rho)^{1/2.3},\ \rho_c/\bar\rho \gtrsim 500 \label{eq:beta_crit_appendix}
\end{cases}
\end{eqnarray}
The range for $\Delta M/M_\star$ is $x \in [0.8,1.0]$ and for $t_{\rm peak}$ and $\dot M_{\rm peak}$ is $x \in [0.8,1.7]$. 
\added{We note that the equations for $\beta_{\rm crit}$ are only approximate.}

Below we provide python code to read in and evaluate the B-spline fits to the disruption quantities $\Delta M/M_\star$, $t_{\rm peak}$, $\dot M_{\rm peak}$, and $n_\infty$. This code will reproduce the blue lines in Figure~\ref{fig:fitting_formulae}. Note that the order of all of the splines is 3. The knots and coefficients have been rounded to 3 decimals, which gives precision indistinguishable from the original fitted splines.

\texttt{
\\
import numpy as np\\
from scipy.interpolate import splev\\
\# deltam/mstar\\
knots = [0.804, 0.804, 0.804, 0.804, 0.859, 0.881, 0.892, 0.903, 0.957, 1.001, 1.001, 1.001, 1.001]; \\
coeffs = [0.028, 0.068, 0.027, 0.159, 0.234, 0.416, 0.745, 0.97, 1.015, 0.0, 0.0, 0.0, 0.0] \\
\# tpeak\\
knots = [0.804, 0.804, 0.804, 0.804, 1.082, 1.638, 1.638, 1.638, 1.638]; \\
coeffs = [40.874, 27.579, 24.097, 39.004, 33.804, 0.0, 0.0, 0.0, 0.0] \\
\# mdotpeak/mstar\\
knots = [0.804, 0.804, 0.804, 0.804, 0.897, 0.943, 1.082, 1.638, 1.638, 1.638, 1.638]; \\
coeffs = [0.167, 0.127, 2.617, 3.477, 1.714, 2.034, 2.024, 0.0, 0.0, 0.0, 0.0] \\
\# ninf\\
knots = [0.804, 0.804, 0.804, 0.804, 0.897, 0.943, 0.99, 1.036, 1.082, 1.175, 1.268, 1.314, 1.499, 1.638, 1.638, 1.638, 1.638]; \\
coeffs = [-2.079, -2.212, -2.19, -2.218, -1.333, -1.637, -1.652, -1.675, -1.73, -1.701, -1.671, -1.656, -1.695, 0.0, 0.0, 0.0, 0.0] \\
\\
\# evaluate B-splines\\
x = np.linspace(min(knots), max(knots), 100); \\
y = splev(x, [knots, coeffs, 3]) \\
}

Note that in order to use the B-splines to retrieve values for a specific star, one needs to use the scaled $x$-axis, and thus input the $\rho_c/\bar\rho$ and $\beta_{\rm crit}$ values for the star. $\rho_c/\bar\rho$ can be obtained as above and $\beta_{\rm crit}$ can be obtained from either Table~\ref{tab:beta_crit} or our simple formula (Eq.~\ref{eq:beta_crit_appendix}).
Then one plugs these into
\begin{equation}
    x=\exp[(\beta/\beta_{\rm crit})^\alpha -1],\ \ \ \alpha=(\rho_c/\bar\rho)^{-1/3}.
\end{equation}
This $x$ is the value to plug into the B-spline in order to retrieve the desired $y$-axis value.

%%%%%%%%%%%%%%%%%%%
\section{MESA profiles vs. polytropes}

\begin{figure}[htp!]
\epsscale{0.45}
\plotone{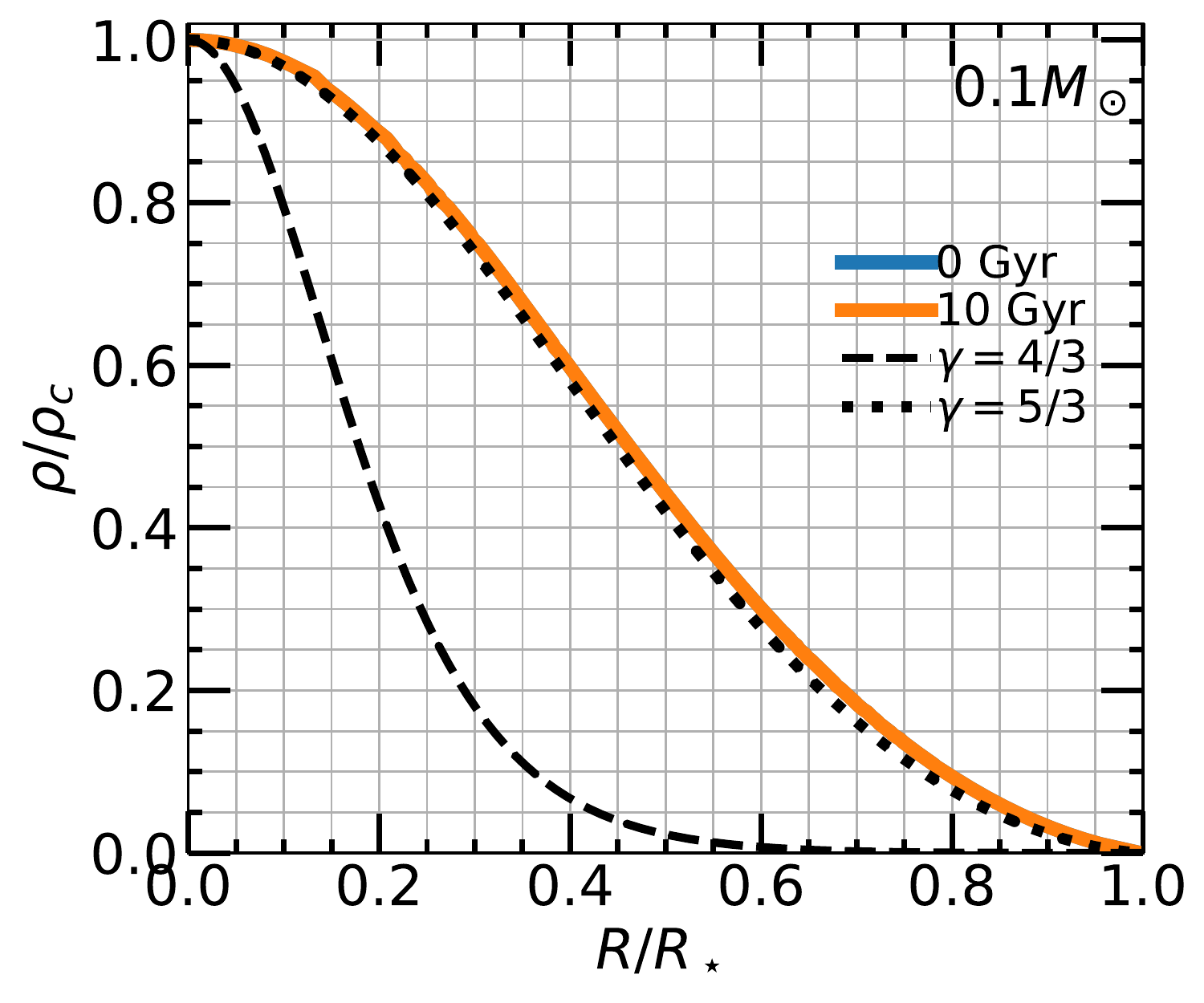}
\plotone{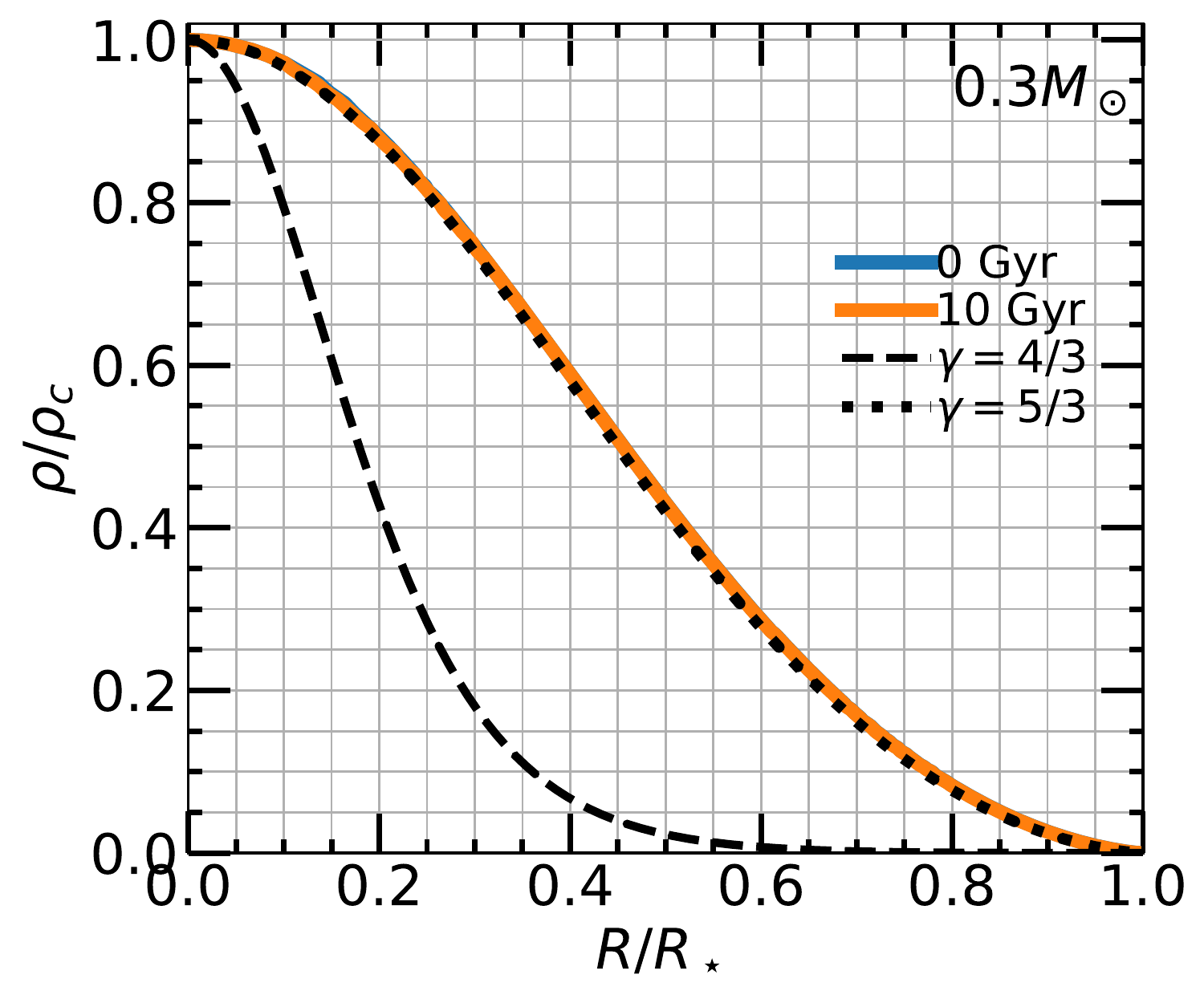}
\plotone{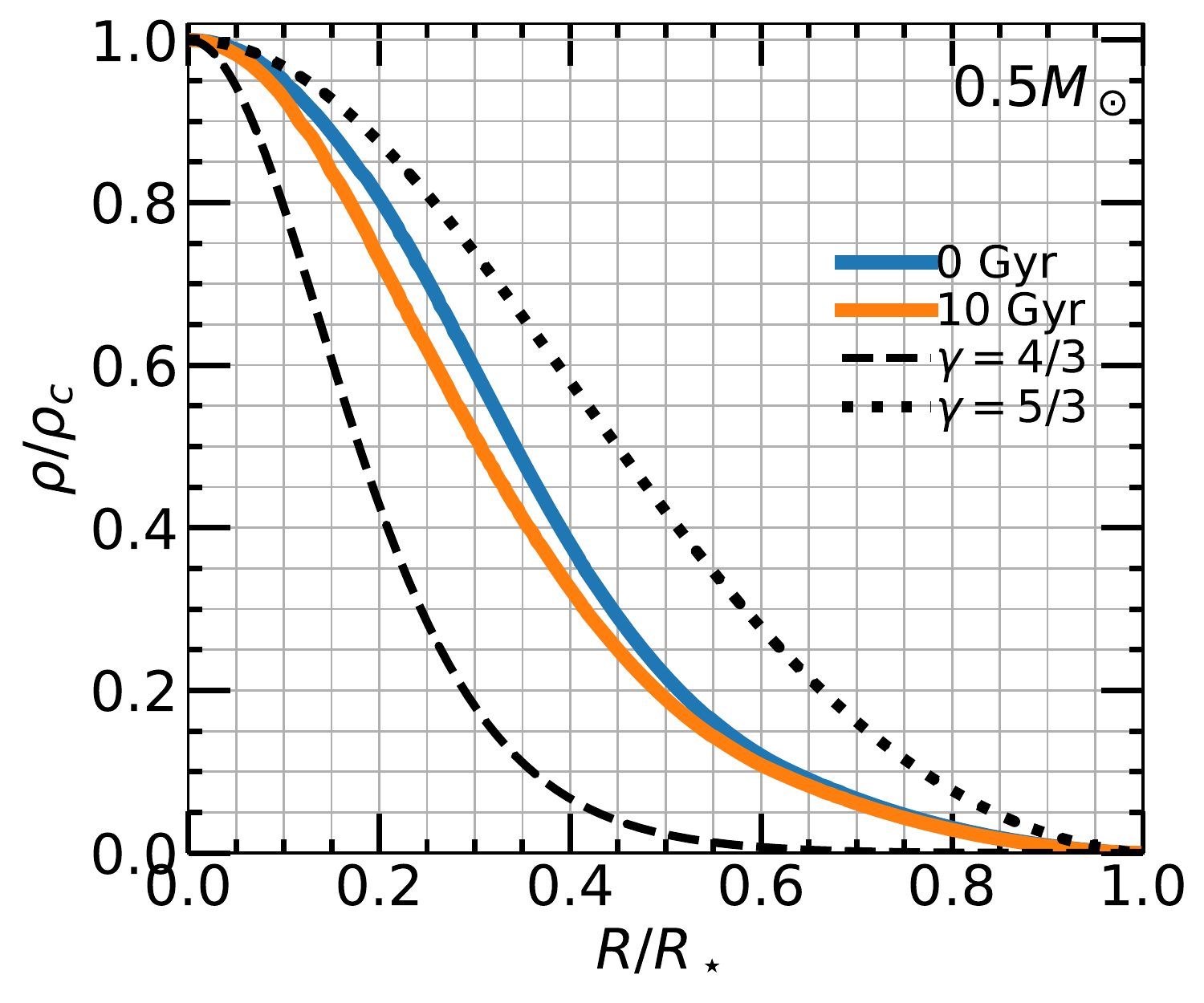}
\plotone{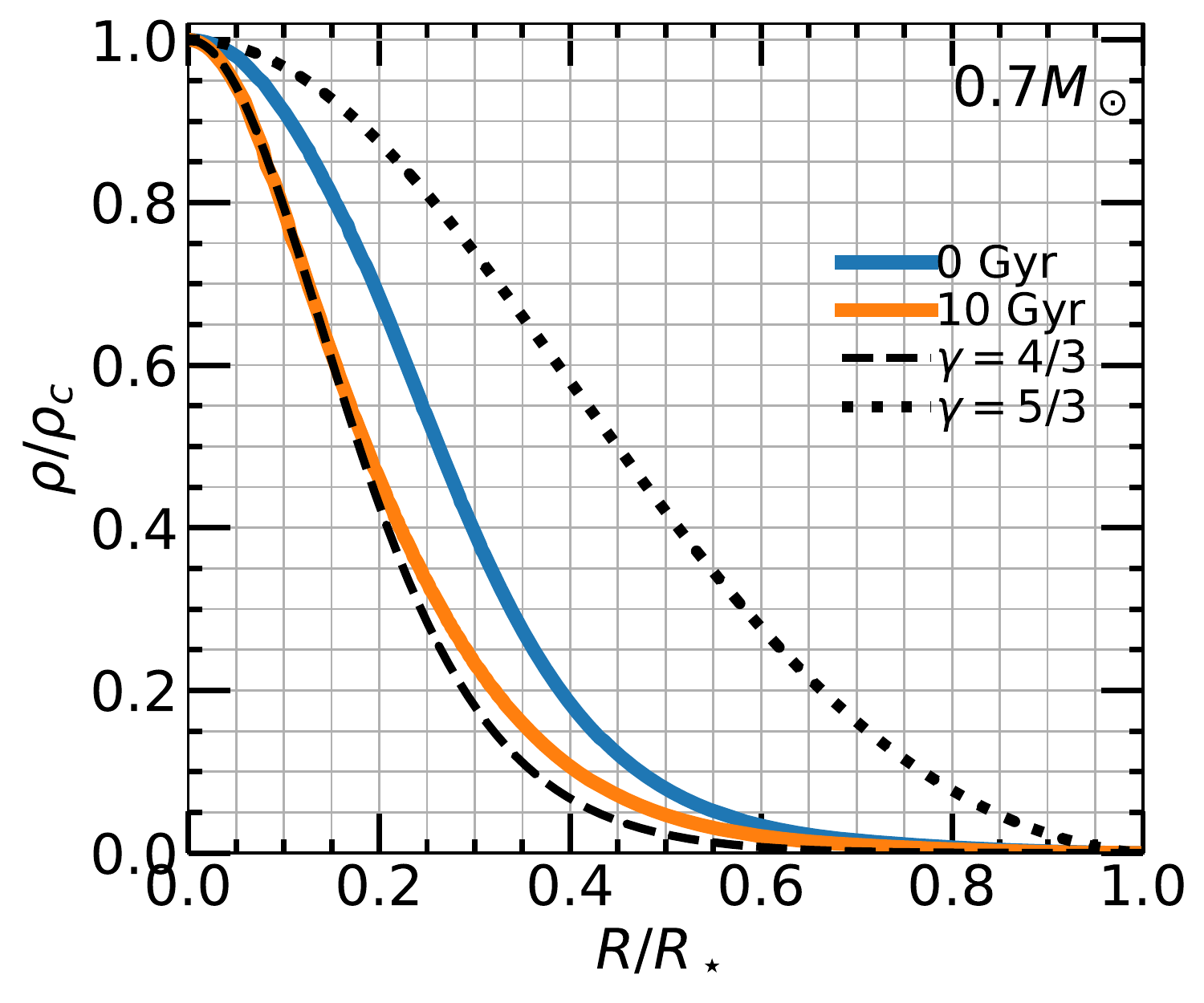}
\plotone{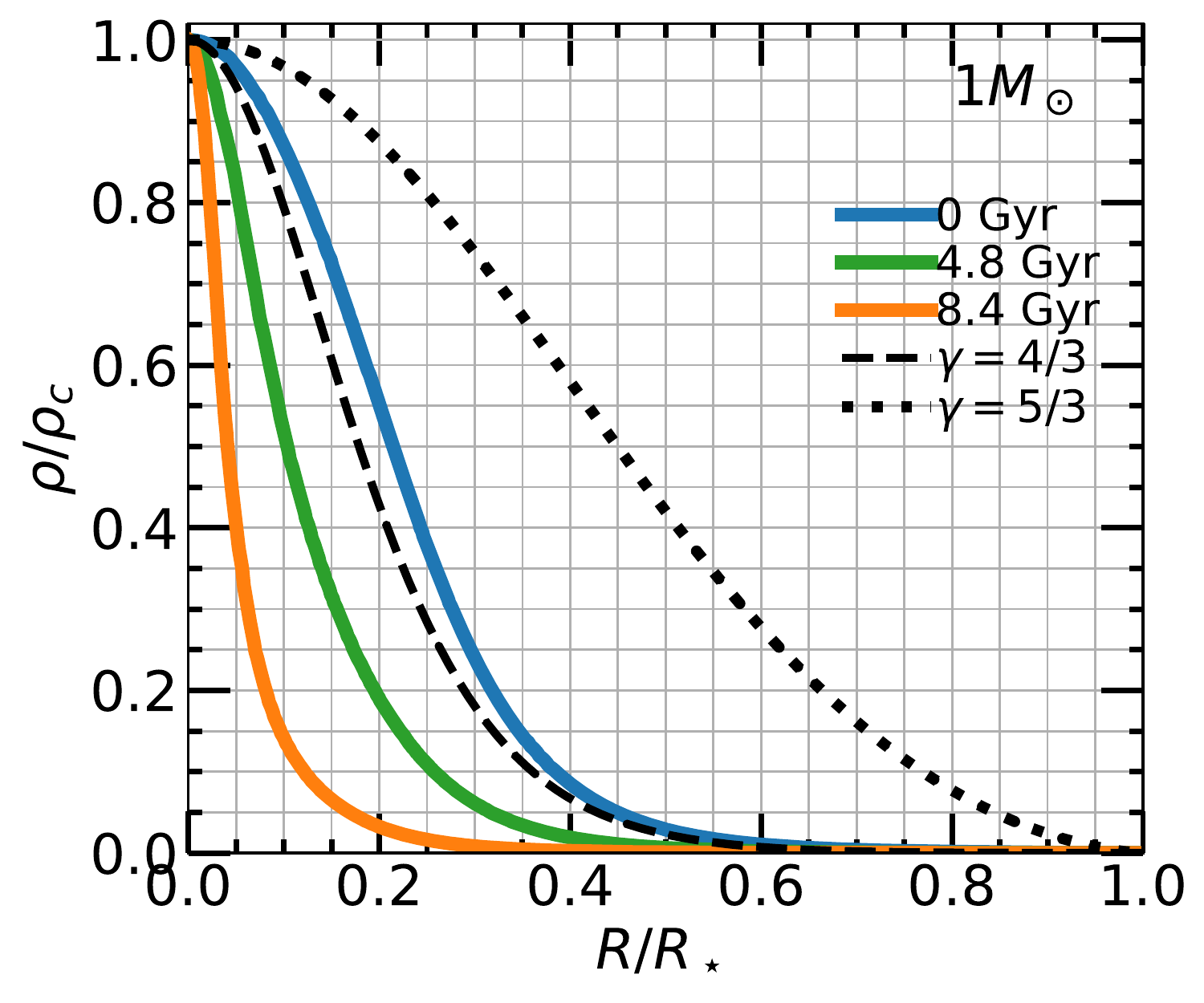}
\plotone{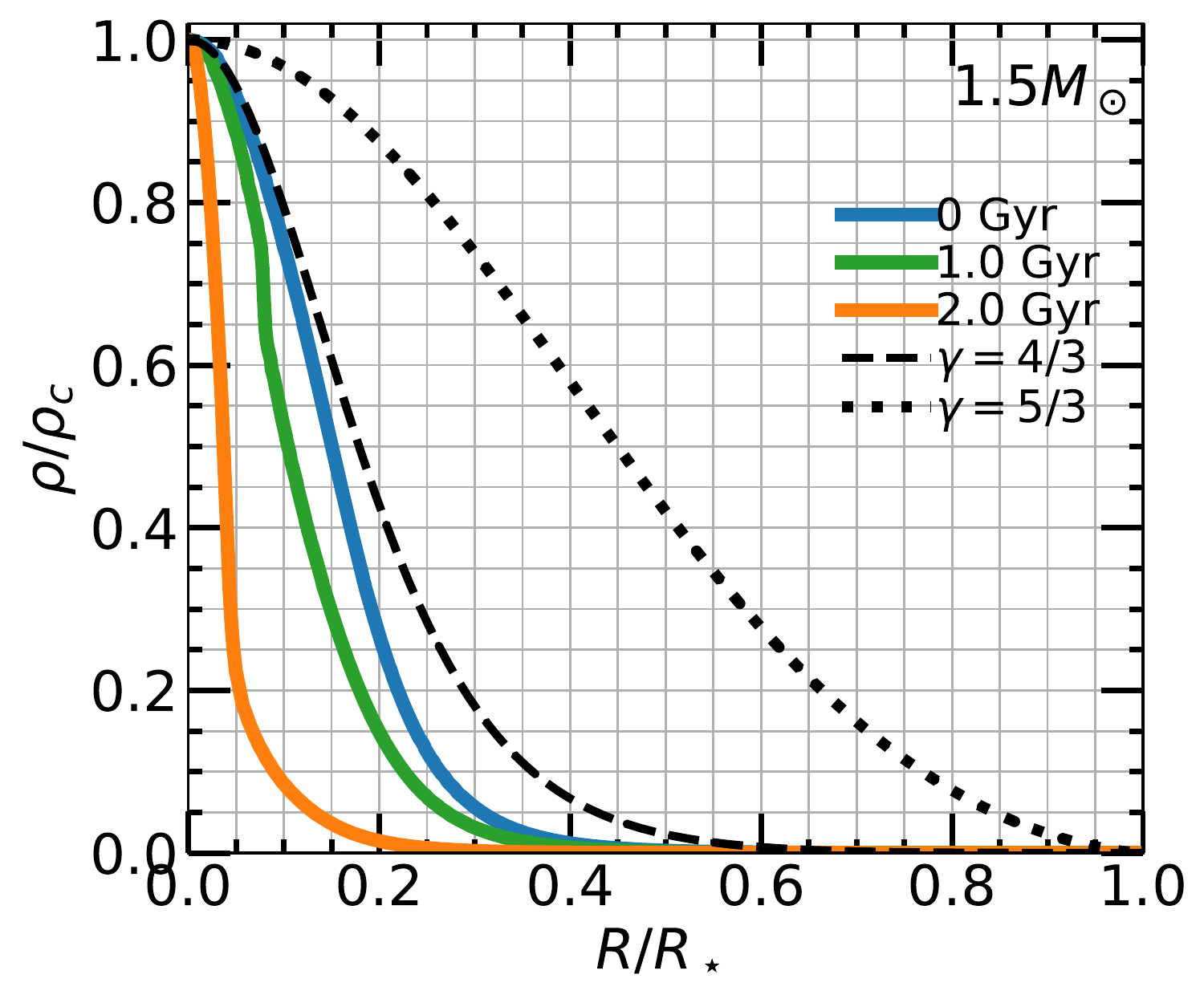}
\plotone{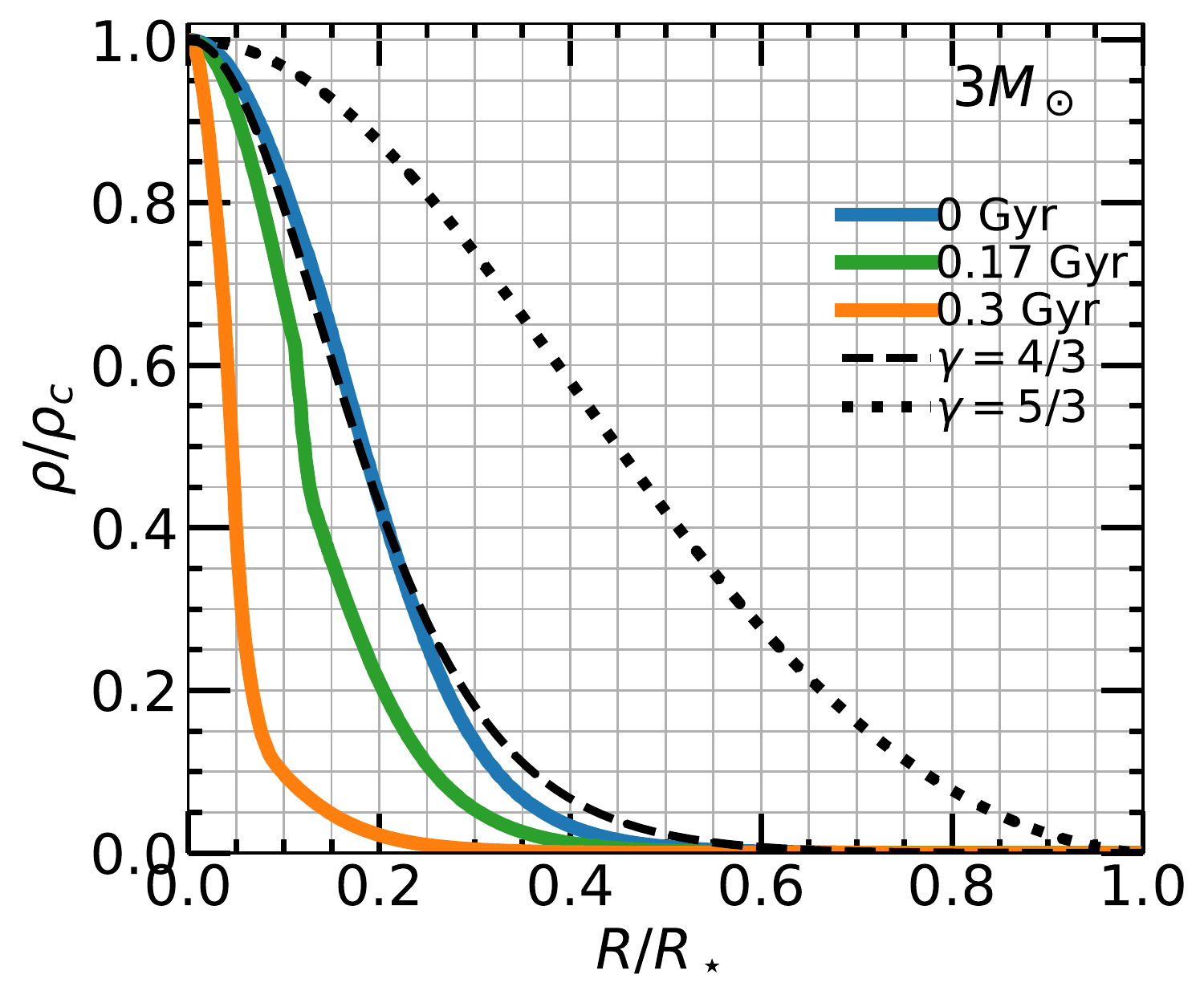}
\plotone{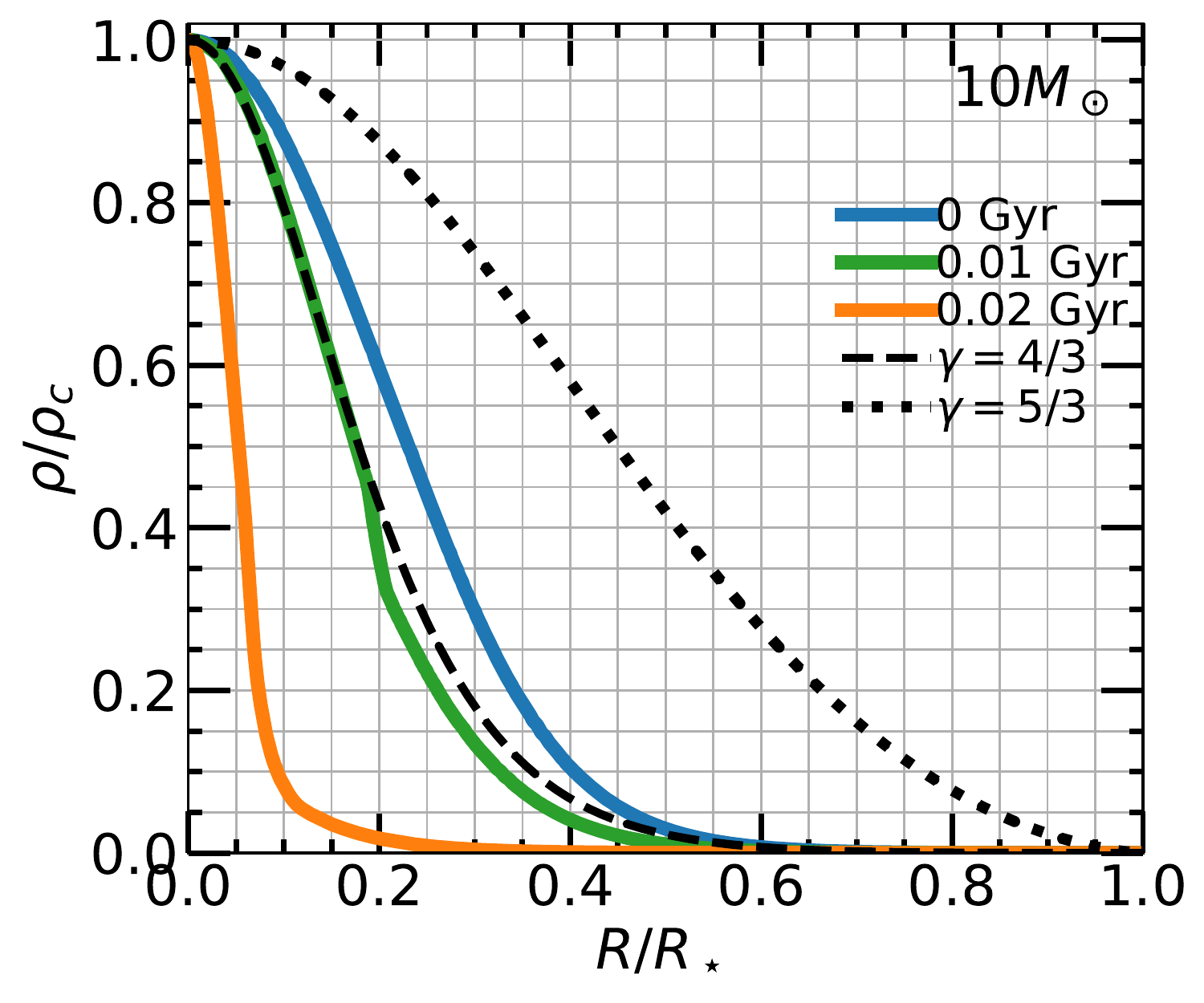}
\caption{
MESA density profiles vs. $\gamma=5/3$ and $\gamma=4/3$ polytropes, normalized to central density and stellar radius. Each panel is a different mass. Blue is ZAMS, \added{green is MAMS (shown for $M_\star \geq 1M_\sun$)}, and orange is min(10 Gyr, TAMS).
\label{fig:MESA_profiles_vs_polytropes_norm}
}
\end{figure}

Figure~\ref{fig:MESA_profiles_vs_polytropes_norm} shows MESA density profiles vs. $\gamma=5/3$ and $\gamma=4/3$ polytropes, normalized to central density and stellar radius. One can see where a polytropic stellar model is sufficient and where we need to use a MESA stellar model as initial conditions for the FLASH simulations. For $M_\star=0.1, 0.3M_\sun$, the profiles are nearly identical to $\gamma=5/3$ polytropic stellar structures. For $M_\star=0.5, 0.7 M_\sun$ the profiles are in between $\gamma=5/3$ $\gamma=4/3$ polytropic stellar structures, and could in principle be simulated with a polytropic simulation using a $\gamma$ that has been matched to this star. For $M_\star \gtrsim 1M_\sun$, as the star evolves off the ZAMS, the profile becomes more centrally concentrated than $\gamma=4/3$ and thus cannot be simulated self-consistently with a polytropic stellar structure, as these are unstable for $\gamma \lesssim 4/3$. One requires a Helmholtz EOS in order to provide hydrodynamic support for these non-ZAMS stars. Note that the ZAMS $M_\star \gtrsim 1M_\sun$ stars are not uniformly well represented by $\gamma=4/3$ polytropes---the only star that is quite well approximated by a $\gamma=4/3$ polytrope is the 3$M_\sun$ ZAMS star.
\added{Though we do not do FLASH simulations for intermediate ages between ZAMS and TAMS for the 1.5$M_\sun$, 3$M_\sun$, and 10$M_\sun$ stars in this work (we interpolate instead; see Section~\ref{sec:stars_lib}), we show a MAMS profile at age $t \approx t_{\rm TAMS}/2$ for these higher mass stars (including for the 1$M_\sun$ star, which we do simulate) for comparison. While the TAMS profiles are clearly much more centrally concentrated than a $\gamma=4/3$ profile, the MAMS profiles are somewhat closer to the polytropic approximation, though still not a good fit.
}

One can also see that $\rho_c/\bar\rho$ does not map 1-1 exactly to the stellar density profile, though it is a very good approximation. Consider the 10Gyr 0.7$M_\sun$ star ($\rho_c/\bar\rho=36$) and the ZAMS 10$M_\sun$ star ($\rho_c/\bar\rho=38$). The 10$M_\sun$ star has a \added{slightly} higher value of $\rho_c/\bar\rho$ but its density profile is \added{slightly} less centrally concentrated than that of the 0.7$M_\sun$ star.

%%%%%%%%%%%%%%%%
\section{Initial profiles after relaxation}

\begin{figure}[htp!]
\epsscale{0.38}
\plotone{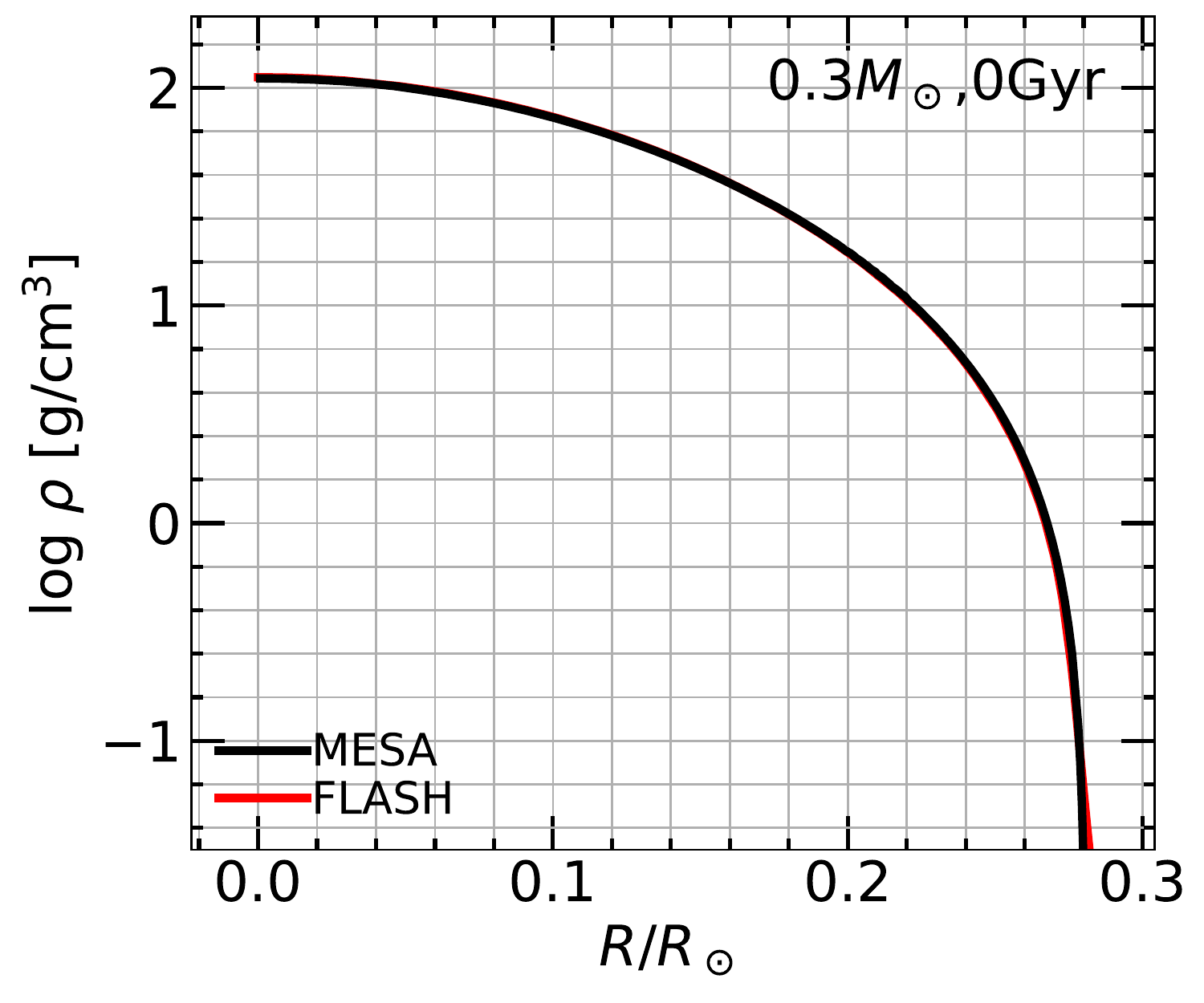}
\plotone{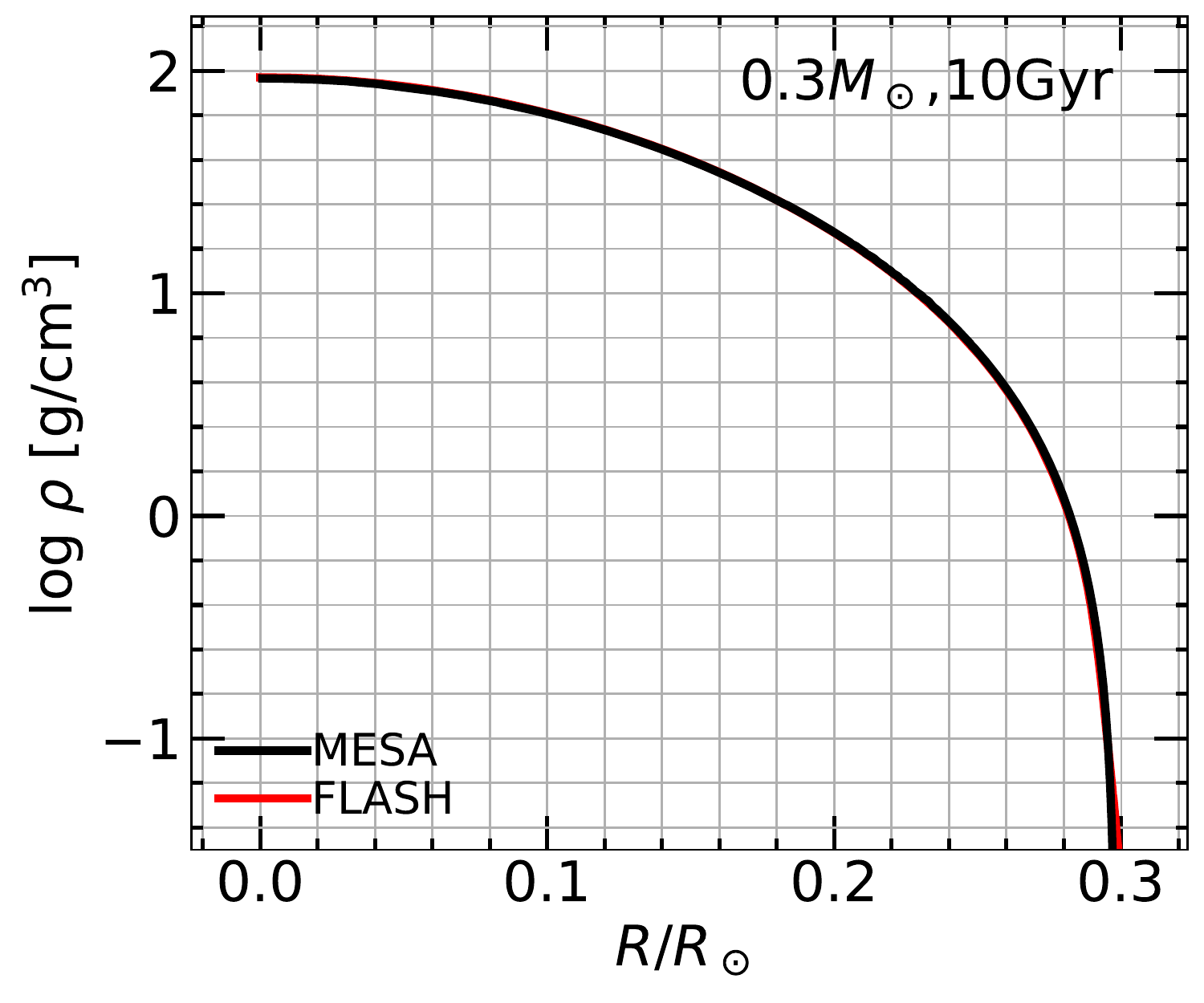}
\plotone{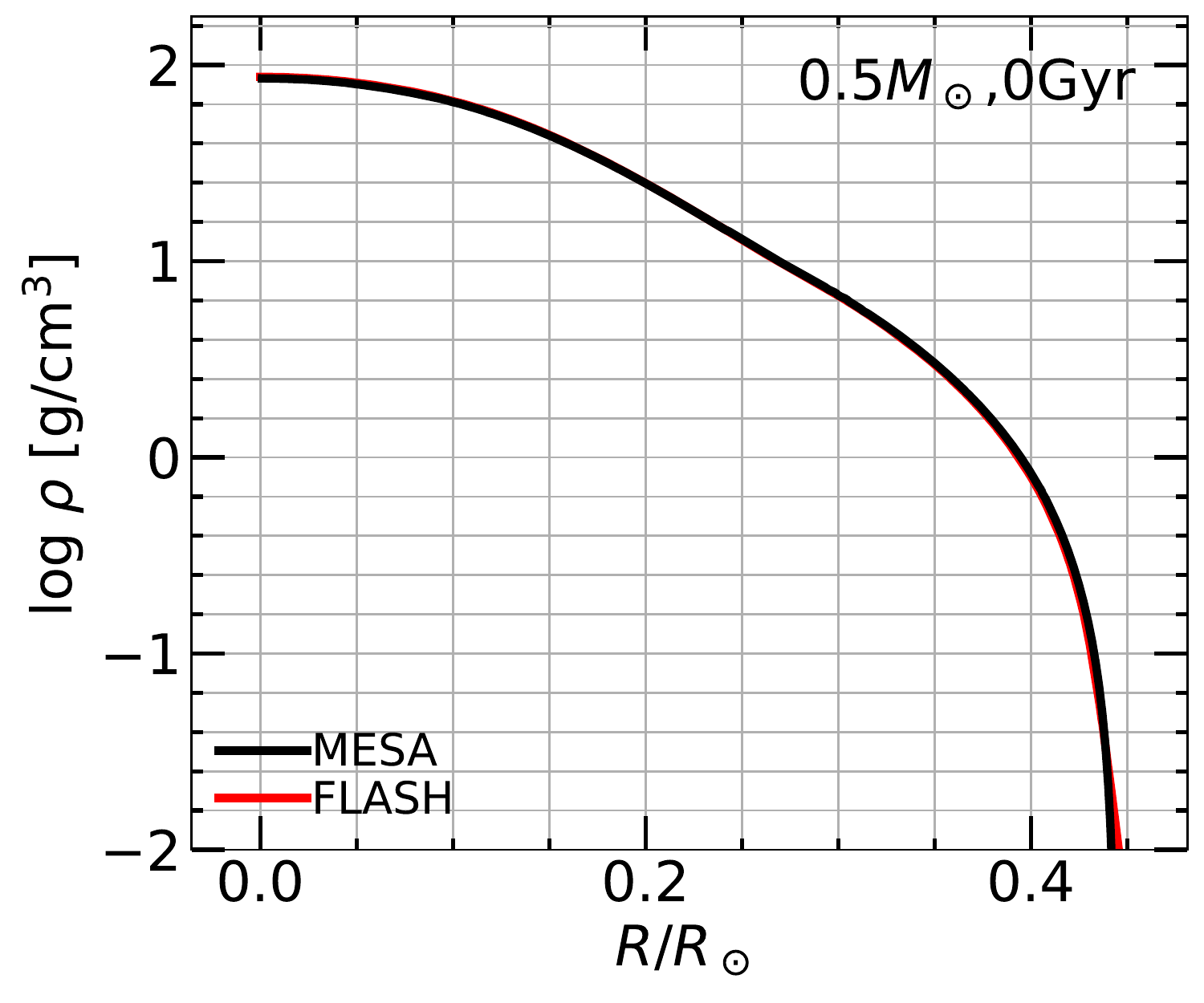}
\plotone{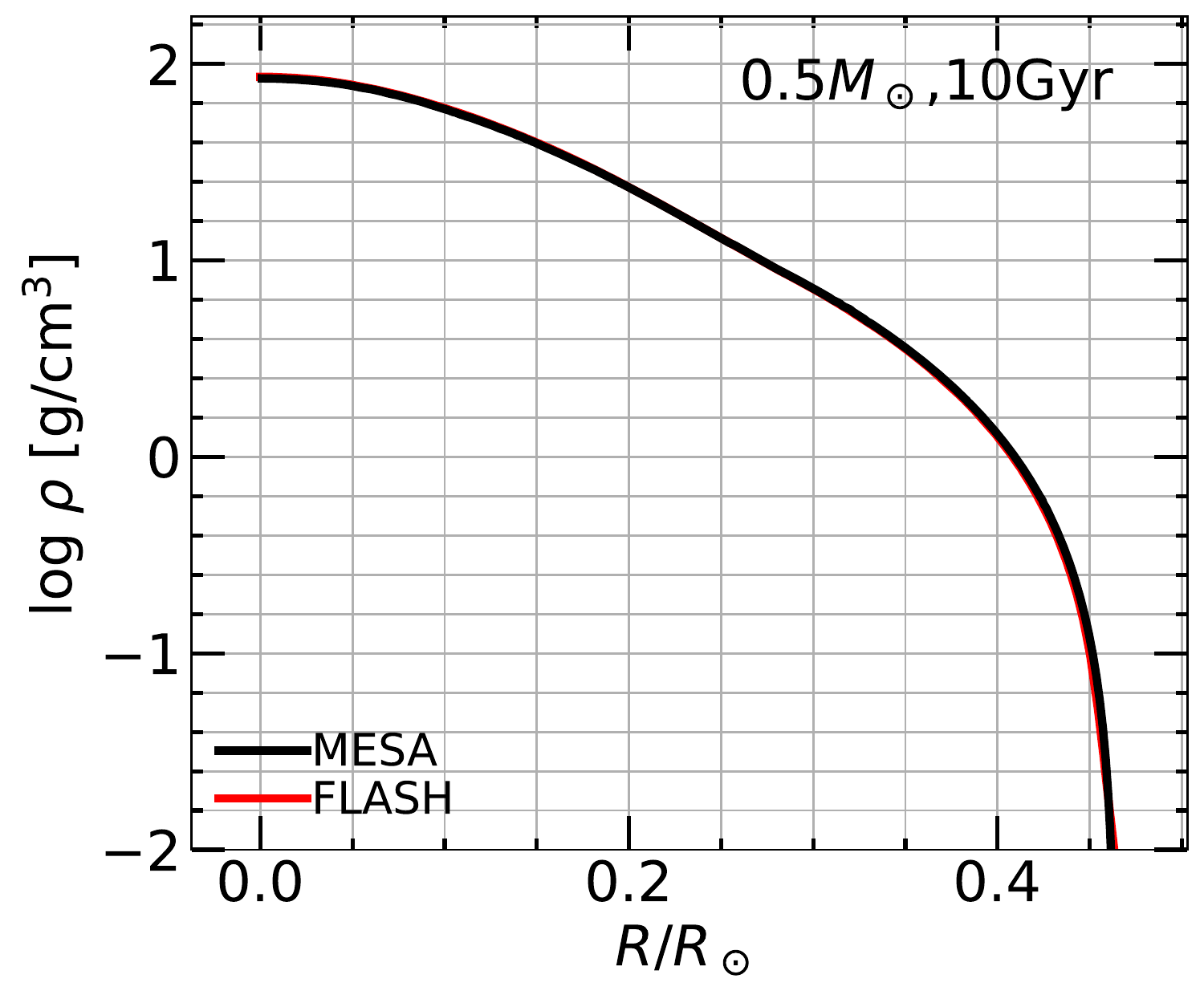}
\plotone{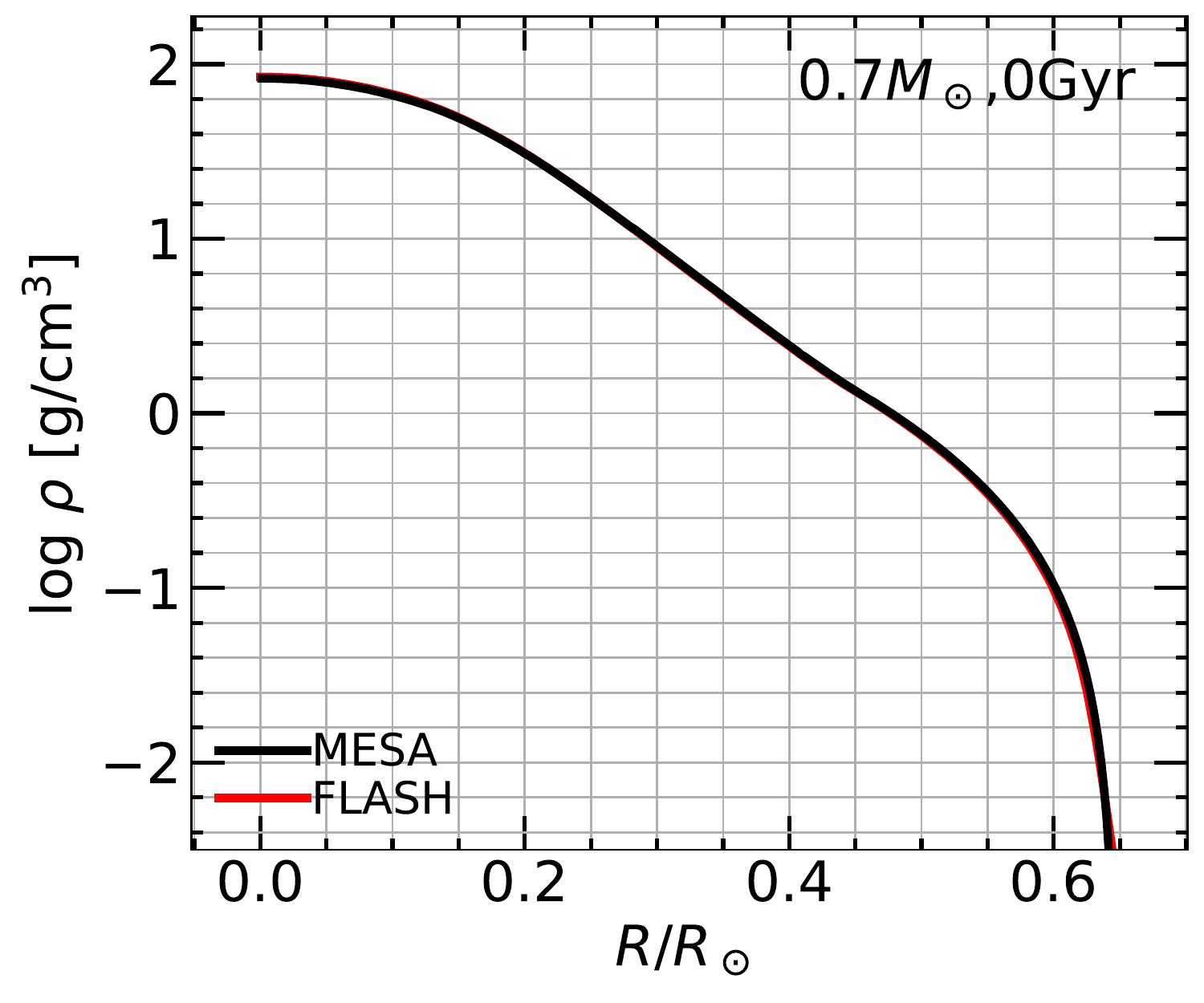}
\plotone{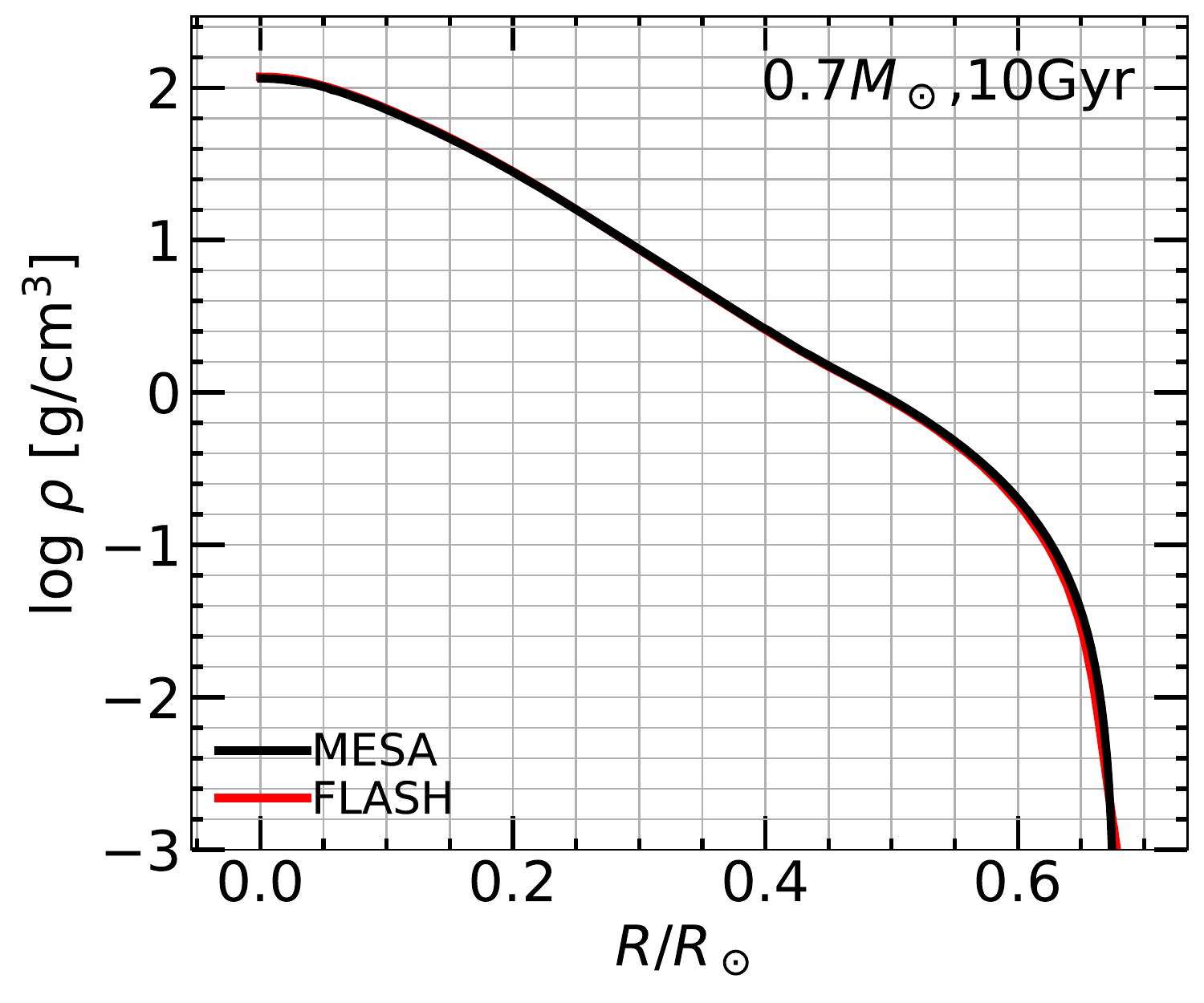}
\plotone{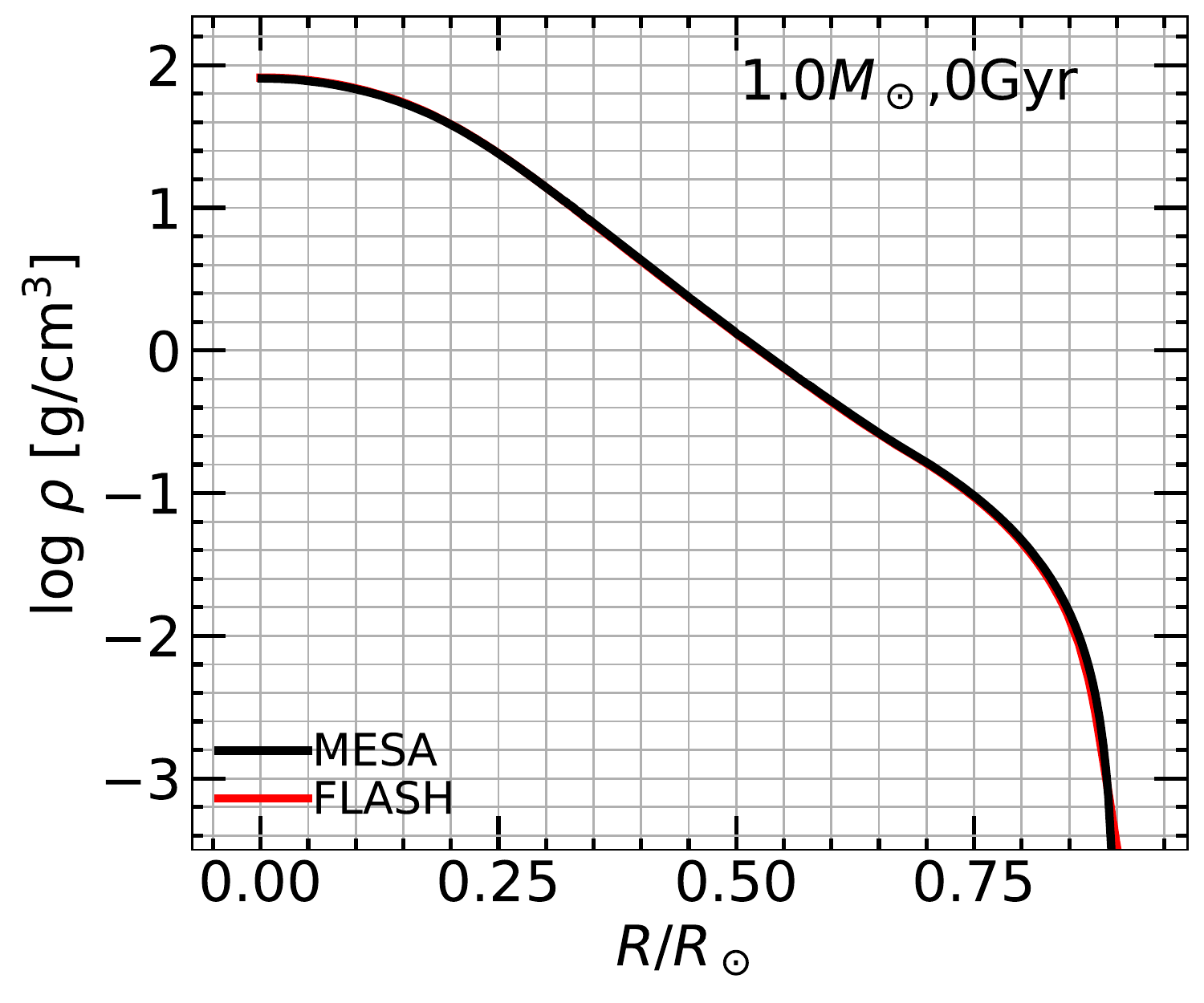}
\plotone{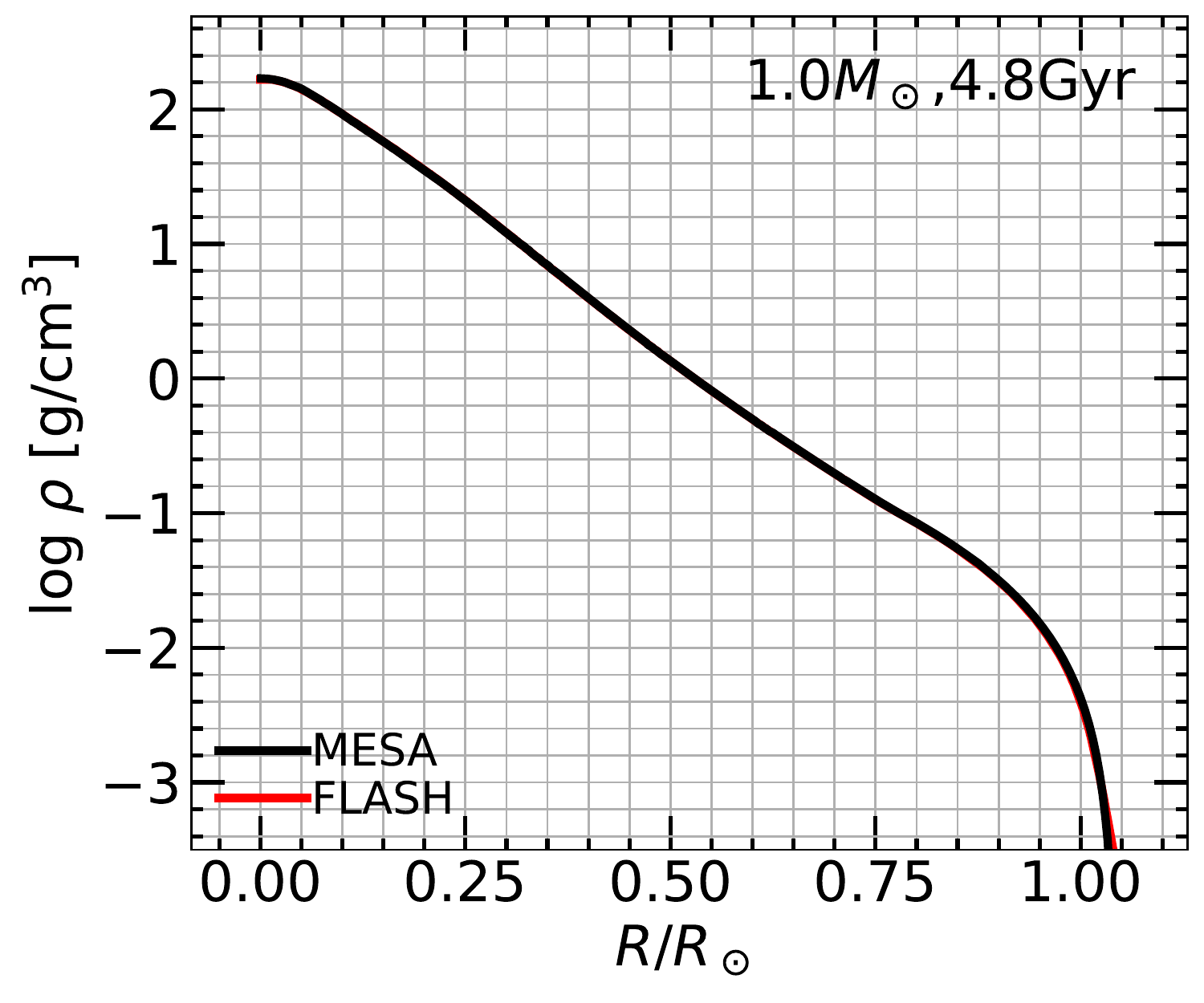}
\plotone{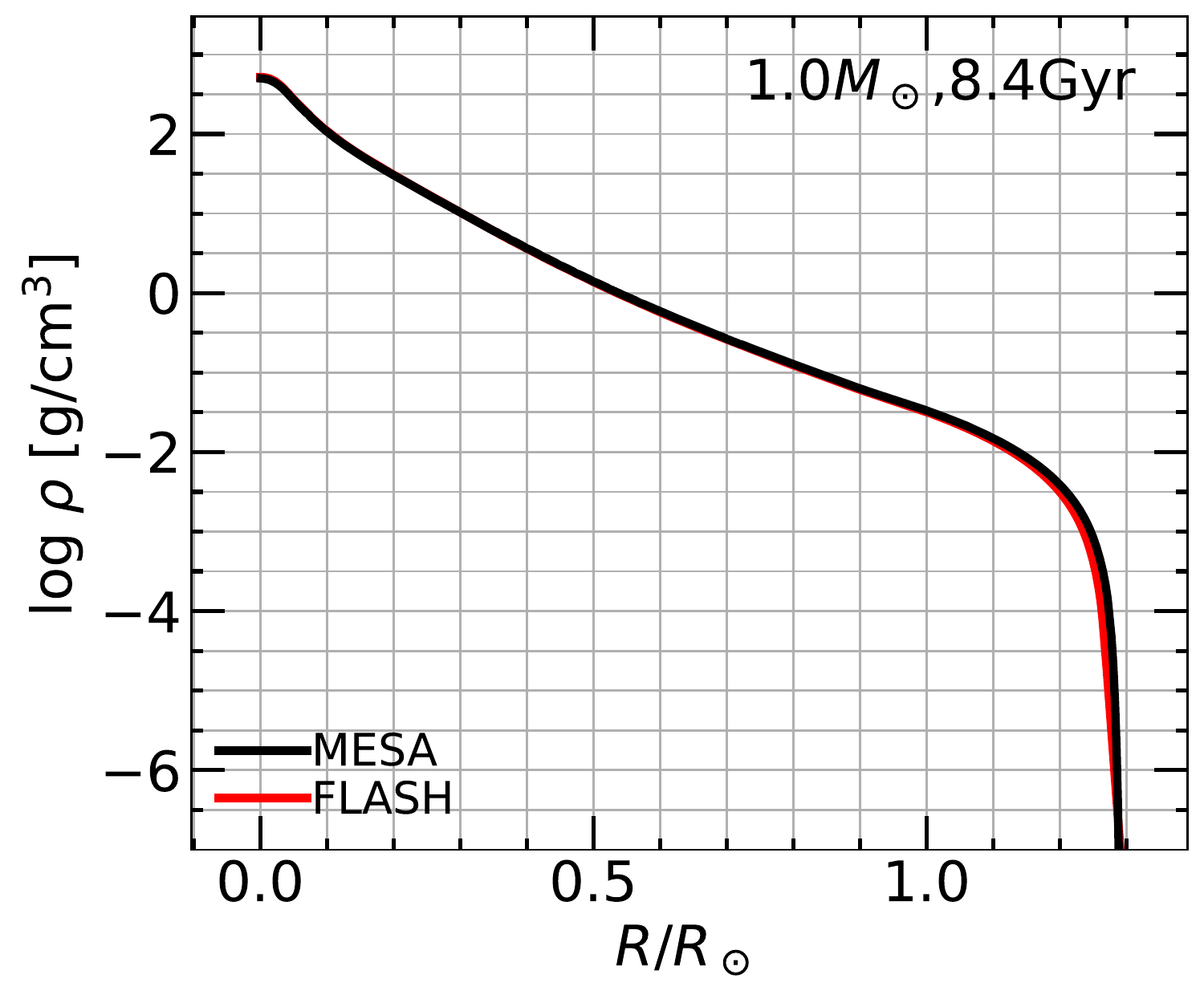}
\plotone{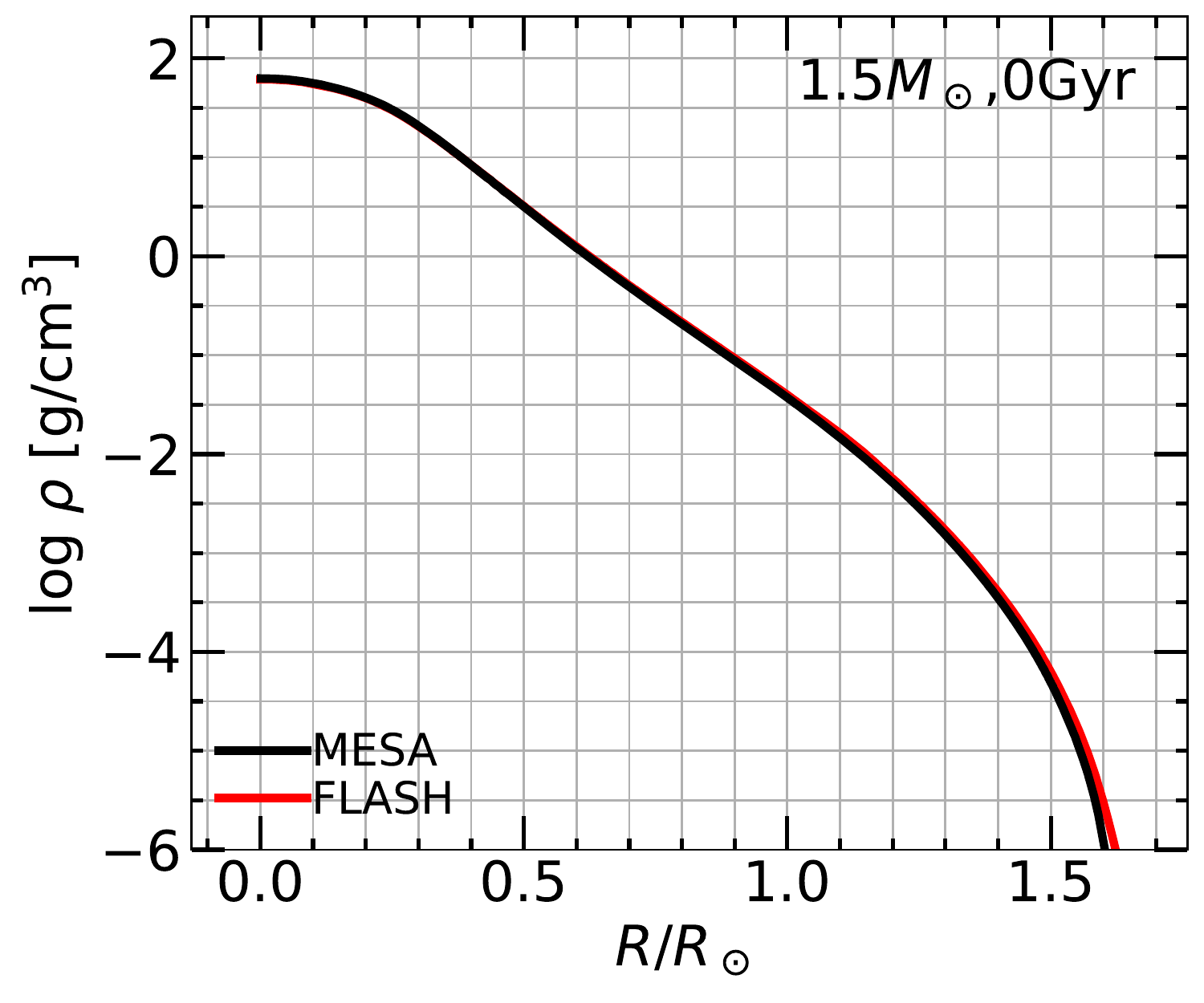}
\plotone{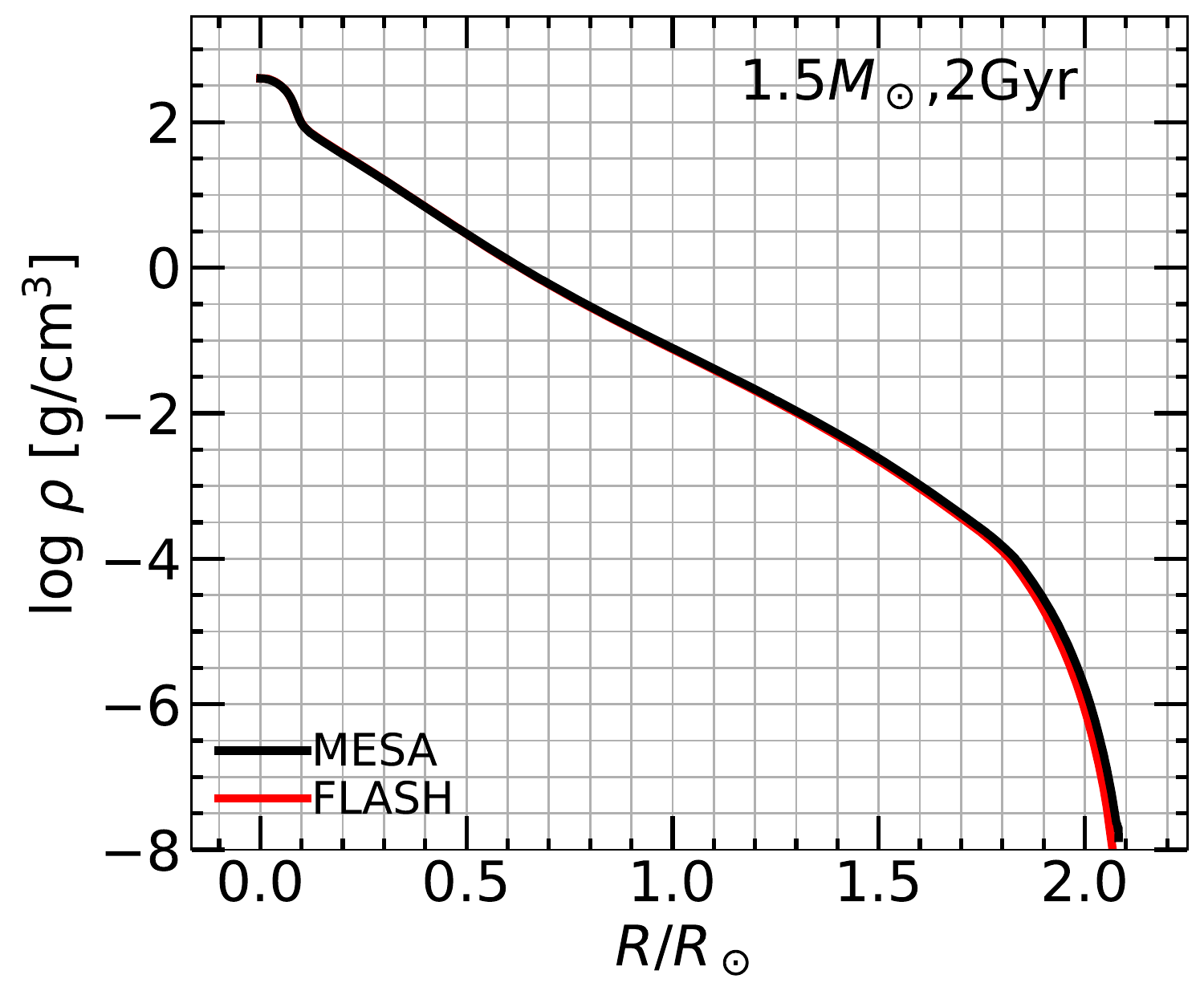}
\plotone{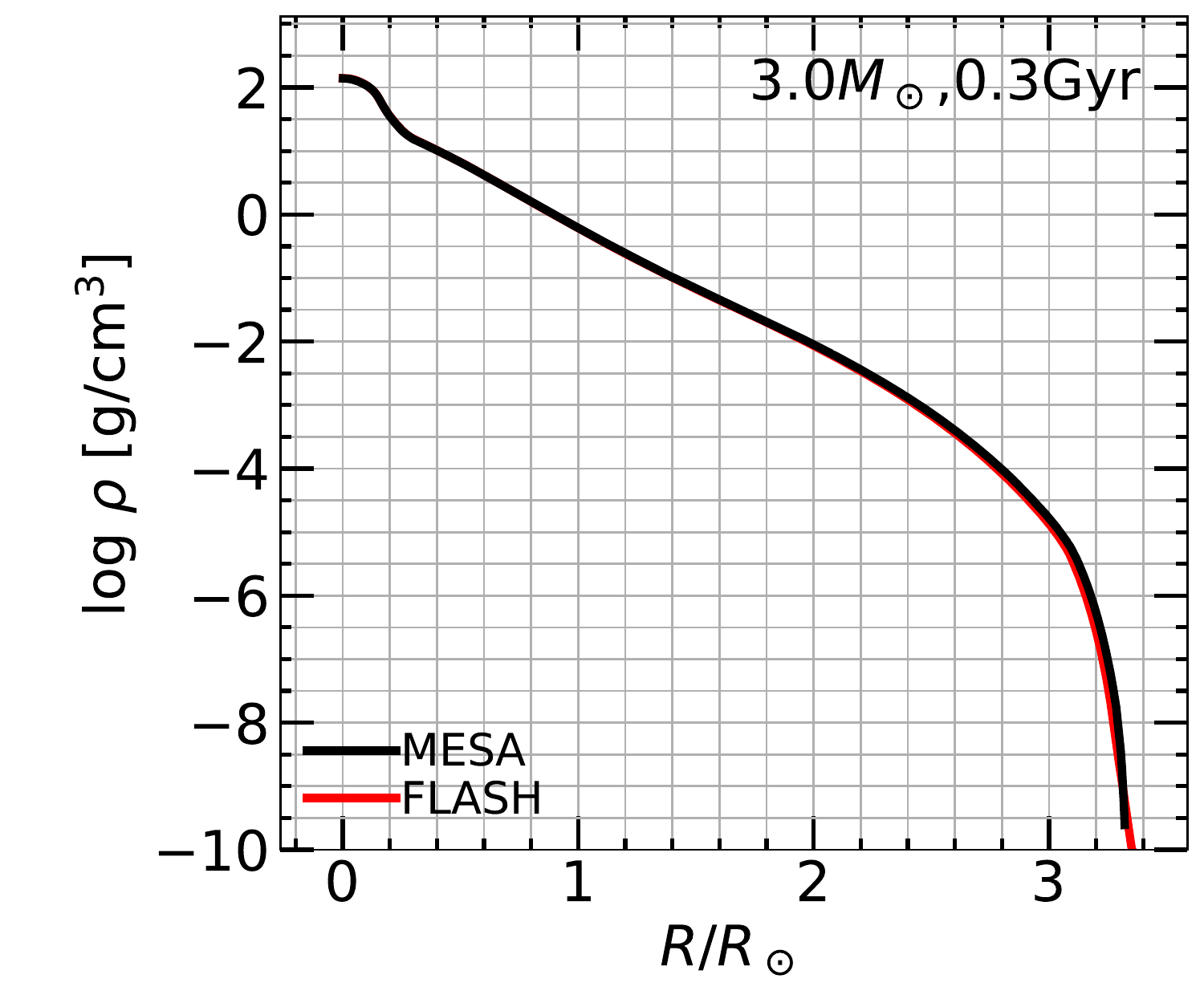}
\caption{
Stellar density profiles after $5 t_{\rm dyn}$ of relaxation onto the hydrodynamical grid in FLASH (red), compared to initial profiles from MESA (black). These two profiles match exactly at the start of the simulation ($t=0$).
\label{fig:initial_profiles}
}
\end{figure}

Figure~\ref{fig:initial_profiles} shows the stellar density profiles after $5 t_{\rm dyn}$ of relaxation onto the hydrodynamical grid in FLASH, compared to initial profiles from MESA. To be clear, these two profiles match exactly at the start ($t=0$) of the FLASH simulation, but here they are shown after $5 t_{\rm dyn}$. Our setup has been tested such that the density profile does not change significantly after $\approx 100 t_{\rm dyn}$ of relaxation onto the hydrodynamical grid (without a BH present) in \citet{2009ApJ...705..844G}.

As mentioned in Section~\ref{sec:methods}, the most centrally concentrated stars, for which $\rho_c/\bar\rho \gtrsim 150$, have a higher initial resolution in FLASH of 512 cells across the diameter of the star. Stars with $\rho_c/\bar\rho \lesssim 150$ are initially resolved by 131 cells across their diameters.

%%%%%%%%%%%%%%%%
\section{Critical impact parameter and pericenter distance vs. stellar mass}

\begin{figure*}[htp!]
\epsscale{0.55}
\plotone{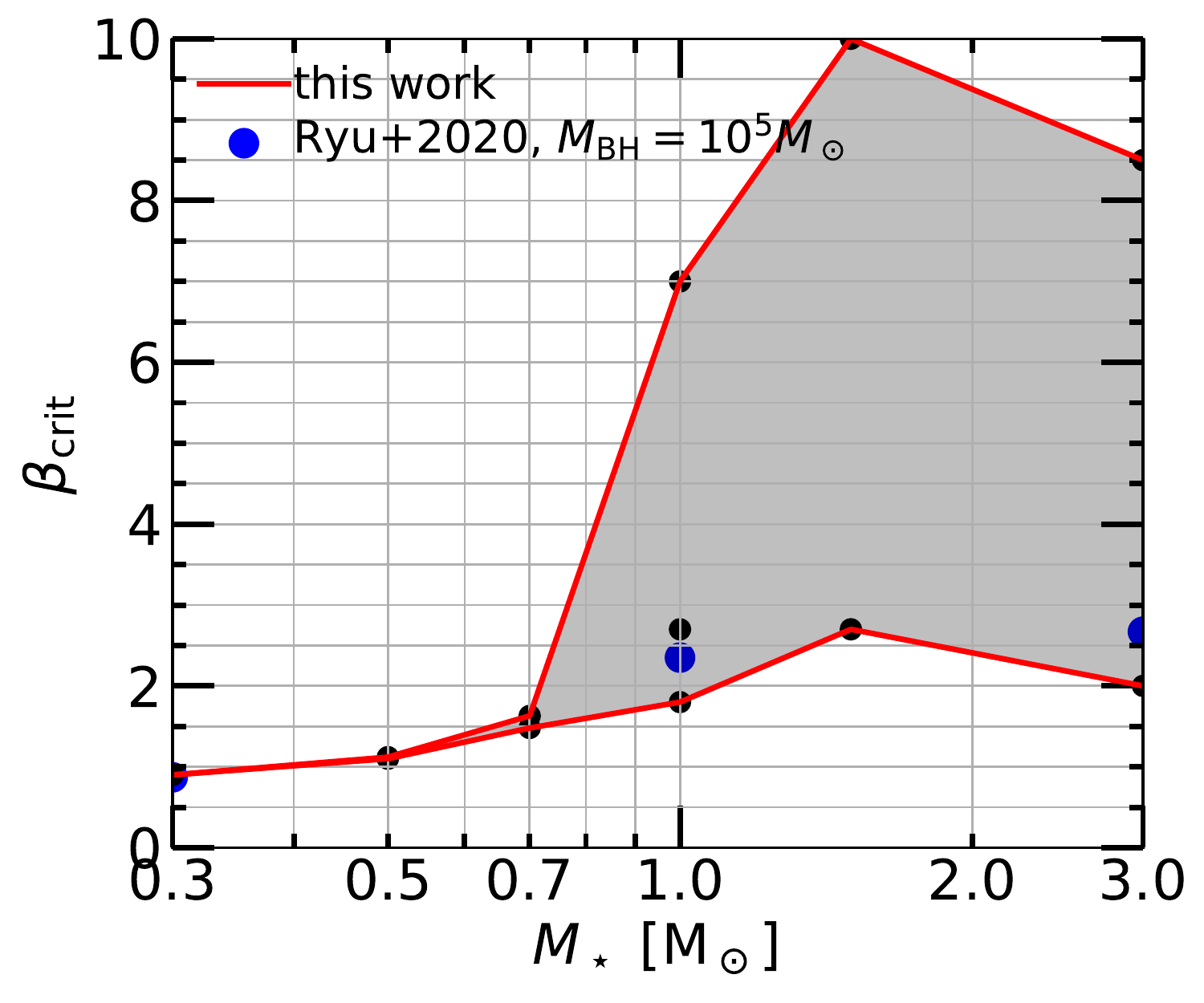}
\plotone{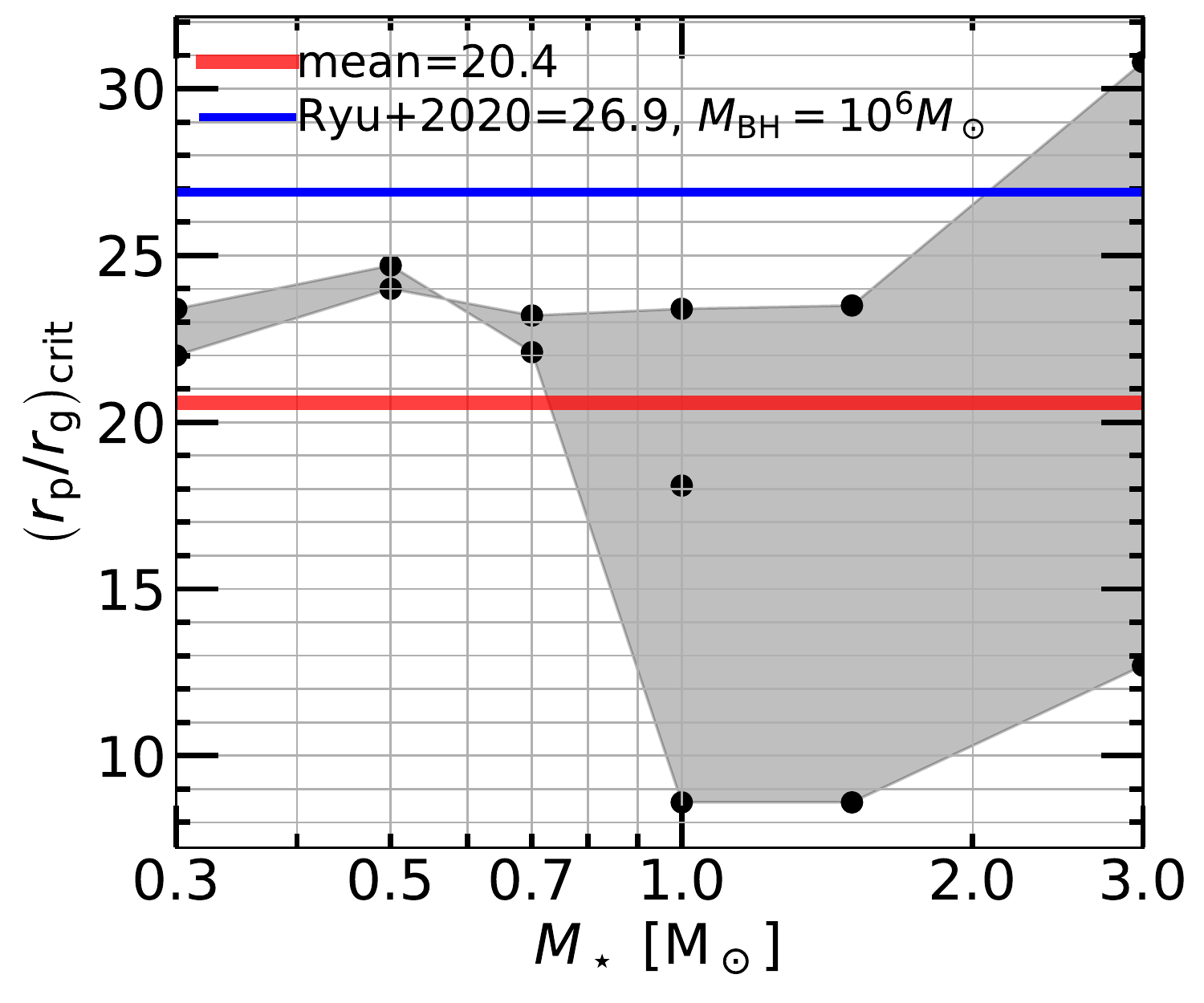}
\caption{
Left: critical $\beta$ vs. stellar mass. See also Figure~\ref{fig:beta_crit_vs_Mstar}. Right: critical pericenter distance over gravitational radius ($r_{\rm p}/r_{\rm g}$) vs. stellar mass. Points at the same stellar mass have different stellar ages. 
Comparison to \citet{2020arXiv200103501R,2020arXiv200103502R,2020arXiv200103503R,2020arXiv200103504R} in blue.
\label{fig:beta_crit_vs_Mstar2}
}
\end{figure*}

Figure~\ref{fig:beta_crit_vs_Mstar2} shows critical $\beta$ and critical pericenter distance over gravitational radius ($r_{\rm p}/r_{\rm g}$) as a function of stellar mass. 
See also Figure~\ref{fig:beta_crit_vs_Mstar} for critical $\beta$ as a function of $\rho_c/\bar\rho$, in which there is a simple power-law relationship.
We find large ranges in critical $\beta$ and critical $r_{\rm p}/r_{\rm g}$, especially with stellar age for $M_\star \gtrsim 0.8 M_\sun$. 
At a fixed stellar mass, $\beta_{\rm crit}$ can vary by a factor of $\approx$4.
We also compare to the results from \citet{2020arXiv200103501R,2020arXiv200103502R,2020arXiv200103503R,2020arXiv200103504R}. 
To avoid repetition, see Sections~\ref{sec:methods} and \ref{sec:conclusion} for discussion of the differences between these works.
We note that these authors studied only a single stellar age, so a direct comparison is often not possible.
For physical pericenter distance, our average from 0.3 to 3$M_\sun$ is $r_{\rm p}/r_{\rm g} \approx 20.4$.
Our results are consistent with those of \citet{2020arXiv200103501R,2020arXiv200103502R,2020arXiv200103503R,2020arXiv200103504R} for a $M_{\rm BH}=10^5 M_\odot$, and for a $M_{\rm BH}=10^6 M_\odot$ \citet{2020arXiv200103501R,2020arXiv200103502R,2020arXiv200103503R,2020arXiv200103504R} find $r_{\rm p}/r_{\rm g} \sim 26.9$, averaging from 0.15$M_\sun$ to 3$M_\sun$.

%%%%%%%%%%%%%%%%
\section{Range of applicability of nonrelativistic TDE hydrodynamics simulations}

\begin{figure}[htp!]
\epsscale{0.55}
\plotone{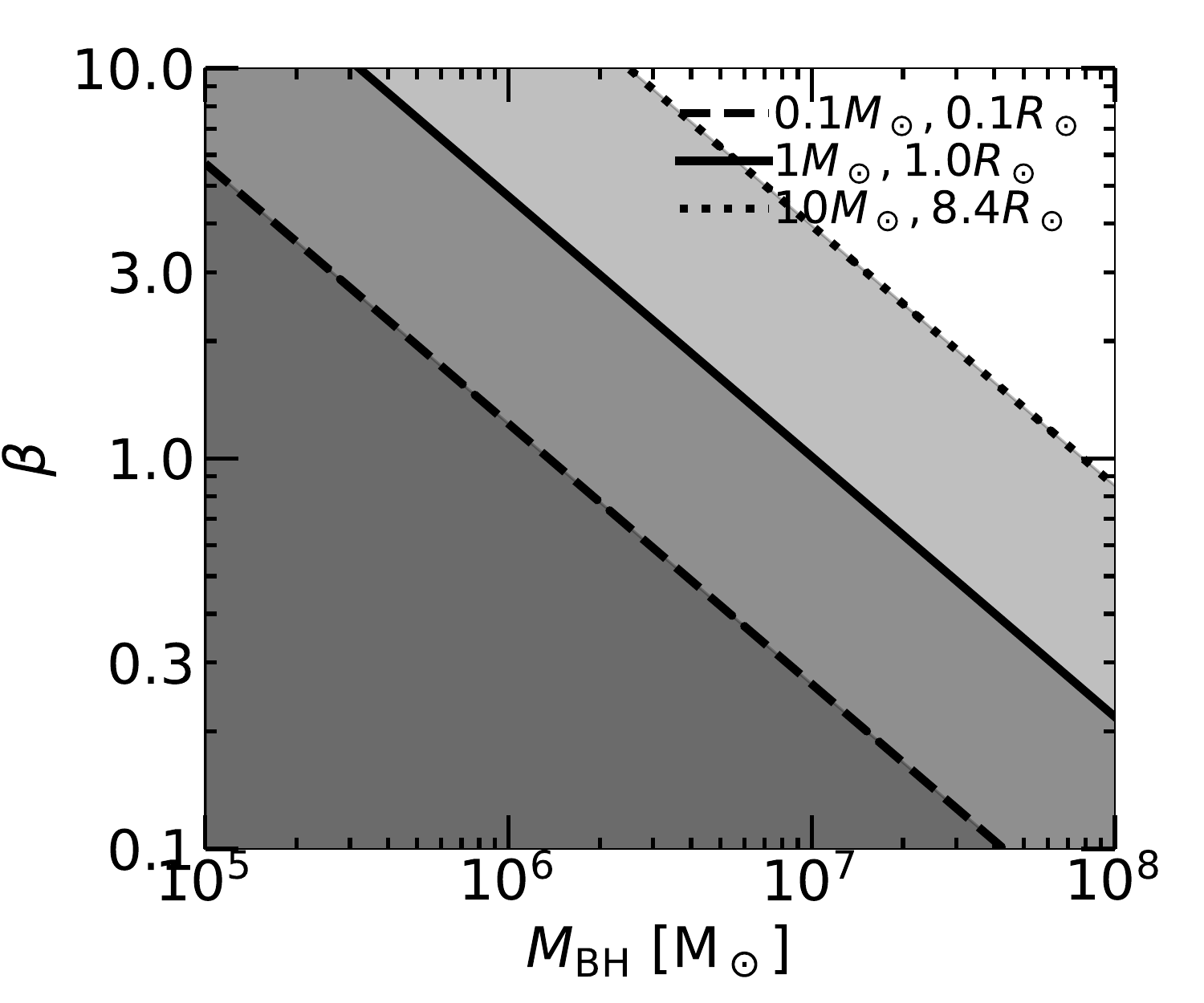}
\caption{
Range of applicability of nonrelativistic TDE fallback-rate simulations in $\beta$ vs. BH mass. Lines are $r_{\rm p}=10 r_{\rm g}$. Shaded regions below lines are where $r_{\rm p}>10 r_{\rm g}$ and we expect nonrelativistic simulations to have $\lesssim$10\% error.
\label{fig:mbh_beta_phase_space}
}
\end{figure}

Figure~\ref{fig:mbh_beta_phase_space} shows regions where $r_{\rm p}>10 r_{\rm g}$ for a few stars (0.1$M_\sun$, 1$M_\sun$, 10$M_\sun$). This is where we expect nonrelativistic tidal disruption simulations to have $\lesssim$10\% error. 
For more detailed discussion of relativistic effects, see Section~\ref{sec:conclusion}.
Note that the BH masses of the host galaxies of observed TDEs, found independently through new observations and the $M$--$\sigma$ relationship \citep{2017MNRAS.471.1694W,2019MNRAS.487.4136W}, lie in the range $3\times10^5 M_\sun \leq M_{\rm BH} \leq 2 \times 10^7 M_\sun$ and peak at $10^6 M_\sun$.
The stellar masses obtained through fitting TDEs \citep{2019ApJ...872..151M} are almost all $M_\star \lesssim 1 M_\sun$.

%%%%%%%%%%%%%%%%
\section{Test of stellar structure vs. EOS}\label{sec:stellar_structure_vs_EOS}

In this paper we claim that stellar structure is the sole determinant of several tidal disruption quantities. 
To make this claim more robust we must first disentangle the effects of the equation of state.
As a test of the effects of stellar structure vs. the EOS, we run simulations of two stars that have very similar density profiles but different energy support (radiative vs. convective). We use a ZAMS 1$M_\sun$ star (36 Myr) and a close-to-ZAMS 10$M_\sun$ star (4 Myr). The 10$M_\sun$ star is at central hydrogen fraction of $X=0.65$, compared the ZAMS value of $X=0.71$. 
The 1$M_\sun$ and 10$M_\sun$ star have very similar (but not identical) normalized density profiles, and $\rho_c/\bar\rho=42$ for the 1$M_\sun$ star and $\rho_c/\bar\rho=45$ for the 10$M_\sun$ star. We chose to have the profiles match nearly exactly, rather than have the $\rho_c/\bar\rho$ values match. We ran one simulation at $\beta=1.5$ for each, a ``solid'' partial disruption.

\begin{figure*}[tp!]
\epsscale{0.55}
\plotone{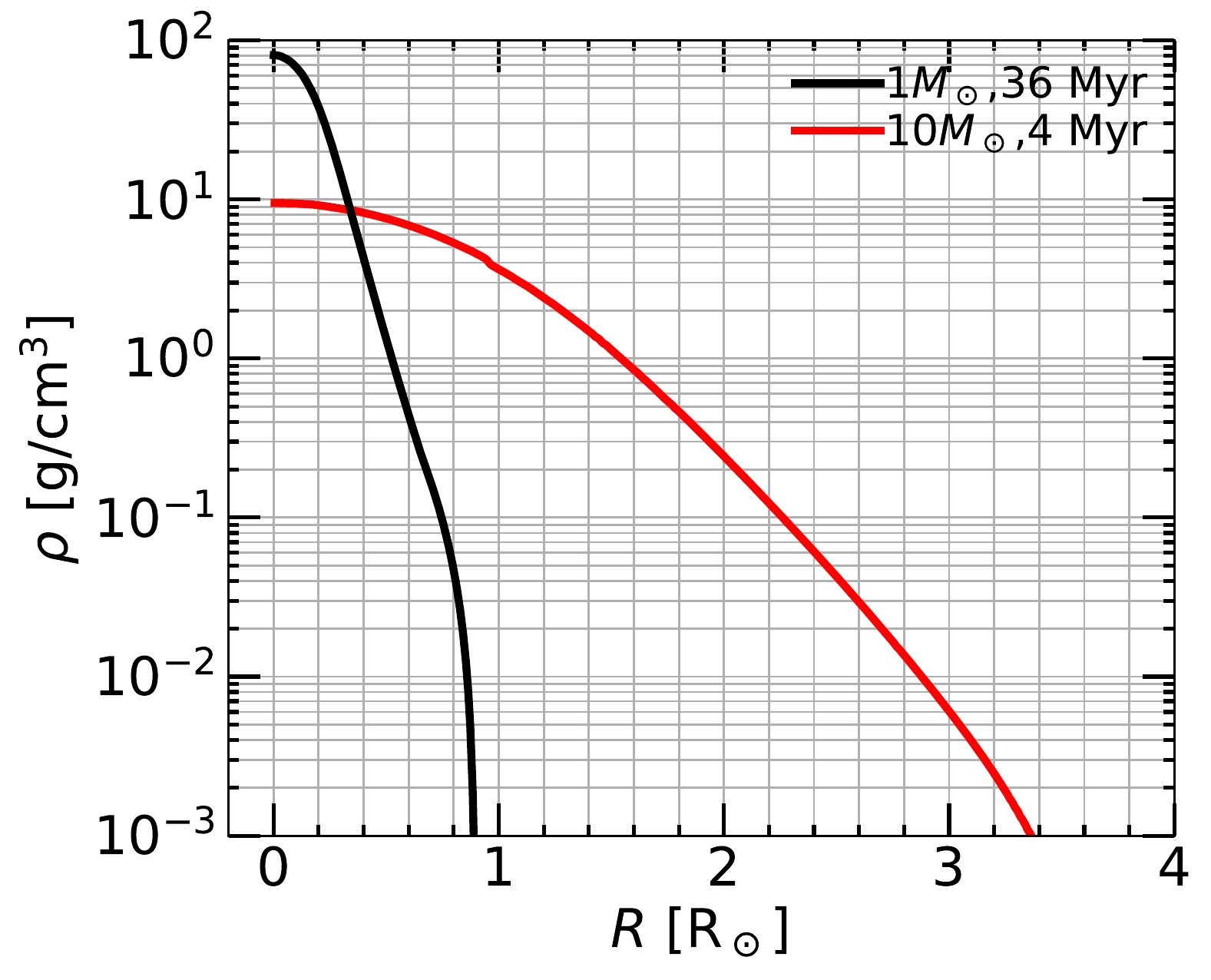}
\plotone{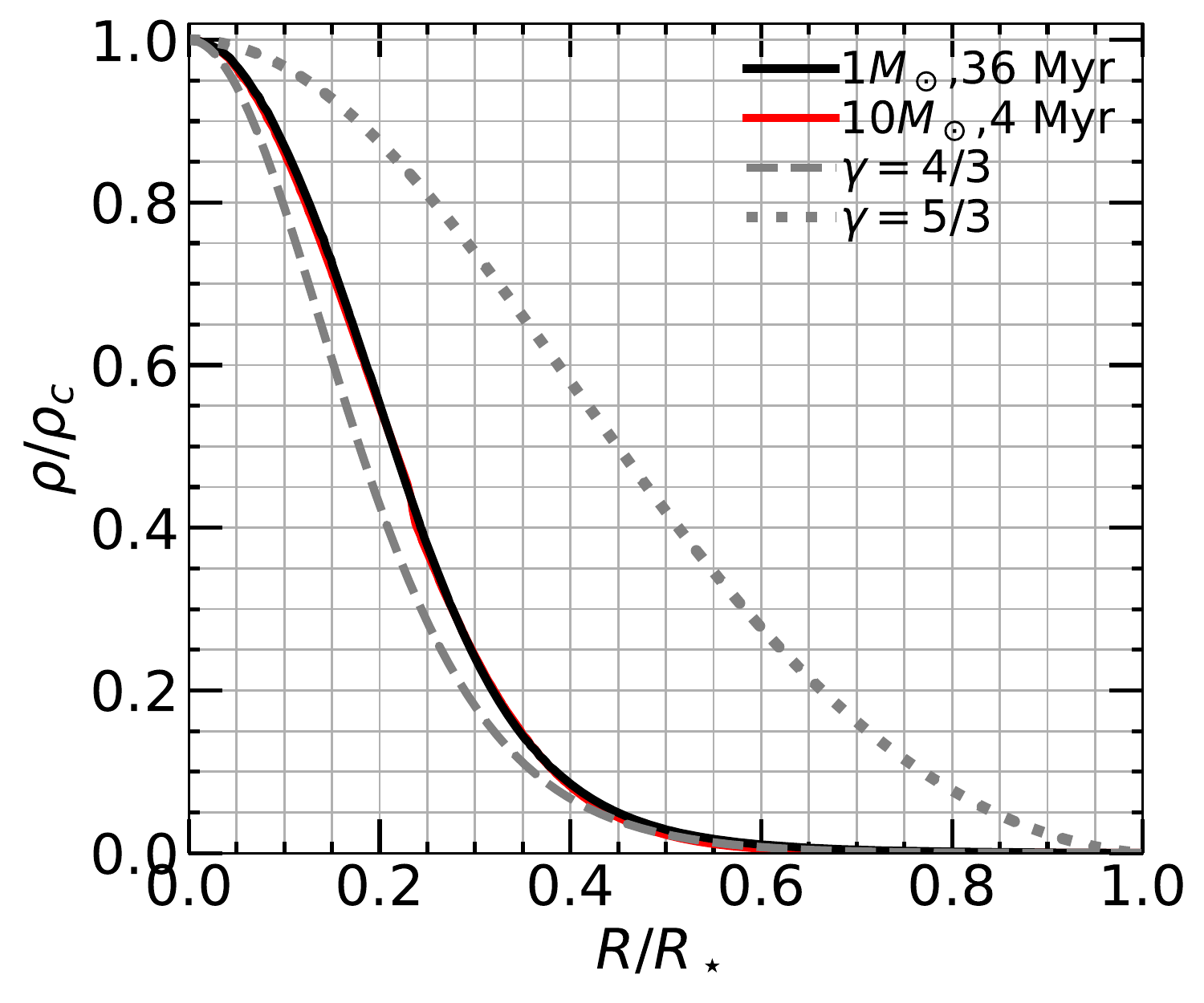}
\caption{
Density profiles for two stars in our ``stellar structure vs. EOS'' study.
Left: absolute; right: normalized.
\label{fig:stellar_structure_vs_EOS_density_profile}
}
\end{figure*}

Figure~\ref{fig:stellar_structure_vs_EOS_density_profile} shows a comparison of the density profiles for these two stars. They are very different in absolute scales, but normalized to $\rho_c$ and $R_\star$, the profiles are nearly indistinguishable.

\begin{figure*}[tp!]
\epsscale{0.55}
\plotone{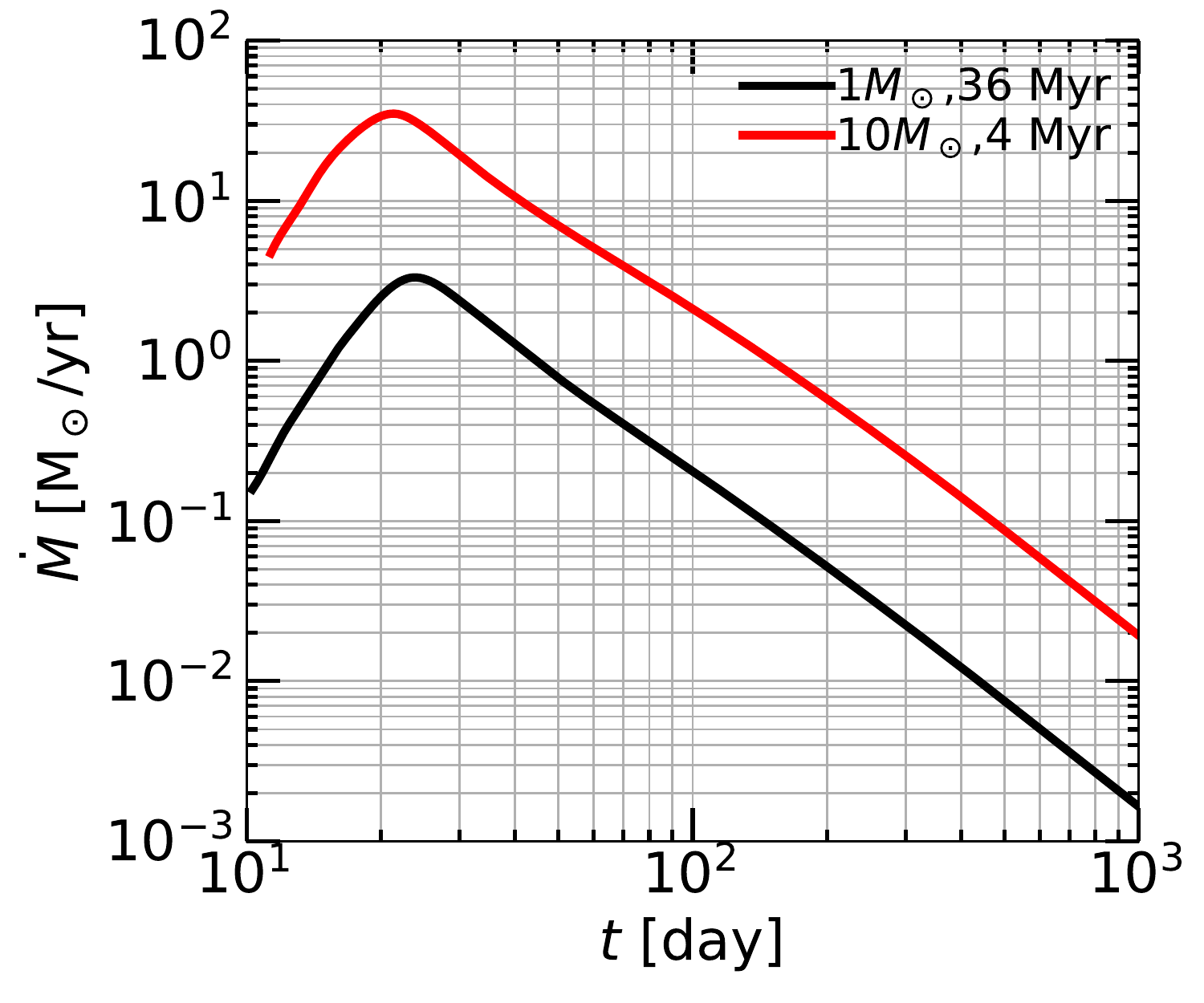}
\plotone{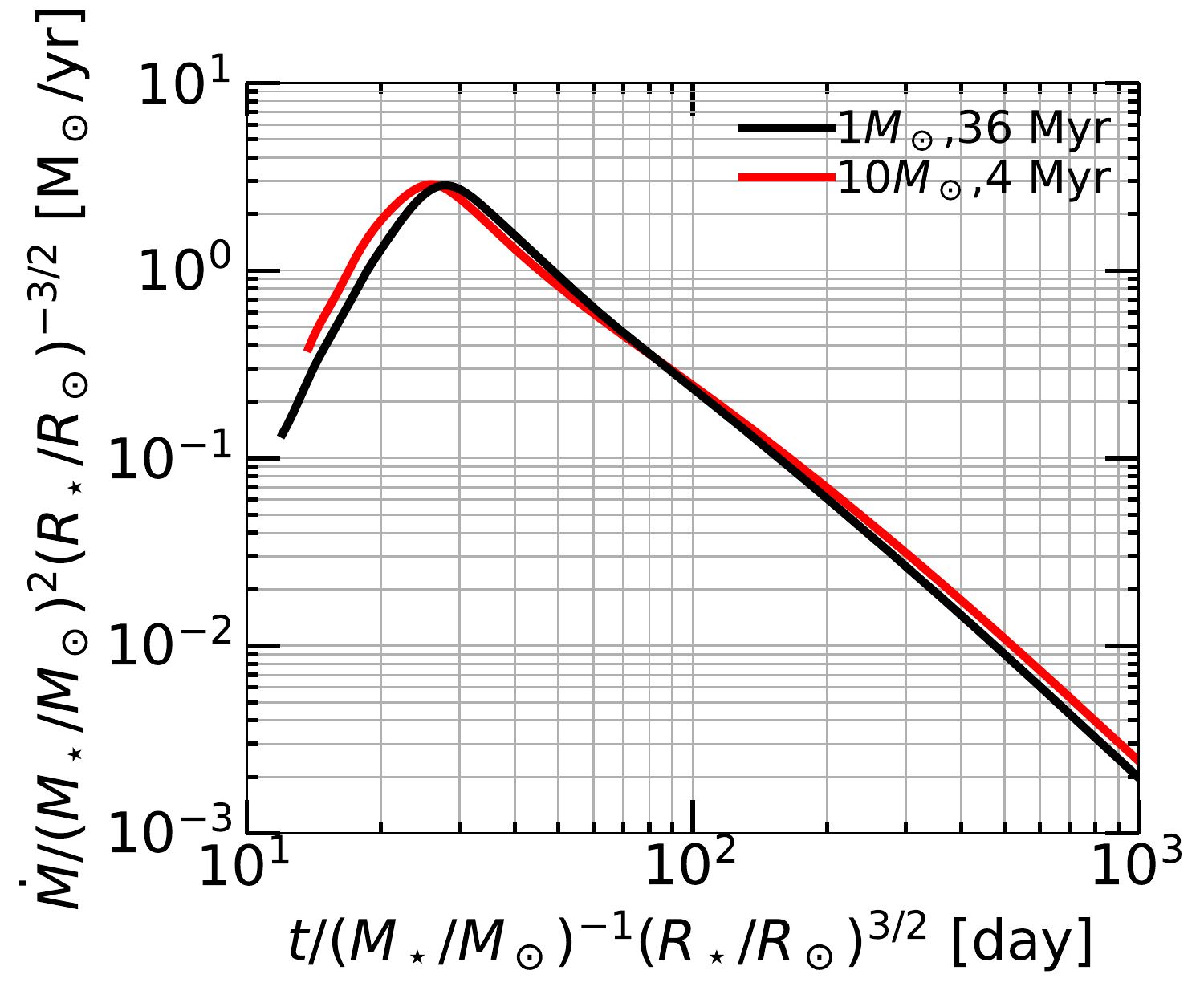}
\caption{
$dM/dt$'s for two stars in our ``stellar structure vs. EOS'' study. 
Left: absolute; right: normalized using stellar mass and radius scalings (Eq. \ref{eq:mdotpeaktpeakscalings}).
\label{fig:stellar_structure_vs_EOS_dmdt}
}
\end{figure*}

Figure~\ref{fig:stellar_structure_vs_EOS_dmdt} shows the $dM/dt$'s, absolute and scaled with $M_\star$ and $R_\star$. 
The 1$M_\sun$ star has $\Delta M/M=0.581$ and the 10$M_\sun$ star has $\Delta M/M=0.583$. 
The $dM/dt$'s are very similar, which supports the argument advanced in this paper that the shape of the $dM/dt$ curve is determined by the stellar density profile, and allows us to provide the fitting formulae as a function of only $\rho_c/\bar\rho$ and $\beta$ for disruption quantities for any stellar mass and age within our simulation grid range. 
In fact, if, from this initial study of stellar structure vs. EOS, we can conclude that these fitting formulae are robust for any star, then we can extend the scalings to stars with masses outside our grid range, as their different EOS's will have only a small effect on the resulting disruption quantities.

The differences in shape between these two $dM/dt$'s could be due partially to the slightly different density profiles (note the small ``notch'' in the profile for the 10$M_\sun$ star in Figure~\ref{fig:stellar_structure_vs_EOS_density_profile}) and values of $\rho_c/\bar\rho$ of these two stars. However---and this is the motivation for this test---it could be due to differences in the EOS of the two stars. In particular, the change in slope at $t\approx100$ days may be due to different transitions between $\gamma=5/3$ and $\gamma=4/3$ pressure support in the $\rho$--$T$ plane for these two stars \citep[e.g. see Figure 7 in][]{2017ApJ...845..173M}.

%%%%%%%%%%%%%%%%
\section{MESA and FLASH parameters}

In Table~\ref{tab:MESA_params} and Table~\ref{tab:FLASH_params} we list a few relevant parameters for the MESA and FLASH simulations. See Section~\ref{sec:methods} for explanations. 
MESA inlists are available upon request. 
We also turn on overshooting (not shown in the table) for $M_\star > 3M_\sun$ using the same parameter choices as in the MIST models \citep{2016ApJ...823..102C}.

\begin{deluxetable}{l l}
\tablecaption{Relevant MESA parameters. \label{tab:MESA_params}}
\tablehead{
\colhead{Parameter} & \colhead{Value}
}
\startdata
\texttt{create\_pre\_main\_sequence\_model} & \texttt{.true.} \\
\texttt{new\_net\_name} & \texttt{`mesa\_49.net'} \\
\texttt{new\_rates\_preference} & \texttt{2 ! jina} \\
\texttt{kappa\_file\_prefix} & \texttt{`a09'} \\
\texttt{initial\_zfracs} & \texttt{6 ! AGSS09} \\
\texttt{kappa\_lowT\_prefix} & \texttt{`lowT\_fa05\_a09p'} \\
\texttt{kappa\_CO\_prefix} & \texttt{`a09\_co'} \\
\texttt{initial\_z} & \texttt{0.0142d0} \\
\texttt{initial\_y} & \texttt{0.2703d0} \\
\texttt{Lnuc\_div\_L\_zams\_limit} & \texttt{0.999d0} \\
\texttt{mixing\_length\_alpha} & \texttt{2} \\
\texttt{delta\_lg\_XH\_cntr\_hard\_limit} & \texttt{0.00432d0} \\
\texttt{do\_element\_diffusion} & \texttt{.true.} \\
\texttt{xa\_central\_lower\_limit\_species(1)} & \texttt{`h1'} \\
\texttt{xa\_central\_lower\_limit(1)} & \texttt{0.001d0} \\
\enddata
\end{deluxetable}

\begin{deluxetable}{l l}
\tablecaption{A few relevant FLASH parameters. \texttt{xmax} is in units of $R_\star$ and \texttt{tmax} is in units of the stellar dynamical timescale $t_{\rm dyn}$.
\label{tab:FLASH_params}}
\tablehead{
\colhead{Parameter} & \colhead{Value}
}
\startdata
\texttt{xmax}   & \texttt{1.e3} \\
\texttt{tmax}   & \texttt{1.e2} \\
\texttt{mpole\_lmax} & \texttt{20} \\
\texttt{eos\_tolerance} & \texttt{1.e-6} \\
\texttt{smalle} & \texttt{1.e7} \\
\texttt{smallt} & \texttt{1.e3} \\
\texttt{smlrho} & \texttt{1.e-12} \\
\texttt{smallp} & \texttt{1.e0} \\
\texttt{sim\_rhoAmbient} & \texttt{1.e-11} \\
\texttt{sim\_pAmbient} & \texttt{1.e1} \\
\texttt{sim\_tAmbient} & \texttt{1.e3} \\
\texttt{order} & \texttt{3} \\
\texttt{slopeLimiter} & \texttt{``hybrid''} \\ 
\texttt{charLimiting} & \texttt{.true.} \\
\texttt{use\_3dFullCTU} & \texttt{.true.} \\
\texttt{use\_hybridOrder} & \texttt{.true.} \\
\texttt{RiemannSolver} & \texttt{``HYBRID''} \\
\texttt{shockDetect} & \texttt{.true.} \\
\enddata
\end{deluxetable}

\added{
\section{Numerical convergence}\label{sec:resolution-convergence}
Figure~\ref{fig:resolution-convergence} shows $dM/dt$'s for different resolution simulations in FLASH, demonstrating numerical convergence for the resolutions we choose in this paper: 131 cells across the diameter of the star for less centrally concentrated stars ($\rho_c/\bar\rho \lesssim 150$) and 524 cells across the diameter of the star for highly centrally concentrated stars ($\rho_c/\bar\rho \gtrsim 150$).
We show results for a ZAMS 0.3 $M_\sun$ star (less centrally concentrated, $\rho_c/\bar\rho=6$), a MAMS 1$M_\sun$ star (moderately centrally concentrated, $\rho_c/\bar\rho=138$) and a TAMS 3$M_\sun$ star (highly centrally concentrated, $\rho_c/\bar\rho=1198$).
We note that past simulations have had difficulty reaching convergence on the low-mass, ``high-energy'', tightly bound tail of debris that returns first to the BH. This tail is of observational interest, and we achieve convergence here.

\begin{figure*}[tp!]
%\epsscale{0.55}
\epsscale{0.38}
\plotone{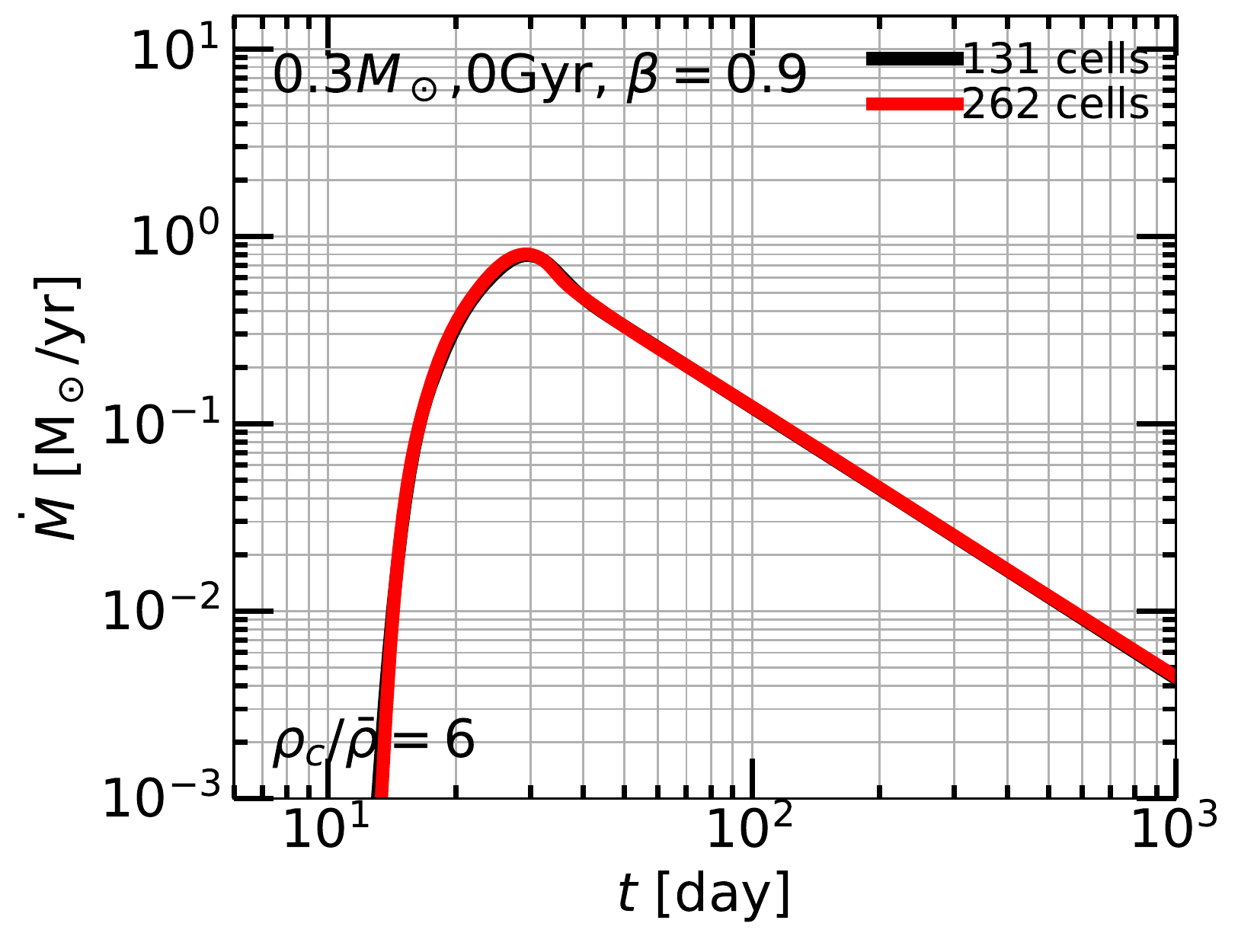}
\plotone{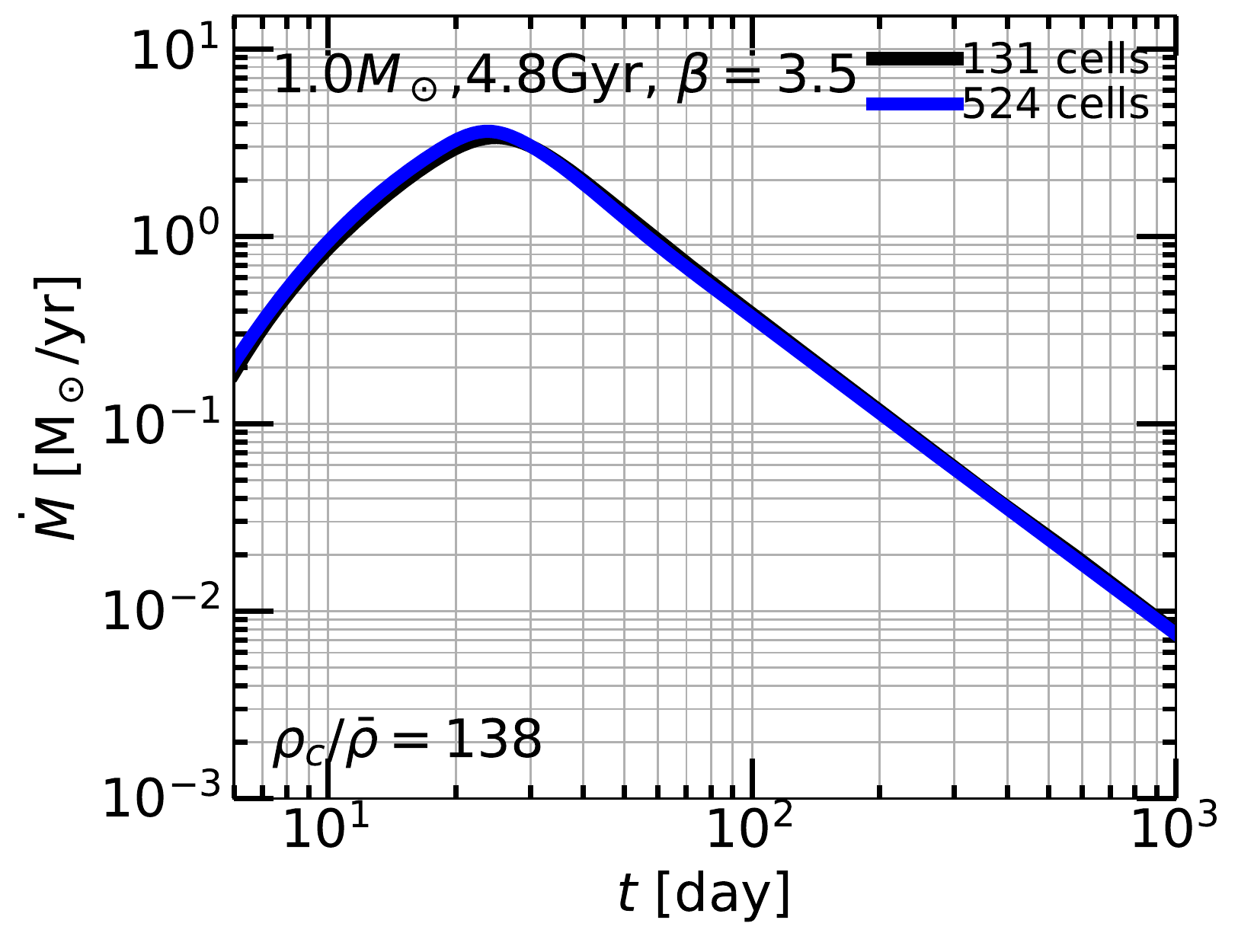}
\plotone{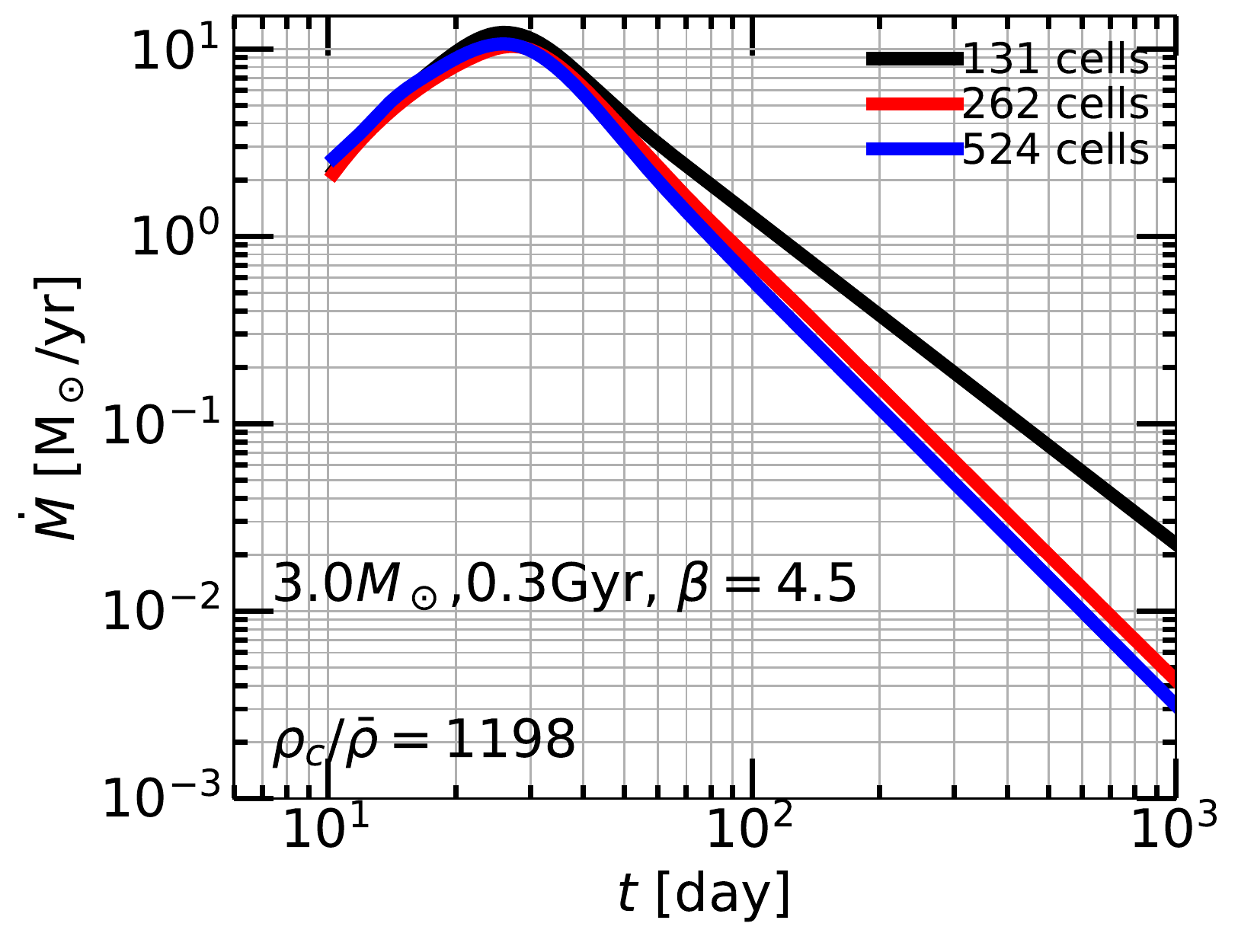}
\caption{
$dM/dt$'s for different resolution simulations in FLASH for a few different examples.
Lines are labeled by the number of cells across the diameter of the star. 
Central concentration (parameterized by $\rho_c/\bar\rho$) is shown in the bottom left.
Left: 0.3$M_\sun$, 0 Gyr (a less centrally concentrated star; we use 131 cells here).
Middle: 1.0$M_\sun$, 4.8 Gyr (a moderately centrally concentrated star; we use 131 cells here).
Right: 3.0$M_\sun$, 0.3 Gyr (a highly centrally concentrated star; we use 524 cells here).
\label{fig:resolution-convergence}
}
\end{figure*}
}

\bibliography{export-bibtex}
\bibliographystyle{aasjournal}

%\listofchanges

\end{document}